\documentclass[12pt]{report} \usepackage{amssymb,latexsym}
\usepackage{amsmath}
 
\columnwidth\textwidth

\begin{document}
 
\newtheorem{df}{Definition} \newtheorem{thm}{Theorem} \newtheorem{lem}{Lemma}
\newtheorem{prop}{Proposition} \newtheorem{assump}{Assumption}
\newtheorem{rl}{Rule}

\begin{titlepage}
 
\noindent
 
\begin{center} {\LARGE Quantum models of classical world} \vspace{1cm}

P. H\'{a}j\'{\i}\v{c}ek \\ Institute for Theoretical Physics \\ University of
Berne \\ Sidlerstrasse 5, CH-3012 Bern, Switzerland \\ hajicek@itp.unibe.ch \\
\vspace{.5cm}

February 2013 \\ \vspace{1cm}
 

\end{center} \vspace*{2cm}
 
\nopagebreak[4]
 
\begin{abstract} This paper is a review of our recent work on three notorious
problems of non-relativistic quantum mechanics: realist interpretation,
quantum theory of classical properties and the problem of quantum
measurement. A considerable progress has been achieved, based on four distinct
new ideas. First, objective properties are associated with states rather than
with values of observables. Second, all classical properties are selected
properties of certain high entropy quantum states of macroscopic systems. Third,
registration of a quantum system is strongly disturbed by systems of the same
type in the environment. Fourth, detectors must be distinguished from
ancillas and the states of registered systems are partially dissipated and
lost in the detectors. The paper has two aims: a clear explanation of all new
results and a coherent and contradiction-free account of the whole quantum
mechanics including all necessary changes of its current textbook version.
\end{abstract}

\end{titlepage}

\tableofcontents

\section{Introduction} Quantum mechanics was originally developed for the
world of atoms and electrons, where it has been very successful. The
understanding of the microscopic world, let us call it "quantum
world", that has developed from this success, seems to be very different
or even incompatible with the understanding of the everyday world of our
immediate experience, which we can call "classical world". This is
unsatisfactory because one of the strongest feelings of a modern physicist is
the belief in the unity of knowledge. It is even paradoxical because the
bodies of the everyday world are composed of atoms and electrons, which ought
to be described by quantum mechanics.

We can distinguish three problems that are met on the way from quantum to
classical physics. Classical theories such as Newtonian mechanics, Maxwellian
electrodynamics or thermodynamics are {\em objective}, in the sense that the
systems they are studying can be considered as real objects, and the values of
their observables, such as position, momentum, field strengths, charge
current, temperature etc. can be ascribed to the systems independently of
whether they are observed or not. If we are going to construct quantum models
of classical systems, the question naturally arises how such an objective
world can emerge from quantum mechanics.

Indeed, in quantum mechanics, the values obtained by most registrations on
microsystems (technically, they are values of observables) cannot be assumed
to exist before the registrations, that is, to be objective properties of the
microsystem on which the registration is made. An assumption of this kind
would lead to contradictions with other assumptions of standard quantum
mechanics and, ultimately, with observable facts (contextuality
\cite{bell4,kochen}, Bell inequalities \cite{bell1}, Hardy impossibilities
\cite{hardy}, Greenberger--Horne--Zeilinger equality \cite{GHZ}).

This property of quantum mechanical observables has lead to growing popularity
of various forms of weakened realism. For example, according to Bohr, realism
applies only to the results of quantum measurements, which can be described by
the relation between objective classical properties of real classical
preparation and registration apparatuses. Various concepts of quantum
mechanics itself, such as electron, wave function, observable, etc., do not
possess any direct counterparts in reality; they are just instruments to keep
order in our experience and to make it ready for application. A rigorous
account of this kind of weakened realism is \cite{ludwig1}. Similar view is
the so-called "statistical" or "ensemble" interpretation
\cite{ball}. This refuses to attribute any kind of reality to the
quantum-mechanical probability amplitudes either at the microscopic or
macroscopic level. According to this view, the amplitudes are simply
intermediate symbols in a calculus whose only ultimate function is to predict
the statistical probability of various directly observed macroscopic outcomes,
and no further significance should be attributed to them. Another example of
even weaker realism is the "constructive empiricism" by van Fraassen
\cite{fraassen}, which proposes to take only empirical adequacy, but not
necessarily "truth", as the goal of science. Even the reality of
classical systems and their properties is then only apparent. The focus is on
describing "appearances" rather than how the world really is
\cite{Zeh}. Thus, quantum mechanics does not seem to allow a realist
interpretation. Let us call this {\em Problem of Realist Interpretation}.

The second problem for construction of quantum models of classical systems can
be called {\em Problem of Classical Properties}. This is the apparent absence
of quantum superpositions, as well as the robustness, of classical properties
(for detailed discussion, see Ref.\ \cite{leggett}). Clearly, classical
properties such as a position do not allow linear superposition. Nobody has
ever seen a table to be in a linear superposition of being simultaneously in
the kitchen as well as in the bedroom. Also, observing the table in the
kitchen does not shift it to the bedroom, while quantum registration changes
properties of registered systems. That is roughly what is meant when one says
that classical properties are {\em robust} (for a better definition, see Ref.\
\cite{leggett}).

Finally, there is a serious problem at the quantum--classical boundary. For
quantum measurements, evidence suggests the assumption that the registration
apparatus always is in a well-defined classical state at the end of any
quantum measurement indicating just one value of the registered
observable. This is called {\em objectification requirement}
\cite{BLM}. However, if the initial state of the registered system is a linear
superposition of different eigenvectors of the observable, then the linearity
of Schr\"odinger equation implies that the end state of the apparatus is
also a linear superposition of eigenstates of its pointer
observable\footnote{This holds, strictly speaking, only if the measured system
${\mathcal S}$ and the apparatus ${\mathcal A}$ constitute together an
isolated system. But if they are not isolated from their environment
${\mathcal E}$ then the composite ${\mathcal S} + {\mathcal A} + {\mathcal E}$
can be considered as isolated and the difficulty reappears.}. Thus, it turns
out that realism in most cases leads to contradictions with the postulate of
linear quantum evolution, see the analysis in \cite{leggett}. Let us call this
{\em Problem of Quantum Measurement}.

There is a vast literature dealing with the three problems containing many
different proposals from various variants of weakened realism to radical
changes of quantum mechanics (for clear reviews see
\cite{d'Espagnat,BLM,schloss}). One also proposes that quantum mechanics is
based on a kind of approximation that ceases to be valid for macroscopic
systems, see e.g.\ \cite{leggett}. An even more radical approach is to look
for the way quantum mechanics could be obtained from a kind of deeper theory with
classical character (see e.g.\ the proceedings \cite{EQM}) These or other
attempts in the literature do not seem to lead to a satisfactory solution. We
shall not discuss this work because we shall look for the solution in a
different and utterly novel direction. Our approach starts directly from
quantum mechanics as it has been formulated by Bohr, Born, Dirac, Heisenberg,
von Neumann, Pauli and Schr\"{o}dinger. Our analysis has shown that they
have delivered enough tools to deal with the problems. The aim of this review
is to prove that our proposals of solutions to the three problems form a
logically coherent whole with the rest of quantum mechanics.

Let us briefly describe the ideas from which our approach starts. First, it is
true that values of observables are not objective. However, in \cite{PHJT}, we
have shown that there are other observable properties of quantum systems that
can be ascribed to the systems without contradictions and there is a
sufficient number of them to describe the state of the systems completely. We
shall introduce and discuss this approach in Section 0.1.3 in general terms
and then, in technical detail, in Section 1.1.2.

Second, many attempts to derive classical theories from quantum mechanics are
based on quantum states of minimal uncertainty (Gaussian wave packets,
coherent states, etc.). But a sharp classical trajectory might be just a
figment of imagination because each measurement of a classical trajectory is
much fuzzier than the minimum quantum uncertainty. Thus, the experimental
results of classical physics do not justify any requirement that we have to
approximate absolutely sharp trajectories as accurately as possible. This was
observed as early as 1822 by Exner \cite{Exner} and evolved further in 1955 by
Born \cite{Born}. Taking this as a starting point, one must ask next what the
quantum states that correspond to realistically fuzzy classical ones are. In
\cite{hajicek}, a new assumption about such quantum states has been formulated
and studied. We shall discuss it in Chapter 3.

Third, in \cite{hajicek2} it is shown that the measurement of observables such
as a position, momentum, spin, angular momentum, energy, etc.\ on a quantum
system must be strongly disturbed by all existing systems of the same type, at
least according to the standard quantum mechanics. To eliminate the
disturbance, and to give an account of what is in fact done during a quantum
measurement process, the standard theory of observables must be rewritten. We
shall do this in Section 2.2 on identical quantum systems.

Fourth, the current theory of quantum measurement describes a registration
apparatus by a microscopic quantum system, the so-called pointer, and assumes
that a reading of the apparatus is an eigenvalue of a pointer observable. For
example, Ref.\ \cite{pauli}, p.\ 64, describes a measurement of energy
eigenvalues with the help of scattering process similar to Stern--Gerlach
experiment, and it explicitly states:
\begin{quote} We can consider the centre of mass [of a microscopic system] as
a 'special' measuring apparatus...
\end{quote} Similarly, Ref.\ \cite{peres}, p. 17 describes Stern--Gerlach
experiment:
\begin{quote} The microscopic object under investigation is the magnetic
moment $\mathbf \mu$ of an atom.\dots The macroscopic degree of freedom to
which it is coupled in this model is the centre of mass position $\mathbf
r$.\dots I call this degree of freedom {\em macroscopic} because different
final values of $\mathbf r$ can be directly distinguished by macroscopic
means, such as the detector.\dots From here on, the situation is simple and
unambiguous, because we have entered the macroscopic world: The type of
detectors and the detail of their functioning are deemed irrelevant.
\end{quote} Paradoxically, this notion of measurement apparatus (mostly called
"meter") is quite useful for the very precise modern quantum
experiments, such as non-demolition measurements or weak measurements, see,
e.g., Refs.\ \cite{WM,svensson}. Quite generally, these experiments utilise
auxiliary microscopic systems called {\em ancillas}. If the pointer is
interpreted as an ancilla, then the old theory works well at least for the
interaction of the measured system with the ancilla. However, the fact that
ancillas themselves must be registered by macroscopic detectors is
"deemed irrelevant". A more detailed analysis is given in Section
4.1.

In fact, detectors are very special macroscopic systems. We shall show that
their role is to justify the state reduction (collapse of wave function) and
to define the so-called "preferred basis" (see \cite{schloss}) that
determines the form of state reduction. Further, the reduction can be assumed
as a process within detectors at the time of the detector's macroscopic
signal. We shall discuss the design and role of detectors in Section 4.3.

The present paper has grown from review \cite{survey} by adding new results,
by correcting many minor errors and by explaining many points in a clearer and
more coherent way. It is a systematic exposition of the non-relativistic
quantum mechanics (the space and time structure is everywhere assumed to be
Newtonian).

\subsection{Examples of quantum systems} To explain what quantum mechanics is
about, this section describes some well-known quantum systems following Ref.\
\cite{survey}. It also introduces some general notions, such as microsystem,
macrosystem, type of system and structural property following Ref.\
\cite{PHJT}.

Quantum mechanics is a theory that describes certain class of properties of
certain class of objects in a similar way as any other physical theory
does. For example, among others, Newtonian mechanics describes bodies that can
be considered as point-like in a good approximation and studies the motion of
the bodies.

Quantum systems that we shall consider are photons\footnote{One can ask
whether there is a non-relativistic limit of photons. In one such limit,
photons may move with infinite velocity and therefore their position does not need to
be very well defined. In another, photons may be represented by a classical
electromagnetic wave.}, electrons, neutrons and nuclei, which we call {\em
particles}, and systems containing some number of particles, such as atoms,
molecules and macroscopic systems, which are called {\em composite}. Of
course, neutrons and nuclei themselves are composed of quarks and gluons, but
non-relativistic quantum mechanics can and does start from some
phenomenological description of neutrons and nuclei.

Let us call particles and quantum systems that are composed of small number
of particles {\em microsystems}. They are extremely tiny and they mostly
cannot be perceived directly by our senses. We can observe directly only
macroscopic quantum systems that are composed of very many
particles\footnote{It is true that the eye can recognize signals of just
several photons, but it can be viewed as a quantum registration apparatus with
macroscopic parts and only these are observed
"directly".}. "Very many" is not too different from
$10^{23}$ (the Avogadro number). Let us call these {\em macrosystems}. Some
properties of most macroscopic systems obey classical theories. For example,
shape and position of my chair belong to Euclidean geometry, its mass
distribution to Newtonian mechanics, chemical composition of its parts to
classical chemistry and thermodynamic properties of the parts such as phase or
temperature to phenomenological thermodynamics. Such properties are called
``classical''. Thus, properties of microsystems can only be observed via
classical properties of macrosystems; if microsystems interact with them and if
this interaction changes their classical properties.

Microsystems are divided into types, such as electrons, hydrogen atoms,
etc. Systems of one type are not distinguishable from each other in a sense
not existing in classical physics. Systems of the same type are often called
{\em identical}. Microsystems exist always in a huge number of identical
copies. The two properties of microsystems, viz. 1) their inaccessibility to
direct observations and 2) utter lack of individuality that is connected with
the existence of a huge number of identical copies, make them rather different
from classical systems or "things". Each classical system can be
observed directly by humans (in principle: for example, the distant galaxies)
and each can be labelled and distinguished from other classical systems,
because it is a quantum system composed of a huge number of particles and
hence it is highly improbable that it has a kin of the same type in the world.

Objective properties that are common to all microsystems of the same type will
be called {\em structural}. Thus, each particle has a mass, spin and electric
charge. For example, the mass of electron is about 0.5 MeV, the spin 1/2 and
the charge about $10^{-19}$ C. In non-relativistic quantum mechanics, any
composite system consists of definite numbers of different particles with
their masses, spins and charges\footnote{We do not view quasiparticles as
particles but as auxiliary entities useful for description of the spectrum of
some composite systems.}. E.g., a hydrogen atom contains one electron and one
proton (nucleus). The composition of a system is another structural
property. The structural properties influence the dynamics of quantum systems;
the way they do it and what dynamics is will be explained later. Only then, it
will be clear what the meaning of these parameters is and that the type of
each system can be recognized, if its dynamics is observed. When we shall know
more about dynamics, further structural properties will emerge.

Structural properties are objective in the sense that they can be assumed to
exist before and independently of any measurement. Such assumption does not
lead to any contradictions with standard quantum mechanics and is at least
tacitly made by most physicists. In fact, the ultimate goal of practically all
experiments is to find structural properties of quantum systems.

From the formally logical point of view, all possible objective properties of
given kind of objects ought to form a Boolean lattice. The structural
properties satisfy this condition: systems with a given structural property
form a subset of all systems. These subsets are always composed of whole
type-classes of quantum systems. Clearly, the intersection of two such subsets
and the complement of any such subset is again a structural property.

Structural properties characterise a system type completely but they are not
sufficient to determine the dynamics of individual systems.

\subsection{Examples of quantum experiments} The topic of this section plays
an important role in understanding quantum mechanics. Specific examples of
typical experiments will be given in some detail following Ref.\
\cite{survey}. In this way, we gain access to the notions of preparation and
registration, which are assumed by the basic ideas of our realist
interpretation of quantum mechanics. Describing the experiments, we shall
already use some of the language of the interpretation, which will be
introduced and motivated in this way.

Let us first consider experiments with microsystems that are carried out in
laboratories. Such an experiment starts at a source of microsystems that are
to be studied. Let us give examples of such sources.
\begin{enumerate}
\item Electrons. One possible source (called field emission, see e.g.\ Ref.\
\cite{merton}, p.\ 38) consists of a cold cathode in the form of a sharp tip
and a flat anode with an aperture in the middle at some distance from the
cathode, in a vacuum tube. The electrostatic field of, say, few kV will enable
electrons to tunnel from the metal and form an electron beam of about $10^7$
electrons per second through the aperture, with a relatively well-defined
average energy.
\item Neutrons can be obtained through nuclear reaction. This can be initiated
by charged particles or gamma rays that can be furnished by an accelerator or
a radioactive substance. For example the so-called Ra-Be source consists of
finely divided RaCl$_2$ mixed with powdered Be, contained in a small
capsule. Decaying Ra provides alpha particles that react with Be. The yield
for 1 mg Ra is about $10^4$ neutrons per second with broad energy spectrum
from small energies to about 13 MeV. The emission of neutrons is roughly
spherically symmetric centred at the capsule.
\item Atoms and molecules. A macroscopic specimen of the required substance in
gaseous phase at certain temperature can be produced, e.g., by an oven. The gas
is in a vessel with an aperture from which a beam of the atoms or molecules
emerges.
\end{enumerate} Each source is defined by an arrangement of macroscopic bodies
of different shapes, chemical compositions, temperatures and by electric or
magnetic fields that are determined by their macroscopic characteristics, such
as average field intensities: that is, by their classical properties. These
properties determine uniquely what type of microsystem is produced. Let us
call this description {\em empirical}. It is important that the classical
properties defining a source do not include time and position so that the
source can be reproduced later and elsewhere. We call different sources that
are defined by the same classical properties {\em equivalent}. Empirical
description is sufficient for reproducibility of experiments but it is not
sufficient for understanding of how the sources work. If a source defined by
an empirical description is set into action, we have an instance of the
so-called {\em preparation}.

Quantum mechanics assumes that these are general features of all sources,
independently of whether they are arranged in a laboratory by humans or occur
spontaneously in nature. For example, classical conditions at the centre of
the Sun (temperature, pressure and plasma composition) lead to emission of
neutrinos that reach the space outside the Sun.

Often, a source yields very many microsystems that are emitted in all possible
directions, a kind of radiation. We stress that the detailed structure of the
radiation as it is understood in classical physics, that is where each
individual classical system exactly is at different times, is not determined
for quantum systems and the question even makes no sense. Still, a fixed
source gives the microsystems that originate from it some properties. In
quantum mechanics, these properties are described on the one hand by the
structural properties that define the prepared type, on the other, e.g., by
the so-called {\em quantum state}. The mathematical entity that is used in
quantum mechanics to describe a state (the so-called state operator) will be
explained in Section 1.1. To determine the quantum state that results from a
preparation with a given empirical description in each specific case requires
the full formalism of quantum mechanics. Hence, we postpone this point to
Chapter 4.

After arranging the source, another stage of the experiment can
start. Generally, only a very small part of the radiation from a source has
the properties that are needed for the planned experiment. The next step is,
therefore, to select the part and to block off the rest. This is done by the
so-called collimator, mostly a set of macroscopic screens with apertures and
macroscopic electric or magnetic fields. For example, the electron radiation
can go through an electrostatic field that accelerates the electrons and
through electron-microscope "lenses", each followed by a suitable
screen. A narrow part of the original radiation, a beam, remains. Another
example is a beam of molecules obtained from an oven. It can also contain
parts of broken molecules including molecules with different degrees of
ionisation. The part with suitable composition can then be selected by a mass
spectrometer and the rest blocked off by a screen. Again, the beam resulting
from a raw source and a collimator consists of individual quantum objects with
a well-defined type and quantum state. The process of obtaining these
individual quantum objects can be viewed as a second stage of the
preparation. Again, there is an empirical description that defines an
equivalence class of preparations and equivalent preparations can be
reproduced.

The final beam can be characterized not only by the quantum state of
individual objects but also by its approximate current, that is how many
individual objects it yields per second. The beam can be made very thin. For
example, in the electron-diffraction experiment \cite{tono}, the beam that
emerges from the collimator represents an electric current of about $10^{-16}
A$, or $10^3$ electrons per second. As the approximate velocity of the
electrons and the distance between the collimator and the detector are known,
one can estimate the average number of electrons that are there simultaneously
at each time. In the experiment, it is less than one. One can understand in
this way that it is an experiment with individual electrons.

Next, the beam can be lead through further arrangement of macroscopic bodies
and fields. For example, to study the phenomenon of diffraction of electrons,
each electron can be scattered by a thin slab of crystalline graphite or by an
electrostatic biprism interference apparatus. The latter consists of two
parallel plates and a wire in between with a potential difference between the
wire on the one hand and the plates on the other. An electron object runs
through between the wire and both the left and right plate simultaneously and
interferes with itself afterwards (for details see \cite{tono}). Again, the
beam from the graphite or the biprism can be viewed as prepared by the whole
arrangement of the source, collimator and the interference apparatus. This is
another example of a reproducible preparation procedure.

Finally, what results from the original beam must be made directly perceptible
by its interaction with another system of macroscopic bodies and fields. This
process is called {\em registration} and the system {\em registration
apparatus}. The division of an experimental arrangement into preparation and
registration parts is not unique. For example, in the electron diffraction
experiment, one example of a registration apparatus begins after the biprism
interference, another one includes also the biprism interference
apparatus. Similarly to preparations, the registrations are defined by an
empirical description of their relevant classical properties in such a way
that equivalent registrations can be reproduced.

An important, even definition, property of a registration is that it is
applicable to an individual quantum system and that each empirical result of a
registration is caused by just one individual system. In the above experiment,
this assumption is made plausible by the extreme thinning of the beam, but it
is adopted in general even if the beam is not thin.

An empirical description of a registration apparatus can determine a quantum
mechanical {\em observable} similarly as preparation determines a quantum
state. Again, more theory is needed for understanding of what are the
mathematical entities (the so-called {\em positive valued operator measures},
see Section 1.2) describing observables and how they are related to
registration devices. Each individual registration performed by the apparatus,
i.e., registration performed on a single quantum object, then gives some value
of the observable. The registration is not considered to be finished without
the registration apparatus having given a definite, macroscopic and classical
signal. This is the objectification requirement.

A part of registration apparatuses for microsystems are {\em detectors}. At
the empirical level, a detector is determined by an arrangement of macroscopic
fields and bodies, as well as by the chemical composition of its sensitive
matter \cite{leo}. For example, in the experiment \cite{tono} on electron
diffraction, the electrons coming from the biprism interference apparatus are
absorbed in a scintillation film placed transversally to the beam. An incoming
electron is thus transformed into a light signal. The photons are guided by
parallel system of fibres to a photo-cathode. The resulting (secondary)
electrons are accelerated and lead to a micro-channel plate, which is a system
of parallel thin photo-multipliers. Finally, a system of tiny anodes enables
to record the time and the transversal position of the small flash of light in
the scintillation film.

In this way, each individual electron coming from the biprism is detected at a
position (two transversal coordinates determined by the anodes and one
longitudinal coordinate determined by the position of the scintillation
film). Such triple $\vec{x}$ of numbers is the result of each registration and
the value of the corresponding observable, which is a coarsened and localised
position operator $\vec{\mathsf q}$ in this case (see Section 2.2.3). Also the
time of the arrival at each anode can be approximately determined. Thus, each
position obtains a certain time.

For our theory, the crucial observation is the following. When an electron
that has been prepared by the source and the collimator hits the scintillation
film, it is lost as an individual system. Indeed, there is no property that
would distinguish it from other electrons in the scintillation matter. Thus,
this particular registration is a process inverse to preparation: while the
preparation has created a quantum system with certain individuality, the
registration entails a loss of the individuality. Our theory of quantum
registration in Chapter 4 will make precise and generalise this observation.

If we repeat the experiment with individual electrons many times and record
the transversal position coordinates, the gradual formation of the electron
interference pattern can be observed. The pattern can also be described by
some numerical values. For example, the distance of adjacent maxima and the
direction of the interference fringes can be such values. Still, the
interference pattern is {\em not} a result of one but of a whole large set of
individual registrations.

In some sense, each electron must be spread out over the whole plane of the
scintillation detector after coming from the biprism but the excitation of the
molecules in the detector matter happens always only within a tiny
well-localized piece of it, which is different for different
electrons\footnote{This is what is sometimes called "the collapse of
wave function".}. Thus, one can say that the interference pattern must be
encoded in each individual electron, even if it is not possible to obtain the
property by a single registration. The interference pattern can be considered
as an objective property of the individual electrons prepared by the source,
the collimator and the biprism interference apparatus. The interference
pattern is not a structural property: preparations that differ in the voltage
at some stage of the experiment (e.g., the accelerating field in the
collimator or the field between the wire and the side electrodes in the
biprism interference apparatus, etc.) will give different interference
patterns. We call such objective properties {\em dynamical}. On the other
hand, the hitting position of each individual electron cannot be considered as
its objective property. Such an assumption would lead to contradictions with
results of other experiments. The position must be regarded as created in the
detection process.

It is a double-slit experiment, a special kind of which is described above,
that provides a strong motivation for considering an individual quantum
particle as an extended object of sorts. Without any mathematical description,
it is already clear that such an extended character of electrons could offer
an explanation for the stability of some states of electrons orbiting atomic
nuclei. Indeed, a point-like electron would necessarily have a time-dependent
dipole momentum and lose energy by radiation. However, an extended electron
can define a stationary charge current around the nucleus.

Some structural properties can be measured directly by a registration (on
individual quantum systems) and their values are real numbers. For example,
mass can be measured by a mass spectrometer. Such structural properties can be
described by quantum observables (see Section 1.2.5)\footnote{These
observables must commute with all other observables (\cite{ludwig1}, IV.8),
and can be associated with the so-called superselection rules, see e.g.\
\cite{BLM}.}. However, there are also structural properties that cannot be
directly measured on individual objects similarly to the interference pattern,
such as cross sections or branching ratios. They cannot be described by
observables.

\subsection{Realist Model Approach to quantum mechanics} In the previous
section, describing specific experiments, we have used certain words that are
avoided in careful textbooks of quantum mechanics such as {\em objective
properties} or {\em quantum object} or
\begin{quote} {\em \dots An electron object runs through between the wire and
both the left and right plate simultaneously and interferes with itself
afterwards \ldots}
\end{quote} The electron is viewed here as a real object that is extended over
the whole width of the biprism apparatus. After this intuitive introduction,
we give now a general and systematic account of our realist interpretation.

A realist interpretation of a physical theory is a more subtle question than
whether the world exists for itself rather than being just a construction of
our mind. This question can always be answered in positive without any danger
of falsification. However, every physical theory introduces some general,
abstract concepts. For example, Newtonian mechanics works with mass points,
their coordinates, momenta and their dynamical trajectories. The truly
difficult question is whether such concepts possess any counterparts in the
real world. On the one hand, it seems very plausible today that mass points
and their sharp trajectories cannot exist and are at most some
idealisations. On the other hand, if we are going to understand a real system,
such as a snooker ball moving on a table, then we can work with a construction
that uses these concepts but is more closely related to the reality. For
example, we choose a system of infinitely many mass points forming an elastic
body of a spherical shape and calculate the motion of this composite system
using Newton's laws valid for its constituent points. Then, some calculated
properties of such a model can be compared with interesting observable
properties of the real system. Thus, even if the general concepts of the
theory do not describe directly anything existing, a suitable model
constructed with the help of the general concepts can account for some aspects
of a real system.

Motivated by this observation, we shall divide any physical theory into two
parts. First, there is a treasure of successful models. Each model gives an
approximative representation of some aspects of a {\em real object}
\cite{giere}. Historically, models form a primary and open part of the
theory. For example, in Newtonian mechanics, the solar system was carefully
observed by Tycho de Brahe and then its model was constructed by
Kepler. Apparently, Newton was able to calculate accelerations and doing so
for Kepler trajectories, he might discover that they pointed towards the
Sun. Perhaps this lead to the Second Law. The hydrogen atom had a similar role
in quantum mechanics.

Second, there is a general language part. It contains the mathematical
structure of state space, conditions on trajectories in the state space, their
symmetries and the form of observables \cite{fraassen}. It is obtained by
generalisation from the study of models and is an instrument of further model
construction and of model unification. For example, in Newtonian mechanics,
the state space is a phase space, the conditions on trajectories is the
general structure of Newton's dynamical equations, symmetries are Galilean
transformations and observables are real functions on the phase space.

A model is constructed as a particular subset of trajectories in a particular
state space as well as a choice of important observables. For example, to
describe the solar system, assumptions such as the number of bodies, their
point-like form, their masses, the form of gravitational force and certain
class of their trajectories can be made if we want to construct a model. The
observed positions of the planets would then match the theoretical
trajectories of the model within certain accuracy. Thus, a model consists of a
language component on the one hand, and an identifiable-object component on
the other. The language component always contains simplifying assumptions,
always holds only for some aspects of the associated object and only within
some approximation. The approximation that is referred to is bounded from
above by the accuracy of performed measurements. This is measurable and can be
expressed numerically by statistical variances.

Clearly, the models of a given theory are not predetermined by the general
part but obtained in the historical evolution and dependent on observation of
real objects. On the one hand, the general part can also be used to construct
language components of models that do not have any real counterparts. On the
other hand, the model part is steadily evolving and never closed. For example, a
satisfactory quantum model of high temperature superconductivity is not yet
known. This is why the treasure of successful models is an independent and, in
fact, the basic part of any theory.

Such philosophy forms a first step of what we call our Realist Model Approach
to quantum mechanics. Thus, the Approach lies somewhat within the recent trend
of the philosophy of science that defines a theory as a class of models (see,
e.g., \cite{suppes,sneed,fraassen,giere,cart}). It can be said that it
combines ideas of the constructive realism by Ronald Giere\footnote{I
enthusiastically adopt Giere's view that philosophy of science is to be
removed from the realm of philosophy and put into the realm of cognitive
sciences.} with van Fraassen notions of state space and symmetries
\cite{fraassen} as a basis of the general language part\footnote{Van Fraassen
also applied his constructive empiricism to quantum mechanics \cite{fraass}
and, adding some further ideas, arrived at his own, the so-called "modal
interpretation" of it. To prevent misunderstanding, it must be stressed
that the account and interpretation of quantum mechanics described here is
different from van Fraassen's.}. It is important that constructive realism is
immune to the usual objections against naive realism. In addition, we add some
further importance to the general part by recognising its unification
role. The effort at unification is without any doubts a salient feature of
scientists that can be observed at any stage of research. For example, Newton
was admired for his unification of such different phenomena as apples falling
from trees and the Moon moving in the sky. Today's endeavour to unify the
theories of quantum fields and gravity is a very well observable historical
fact. From the point of view of Giere, this bent might, perhaps, be understood
as one of cognitive instincts.

The focus on models allows to define the task of quantum-explaining the
classical world in the following way. Instead of trying to find a direct
relation between the general language parts of, say, quantum and Newtonian
mechanics or a universal correspondence between states of Newtonian and
quantum mechanics such as Wigner--Weyl--Moyal map \cite{schroeck}, p.\ 85, one
ought to build quantum models of real macroscopic systems and their aspects
for which there are models in Newtonian mechanics striving for approximate
agreement between the two kinds of models on those aspects for which the
Newtonian models are successful. For example, we shall not attempt to obtain
from quantum mechanics the sharp trajectories that is a concept of general
language part of Newtonian mechanics, but rather try to model the observed
fuzzy trajectories of specific classical systems (Chapter 3), or to analyse
different specific registration apparatuses first and then try to formulate
some features common to all (Chapter 4).

However, the Realist Model Approach is not so easily applied to quantum
mechanics as it is to Newtonian mechanics. A question looms large at the very
start: What is a real quantum object? Of course, such objects are met
"empirically" in preparations and registrations. However, we would
like to subscribe to the notion that the language component of a model must
ascribe to its real object a sufficient number of objective
properties. Objective means that the properties can be ascribed to the object
alone. Sufficient means that the dynamics of any object as given by its model
is uniquely determined by initial data defined by values of a minimal set of
its objective properties. For example, in Newtonian mechanics, the values of
coordinates and momenta determine a unique solution of Newton's dynamical
equations.

In Newtonian mechanics, coordinates and momenta are observables, and values of
observables can be viewed as objective without any danger of
contradictions. Using this analogy, one asks: Can values of observables be
viewed as objective properties of quantum systems? As is well known, the
answer is negative (see Section 1.2.4). If we assume that values of
observables are the only properties of quantum systems that are relevant to
their reality, then there are no real quantum systems. For rigorous no-go
theorems concerning such objective properties see, e.g., Ref.\ \cite{ludwig1}.

Our approach to properties of quantum systems is therefore different from
those that can be found in literature. First, we extend the notion of
properties to include complex ones in the following sense \cite{survey}:
\begin{enumerate}
\item Their values may be arbitrary mathematical entities (sets, maps between
sets, etc.). For example, the Hamiltonian of a closed quantum system involves
a relation between energy and some other observables of the system. This
relation is an example of such a complex property.
\item Their values do not need to be directly obtained by individual
registrations. For example, to measure a cross-section a whole series of
scattering experiments must be done. Thus, their values do not necessarily possess
probability distribution but may be equivalent to, or derivable from,
probability distributions.
\end{enumerate}

Point 2 is usually not clearly understood and we must make it more precise. A
real system of Newton mechanics is sufficiently robust so that we can do many
experiments with it and perform many measurements on it without changing
it. Moreover, any such system is sufficiently different from other systems
anywhere in the world (even two cars from one factory series can be
distinguished from each other). Any physical experiment on a given classical
system can then be repeated many times and only then the results can be
considered reliable. The results are then formulated in statistical terms
(e.g., as averages, variances, etc.). One can, therefore, feel that it might
be more precise account of what one does generally in physics if one spoke of
ensembles of equivalent experiments done on equivalent object systems in terms
of equivalent experimental set-ups and of the statistics of these ensembles.

This is of course a well-known idea. We shall apply it consequently to
Newtonian mechanics in the theory of classical limit in Chapter 3. However,
one ought not to forget that each ensemble must consist of some
elements. Indeed, to get a statistics, one has to possess a sufficiently large
number of different individual results. Hence, these individual elements must
always be there independently of how large the ensemble is. Then, we can ask
the question: What do the statistical properties of an ensemble tell us about
properties of the individual object systems used in each individual
experiment? In the classical physics, at least, the answer to this question is
considered quite obvious and one interprets the experimental results as
properties of the individual objects.

Quantum microsystems are never robust in the above sense. After a single
registration, the microsystem is usually lost. Then we can repeat the
experiment only if we do it with another system. Here, we can utilise another
property of quantum microsystems that is different from classical ones: there
is always a huge number of microscopic systems of the same type, which are
principally indistinguishable from each other. Thus, we can apply the same
preparation together with the same registration many times. In quantum
mechanics, the thought set of all such experiments is called {\em ensemble} of
experiments, and similarly ensembles of prepared systems and of obtained
results. The elements of the ensembles are again called {\em
individuals}. Suppose that each individual result of an ensemble of
measurements is a real number. Then we can e.g.\ calculate an average of the
results ensemble. The average can then surely be considered as a property of
the ensemble.

However, as in classical physics, one can also understand the average as a
property of each individual system of the ensemble. In any case, the fact that
a given individual belongs to a given ensemble is a property of the
individual. It is a crucial step in our theory of properties that we consider
a property of an ensemble as a property of each individual element of the
ensemble. In fact, this is the only way of how the logical union or
intersection of two properties can be understood. For example, the logical
union, $A \vee B$, of properties $A$ and $B$ of system $\mathcal S$ is the
property, that $\mathcal S$ has either property $A$ or $B$.

In our theory, we shall use both notions, individual object and ensemble. The
notion of system ensemble is defined as usual (see, e.g., \cite{peres}, p.\
25): it is the thought set of all systems obtained through equivalent
preparations.

Returning to objectivity of observables in quantum mechanics, the problem is
that a registration of an eigenvalue $a$ of an observable $\mathsf A$ of a
quantum system $\mathcal S$ by an apparatus $\mathcal A$ disturbs the
microsystem and that the result of the registration is only created during the
registration process. The result of the individual registration cannot thus be
assumed to be an objective property of $\mathcal S$ before the
registration. It can however be assumed, as we shall do, that it is an
objective property of the composite $\mathcal S + \mathcal A$ after the
registration. This is the objectification requirement \cite{BLM}.

It seems therefore that the objective properties of quantum systems, if there
are any, cannot be directly related to individual registrations, as they can
in classical theories. (Paradoxically, most of the prejudices that hinder
construction of quantum models of classical theories originate in the same
classical theories.) However, there are observable properties in quantum
mechanics that are different from values of observables \cite{survey}:
\par\vspace{.5cm} \noindent {\bf Basic Ontological Hypothesis of Quantum
Mechanics} {\it A sufficient condition for a property to be objective is that
its value is uniquely determined by a preparation according to the rules of
standard quantum mechanics. The "value" is the value of the
mathematical expression that describes the property and it may be more general
than just a real number. To observe an objective property, many registrations
of one or more observables are necessary.}\par \vspace{.5cm}

In fact, the Hypothesis just states explicitly the meaning that is tacitly
given to preparation by standard quantum mechanics. More discussion on the
meaning of preparation is in \cite{hajicek2,hajicek4}. In any case, prepared
properties can be assumed to be possessed by the prepared system without
either violating any rule of standard quantum mechanics or contradicting
possible results of any registration performed on the prepared system. The
relation of registrations to such objective properties is only indirect: an
objective property entails limitations on values of observables that will be
registered. In many cases, we shall use the Hypothesis as a heuristic
principle: it will just help to find some specific properties and then it will
be forgotten, that is, an independent assumption will be made that these
properties can be objective and each of them will be further studied.

We shall divide objective properties into structural (see Section 0.1.1) and
{\em dynamical} and describe the dynamical ones mathematically in Section
1.1.2. Examples of dynamical properties are a state, the average value and the
variance of an observable. We shall define so-called simple objective
properties and show first, that there is enough simple objective properties to
characterise quantum systems completely (at least from the standpoint of
standard quantum mechanics) and second, that the logic of simple properties
satisfies Boolean lattice rules. Thus, a reasonable definition of a real
object in quantum mechanics can be given (see Sec.\ 1.1.2).

Often, the Hypothesis meets one of the following two questions. First, how can
the Hypothesis be applied to cosmology, when there was nobody there at the Big
Bang to perform any state preparation? Second, a state preparation is an
action of some human subject; how can it result in an objective property? Both
objections result from a too narrow view of preparations (see Section
0.1.2). Moreover, the second objection is not much more than a pun. It is not
logically impossible that a human manipulation of a system results in an
objective property of the system. For example, pushing a snooker ball imparts
it a certain momentum and angular momentum that can then be assumed to be
objective properties of the ball.

One may wonder how the average of an observable in a state can be objective
while the individual registered value of the observable are not because an
average seems to be defined by the individual values. However, the average is
a property of a prepared state and is, therefore, defined also by the
preparation. The results of a huge number of individual registrations must add
to their predetermined average. This can be seen very well for averages with
small variances. In our theory of classical properties (Chapter 3), the
explanation of classical realism will be based on the objectivity of averages
because some important classical observables will be defined as (quantum)
averages of (quantum) observables in a family of specific (quantum) states
that will be called {\em classicality states}.

Let us compare our Realist Model Approach with what is usually understood as
the Realism of Classical Theories. This is the philosophy that extends some
successful features of classical theories, especially Newtonian mechanics, to
the whole real world. There are three aspects of the Realism of Classical
Theories that are not included in our Realist Model Approach. First, classical
physics is deterministic, assuming that every event has a cause but quantum
theory does not tell us the causes of some of its events. Second, each
interaction of classical physics is local in the sense that the mutual
influence of two interacting systems asymptotically vanishes if the systems
are separated by increasing spatial distance (cluster separability). But, in
addition to local interactions, quantum theory contains mutual influence that
is independent of distance (entanglement, mutual influence between particles
of the same type). Third, classical physics requires a causal explanation for
every correlation. This can be rigorously expressed by Reichennbach's
condition of {\em common cause} \cite{reichenbach}. The existence of the
common cause for some quantum correlation is incompatible with experiments
(for discussion, see \cite{fraass}).

Our Realist Model Approach just states which ontological hypotheses can
reasonably be made under the assumption that quantum mechanics is valid. A few
words have to be said on ontological hypotheses. As is well-known, the
objective existence of anything cannot be proved (even that of the chair on
which I am now sitting, see, e.g., Ref.\ \cite{d'Espagnat}, where this old
philosophical tenet is explained from the point of view of a physicist). Thus,
all such statements are only hypotheses, called ontological.

It is clear, however, that a sufficiently specific ontological hypothesis may
lead to contradictions with some observations. Exactly that happens if one
tries to require objectivity of quantum observables\footnote{More precisely,
the existence in question is that of systems with sufficient number of
properties defined by values of observables.}. Moreover, hypotheses that do
not lead to contradictions may be useful. For example, the objective existence
of the chair nicely explains why we all agree on its properties. Similarly,
the assumption that quantum systems possess certain objective properties will
be useful for the quantum theory of classical properties or for a solution of
the problem of quantum measurement. The usefulness of ontological hypotheses
in the work of experimental physicists has been analysed by Giere
\cite{giere}, p.\ 115. The hypothesis in question is the existence of certain
protons and neutrons. It explains, and helps to perform, the production, the
manipulations, the control and the observations of proton and neutron beams in
an experiment at Indiana University Cyclotron Facility. From the point of view
of van Fraassen, the ontological hypotheses of the kind used by this paper
might perhaps be considered as a part of theoretical models: such a hypothesis
may or may not be "empirically adequate".

The position on ontological hypotheses taken here is, therefore, rather
different from what has been called "metaphysical realism" by
Hillary Putnam \cite{putnam}: "There is exactly one true and complete
description of 'the way the world is'\,".

The Realist Model Approach enables us to characterise the subject of quantum
mechanics as follows:
\begin{quote} Quantum mechanics studies objective properties of existing
microscopic objects.
\end{quote} This can be contrasted with the usual cautious characterisation of
the subject, as e.g.\ \cite{peres}, p.\ 13:
\begin{quote} \dots quantum theory is a set of rules allowing the computation
of {\em probabilities} for the outcomes of tests [registrations] which follow
specific preparations.
\end{quote}

\subsection{Probability and information} Let us return to Tonomura
experiment. At each individual registration, a definite value $\vec{x}$ of the
observable $\vec{\mathsf x}$ is obtained. Quantum mechanics cannot predict
which value it will be, but it can give the probability $p(\vec{x})$ that the
value $\vec{x}$ will be obtained. This is a general situation for any
registration. In this way, registrations introduce a specific statistical
element into quantum mechanics.

A correct understanding of probability and information is an important part in
the conceptual framework of the theory. The discussion whether probability
describes objective properties that can be observed in nature or subjective
states of the knowledge of some humans has raged since the invention of
probability calculus by Jacob Bernoulli and Pierre-Simon Laplace
\cite{Jaynes}. The cause of this eternal argument might be that the dispute
cannot be decided: probability has both aspects, ontological and epistemic
\cite{survey}.

Probability is a function of a proposition $A$ and its value, $p(A)$, is a
measure of the degree of certainty that $A$ is true. As a function on a
Boolean lattice of propositions, it satisfies Cox's axioms \cite{Jaynes}. Then
it becomes a real additive measure on the lattice. Whether a proposition is
true or false must be decided by observation, at least in principle. Hence,
the probability always concerns objective events, at least indirectly.

As an example, consider the Tonomura experiment. The probability $p(\vec{x})$
concerns the proposition that the value of observable $\vec{\mathsf x}$ is
individually registered on an electron is $\vec{x}$. The value of the
probability can be verified by studying an ensemble of such
registrations. Indeed, if we perform a huge number of such registrations, we
obtain an interference pattern that approximately reproduces the smooth
probability distribution obtained by calculating the quantum mechanical model
of an electron in Tonomura's apparatus. It is a real interference pattern
shown by the apparatus.

We have used the term "ensemble" with the meaning of a statistical
ensemble of real events or objects. Here, the objectification requirement is
involved. The value $\vec{x}$ obtained in each individual registration is
considered as a real property of the system consisting of the electron and the
apparatus. Then, the probability concerns both the lack of knowledge of what
objectively happens and properties of real systems. We emphasise that
$p(\vec{x})$ is not the probability that the prepared electron possesses value
$\vec{x}$ of observable $\vec{\mathsf x}$ but the probability that a
registration will give such a value.

Another question is whether the individual outcomes are in principle
predictable from some more detailed initial conditions (the so-called
"hidden variables") on the electron that we do not know. Quantum
mechanics does not contain information on any such conditions. It does not
deliver these predictions and it would even be incompatible with any 'deeper'
theory that did (see Section 1.2.4). We shall, therefore, assume that they are
objectively unpredictable.

Let us describe the general framework that is necessary for any application of
probability theory. First, there is a system, denote it by $\mathcal
S$. Second, certain definite objective conditions are imposed on the system,
e.g., it is prepared as in the example above and observable $\vec{\mathsf x}$
is registered. In general, there must always be an analogous set of
conditions, let us denote it by $\mathcal C$. To each system $\mathcal S$
subject to condition $\mathcal C$, possibilities in some range are open. These
possibilities are described by a set of propositions that form a Boolean
lattice $\mathbf F$. A probability distribution $p : {\mathbf F} \mapsto
[0,1]$ is a real additive measure on $\mathbf F$.

In quantum mechanics, $\mathbf F$ is usually constructed with the help of some
observable, ${\mathsf E}$ say. If ${\mathsf E}$ is a discrete observable, its
value set ${\mathbf \Omega}$ is at most countable, ${\mathbf \Omega} =
\{\omega_1, \omega_2 \cdots \}$. Then the single-element sets $\{\omega_k\}, k
= 1, 2, \cdots $, are atoms of $\mathbf F$ that generate $\mathbf F$ and the
probability distribution $p(X)$, $X \in {\mathbf F}$, can be calculated from
$p(\omega_k)$ by means of Cox's axioms. The atoms are called outcomes. If
${\mathsf E}$ is a continuous observable, then there are no atoms but
continuous observables can be considered as idealisations of more realistic
discrete observables with well-defined atoms. For example, in Tonomura's
experiment, observable $\vec{\mathsf x}$ is defined by the photo-multiplier
cells in the micro-channel plate, and not only is discrete, but also even has a
finite number of values.

If condition $\mathcal C$ is reproducible or it obtains spontaneously
sufficiently often (which is mostly the case in physics), anybody can test the
value of the probabilities because probability theory enables us to calculate
the frequencies of real events starting from any theoretical probability
distribution and the frequencies are measurable \cite{Jaynes}. The probability
distribution $p(X)$ is therefore an objective property of condition $\mathcal
C$ on $\mathcal S$. It is so even in cases when the outcomes can be in
principle predicted if occurrence of more detailed conditions is observable and
can in principle be known. More precisely, this would depend on whether
condition $\mathcal C$ can be decomposed into other conditions ${\mathcal
C}_i$, $i=1,\dots,N$ in the following sense. Condition $\mathcal C$ can be
viewed as a logical statement '$\mathcal S$ satisfies $\mathcal C$'. Let
\begin{enumerate}
\item $\mathcal C = {\mathcal C}_1 \vee \ldots \vee {\mathcal C}_N$, where
$\vee$ is the logical union (disjunction),
\item each ${\mathcal C}_i$ be still recognisable and reproducible,
\item each outcome allowed by $\mathcal C$ is uniquely determined by one of
${\mathcal C}_i$'s.
\end{enumerate} It follows that ${\mathcal C}_i \wedge {\mathcal C}_k =
\emptyset$ for all $k \neq i$, where $\wedge$ is logical intersection
(conjunction). Even in such a case, condition $\mathcal C$ itself leaves the
system a definite amount of freedom that can be described in all detail by the
probability distribution $p(X)$ and it is an objective, verifiable property of
$\mathcal C$ alone. And, if we know only that condition $\mathcal C$ obtains
and that a probability distribution is its objective property, then this
probability distribution describes the state of our knowledge, independently
of whether the conditions $\mathcal C_k$ do exist and we just do not know
which of them obtains in each case or not. Examples of these two different
situations are given by standard quantum mechanics, which denies the existence
of $\mathcal C_k$'s, and the Bohm--de Broglie pilot wave theory, which
specifies such $\mathcal C_k$'s.

Of course, there are also cases where some condition, $\mathcal C$, say,
occurs only once so that a measurement of frequencies is not possible. Then,
no probability distribution associated with $\mathcal C$ can be verified so
that our knowledge about $\mathcal C$ is even more incomplete. However, in
some such cases, one can still give a rigorous sense to the question
\cite{Jaynes}, p.\ 343 (see also the end of this section): "What is the
most probable probability distribution associated with $\mathcal C$?" One
can then base one's bets on such a probability distribution. Such a
probability distribution can be considered as an objective property of
$\mathcal C$ and again, there is no contradiction between the objective and
subjective aspects of probability.

In quantum mechanics, it is also possible to mix preparations in a random
way. Suppose we have two preparations, ${\mathcal P}_1$ and ${\mathcal P}_2$,
and can mix them randomly by e.g.\ mixing the resulting particle beams in
certain proportions $c_1$ and $c_2$, $c_1 + c_2 = 1$. Then, each particle in
the resulting beam is either prepared by ${\mathcal P}_1$, with probability
$c_1$, or by ${\mathcal P}_2$, with probability $c_2$. In this way, another
kind of statistical element can be introduced into quantum theory. This
element will be discussed in Section 1.1.2.

Condition ${\mathcal C}$ is defined by the two beams and their mixing so that
the ensemble has, say, $N$ particles. Then, ${\mathcal C}$ can be decomposed
in $N_1{\mathcal C}_1 \vee N_2{\mathcal C}_2$, where ${\mathcal C}_k$ is the
preparation by ${\mathcal P}_k$ and $c_k = N_k/N$, $k = 1,2$. We can know
$c_1$ and $c_2$ because we know the intensity of the corresponding beams but
it is unlikely that we also know whether a given element of the ensemble has
been prepared by ${\mathcal P}_1$ or ${\mathcal P}_2$.

An important role in probability theory is played by entropy. Entropy is a
certain functional of $p(X)$ that inherits both objective and subjective
aspects from probability. Discussions similar to those about probability spoil
the atmosphere about entropy. The existence of a subjective aspect of
entropy---the lack of information---seduces people to ask confused questions
such as: Can a change of our knowledge about a system change the system
energy?

The general definition of entropy as a measure of missing information has been
given by Shannon \cite{shannon} and its various applications to communication
theory are, e.g., described in the book by Pierce \cite{pierce}. Our version
is:
\begin{df} Let $p : {\mathbf F} \mapsto [0,1]$ be a probability distribution
and let $X_k$ be the atoms of ${\mathbf F}$. Then the {\em entropy} of $p(X)$ is
\begin{equation}\label{entropy} S = -\sum_k p(X_k)\ln(p(X_k))\ .
\end{equation}
\end{df}

Let us return to the quantum-mechanical example in order to explain which
information is concerned. After the choice of preparation and registration
devices, we do not know what will be the outcome of a registration but we just
know that any outcome $X_k$ of the registration has probability
$p(X_k)$. After an individual registration, one particular outcome will be
known with certainty. The amount of information gained by the registration is
the value of $S$ given by Eq.\ (\ref{entropy}). For more detail, see Ref.\
\cite{peres}. Thus, the value of $S$ measures the lack of information before
the registration.

The entropy and the so-called Maximum Entropy Principle\footnote{There is also
a principle of statistical thermodynamics that carries the same name but ought
not to be confused with the mathematical MEP.} (MEP) have become important
notions of mathematical probability calculus, see, e.g., Ref.\
\cite{Jaynes}. The mathematical problem MEP solves can be generally
characterised as follows. Let system $\mathcal S$, condition $\mathcal C$ and
lattice $\mathbf F$ with atoms $X_k$ be given. Let there be more than one set
of $p(X_k)$'s that appears compatible with $\mathcal C$. How the probabilities
$p(X_k)$ are to be assigned so that condition $\mathcal C$ is properly
accounted for without any additional bias? Such $p(X_k)$'s yield the maximum
of $S$ as given by (\ref{entropy}). MEP clearly follows from the meaning of
entropy as a measure of lack of information. We shall use this kind of MEP in
Section 3.2.

\part{Corrected language of quantum mechanics} The origin of quantum mechanics
can be traced back to the study of a few real systems: hydrogen atom and
black-body radiation. The resulting successful models used some new concepts
and methods that were readily generalised so that they formed a first version
of a new theoretical language. This language was then used to construct models
of some aspects of further real objects, such as atoms and molecules, solid
bodies, etc., and this activity lead in turn to the refinement of the language.

This evolution does not seem to be finished. A large number of real objects,
the so-called classical world, have as yet no satisfactory quantum models. Our
own attempts \cite{PHJT} and \cite{hajicek2} at constructing such models have
lead to some changes in the quantum language. The first part of this review
starting here is an attempt at a systematic formulation of this new language.

The general notions of a theoretical language are imported from some
mathematical theory and satisfy the corresponding relations given by the {\em
axioms} and {\em theorems} of the mathematical theory. They are rather
abstract and by themselves, they do not possess any direct connection to real
physical systems. However, as building blocks of various models that do
possess such connections, some of them acquire physical meaning. Such a
model-mediated physical interpretation can be postulated for most of the
mathematical notions by basic assumptions that will be called {\em rules} to
distinguish them from the axioms of the mathematical theory. What can be
derived from these rules and axioms will be called {\em propositions}. We
shall however formulate only most important theorems and propositions
explicitly as such in order to keep the text smooth.

A state of a quantum system is determined by a preparation while a value of an
observable is determined by a registration. The notions of preparation and
registration are used in their empirical (see Section 0.1.2) meaning first,
just to catch the model-mediated significance of mathematical notions, and the
quantum mechanical models explaining relevant aspects of preparation and
registration processes will be constructed later. The calculation of a state
from classical conditions defined by the preparation needs a sophisticated
model of the mature quantum mechanics. Similarly, to calculate an observable
from classical properties of the registration device, a quantum model must be
used. Only in Chapters 3 and 4 shall we be able to find the way from the
empirical description of preparation and registration to a particular
mathematical state or observable.

\chapter{States, observables and symmetries} This chapter introduces the
notions and most important properties of quantum states, quantum observables
and their relation to symmetry transformations. The states and observables are
described by specific mathe\-matical entities. We study the mathematical
aspects first and then discuss the physical interpretation.

\section{States} Quantum-mechanical states are often described by wave
functions. However, this leads to some confusion of the preparation statistics
with the statistics of registered values, which is a hindrance to understanding
quantum measurement (see Chapter 4). Moreover, it is not adequate for 
studying the states of macroscopic bodies that we meet in our everyday life (see
Chapter 3), which are very different from wave functions. We start, therefore,
with the general notion of quantum states, called either density matrices
\cite{peres} or state operators \cite{BLM}.

In the mathematical part of this section, the construction of the space of
states from the Hilbert space of a system is described and the most important
general properties of states are listed. In the interpretation part, the
Realist Model Approach introduced in Section 0.1.3 is described in detail.

\subsection{Mathematical preliminaries} This subsection lists briefly all
necessary definitions and theorems, stating some explicitly and giving
reference to literature for others. Good textbooks are \cite{BEH,ludwig1,RS}.

Let ${\mathbf H}$ be a complex separable Hilbert space with inner product
$\langle\cdot|\cdot\rangle$ satisfying $\langle a\phi|b\psi\rangle =
a^*b\langle\phi|\psi\rangle$, where '$*$' denotes complex conjugation. An
element $\phi\in{\mathbf H}$ is a unit vector if its norm defined by
$$\| \phi \| = \langle\phi|\phi\rangle$$
equals one,
$$\| \phi \| = 1\ ,$$
and the non-zero vectors $\phi,\ \psi\in{\mathbf H}$ are orthogonal if
$$\langle\phi|\psi\rangle = 0\ .$$
A set $\{\phi_k\}\subset{\mathbf H}$ is orthonormal if the vectors $\phi_k$
are mutually orthogonal unit vectors. $\{\phi_k\}\subset{\mathbf H}$,
$k=1,2,\cdots$, is an orthomormal basis of ${\mathbf H}$ if any
$\psi\in{\mathbf H}$ can be expressed as a series
$$
\psi = \sum_k\langle\phi_k|\psi\rangle\phi_k
$$
with
$$
\| \psi\|^2 = \sum_k|\langle\phi_k|\psi\rangle|^2\ .
$$
Separability means that there is at least one countable basis.

Let ${\mathbf H}$ and ${\mathbf H}'$ be two separable Hilbert spaces,
$\{\phi_k\}$ and $\{\phi'_k\}$ two orthonormal bases, $\{\phi_k\} \subset
{\mathbf H}$ and $\{\phi'_k\} \subset {\mathbf H}'$. Let us define map
${\mathsf U} : \{\phi_k\} \mapsto \{\phi'_k\}$ by
$$
{\mathsf U}\phi_k = \phi'_k
$$
for each $k$. Then, ${\mathsf U}$ can be extended by linearity and continuity
to the whole of ${\mathbf H}$ and it maps ${\mathbf H}$ onto ${\mathbf
H}'$. The map ${\mathsf U}$ is called {\em unitary}. Unitary maps preserve
linear superposition,
$$
{\mathsf U}(a\psi + b\phi) = a{\mathsf U}\psi + b{\mathsf U}\phi
$$
and inner product,
\begin{equation}\label{sp} \langle {\mathsf U}\psi|{\mathsf U}\phi\rangle =
\langle \psi|\phi\rangle\ .
\end{equation} They can be defined by these properties for general Hilbert
spaces and used as equivalence morphisms in the theory of Hilbert spaces. Each
two separable Hilbert spaces are thus unitarily equivalent.

Any unit vector $\phi\in{\mathbf H}$ determines a one-dimensional (orthogonal)
projection operator ${\mathsf P}[\phi]$ by the formula
$${\mathsf P}[\phi]\psi = \langle\phi|\psi\rangle\phi$$
for all $\psi\in{\mathbf H}$. We also use the Dirac notation
$|\phi\rangle\langle\phi|$ for this projection. If $\{\phi_k\}$ is an
orthonormal basis of ${\mathbf H}$, then the projection operators ${\mathsf
P}[\phi_k]$ satisfy
$$
{\mathsf P}[\phi_k]{\mathsf P}[\phi_l] = 0
$$
for all $k\neq l$---we say they are mutually orthogonal---and
$$
\sum_k {\mathsf P}[\phi_k] = {\mathsf 1}\ ,
$$
where ${\mathsf 1}$ is the identity operator on ${\mathbf H}$.

An operator ${\mathsf A} : {\mathbf H} \mapsto {\mathbf H}$ is called {\em
bounded} if its norm
\begin{equation}\label{opnorm} \|{\mathsf A}\| = \sup_{\|\psi\| = 1}\|{\mathsf
A}\psi\|\
\end{equation} is finite. The domain of a bounded operator is clearly the
whole Hilbert space ${\mathbf H}$.  A linear operator ${\mathsf A}$ is defined
by the property
$$
{\mathsf A}(a\phi + b\psi) = a{\mathsf A}\phi + b{\mathsf A}\psi
$$
for all $\phi, \psi \in {\mathbf H}$. Multiplication ${\mathsf A}{\mathsf B}$
of two linear operators is defined by
$$
{\mathsf A}{\mathsf B}\phi = {\mathsf A}({\mathsf B}\phi)
$$
and linear combination $a{\mathsf A} + b{\mathsf B}$ by
$$
(a{\mathsf A} + b{\mathsf B})\phi = a{\mathsf A}\phi + b{\mathsf B}\phi
$$
for all $\phi \in {\mathbf H}$ and for any $a,b \in {\mathbb C}$. Let us
denote the algebra of all bounded linear operators on ${\mathbf H}$ by
${\mathbf L}({\mathbf H})$.

The adjoint ${\mathsf A}^\dagger$ of operator ${\mathsf A} \in {\mathbf
L}({\mathbf H})$ is defined by
$$
\langle {\mathsf A}^\dagger \phi|\psi\rangle = \langle \phi| {\mathsf A}
\psi\rangle
$$
for all $\phi,\ \psi\in$, and ${\mathsf A}$ is self-adjoint (s.a.) if
$$
{\mathsf A}^\dagger = {\mathsf A}\ .
$$
Let us denote the set of all bounded s.a.\ operators by ${\mathbf
L}_r({\mathbf H})$. For self-adjoint operators, the spectral theorem holds
(see Sec.\ 1.2.1).

Unitary maps ${\mathsf U} : {\mathbf H} \mapsto {\mathbf H}$ are bounded
operators and we obtain from Eq.\ (\ref{sp})
$$
{\mathsf U}^\dagger\cdot{\mathsf U} = {\mathsf U}\cdot{\mathsf U}^\dagger =
{\mathsf 1}\ .
$$
Let ${\mathbf H}$ and ${\mathbf H}'$ be two separable Hilbert spaces and
${\mathsf U} : {\mathbf H} \mapsto {\mathbf H}'$ be a unitary map. Then
${\mathsf U}$ defines a map of ${\mathbf L}({\mathbf H})$ onto $ {\mathbf
L}({\mathbf H'})$ by ${\mathsf A} \mapsto {\mathsf U}{\mathsf A}{\mathsf
U}^\dagger$. This map preserves operator action,
$$
({\mathsf U}{\mathsf A}{\mathsf U}^\dagger)({\mathsf U}\phi) = {\mathsf
U}({\mathsf A}\phi)\ ,
$$
linear relation,
$$
{\mathsf U}(a{\mathsf A} + b{\mathsf B}){\mathsf U}^\dagger = a{\mathsf
U}{\mathsf A}{\mathsf U}^\dagger + b{\mathsf U}{\mathsf B}{\mathsf U}^\dagger\
,
$$
operator product
$$
({\mathsf U}{\mathsf A}{\mathsf U}^\dagger)({\mathsf U}{\mathsf B}{\mathsf
U}^\dagger) = {\mathsf U}({\mathsf A}{\mathsf B}){\mathsf U}^\dagger
$$
and norm,
$$
\|{\mathsf U}{\mathsf A}\| = \|{\mathsf A}\|\ .
$$

An operator ${\mathsf A} \in {\mathbf L}_r({\mathbf H})$ is positive,
${\mathsf A}\geq {\mathsf 0}$, where ${\mathsf 0}$ is the null operator, if
$$
\langle\phi|{\mathsf A}\phi\rangle \geq 0
$$
for all vectors $\phi\in{\mathbf H}$. The relation ${\mathsf A}\geq {\mathsf
B}$ is defined by
$$
{\mathsf A}-{\mathsf B}\geq {\mathsf 0}\ .
$$
The order relation is preserved by unitary maps,
$$
{\mathsf U}{\mathsf A}{\mathsf U}^\dagger \geq {\mathsf U}{\mathsf B}{\mathsf
U}^\dagger \quad \text{if} \quad {\mathsf A} \geq {\mathsf B}\ .
$$

Let $\{\phi_k\}$ be any orthonormal basis of ${\mathbf H}$. For any ${\mathsf
A} \in{\mathbf L}_r({\mathbf H})$, we define the trace by
$$
tr[{\mathsf A}] = \sum_k\langle\phi_k|{\mathsf A}\phi_k\rangle\ .
$$
Trace is independent of basis and invariant with respect to unitary maps,
$$
tr[{\mathsf U}{\mathsf A}{\mathsf U}^\dagger] = tr[{\mathsf A}]\ .
$$.

\begin{thm}\label{propold1} Trace defines the norm $\|{\mathsf A}\|_s$ on
${\mathbf L}_r({\mathbf H})$ by
\begin{equation}\label{trnorm} \|{\mathsf A}\|_s = tr\left[\sqrt{{\mathsf
A}^2}\right]
\end{equation} satisfying
$$
\|{\mathsf A}\|_s \geq \|{\mathsf A}\|
$$
for all ${\mathsf A} \in {\mathbf L}_r({\mathbf H})$
\end{thm} For proof, see Ref.\ \cite{ludwig1}, Appendix IV.11.
\begin{df}\label{dfold2} The norm (\ref{trnorm}) is called {\em trace norm} and all
elements of ${\mathbf L}_r({\mathbf H})$ with finite trace norm are called
{\em trace-class}. The set of all trace-class operators is denoted by ${\mathbf
T}({\mathbf H})$.
\end{df} Trace norm is preserved by unitary maps.
\begin{thm}\label{prop2}\label{propold2} ${\mathbf T}({\mathbf H})$ with the
operation of linear combination of operators on ${\mathbf H}$, partial
ordering $\geq$ defined above and completed with respect to the norm
(\ref{trnorm}) is an ordered Banach space. A trace-class operator is bounded,
its trace is finite and its spectrum is discrete.
\end{thm} For proof, see Ref.\ \cite{ludwig1}, Appendix IV.11.

Let ${\mathbf T}({\mathbf H})^+_1$ be the set of all positive elements of
${\mathbf L}_r({\mathbf H})$ with trace 1. As these operators are positive,
their trace is equal to their trace norm and they lie on the unit sphere in
${\mathbf T}({\mathbf H})$. ${\mathbf T}({\mathbf H})^+_1$ is not a linear
space but a convex set: let ${\mathsf T}_1, {\mathsf T}_2 \in {\mathbf
T}({\mathbf H})^+_1$, then
$$
{\mathsf T} = w{\mathsf T}_1 + (1-w){\mathsf T}_2 \in {\mathbf T}({\mathbf
H})^+_1
$$
for all $0<w<1$. The sum is called {\em convex combination} and states
${\mathsf T}_1$ and ${\mathsf T}_2$ are called {\em convex components} of
${\mathsf T}$.

It follows that any convex combination
\begin{equation}\label{properm'} {\mathsf T} = \sum_k w_k{\mathsf T}_k
\end{equation} of at most countable set of ${\mathsf T}_k\in {\mathbf
T}({\mathbf H})^+_1$ with weights $w_k$ satisfying
\begin{equation}\label{weight} 0\leq w_k \leq 1\ ,\quad \sum w_k = 1
\end{equation} and the series converging in the trace-norm topology also lies
in ${\mathbf T}({\mathbf H})^+_1$.

In general, elements of ${\mathbf T}({\mathbf H})^+_1$ can be written in
(infinitely) many ways as a convex combinations of other elements.
\begin{df}\label{dfface} {\em Face} ${\mathbf W}$ is a (norm) closed
subset of ${\mathbf T}({\mathbf H})^+_1$ that is invariant with respect to
convex combinations and contains all convex components of any ${\mathbf T} \in
{\mathbf W}$.
\end{df} Then, ${\mathbf T}({\mathbf H})^+_1$ itself is a
face. "Face" is an important notion of the mathematical theory of
convex sets.
\begin{thm}\label{face1} Every face ${\mathbf W} \subset {\mathbf T}({\mathbf
H})^+_1$ can be written as ${\mathbf W}({\mathsf T})$ for a suitably chosen
${\mathsf T} \in {\mathbf W}({\mathsf T})$ where ${\mathbf W}({\mathsf T})$ is
the smallest face that contains ${\mathsf T}$.
\end{thm} For proof, see \cite{ludwig1}, p.\ 76. There is a useful relation
between faces and projections:
\begin{thm}\label{face2} To each face ${\mathbf W}$ of ${\mathbf T}({\mathbf
H})^+_1$ there is a unique projection ${\mathsf P} : {\mathbf H} \mapsto
{\mathbf H}'$, where ${\mathbf H}'$ is a closed subspace of ${\mathbf H}$, for
which ${\mathsf T} \subset {\mathbf W}$ is equivalent to
$$
{\mathsf T} = {\mathsf P}{\mathsf T}{\mathsf P}\ .
$$
The map so defined between the set of faces and the set of projections is an
order isomorphism, i.e., it is invertible and ${\mathsf P}' < {\mathsf P}$ is
equivalent to ${\mathbf W}' \subset {\mathbf W}$.
\end{thm} For proof, see \cite{ludwig1}, p.\ 77. We shall denote the face that
corresponds to a projection ${\mathsf P}$ by ${\mathbf W}_{\mathsf P}$.

Clearly, intersection of two faces, if non-empty, is a face, and a unitary map
of a face is a face. The next theorem shows that ${\mathbf W}({\mathsf T})$ is
not necessarily the set of all convex components of ${\mathsf T}$.
\begin{thm}\label{face4} Let ${\mathsf P}({\mathbf H})$ be
infinite-dimensional and let ${\mathsf T}_1, {\mathsf T}_2 \in {\mathbf
W}_{\mathsf P}$ be positive definite on ${\mathsf P}({\mathbf H})$. Then
$$
{\mathbf W}({\mathsf T}_1) = {\mathbf W}({\mathsf T}_2) = {\mathbf W}_{\mathsf
P}\ .
$$
Let $\{|k\rangle\}$ be an orthonormal basis of ${\mathsf P}({\mathbf H})$ and
let
\begin{equation}\label{suprem} \sup_k \frac{\langle k|{\mathsf
T}_1|k\rangle}{\langle k|{\mathsf T}_2|k\rangle} = \infty\ .
\end{equation} Then ${\mathsf T}_1$ is not a convex component of ${\mathsf
T}_2$.
\end{thm} {\bf Proof} Suppose that ${\mathsf T}_1$ is a component of ${\mathsf
T}_2$. Then, there is ${\mathsf T}_3$ and $w \in (0,1)$ such that
$$
w{\mathsf T}_1 + (1-w){\mathsf T}_3 = {\mathsf T}_2\ .
$$
Hence ${\mathsf T}_2 - w{\mathsf T}_1 > 0$ and
$$
\langle k|{\mathsf T}_2|k\rangle - w\langle k|{\mathsf T}_1|k\rangle > 0
$$
for some positive $w$ and all $k$, which contradicts Eq.\ (\ref{suprem}), QED.
\begin{df}\label{extrem} An element ${\mathsf T}$ is called {\em extremal} element
of ${\mathbf T}({\mathbf H})^+_1$ if ${\mathbf W}({\mathsf T})$ is
zero-dimensional, i.e., if the condition
$$
{\mathsf T} = w{\mathsf T}_1 + (1-w){\mathsf T}_2
$$
with ${\mathsf T}_1, {\mathsf T}_2 \in {\mathbf T}({\mathbf H})^+_1$ and
$0\leq w \leq 1$, implies that ${\mathsf T} = {\mathsf T}_1 = {\mathsf T}_2$.
\end{df} For extremal states, we have:
\begin{thm}\label{face3} ${\mathsf T}$ is extremal iff ${\mathsf T} =
|\psi\rangle \langle \psi |$, where $\psi$ is a unit vector of ${\mathbf H}$.
\end{thm} For proof, see \cite{ludwig1}, p.\ 78. The set of all extremal
elements of ${\mathbf T}({\mathbf H})^+_1$ generates ${\mathbf T}({\mathbf
H})^+_1$ in the sense that any ${\mathsf T} \in {\mathbf T}({\mathbf H})^+_1$
can be expressed as countable convex combination of some extremal elements,
$$
{\mathsf T} = \sum_k w_k{\mathsf P}[\phi_k]\ .
$$
Such a decomposition can be obtained, in particular, from the spectral
decomposition
$$
{\mathsf T} = \sum_k t_k {\mathsf P}_k\ .
$$
In that case ${\mathsf T}$ is decomposable into mutually orthogonal projectors
${\mathsf P}[\phi_l]$ onto elements of a basis, with weights $w_l =
t_l/\sqrt{n_l}$, where $n_l$ is the degeneracy of the eigenvalue $t_l$ (the
degeneracy subspaces of trace-class operators have finite dimensions).

A unit vector $\phi$ of ${\mathbf H}$ defines a unique extremal element
${\mathsf P}[\phi] \in {\mathbf T}({\mathbf H})^+_1$ but ${\mathsf P}[\phi]$
determines $\phi$ only up to a phase factor $e^{i\alpha}$. Due to the
(complex) linear structure of ${\mathbf H}$, there is an operation on vectors
called linear superposition. Linear superposition $\psi = \sum c_k\phi_k$ of
unit vectors with complex coefficients satisfying
$$
\sum_k |c_k|^2 = 1
$$
is another unit vector and the resulting projector ${\mathsf P}[\psi]$,
\begin{equation}\label{pures} {\mathsf P}[\psi] = \left|\sum_k
c_k\phi_k\right\rangle\left\langle\sum_l c_l\phi_l\right| =
\sum_{kl}c_l^*c_k|\phi_k\rangle\langle\phi_l| \neq \sum_k
|c_k|^2|\phi_k\rangle\langle\phi_k|\ .
\end{equation} is different from the corresponding convex combination,
\begin{equation}\label{npures} \sum_k |c_k|^2|\phi_k\rangle\langle \phi_k|\ .
\end{equation} Observe that ${\mathsf P}[\psi]$ is not determined by the
projections ${\mathsf P}[\phi_k]$ because it depends on the relative phases of
vectors $\phi_k$.

\subsection{General rules} The preceding subsection has introduced technical
tools that will now be used to further develop Realist Model Approach (see
Section 0.1.3) so that it can serve as a basis of the general quantum
language.

\begin{rl}\label{hilbert} With each quantum system ${\mathcal S}$ of type
$\tau$, a complex separable Hilbert space ${\mathbf H}_\tau$ is
associated. ${\mathbf H}_\tau$ is a representation space of certain group
associated with Galilean group and $\tau$ determines the representation (see
Section 1.3).
\end{rl} Thus, every system has its own copy of a Hilbert space and the
structure of the space depends only on the type of the system. Starting from
the Hilbert space, all important entities concerning ${\mathcal S}$ such as
state space or algebra of observables are constructed. The word
"system" can be used in two different ways: it may represent a real,
i.e., a prepared quantum object, or an idealised entity used in the
construction of a theoretical model of another prepared object. This idealised
entity can be described by, or identified with, the corresponding Hilbert
space because the construction uses only this Hilbert space\footnote{If the
model contains more identical subsystems, then none of these subsystems has an
individual existence and none is in a state of its own (see Section
2.2.1).}. However, we can assume:
\begin{rl}\label{rlold1} The state space of ${\mathcal S}$ is ${\mathbf
T}({\mathbf H}_\tau)^+_1$. For each ${\mathsf T} \in {\mathbf T}({\mathbf
H}_\tau)^+_1$, there is a preparation ${\mathcal P}$ that prepares a system
${\mathcal S}$ of type $\tau$ in state ${\mathsf T}$. ${\mathsf T}$ is then an
objective property of ${\mathcal S}$.
\end{rl}

If a state is not extremal, then it can always be written as a convex
combination of other states. What is the physical meaning of this mathematical
operation?

\begin{df}\label{dfstatprep} Let ${\mathcal P}_1$ and ${\mathcal P}_2$ be two
preparation of ${\mathcal S}$ and $w \in [0,1]$. {\em Statistical
mixture}
\begin{equation}\label{statmixP} {\mathcal P} = \{(w,{\mathcal P}_1) ,
((1-w),{\mathcal P}_2)\}
\end{equation} of the two preparations is a new preparation constructed as
follows. Let system ${\mathcal S}$ be prepared either by ${\mathcal P}_1$ or
by ${\mathcal P}_2$ in a random way so that ${\mathcal P}_1$ is used with
probability $w$ and ${\mathcal P}_1$ with probability $1-w$.
\end{df} This definition can easily be extended to any number of preparations.
Examples of statistical mixture are given in Sections 0.1.4 and 4.3.1. Then,
we assume:
\begin{rl}\label{rlstatmixT} Let ${\mathcal P}_1$ and ${\mathcal P}_2$ be two
preparation of ${\mathcal S}$ and let the corresponding states be ${\mathsf
T}_1$ and ${\mathsf T}_2$. Then the statistical mixture (\ref{statmixP})
prepares state
\begin{equation}\label{physdec} {\mathsf T} = w{\mathsf T}_1\ (+)_p\
(1-w){\mathsf T}_2\ .
\end{equation}
\end{rl} The purpose of sign "$(+)_p$" on the right-hand side is to
distinguish statistical decompositions from convex combinations. This
distinction is very important in the theory of quantum measurement. For
example, the theory of quantum decoherence can achieve that the final state of
the apparatus is a convex combination of pointer states but cannot conclude
that it is a statistical decomposition and must, therefore, resort to further
assumptions such as Everett interpretation \cite{schloss}. Let us stress that
a statistical decomposition of state ${\mathsf T}$ is not determined by the
mathematical structure of state operator ${\mathsf T}$ but by the preparation
of ${\mathsf T}$.

Sometimes, one meets the objection that the states $w{\mathsf T}_1\ (+)_p\
(1-w){\mathsf T}_2$ and $w{\mathsf T}_1 + (1-w){\mathsf T}_2$ of system
${\mathcal S}$ cannot be distinguished by any measurement. But this is only
true if the measurements are limited to registrations of observables of
${\mathcal S}$. If observables of arbitrary composite systems containing
${\mathcal S}$ are also admitted, then the difference of the two states can be
found by measurements \cite{survey}. This is exactly the argument against the
decoherence theory described in \cite{d'Espagnat}, p.\ 171. Let us also
emphasise that quantum state statistics has nothing to do with the statistics
of values of observables.

As a mathematical operation, $(+)_p$ is commutative, associative, and state
statistics is invariant with respect to state composition and unitary
evolution \cite{hajicek2}. Thus, the definitions and assumptions can be
generalised to more than two preparations and states (see Rules \ref{rlold9}
and \ref{rlold11}).

Some comment is in order. Rules \ref{rlold1} and \ref{rlstatmixT} imply, on
the one hand, that ${\mathsf T}$ is an objective property of ${\mathcal S}$
because it has been prepared by ${\mathcal P}$ and, on the other hand, that
${\mathcal S}$ is objectively either in state ${\mathsf T}_1$ or ${\mathsf
T}_2$ at the same time because it has been also prepared either by ${\mathcal
P}_1$ or ${\mathcal P}_2$. There seems to be a contradiction: a particular
copy of ${\mathcal S}$ is announced to be in two different states
simultaneously. However, this contradiction is only due to a too narrow
understanding of the term state. To explain this with the help of a simple
example, let us consider a large box filled with small balls, some red and
some green, in a random mixture. Each ball chosen blindly from the box is
either red, with probability 1/2, or green, with probability 1/2, and each of
them is also coloured, with certainty. Ball properties red, green and coloured
in this example can be considered as objective. Thus, two different objective
properties concerning the same thing, namely the colour, e.g., red and
coloured, can exist simultaneously.
\begin{df}\label{dfphysdec} Let ${\mathcal P}$ of the form (\ref{statmixP})
prepare ${\mathcal S}$ in state ${\mathsf T}$. Then we write equation
(\ref{physdec}) and call the right-hand side of equation (\ref{physdec}) the
{\em statistical decomposition} of ${\mathsf T}$. States that have a
non-trivial statistical decomposition ($w \in (0,1)$) will be called
{\em decomposable}, otherwise {\em indecomposable}.
\end{df}

Observe that the statistical decomposition of quantum states has nothing to do
with the statistical structure of values of observables and that it is usually
the latter that is understood as implying the statistical character of quantum
mechanics.

The properties "decomposable", "indecomposable" and
statistical decomposition are determined uniquely by a preparation, hence they
are objective according to Basic Ontological Hypothesis. As explained above,
they are in principle observable. If a preparation is not completely known, it
can prepare a state without determining whether it is decomposable or
indecomposable.

State operator ${\mathsf T}$ does not, by itself, determine the statistical
decomposition of a prepared state described by it, unless ${\mathsf T}$ is
extremal so that every convex decomposition of it is trivial. One works
usually with wave functions, which do represent extremal states. There are
also some mathematical properties of statistical decomposition that can be
helpful for deciding if a state is decomposable that will be introduced later
(conservation under unitary time evolution and under system
composition). However, in many cases, possible statistical decomposition of a
state is not important because sufficiently many properties of the state are
independent of its statistical decomposition so that everything one needs can
be obtained from the state operator.

There are examples of prepared non-extremal states that are
indecomposable. Consider system ${\mathcal S}$ prepared in the EPR experiment
\cite{peres}. ${\mathcal S}$ is composite, ${\mathcal S} = {\mathcal S}_1 +
{\mathcal S}_2$ so that the spins of the two subsystems are
correlated. ${\mathcal S}$ is prepared in extremal state $|\psi\rangle
\langle\psi|$. Then the state of ${\mathcal S}_1$ is $tr_{{\mathcal
S}_2}[{|\psi\rangle \langle\psi|}]$, which is well-known to be indecomposable
but not extremal.

We observe that a quantum state is conceptually very different from a state in
Newtonian mechanics. It may be helpful to look at some important
differences. Let us define a state of a Newtonian system as a point of the
phase space of the system, ${\mathbf \Gamma}$. Newtonian state defined in this
way is generally assumed to satisfy:
\begin{enumerate}
\item {\em objectivity}: a state of a system is an objective property of the
system,
\item {\em universality}: any system is always in some state,
\item {\em exclusivity}: a system cannot be in two different states
simultaneously,
\item {\em completeness}: any state of a system contains maximum information
that can exist about the system.
\item {\em locality} the state of a system determines the position of the
system.
\end{enumerate} It follows that incomplete information about the state of a
system can be described by a probability distributions on ${\mathbf
\Gamma}$. Indeed, it is always at a particular point, but we do not know at
which. Such a distribution is sometimes called {\em statistical state}. In any
case, we distinguish a state from a statistical state.

A quantum state is an element of ${\mathbf T}({\mathbf H})^+_1$ and the
comparison with Newtonian states shows that it satisfies just the objectivity in
the form that a prepared state is an objective property, at least according to
our interpretation. However, a quantum system does not need to be in any state (an
example is a particle ${\mathcal S}$ in a system ${\mathcal S}'$ of identical
particles and we assume that a state of ${\mathcal S}'$ has been prepared but
that of ${\mathcal S}$ has not \cite{hajicek4}). Also, a system can be in
several states simultaneously, such as ${\mathsf T}_1$ and $w {\mathsf T}_1\
(+)_p\ (1-w) {\mathsf T}_2$ in formula (\ref{physdec}). Moreover, a state
operator alone does not contain any information on the statistical
decomposition of a prepared state. However, if an indecomposable state of a
system is given, no more knowledge on the system can objectively exist than
that given by the state\footnote{It follows that a collapse of wave function
or analogous processes are not just changes of our information about the
system bur genuine physical processes, see Chapter 4 and
\cite{hajicek2,hajicek4}.}. Finally, quantum states are non-local: most states
of a single particle do not determine its position, but simultaneous
registrations by two detectors at different positions will give
anti-correlated results (see Chapter 4).

In particular, disregarding all differences, a Newtonian state is, in a sense,
analogous to indecomposable state in quantum mechanics and probability
distributions on Newtonian phase space are, in the same sense, analogous to
decomposable states. The following comparison is amusing. If the knowledge of
a state of a Newtonian system is not complete, then we can describe it by a
probability distribution on the phase space. If a knowledge of the preparation
of a quantum state is incomplete, then the state is known but its statistical
decomposition does not need to be.

At this stage, we can understand a different approach to reality of states
\cite{spekkens,PBR} and to compare it to ours. In this approach, the existence
of a theory with realist interpretation is assumed, the states of which have
the features 1--4 above and such that quantum states can be considered as
distributions over these real states. Thus, the information contained in
quantum states is never complete. Then, the question is asked, whether
different extremal quantum states can be non-overlapping distributions, so
that each real state defines at most one extremal quantum state. This is
interpreted as reality of quantum states. (Under some reasonable conditions,
one can prove that the answer to the question is positive.) One can then try
to understand the collapse of wave function as the change of information
resulting from a registration together with, or rather than, a physical change
in a real state. Clearly, we do not assume the existence of any further theory
and directly consider all quantum states to be objective properties without
being probability distributions over some different kind of real
states. Rather, we say that indecomposable states are not distributions over
any real states (and represent a complete information) while decomposable
states can be considered as distribution over real indecomposable states. The
collapse of wave function is then a physical process.

Any unit vector $\phi \in {\mathbf H}_\tau$ defines state ${\mathsf
P}[\phi]$. Often, such states are called "pure" while general state
operators are called "mixed". In fact, the common use of the terms
"pure" and "mixed" states is misleading. From the point of
view of statistical physics, pure ought to be the indecomposable states and
mixed the decomposable ones. The confusion is aggravated by the fact that
decomposable states have various names in literature: {\em direct mixture}
\cite{ludwig1} or {\em proper mixture} \cite{d'Espagnat} or {\em Gemenge}
\cite{BLM}.

In Newtonian mechanics, simple physical properties are constructed from real
functions on the phase space. For example energy is such a function and the
proposition "Energy has value $E$" is a property. Such propositions
form a Boolean lattice with the logical operation union and intersection. The
values of the functions also define subsets of the phase space: e.g., the set
of all points of the phase space where the energy has value $E$. The Boolean
lattice of the propositions can be isomorphically mapped onto the Boolean
lattice of the subsets. We can give analogous definitions in quantum
mechanics:
\begin{df}\label{dfphysprop} Let ${\mathcal S}$ be system of type $\tau$ and
$f : {\mathbf T}({\mathbf H}_\tau)^+_1 \mapsto {\mathbb R}$ a function. Then
the proposition "$f = a$" is a {\em simple property}. Simple
properties form a Boolean lattice with the logical operation union and
intersection. Each simple property defines a subset $\{{\mathsf T} \in
{\mathbf T}({\mathbf H}_\tau)^+_1\ |\ f({\mathsf T}) = a\}$. The Boolean
lattice of simple properties is isomorphic to the Boolean lattice of the
subsets.
\end{df} According to Basic Ontological Hypothesis, simple properties are
objective because they are uniquely defined by preparations. Indeed, a
preparation defines a unique state ${\mathsf T}$ and the state a unique simple
property $f = f({\mathsf T})$.

The statistical decomposition of a state is an example of an objective
property that is not simple. Next section will give important examples of
simple properties. Hence, in both theories, Newtonian and quantum mechanics,
the values of simple properties are real numbers and the logics of the
properties are the same.

Let us briefly compare our logic of simple properties with the so-called
quantum logic (for more details, see Section 1.2.4). Properties of quantum
logic are described mathematically by orthogonal projections onto subspaces of
Hilbert space ${\mathbf H}_\tau$. Any such projection ${\mathsf P}$ is an
observable and the corresponding property is the proposition "Observable
${\mathsf P}$ has value $\eta$", where $\eta = 0,1$. The properties are
not objective and do not form a Boolean lattice. The basic difference is that
our properties are associated with preparations while the properties of
quantum logic are associated with registrations.

Extremal states allow another mathematical operation, a linear superposition
(\ref{pures}), which is different from a convex combination. The non-diagonal
(cross) terms in sum (\ref{pures}) lead to {\em interference} phenomena (such
as the electron interference in the Tonomura \cite{tono} experiment, see
Section 0.1.2) that are purely quantum and unknown in the Newtonian mechanics.

\section{Observables} The popular description of observables is by
self-adjoint operators. However, this notion is not adequate for accounts of
registrations that use ancillas (see Section 4.1) and our construction of
$D$-local observables (see Section 2.2.3) also needs a more general notion of
quantum observable. In the present section, the general theory of observables
will be briefly described. First, we give the mathematical construction of
observables from the Hilbert space, then their general relation to
registration. For this purpose, the empirical notion of registration that was
explained in the Introduction is sufficient. Also, some important properties
of observable such as joint measurability, contextuality and superselection
rules are discussed

\subsection{Mathematical preliminaries} This section is a brief review of most
important definitions and theorems. More details and proofs can be found e.g.\
in \cite{ludwig1}.  According to Rule \ref{hilbert}, every system ${\mathcal
S}$ is associated with a Hilbert space ${\mathbf H}$. Starting from ${\mathbf
H}$, we can perform the following constructions.

The set of all bounded linear operators denoted by ${\mathbf L}({\mathbf H})$
in Section 1.1.1 is closed with respect to both multiplication and linear
combination of its elements.  One can show that ${\mathbf L}({\mathbf H})$ is
a Banach algebra with norm (\ref{opnorm}). That is, ${\mathbf L}({\mathbf H})$
is complete with respect to the norm and its multiplication satisfies
$$
\|{\mathsf A}{\mathsf B}\| \leq \|{\mathsf A}\|\,\|{\mathsf B}\|\ .
$$
For proof, see \cite{BEH}, p.\ 75. Moreover, the identity operator ${\mathsf
1}$ defined by ${\mathsf 1}\phi = \phi$ for all $\phi$ satisfies ${\mathsf
1}{\mathsf A} = {\mathsf A}$ for any ${\mathsf A} \in {\mathbf L}({\mathbf
H})$ and $\|{\mathsf 1}\| = 1$, hence ${\mathbf L}({\mathbf H})$ is a Banach
algebra with unity \cite{BEH}.

Linear combinations of bounded s.a.\ operators with real coefficients are
again s.a., but their products are not, in general. Hence, the set of all
bounded s.a.\ operators forms a real linear space. If completed with respect
to the norm (\ref{opnorm}), it is an ordered Banach space denoted by ${\mathbf
L}_r({\mathbf H})$.

\begin{df}\label{dfold4} Let ${\mathbf F}$ be the Boolean lattice of all Borel
subsets of ${\mathbb R}^n$. A {\em positive operator valued (POV) measure}
$$
{\mathsf E} : {\mathbf F} \mapsto {\mathbf L}_r({\mathbf H})
$$
is defined by the properties
\begin{enumerate}
\item positivity: ${\mathsf E}(X) \geq 0$ for all $X\in {\mathbf F}\ ,$
\item $\sigma$-additivity: if $\{X_k\}$ is a countable collection of disjoint
sets in ${\mathbf F}$ then
$$
{\mathsf E}(\cup_k X_k) = \sum_k {\mathsf E}(X_k)\ ,
$$
where the series converges in weak operator topology, i.e., averages in any
state converge to an average in the state.
\item normalisation:
$$
{\mathsf E}({\mathbb R}^n) = {\mathsf 1}\ .
$$
\end{enumerate} The number $n$ is called {\em dimension} of ${\mathsf E}$. Let
us denote the support of the measure ${\mathsf E}$ by ${\mathbf \Omega}$. The
set ${\mathbf \Omega}$ is called the {\em value space} of ${\mathsf E}$. The
operators ${\mathsf E}(X)$ for $X\in {\mathbf F}$ are called {\em effects}.
\end{df} The support ${\mathbf \Omega}$ of measure ${\mathsf E}$ is defined as
follows
$$
{\mathbf \Omega} = \{\vec{x}\in {\mathbb R}^n| \mathsf E(\{\vec{y}\ |\
|\vec{x}-\vec{y}| < \epsilon\}) \neq {\mathsf 0}\quad \forall \epsilon > 0\}\
.
$$

Let ${\mathbf H}$ and ${\mathbf H}'$ be two separable Hilbert spaces and
${\mathsf U} : {\mathbf H} \mapsto {\mathbf H}'$ a unitary map. Then ${\mathsf
U}{\mathsf E}{\mathsf U}^\dagger : {\mathbf F} \mapsto {\mathbf L}({\mathbf
H}')$ is a POV measure on ${\mathbf H}'$.

We denote by ${\mathbf L}_r({\mathbf H})^+_{\leq 1}$ the set of all effects.
\begin{thm}\label{propold4} ${\mathbf L}_r({\mathbf H})^+_{\leq 1}$ is the set
of elements of ${\mathbf L}_r({\mathbf H})$ satisfying the inequality
\begin{equation}\label{effect} {\mathsf 0}\leq {\mathsf E}(X)\leq {\mathsf 1}\
.
\end{equation}
\end{thm} That effects must satisfy (\ref{effect}) follows from the positivity
and normalisation of a POV measure \cite{ludwig1}. On the other hand, each
element of the set defined in Theorem \ref{propold4} is an effect. For
example, if ${\mathsf E}$ is such an element, we can define ${\mathsf
E}(\{1\}) = {\mathsf E}$ and ${\mathsf E}(\{-1\}) = {\mathsf 1}-{\mathsf
E}$. The two operators satisfy Eq.\ (\ref{effect}) and they sum to ${\mathsf
1}$, so they determine a POV measure with the value set $\Omega = \{-1,+1\}$.

Theorem \ref{propold4} implies that the spectrum of each effect is a subset of
$[0,1]$. An effect is a projection operator (${\mathsf E}(X)^2 = {\mathsf
E}(X)$) if an only if its spectrum is the two-point set $\{0,1\}$.
\begin{df}\label{df4} Let ${\mathsf E} : {\mathbf F} \mapsto {\mathbf
L}({\mathbf H})$ and ${\mathsf E}' : {\mathbf F}' \mapsto {\mathbf L}({\mathbf
H})$ be two POV measures that satisfy
$$
{\mathsf E}(X){\mathsf E}'(X') = {\mathsf E}'(X'){\mathsf E}(X)
$$
for all $X \in {\mathbf F}$ and $X' \in {\mathbf F}'$\ .  Then we say that the
two POV measures {\em commute}.
\end{df}

\begin{thm}\label{propold5} For any POV measure ${\mathsf E} : {\mathbf F}
\mapsto {\mathbf L}({\mathbf H})$ the following two conditions are equivalent
$$
{\mathsf E}(X)^2 = {\mathsf E}(X)
$$
for all $X \in {\mathbf F}$ and
$$
{\mathsf E}(X\cap Y) = {\mathsf E}(X){\mathsf E}(Y)
$$
for all $X, Y\in {\mathbf F}$\ .
\end{thm} Thus, a POV measure is a {\em projection valued} (PV) measure
exactly when it is multiplicative. In this case, all its effects commute with
each other.

PV measures for $n=1$ are equivalent to s.a.\ operators are not necessarily
bounded. Section 1.1.1 dealt only with bounded s.a.\ operators, but now we
need more general entities.

Let ${\mathsf A}$ be an operator on Hilbert space ${\mathbf H}$ that is not
necessarily bounded. Then it is not defined on the whole of ${\mathbf H}$ but
only on linear subspace ${\mathbf D}_{\mathsf A}$ that is dense in ${\mathbf
H}$ and is called {\em domain} of ${\mathsf A}$. The definition of adjoint has
two steps: first, ${\mathsf A}^\dagger$ is defined on some domain associated
with ${\mathbf D}_{\mathsf A}$ and then possibly extended to a larger
domain. The self-adjoint operator is defined similarly, for details,
see e.g.\ Ref.\ \cite{RS}.

To define a sum and product of two unbounded operators ${\mathsf A}$ and
${\mathsf B}$, their domains are used as follows. If
$$
{\mathbf D}_{\mathsf A} = {\mathbf D}_{\mathsf B}
$$
then we say that ${\mathsf A}$ and ${\mathsf B}$ have a common domain. If,
moreover, the common domain ${\mathbf D}$ is invariant with respect to both
operators, i.e.,
$$
{\mathsf A}{\mathbf D}\subset {\mathbf D}\ ,\quad {\mathsf B}{\mathbf
D}\subset {\mathbf D}
$$
then the sum ${\mathsf A} + {\mathsf B}$, product ${\mathsf A}{\mathsf B}$ and
commutator $[{\mathsf A},{\mathsf B}] = {\mathsf A}{\mathsf B} - {\mathsf
B}{\mathsf A}$ are well defined on ${\mathbf D}$ and can be possibly extended.

For s.a.\ operators, bounded or unbounded, the so-called spectral theorem
holds, see Refs.\ \cite{RS}, \cite{BEH}, Chap.\ 10. This says that any s.a.\
operator ${\mathsf A}$ is equivalent to a PV measure, which is called, in this
case, the spectral measure of ${\mathsf A}$. Let ${\mathsf E}$ be a PV
measure, then ${\mathsf E}$ determines a unique self-adjoint operator
$\int_{\mathbb R}\iota d{\mathsf E}$, where $\iota$ denotes the identity
function on ${\mathbb R}$. Conversely, each s.a.\ operator $\mathsf A$ on
${\mathbf H}$ determines a unique PV measure ${\mathsf E} : {\mathbf F}\mapsto
{\mathbf L}({\mathbf H})$ such that
\begin{equation}\label{sharp} {\mathsf A} = \int_{\mathbb R}\iota d{\mathsf
E}\ .
\end{equation} If PV measure ${\mathsf E}$ is equivalent to a s.a.\ operator
${\mathsf A}$, we shall denote it ${\mathsf E}^{\mathsf A}$. Thus, POV measure
is a generalisation of a self-adjoint operator. Clearly, two s.a.\ operators
${\mathsf A}$ and ${\mathsf B}$ commute if PV measures ${\mathsf E}^{\mathsf
A}$ and ${\mathsf E}^{\mathsf B}$ commute.

Expression $tr[{\mathsf T},{\mathsf E}]$ is well-defined for all ${\mathsf T}
\in {\mathbf T}({\mathbf H})$ and ${\mathbf E} \in {\mathbf L}_r({\mathbf H})$
and it is a real-valued bilinear form with respect to which the two Banach
spaces are dual to each other \cite{ludwig1}, p.\ 413.  We even have:
\begin{thm}\label{propold6} If ${\mathsf T}_1$ and ${\mathsf T}_2$ from
${\mathbf T}({\mathbf H}_\tau)^+_1$ satisfy
$$
tr[{\mathsf T}_1{\mathsf E}] = tr[{\mathsf T}_2{\mathsf E}]
$$
for all ${\mathsf E} \in {\mathbf L}_r({\mathbf H}_\tau)^+_{\leq 1}$ then
${\mathsf T}_1 = {\mathsf T}_2$; if ${\mathsf E}_1$ and ${\mathsf E}_2$ from
${\mathbf L}_r({\mathbf H}_\tau)^+_{\leq 1}$ satisfy
$$
tr[{\mathsf T}{\mathsf E}_1] = tr[{\mathsf T}{\mathsf E}_2]
$$
for all ${\mathsf T} \in {\mathbf T}({\mathbf H}_\tau)^+_1$ then ${\mathsf
E}_1 = {\mathsf E}_2$.
\end{thm} An important further property of the bilinear form is:
\begin{thm}\label{propold7} For each ${\mathsf T} \in {\mathbf T}({\mathbf
H}_\tau)^+_1$ and ${\mathsf E} \in {\mathbf L}_r({\mathbf H}_\tau)^+_{\leq
1}$, the condition $tr[{\mathsf T}{\mathsf E}] = 1$ is equivalent to
$$
{\mathsf E}{\mathsf T} = {\mathsf T}\ .
$$
\end{thm} In general, we call state ${\mathsf T}$ satisfying
$$
{\mathsf E}{\mathsf T} = a{\mathsf T}
$$
with some real $a$ {\em eigenstate} of ${\mathsf E}$ to {\em eigenvalue}
$a$. If ${\mathsf T} = {\mathsf P}[\psi]$, then $\psi$ is called {\em
eigenvector}.

An example is a discrete POV measure. A POV measure ${\mathsf E}$ is {\em
discrete} if its value set ${\mathbf \Omega}$ is an at most countable subset
of $\mathbb R$. Let ${\mathsf E}$ be a discrete POV measure and let ${\mathbf
\Omega} = \{a_k, k\in {\mathbb N}\}$, where ${\mathbb N}$ is the set of
positive integers. Then ${\mathsf E}$ is a PV measure, ${\mathsf E} = {\mathsf
E}^{\mathsf A}$, ${\mathsf A} = \sum_k a_k {\mathsf E}_k$, $a_k$ are
eigenvalues of ${\mathsf A}$ and
$$
{\mathsf E}_k = {\mathsf E}(\{a_k\})
$$
are projections on the corresponding eigenspaces of the s.a.\ operator
${\mathsf A}$ that is defined in this way.

\begin{thm}\label{proppm} For any POV measure ${\mathsf E} : {\mathbf F}
\mapsto {\mathbf L}({\mathbf H})$ and any ${\mathsf T}\in {\mathbf T}({\mathbf
H})^+_1$, the mapping
$$
p^{\mathsf E}_{\mathsf T} : {\mathbf F} \mapsto [0,1]
$$
defined by
\begin{equation}\label{probab} p^{\mathsf E}_{\mathsf T}(X) = tr[{\mathsf
T}{\mathsf E}(X)]
\end{equation} for all $X\in {\mathbf F}$ is a real $\sigma$-additive
probability measure with values in $[0,1]$.
\end{thm} This follows from the defining properties of ${\mathsf E}$ and the
continuity and linearity of the trace, see e.g.\ \cite{BLM}. The measure is
preserved by unitary maps,
$$
tr[({\mathsf U}{\mathsf T}{\mathsf U}^\dagger)({\mathsf U}{\mathsf
E}(X){\mathsf U}^\dagger)] = tr[{\mathsf T}{\mathsf E}(X)]\ .
$$

We note that a convex combination of states induces a convex combination of
measures,
\begin{equation}\label{comes} {\mathsf T} = \sum_k w_k {\mathsf T}_k \mapsto
p^{\mathsf E}_{\mathsf T} = \sum_k w_k p^{\mathsf E}_{{\mathsf T}_k}\ .
\end{equation}

More about mathematical properties of states and effects and the corresponding
spaces ${\mathbf T}(\mathbf H)^+_1$ and ${\mathbf L}_r(\mathbf H)^+_{\leq 1}$
can be found in Ref.\ \cite{ludwig1}.

\subsection{General rules} The physical interpretation of POV measures is
given by:
\begin{rl}\label{rlold2} Any quantum mechanical observable for system
${\mathcal S}$ of type $\tau$ is mathematically described by some POV measure
${\mathsf E} : {\mathbf F} \mapsto {\mathbf L}_r({\mathbf H}_\tau)$. Each
outcome of an individual registration of the observable ${\mathsf E}(X)$
performed on quantum object ${\mathcal S}$ yields an element of ${\mathbf
F}$. Each registration apparatus that interacts with ${\mathcal S}$ determines
a unique observable of ${\mathcal S}$.
\end{rl}
\begin{df} If a PV measure is an observable, the observable is called
{\em sharp}.
\end{df} Often, a stronger assumption than Rule \ref{rlold2} is made, namely
that each s.a.\ operator on ${\mathbf H}_\tau$ is a sharp observable of system
${\mathcal S}$ of type $\tau$. Such systems are called {\em proper} quantum
systems \cite{BLM}. In Sections 2.2.2 and 4.2, we shall show that, in strict
sense, there is no proper quantum system. This does not seem to represent any
genuine difficulty. Usually, the construction of a model needs only few
observables and, for a given object, our theory of quantum measurement will
itself determine which POV measures cannot be observables\footnote{These facts
do not invalidate all applications of $C^*$-algebras (e.g.\ \cite{haag}) as
powerful mathematical method in quantum mechanics. In particular,
$C^*$-algebras are useful in quantum field theory, where they are generated by
local operators, which are measurable according to our theory. However, some
of the physical interpretations that are sometimes \cite{Primas} given to
$C^*$-algebras in non-relativistic quantum mechanics are difficult to
maintain.}.

An important assumption of quantum mechanics is the following generalisation
of Born's rule
\begin{rl}\label{rlold3} The number $p^{\mathsf E}_{\mathsf T}(X)$ defined by
Eq.\ (\ref{probab}) is the probability that a registration of the observable
${\mathsf E}(X)$ performed on object ${\mathcal S}$ in the state ${\mathsf T}$
leads to a result in the set $X$.
\end{rl} Using the same preparation $N$ times results in a set of systems with
the same state ${\mathsf T}$. Performing on each element the registration of
${\mathsf E}(X)$ gives a value (or an approximate value) $y\in {\mathbf
\Omega}$. The relative frequency of finding $y \in X$ approaches $p^{\mathsf
E}_ {\mathsf T}(X)$ if $N \rightarrow\infty$ on the set. This is equal to the
probability measure given by equation (\ref{comes}). Thus, the relative
frequency of a measurement result for a registration on elements of an
ensemble can be directly calculated from the state operator. The relative
frequencies of a registration result on the so constructed ensemble are
approximately measurable and this gives the physical meaning to the
probability distribution $p^{\mathsf E}_{\mathsf T}(X)$. This does not mean
that we {\em define} probability as a frequency, see Section 0.1.4.

Theorem \ref{propold6} implies that a state ${\mathsf T}$ can be determined
uniquely, if it is prepared many times and a sufficient number of different
registrations can be performed on it. On the other hand, the question of what
is its statistical decomposition cannot be decided in this way because
$tr[{\mathsf E}{\mathsf T}]$ is generally independent of the statistical
decomposition of $\mathsf T$. There is one exception: extremal states.

Observe that the probability is only meaningful if it concerns really existing
(but maybe unknown) outcomes, see Section 0.1.4. Thus, it can be only
associated with registration. Hence, in Rule \ref{rlold3}, the probability
$p^{\mathsf E}_{\mathsf T}(X)$ refers to registrations on an object and not to
the object itself. In general, the property that value $x\in {\mathbf \Omega}$
lies within subset $X\in {\mathbf F}$ is not an objective property of objects
in the sense that it could be attributed to an object itself and that the
measurement would just reveal it. Such assumption would lead to
contradictions, see Section 1.2.4. Thus, there is an asymmetry between states
and observables: states are objective properties but observables are not; all
elements of ${\mathbf T}({\mathbf H}_\tau)^+_1$ can be prepared but not all
effects of the dual space are observables.

Special cases in which the outcome of a registration is predictable are
described as follows. Let ${\mathsf E} : {\mathbf F} \mapsto {\mathbf
L}_r({\mathbf H}_\tau)$ be an observable and let ${\mathsf T} \in {\mathbf
T}({\mathbf H}_\tau)^+_1$ be such that
$$
tr[{\mathsf T}{\mathsf E}(X)] = 1
$$
for some $X\in {\mathbf F}$. Then the probability that the registration of
${\mathsf E}(X)$ on ${\mathsf T}$ will give a value in $X$ is 1 and we can say
that the prepared object in the state ${\mathsf T}$ possesses the property
independently of any measurement. Theorem \ref{propold7} implies that
${\mathsf T}$ is an eigenstate of ${\mathsf E}(X)$ to eigenvalue 1.

Rule \ref{rlold3} implies useful formulae for averages and higher moments of
observable ${\mathsf E}$ that generalise the well-known formulae for sharp
observables. Let us restrict the dimension of POV measures to $n=1$. For the
average $\langle{{\mathsf E}}\rangle_{\mathsf T}$ of observable ${\mathsf E}$
in the state ${\mathsf T}$, we have
$$
\langle{{\mathsf E}}\rangle_{\mathsf T}= \int_{\mathbb R} \iota dp^{\mathsf
E}_{\mathsf T}\ .
$$
Using Eq.\ (\ref{probab}), we obtain
$$
\langle{{\mathsf E}}\rangle_{\mathsf T}= \int_{\mathbb R}
\iota\,d\left(tr[{\mathsf T}{\mathsf E}]\right) = tr\left[{\mathsf T}
\int_{\mathbb R} \iota d{\mathsf E}\right]\ .
$$
For a sharp observable ${\mathsf E}^{\mathsf A}$, Eq. (\ref{sharp}) yields the
usual relation
$$
\langle{{\mathsf A}}\rangle_{\mathsf T}= tr[{\mathsf T}{\mathsf A}]\ .
$$
In the case of extremal state, ${\mathsf T} = {\mathsf P}[\phi]$, we obtain
$\langle{{\mathsf A}}\rangle_{\mathsf T} = \langle \phi|{\mathsf A}\phi
\rangle$. For higher moments, we just have to substitute suitable power of
$\iota$ in the integral and the proof is analogous.
\begin{df}\label{dfold9} Let ${\mathsf A}$ be a sharp observable and ${\mathsf
T}$ be a state. Then
\begin{equation}\label{variance} \Delta_ {\mathsf T}{\mathsf A} =
\sqrt{\langle{\mathsf A}^2\rangle_ {\mathsf T}-\langle{{\mathsf
A}}\rangle_{\mathsf T}^2}
\end{equation} is called {\em variance} of ${\mathsf A}$ in ${\mathsf
T}$.
\end{df} Next,
\begin{df}\label{dfold10} The {\em normalised correlation} in a state
${\mathsf T}$ of any two commuting sharp observables ${\mathsf A}$ and
${\mathsf B}$ is defined by
\begin{equation}\label{norcorr} C({\mathsf A},{\mathsf B}, {\mathsf T}) =
\frac{\langle{\mathsf A}{\mathsf B}\rangle_{\mathsf T} - \langle{\mathsf
A}\rangle_{\mathsf T}\langle{\mathsf B}\rangle_{\mathsf T}}{\Delta_{\mathsf T}
{\mathsf A}\Delta_{\mathsf T} {\mathsf B}}\ .
\end{equation}
\end{df} The normalised correlation satisfies
$$
-1 \leq C({\mathsf A},{\mathsf B}, {\mathsf T}) \leq 1
$$
for all commuting sharp observables ${\mathsf A}$ and ${\mathsf B}$ and for
all states ${\mathsf T}$ because of Schwarz' inequality. The proof is based on
the facts that $Re\,\langle{\mathsf A}{\mathsf B}\rangle_{\mathsf T}$ is a
positive symmetric bilinear form on real linear space ${\mathbf L}_r({\mathbf
H}_\tau)$ and that $Re\,\langle{\mathsf A}{\mathsf B}\rangle_{\mathsf T} =
\langle{\mathsf A}{\mathsf B}\rangle_{\mathsf T}$ if ${\mathsf A}$ and
${\mathsf B}$ commute. If $C({\mathsf A},{\mathsf B}, {\mathsf T}) = 1$, the
observables are strongly correlated, if $C({\mathsf A},{\mathsf B}, {\mathsf
T}) = 0$, they are uncorrelated and if $C({\mathsf A},{\mathsf B}, {\mathsf
T}) = -1$, they are strongly anticorrelated (see \cite{BLM}, p. 50).

The values of an observable are not objective properties of an object but its
states are. As averages and moments of a given observable are uniquely
determined by states, we have
\begin{prop}\label{propold8} The average $\langle{\mathsf A}\rangle_{\mathsf
T}$ and variance $\Delta_{\mathsf T}{\mathsf A}$ of any sharp observable
${\mathsf A}$ and correlation $C({\mathsf A},{\mathsf B}, {\mathsf T})$ of any
two commuting sharp observables ${\mathsf A}$ and ${\mathsf B}$ in state
${\mathsf T}$ of object $\mathcal S$ are an objective (dynamical) property of
$\mathcal S$ that has been prepared in state ${\mathsf T}$.
\end{prop} Proposition \ref{propold8} gives the most important examples of
simple properties (they have been defined in Section 1.1.2).

Let us consider a discrete observable ${\mathsf A}$ (which is always sharp)
with spectrum $\{a_k\}$ that is non-degenerate, ${\mathsf P}_k = {\mathsf
P}[\psi_k]$ for all $k$. $\psi_k$ are eigenvectors of ${\mathsf A}$. If we
prepare the state ${\mathsf P}[\psi_k]$ and then register ${\mathsf A}$, the
result must be $a_k$ with probability 1. Next suppose that we prepare a linear
superposition ${\mathsf P}[\Psi]$,
$$
\Psi = \sum_1^\infty c_k\psi_k\ ,
$$
with $\sum |c_k|^2 = 1$. Before the registration, the object in this state
does not possess any of the values $a_k$ but one can say that it does possess
all those $a_k$ simultaneously for which $c_k \neq 0$. The probability $p_k$
that the registration of ${\mathsf A}$ will give the result $a_k$ is
\begin{equation}\label{probrep} p_k = tr[{\mathsf P}[\psi_k]{\mathsf P}[\Psi]]
= |c_k|^2\ .
\end{equation} Eq.\ (\ref{probrep}) gives the physical meaning to the absolute
value of the coefficients in the linear superposition and is called the Born rule.

There is also some physical meaning of the relative phases of the coefficients
of decomposition (\ref{probrep}). This can be seen experimentally by means of
the interference phenomena and correlations. As for interference, the average
in state ${\mathsf P}[\Psi]$ of observable ${\mathsf B}$ that does not commute
with ${\mathsf A}$ does depend on matrix elements of ${\mathsf B}$ between
different states $\psi_k$:
\begin{equation}\label{interfer} tr[{\mathsf B}{\mathsf P}[\Psi]] =
\sum_{kl}c^*_kc_l\langle \psi_k |{\mathsf B}| \psi_l \rangle\ .
\end{equation} Registering values of ${\mathsf B}$ many times on systems in
state ${\mathsf P}[\Psi]$ and knowing that they add to average
(\ref{interfer}), we can see that each individual system must "know"
the values of $c^*_kc_l\langle \psi_k |{\mathsf B}| \psi_l \rangle$ for all
pairs $\{k,l\}$. Similarly, the normalised correlations of a suitable pair of
observables can depend on the cross terms in equation (\ref{pures}).

Let $\{\psi_n\}$ be a basis of Hilbert space ${\mathbf H}_\tau$ and $a(n)$ a
monotonous function of $n$. Then, the s.a.\ operator
$$
{\mathsf A} = \sum_n a(n) |\psi_n \rangle \langle \psi_n|
$$
is a discrete non-degenerate observable. The registration of any discrete
non-degener\-ate observable is called {\em complete test} (see, e.g.,
\cite{peres}, p.\ 29). If state ${\mathsf T}$ is prepared, then the
probability distribution of a complete test ${\mathsf A}$ is
$$
p_n = tr[{\mathsf T}|\psi_n \rangle \langle \psi_n|]
$$
In quantum mechanics, probability distributions of registration outcomes are
generally associated with both preparations and registrations. This has to do
with the non-objective nature of observables. The amount of information that
can be gained in a complete test is given by the Shannon entropy
(\ref{entropy}) of the distribution $p_n$ and depends both on state $\psi$ and
observable ${\mathsf A}$.

Still, can there be a general measure of information lack associated with a
state alone? Clearly, such a question is meaningful and the answer is the
minimum of entropies, each associated with a complete test. This minimum,
$S({\mathsf T})$ is a well-defined function of state ${\mathsf T}$ and is
called von Neumann entropy. One can show (see, e.g., Ref.\ \cite{peres}) that,
up to a constant factor,
\begin{equation}\label{vNentropy} S(\mathsf T) = -tr[{\mathsf T}\ln({\mathsf
T})].
\end{equation} As each ${\mathsf T}$ must have a discrete spectrum with
positive eigenvalues $t_k$, we have
$$
S({\mathsf T}) = -\sum_k t_k \ln(t_k)\ .
$$

The lack of information $S({\mathsf T})$ is associated with the state $\mathsf
T$ of object $\mathcal S$ and consequently with the (classical) condition
$\mathcal C$ that defines the preparation. Hence, according to our criterion
of objectivity, von Neumann entropy is an objective property of object
$\mathcal S$ prepared in state $\mathsf T$. The objective property of a state
that is given by a value of von Neumann entropy can be called its {\em
fuzziness}.

The only states that are not fuzzy are the extremal ones. Indeed, the complete
test on extremal state $|\psi \rangle \langle \psi|$ defined by any basis with
$\psi_1 = \psi$ has a trivial probability distribution $p_n = \delta_{n1}$,
for which the entropy is zero, and entropy cannot be negative. A very
important observation is that a quantum mechanical state can be both fuzzy and
indecomposable. An example is given in Section 2.1.2.

\subsection{Joint measurability} This section adapts the theory of joint
measurability as given, e.g., in \cite{ludwig1}, p.\ 84., to our changes in
general language of quantum mechanics.
\begin{prop} Let ${\mathsf A}$ and ${\mathsf B}$ be two s.a.\ operators with a
common invariant domain, let $i[{\mathsf A},{\mathsf B}]$ have a s.a.\
extension and let ${\mathsf T}$ be an arbitrary state. Then
\begin{equation}\label{uncert} \Delta {\mathsf A}\Delta {\mathsf B} \geq
\frac{|\langle[{\mathsf A},{\mathsf B}]\rangle|}{2}\ .
\end{equation}
\end{prop} Equation (\ref{uncert}) is called {\em uncertainty relation}.

The interpretation of the uncertainty relation is as follows. If we prepare
many copies of an object in state ${\mathsf T}$ and register either ${\mathsf
A}$ or ${\mathsf B}$ or $i[{\mathsf A},{\mathsf B}]$ on each, then the average
of $[{\mathsf A},{\mathsf B}]$ and variances of ${\mathsf A}$ as well as of
${\mathsf B}$ will satisfy Eq. (\ref{uncert}). It is not necessary to register
all observables jointly.

An important notion of quantum mechanics is that of joint
measurability. Often, it is also called simultaneous measurability but the
notion has nothing to do with time. It simply means that there is one
registration device that can measure both quantities.
\begin{df} Two elements ${\mathsf E}_1$ and ${\mathsf E}_2$ of ${\mathbf
L}_r({\mathbf H})^+_{\leq 1}$ are {\em jointly measurable} if there is a POV measure
${\mathsf E} : {\mathbf F} \mapsto {\mathbf L}_r({\mathbf H})^+_{\leq 1}$ such
that ${\mathsf E}_1 = {\mathsf E}(X_1)$ and ${\mathsf E}_2 = {\mathsf E}(X_2)$
for some $X_1$ and $X_2$ in ${\mathbf F}$.
\end{df} This is a mathematical property that, in fact, is only a necessary
condition for existence of the required registration device. Even if ${\mathsf
E}$ existed, the question whether it is an observable is non-trivial.
\begin{prop}\label{propold16} Two effects ${\mathsf E}_1$ and ${\mathsf E}_2$
are jointly measurable if and only if there are three elements ${\mathsf
E}'_1$, ${\mathsf E}'_2$ and ${\mathsf E}'_{12}$ in ${\mathbf L}_r({\mathbf
H})^+_{\leq 1}$ such that
$$
{\mathsf E}'_1 +{\mathsf E}'_2 + {\mathsf E}'_{12} \in {\mathbf L}_r({\mathbf
H})^+_{\leq 1}
$$
and
$$
{\mathsf E}_1 = {\mathsf E}'_1 + {\mathsf E}'_{12}\ ,\quad {\mathsf E}_2 =
{\mathsf E}'_2 + {\mathsf E}'_{12}\ .
$$
\end{prop} For proof, see \cite{ludwig1}, p. 89. Proposition \ref{propold16}
has two corollaries: (1) two projections are jointly measurable if they commute
and (2) a sufficient condition for joint measurability of ${\mathsf E}_1$ and
${\mathsf E}_2$ is that
$$
{\mathsf E}_1 +{\mathsf E}_2 \in \ {\mathbf L}_r({\mathbf H})^+_{\leq 1}.
$$

If an observable is sharp then its effects commute, but the effects of a
general observable does not necessarily commute. Thus, even non commuting operators can be
jointly measurable. This is completely compatible with standard quantum
mechanics. Indeed, any POV measure that is an observable of an object
${\mathcal S}$ can be measured as a sharp observable of a composite system
containing ${\mathcal S}$ and another object called {\em ancilla} as
subsystems. For proof, see e.g. \cite{peres}, p.\ 285. An important example
will be given in Section 3.7.

Two observables ${\mathsf E}_1 : {\mathbf F}_1 \mapsto {\mathbf L}_r({\mathbf
H})$ and ${\mathsf E}_2 : {\mathbf F}_2 \mapsto {\mathbf L}_r({\mathbf H})$
with ${\mathbf F}_1$ comprising the Borel subsets of ${\mathbb R}^{n_1}$ and
${\mathbf F}_2$ those of ${\mathbb R}^{n_2}$ are called jointly measurable, if
each two effects ${\mathsf E}_1(X_1)$ and ${\mathsf E}_2(X_2)$ are jointly
measurable. In this case, there is an effect ${\mathsf E}_1(X_1)\wedge
{\mathsf E}_2(X_2)$ in ${\mathbf L}_r (\mathbf H)^+_{\leq 1}$ giving the
probability that observable ${\mathsf E}_1$ has value in $X_1$ and observable
${\mathsf E}_2$ has value in $X_2$. Then, there is a unique observable
$$
{\mathsf E} : {\mathbf F} \mapsto {\mathbf L}_r({\mathbf H})\ ,
$$
where ${\mathbf F}$ is the set of Borel subsets of ${\mathbb R}^{n_1+n_2}$ and
$$
{\mathsf E}(X_1\times X_2) = {\mathsf E}_1(X_1)\wedge {\mathsf E}_2(X_2)\ .
$$
We call ${\mathsf E}$ {\em compound} of ${\mathsf E}_1$ and ${\mathsf E}_2$

An important example are two sharp observables ${\mathsf A}$ and ${\mathsf
B}$. They are jointly measurable, if and only if all their effects (which are
projections) commute with each other, i.e., the operators commute. In this
case, the wedge is just the ordinary operator product,
$$
{\mathsf E}^{\mathsf A}(X_1)\wedge {\mathsf E}^{\mathsf B}(X_2) = {\mathsf
E}^{\mathsf A}(X_1) {\mathsf E}^{\mathsf B}(X_2)\ .
$$
The compound ${\mathsf E}^{{\mathsf A}\wedge{\mathsf B}}$ of ${\mathsf
E}^{\mathsf A}$ and ${\mathsf E}^{\mathsf B}$ is an observable that represents
registration of a pair of values, one of ${\mathsf A}$, the other of ${\mathsf
B}$.

If two sharp observables ${\mathsf A}$ and ${\mathsf B}$ do not commute then
they can be jointly measurable at best with the inaccuracy corresponding to
their uncertainty relation. This means that only some effects of ${\mathsf A}$
are jointly measurable with only some effects of ${\mathsf B}$. An example
will be given in Section 3.7.

\subsection{Contextuality} The investigations in the field of contextuality
were motivated by the following problem. In Newton mechanics, a statistical
state of any system $\mathcal S$ is described by probability distributions
$\rho : Z \mapsto [0,1]$, $Z \in {\mathbf \Gamma}$ on its phase space
${\mathbf \Gamma}$. Any such distribution results from a fixed preparation and
describes the ensemble of individual systems $\mathcal S$ prepared in this
way. Each individual system of the ensemble is, however, always assumed, at
least in Newton mechanics, to be in some state given by a point $Z$ of
${\mathbf \Gamma}$. We just do not know which point and the distribution
$\rho(Z)$ describes the state of our incomplete knowledge. Thus, all
observables, which are functions on ${\mathbf \Gamma}$, have determinate
values for each element of the ensemble.

Can anything analogous be assumed for quantum mechanics making a general state
operator analogous to a statistical state of Newtonian mechanics? Could a
state operator of a quantum object $\mathcal S$ describe our knowledge of an
ensemble defined by the preparation and could the individual elements of the
ensemble each have determinate values of all observables, which are just not
known, or even, be it for whatever reason, could not be known? To solve this
problem, it is sufficient to restrict the observables to sharp observables,
because if the question had negative answer for a restricted class, it would
have negative answer for any class containing the restricted one.

It is technically advantageous to restrict the problem even further by
limiting oneself to (orthogonal) projection operators to subspaces of
${\mathbf H}_\tau$. Let us denote the set of all such projections by ${\mathbf
P}({\mathbf H}_\tau) \subset {\mathbf L}_r({\mathbf H}_\tau)$. The problem can
then be formulated as follows. Is there a dispersion-free probability
distribution $h : {\mathbf P}({\mathbf H}_\tau) \mapsto \{0,1\}$?
Dispersion-free means that its values are just zero and one so that such a
distribution determines values of all projections.

The dispersion-free distribution $h$ has to fulfil certain conditions, or else
it could not be interpreted as concerning properties. Let us first describe
the structure of ${\mathbf P}({\mathbf H}_\tau)$.

Each projection ${\mathsf a} \in {\mathbf P}({\mathbf H}_\tau)$ can be mapped
on linear subspace ${\mathsf a}({\mathbf H}_\tau)$ of ${\mathbf H}_\tau$ and
this bijective map allows to define the lattice relations on ${\mathbf
P}({\mathbf H}_\tau)$. First, ${\mathsf a} \leq {\mathsf b}$ if ${\mathsf
a}({\mathbf H}_\tau) \subset {\mathsf b}({\mathbf H}_\tau)$. This defines a
partial ordering on ${\mathbf P}({\mathbf H}_\tau)$. Second, ${\mathsf a} =
{\mathsf b}^\bot$ if ${\mathsf a}({\mathbf H}_\tau)$ contains all vectors
orthogonal to ${\mathsf b}({\mathbf H}_\tau)$. Projection ${\mathsf a}$ is
called the orthocomplement of ${\mathsf b}$. Third, ${\mathsf c} = {\mathsf a}
\wedge {\mathsf b}$ if ${\mathsf c}({\mathbf H}_\tau)$ is the set-theoretical
intersection of ${\mathsf a}({\mathbf H}_\tau)$ and ${\mathsf b}({\mathbf
H}_\tau)$. Observe that ${\mathsf a} \wedge {\mathsf b} = {\mathsf a} {\mathsf
b}$ (operator product of the projections) only if ${\mathsf a}$ and ${\mathsf
b}$ commute (are orthogonal). Finally, ${\mathsf c} = {\mathsf a} \vee
{\mathsf b}$ if ${\mathsf c}({\mathbf H}_\tau)$ is the linear hull of
${\mathsf a}({\mathbf H}_\tau)$ and ${\mathsf b}({\mathbf H}_\tau)$.

${\mathbf P}({\mathbf H}_\tau)$ with these relations forms the so-called
orthocomplemented lattice (for proof, see, e.g., Refs.\ \cite{BvN,bub}), but
not a Boolean lattice\footnote{In the so-called {\em quantum logic},
properties of ${\mathcal S}$ are described by elements of ${\mathbf
P}({\mathbf H}_\tau)$. They re\-present the mathematical counterpart of the
so-called YES-NO registrations \cite{Piron}. If we pretend that the values
obtained in the YES-NO experiments are properties of a well-defined single
quantum system, then we are forced to replace the Boolean lattice of ordinary
logic by the orthocomplemented lattice of quantum logic. But this pretence is
against all logic because these values are not properties of one but of many
different systems each consisting of ${\mathcal S}$ plus some registration
apparatus.}. It has however Boolean sublattices, which represent sets of
jointly measurable sharp effects, and must, therefore contain only mutually
commuting projections. On these sublattices, $h$ is, in fact, an assignment of
truth values 0, 1 and it has to satisfy the usual logical conditions
$$
h({\mathsf a} \vee {\mathsf b}) = h({\mathsf a}) \vee h({\mathsf b})\ ,\quad
h({\mathsf a} \wedge {\mathsf b}) = h({\mathsf a}) \wedge h({\mathsf b})\
,\quad h({\mathsf a^\bot}) = h({\mathsf a})^\bot\ .
$$
In these relations, we consider the set $\{0,1\}$ as a Boolean lattice of one
empty set and one arbitrary non-empty set $\mathbf A$ with $\bot$ the
set-theoretical complement in $\mathbf A$, $\vee$ the set-theoretical union
and $\wedge$ the set-theoretical intersection. Thus, on each Boolean
sublattice, $h$ must also be a Boolean lattice homomorphism.

The first result relevant to the question of existence of such maps is the
Gleason's theorem \cite{gleason}. It states that the set of all probability
distributions on ${\mathbf P}({\mathbf H}_\tau)$ for Hilbert space ${\mathbf
H}_\tau$ of dimensions greater than 2 is just ${\mathbf T}({\mathbf
H}_\tau)^+_1$. Thus, our dispersion-free distributions are to be found in
${\mathbf T}({\mathbf H}_\tau)^+_1$. It is easy to show that there are none
there.

One can object that ${\mathbf P}({\mathbf H}_\tau)$ is an idealisation
containing an infinity of observables. This leads to the question whether
there is a finite subset of ${\mathbf P}({\mathbf H}_\tau)$ that does not
admit such distributions, either. This is the next relevant result,
Kochen--Specker no-go theorem \cite{kochen}, in which an example of such a
subset is given. There are more examples now provided by different physicists
(see, e.g., Ref.\ \cite{bub}). The reason why $h$ does not exist is that the
assignment of a truth value to projection ${\mathsf a}$, say, depends on to
which Boolean sublattice ${\mathsf a}$ belongs. This can be understood as
"context": in one case, ${\mathsf a}$ is registered jointly with
elements of one Boolean sublattice, in another case with those of another
sublattice.

The definitive result in this field seems to be Bub--Clifton--Goldstein theorem
\cite{bub}, which lists all maximal sub-lattices of ${\mathbf P}({\mathbf
H}_\tau)$ that admit dispersion-free probability distributions. They do not need to
be Boolean but they are always only proper sublattices of ${\mathbf
P}({\mathbf H}_\tau)$. Hence, only a limited number of projections can be
assumed to have determinate values before registration, and this limits
possible "non-collapse" interpretations and modifications of quantum
mechanics such as Bohm--de Broglie or modal interpretations. In fact, each of
these interpretations or modifications is based on a unique
Bub--Clifton--Goldstein sublattice so that these theories can be classified
according to these sublattices \cite{bub}.

\subsection{Superselection rules} A preparation that deals only with systems
of one and the same type prepares so-called {\em one-type systems}. Not every
preparation is such. For example, we can randomly mix electrons in some states
with protons in some states. Such systems, and some generalisations of them,
are called {\em mixed systems}\footnote{This has nothing to do with the term
"mixed state", which is sometimes used for non-vector states.}. In
this section, the mathematical description of mixed systems will be
explained. We shall mix just two one-type systems, ${\mathcal S}_1$ and
${\mathcal S}_2$, but the generalisation to any number is straightforward.

Let ${\mathbf H}_1$ and ${\mathbf H}_2$ be the Hilbert spaces of ${\mathcal
S}_1$ and ${\mathcal S}_2$. Then the Hilbert space ${\mathbf H}$ of the system
${\mathcal S}$ mixing ${\mathcal S}_1$ and ${\mathcal S}_2$ is
$$
{\mathbf H} = {\mathbf H}_1 \oplus {\mathbf H}_2\ .
$$
The direct sum on the right-hand side is the space of pairs,
$(\psi_1,\psi_2)$, $\psi_1 \in {\mathbf H}_1$ and $\psi_2 \in {\mathbf H}_2$
with linear superposition defined by
$$
a(\psi_1,\psi_2) + b(\phi_1,\phi_2) = (a\psi_1 + b\phi_1,a\psi_2 + b\phi_2)
$$
and the inner product defined by
$$
\langle (\psi_1,\psi_2),(\phi_1,\phi_2)\rangle = \langle
\psi_1,\phi_1\rangle_1 + \langle \psi_2,\phi_2\rangle_2\ ,
$$
where $\langle \cdot,\cdot\rangle_i$ is the inner product of ${\mathbf
H}_i$. There are then embeddings $\iota_i : {\mathbf H}_i \mapsto {\mathbf H}$
defined by
$$
\iota_1(\psi_1) = (\psi_1,0)\ ,\quad \iota_2(\psi_2) = (0,\psi_2)
$$
for any $\psi_i \in {\mathbf H}_i$ and $i=1,2$.

Let ${\mathbf L}({\mathbf H})$ be the algebra of bounded linear operators on
${\mathbf H}$ and ${\mathbf L}({\mathbf H}_i)$ that of ${\mathbf H}_i$. If two
operators ${\mathsf A}_i \in {\mathbf L}({\mathbf H}_i)$, $i=1,2$, are given,
then an operator $({\mathsf A}_1 \oplus {\mathsf A}_2) \in {\mathbf
L}({\mathbf H})$, called direct sum of ${\mathsf A}_1$ and ${\mathsf A}_2$, is
defined by
$$
({\mathsf A}_1 \oplus {\mathsf A}_2) (\psi_1,\psi_2) = ({\mathsf
A}_1(\psi_1),{\mathsf A}_2(\psi_2))
$$
for any $(\psi_1,\psi_2) \in {\mathbf H}$. The special property of the
direct-sum operators is that the subspaces $\iota_i({\mathbf H}_i) \subset
{\mathbf H}$ are invariant with respect to them. Clearly, not all operators in
${\mathbf L}({\mathbf H})$ are of this form. Effects of the form ${\mathsf
A}_1 \oplus {\mathsf 0}_2$ (${\mathsf 0}_1 \oplus {\mathsf A}_2$) can be
interpreted as representing registrations done on ${\mathcal S}_1$ (${\mathcal
S}_2$) alone.

The embeddings $\iota_i$ define maps on projections ${\mathsf P}[\psi_i]$,
which can be denoted by the same symbol, viz.
$$
\iota_i({\mathsf P}[\psi_i]) = {\mathsf P}[\iota_i(\psi_i)]
$$
for all $\psi_i \in {\mathbf H}_i$. $\iota_i$ can be extended to the whole
spaces ${\mathbf T}({\mathbf H}_i)^+_1$, $\iota_i : {\mathbf T}({\mathbf
H}_i)^+_1 \mapsto {\mathbf T}({\mathbf H})^+_1$ as follows. Let ${\mathsf T}_i
\in {\mathbf T}({\mathbf H}_i)^+_1$, then
$$
\iota_1({\mathsf T}_1) = {\mathsf T}_1 \oplus {\mathsf 0}\ ,\quad
\iota_2({\mathsf T}_2) = {\mathsf 0} \oplus {\mathsf T}_2\ .
$$
The convex combinations of states from ${\mathbf T}({\mathbf H}_1)^+_1$ and
${\mathbf T}({\mathbf H}_2)^+_1$ are states of mixed systems defined at the
beginning of this section. However, the states ${\mathbf T}({\mathbf H})^+_1$
are not exhausted by convex combinations of states from ${\mathbf T}({\mathbf
H}_1)^+_1$ and ${\mathbf T}({\mathbf H}_2)^+_1$.

The structural properties in which the systems ${\mathcal S}_1$ and ${\mathcal
S}_2$ differ from each other can now be viewed as non-trivial operators on
${\mathbf H}$. Let for example the masses $\mu_1$ and $\mu_2$ satisfy $\mu_1
\neq \mu_2$. Then the s.a.\ operator $\mu_1{\mathsf 1}_1 \oplus \mu_2{\mathsf
1}_2$ defines the mass operator ${\mathsf m}$ on ${\mathbf H}$ with two
eigenspaces $\iota_i({\mathbf H_i})$ and eigenvalues $\mu_i$,
$i=1,2$. Similarly for charges, spins, etc. Now, it is easy to see that all
operators of ${\mathbf L}({\mathbf H})$ that commute with ${\mathsf m}$ have
the form ${\mathsf A}_1 \oplus {\mathsf A}_2$ and all operators of this form
commute with ${\mathsf m}$. Clearly, s.a.\ operators of the form ${\mathsf
A}_1 \oplus {\mathsf A}_2$ are observables of system ${\mathcal S}$. The
nature of mixing systems suggests the basic rule
\begin{rl} All effects that can be registered on system ${\mathcal S}$ mixing
${\mathcal S}_1$ and ${\mathcal S}_2$ have the form ${\mathsf A}_1 \oplus
{\mathsf A}_2$ where ${\mathsf A}_1 \in {\mathbf L}_r({\mathbf H}_1)^+_{\leq
1}$ and ${\mathsf A}_2 \in {\mathbf L}_r({\mathbf H}_2)^+_{\leq 1}$.
\end{rl} Hence, the space of all observables of ${\mathcal S}$ is not
${\mathbf L}_r({\mathbf H})^+_{\leq 1}$ but the subset of all observables that
commute with ${\mathsf m}$.

In general, we define
\begin{df} A discrete sharp observable ${\mathsf Z}$ of a system ${\mathcal
S}$ with Hilbert space $\mathbf H$ is called {\em superselection observable}
if ${\mathsf Z}$ commutes with all observables of ${\mathcal S}$, sharp or not
sharp. The existence of such an observable is called {\em superselection
rule}. The eigenspaces of ${\mathsf Z}$ are called {\em superselection
sectors}. All superselection observables form the centre ${\mathbf Z}$ of the
algebra of all sharp observables of ${\mathcal S}$.
\end{df}

A restriction of the set of observables for system ${\mathcal S}$ suggests the
introduction of an equivalence relation on the set of states ${\mathbf
T}({\mathbf H}_\tau)^+_1$.
\begin{df} Two states ${\mathsf T}_1$ and ${\mathsf T}_2$ are {\em equivalent} with
respect to a set ${\mathbf O}$ of observables if these states assign the same
probability measures to each observable of ${\mathbf O}$, that is ,
$p_{{\mathsf T}_1}^{\mathsf E} = p_{{\mathsf T}_2}^{\mathsf E}$ for each
${\mathsf E} \in {\mathbf O}$. In that case, we write ${\mathsf T}_1
\cong_{\mathbf O} {\mathsf T}_2$.
\end{df} For example, linear superposition
$$
\Psi = a(\psi_1,0) + b(0,\psi_2)
$$
for any $\psi_i \in {\mathbf H}_i$, $i=1,2$, $|a|^2 + |b|^2 = 1$, and convex
combination
$$
{\mathsf T} = |a|^2\iota_1(|\psi_1\rangle \langle\psi_1|) +
|b|^2\iota_2(|\psi_2\rangle \langle\psi_2|)
$$
are equivalent with respect to the set of superselection observables ${\mathbf
O}$,
$$
|\Psi\rangle \langle\Psi |\cong_{\mathbf O} {\mathsf T}\ .
$$

For a given set of observables ${\mathbf O}$, $\cong_{\mathbf O}$ is, indeed,
an equivalence relation on the set of states ${\mathbf T}({\mathbf
H})^+_1$. Let us denote the set of equivalence classes of states in ${\mathbf
T}({\mathbf H})^+_1$ with respect to the set of observables $\mathbf O$ by
${\mathbf T}_{\mathbf O}({\mathbf H})^+_1$. No two states of the same class
can be distinguished by any registrations. No two different states from
${\mathbf T}({\mathbf H})^+_1$ are equivalent with respect to all observables
unless the set of effects ${\mathbf O}$ is restricted.
\begin{prop} Given a system ${\mathcal S}$ with the sets $\mathbf O$ and
${\mathbf T}_{\mathbf O}({\mathbf H})^+_1$ of observables and state classes,
respectively. Then the following statements are equivalent:
\begin{description}
\item[A]${\mathsf Z}$ is a superselection observable.
\item[B]Each state is equivalent to a unique convex combination of eigenstates
of ${\mathsf Z}$.
\end{description}
\end{prop} See Ref.\ \cite{BLM}, p. 18.
\begin{rl}\label{rlssrules} The states of system ${\mathcal S}$ mixing
${\mathcal S}_1$ and ${\mathcal S}_2$ are the equivalence classes with respect
to the set of observables $\mathbf O$ of the form ${\mathsf A}_1 \oplus
{\mathsf A}_2$. The unique convex combination ${\mathsf C}_{\mathsf
Z}({\mathsf T})$ of eigenstates of ${\mathsf Z}$ to which each element
${\mathsf T}$ of ${\mathbf T}({\mathbf H})^+_1$ is equivalent describes the
physical meaning of the class.
\end{rl} Clearly, state ${\mathsf C}_{\mathsf Z}({\mathsf T})$ does not
contain any unmeasurable correlation.

Attempts at a solution of the problem of classical properties and of quantum
measurement \cite{Hepp,Bell3,Bona,Sewell,Primas,wanb} utilise properties of
superselection rules. For example, in the theory of measurement, one would
like to evolve extremal states to convex combinations and this can indeed be
achieved by suitable superselection rules. However, the method might work only
if a stronger assumption than Rule \ref{rlssrules} were made: the words
"convex combination" had to be replaced by "statistical
decomposition".

\section{Galilean group} In quantum mechanics, {\em Galilean relativity
principle} holds in the following form:
\begin{rl}\label{rlold4} The same experiments performed in two different
Newtonian inertial frames have the same results, i.e., give the frequencies of
observable values.
\end{rl} "The same experiment" means that the first experiment has
the same empirical description with respect to the first inertial frame as the
second experiment has with respect to the second frame. In the present
section, we restrict ourselves to the proper Galilean group and work out some
consequences of the principle. For example, the most important PV measures of
quantum mechanics will be defined and the time evolution equation
formulated. We keep the exposition brief; for more detail and references see
e.g.\ Ref.\ \cite{peres,KP}.

The group of transformations that leave the geometric structure of Newtonian
spacetime invariant\footnote{We understand symmetry as transformation that
leaves some well-defined structure invariant. For example, Galilean group
contains all transformations that leave Newtonian spacetime geometry invariant
and a symmetry of a quantum system leaves its Hamiltonian invariant.} (see,
e.g., Ref.\ \cite{MTW}, p. 296) is called Galilean group ${\mathbf G}$. It is
also the group of transformations between inertial frames. We shall restrict
ourselves just to the component of unity of ${\mathbf G}$ called {\em proper
Galilean group}, ${\mathbf G}^+$ and shall ignore the improper transformations
such as space inversions and time reversal. More about them can be found in
Ref.\ \cite{wigner}.

Let $(x^1,x^2,x^3,t)$ be a Newtonian inertial frame so that $x^1,x^2,x^3$ are
Cartesian coordinates. A general element $g(\lambda)$ of ${\mathbf G}^+$ can
be written in the form
\begin{equation}\label{actiong1} \vec{x}^\top \mapsto {\mathsf O}\vec{x}^\top
+ \vec{a}^\top + \vec{v}^\top t
\end{equation} and
\begin{equation}\label{actiong2} t \mapsto t + \lambda^{10}\ ,
\end{equation} where ${\mathsf O}$ is a proper orthogonal matrix (element of
rotation group $SO(3)$) determined by three parameters
$\lambda^1,\lambda^2,\lambda^3$, $\vec{a}$ is a vector of space shift with
components $\lambda^4,\lambda^5,\lambda^6$, $\vec{v}$ is a boost velocity with
components $\lambda^7,\lambda^8,\lambda^9$ and $\lambda^{10}$ is a time
shift. The relations are written in matrix notation so that, e.g.,
$\vec{x}^\top$ is a column matrix with components of vector $\vec{x}$.

The group product $g(\lambda_3) = g(\lambda_2)g(\lambda_1)$ is defined as the
composition of the transformations $g(\lambda_2)$ and $g(\lambda_1)$ so that
$g(\lambda_1)$ is performed first and $g(\lambda_2)$ second. Then,
$$
{\mathsf O}_3 = {\mathsf O}_2{\mathsf O}_1\ ,
$$
$$
\vec{a}_3^\top = \vec{a}_2^\top + {\mathsf O}_2\vec{a}_1^\top +
\vec{v}_2^\top\lambda_{10}^1\ ,
$$
$$
\vec{v}_3^\top = \vec{v}_2^\top + {\mathsf O}_2\vec{v}_1^\top\ ,
$$
and
$$
\lambda^{10}_3 = \lambda^{10}_2 + \lambda^{10}_1\ .
$$
Group ${\mathbf G}^+$ is not simply connected because its subgroup $SO(3)$ is
not. The universal covering of $SO(3)$ is group $SU(2)$ so that there are
always two elements of $SU(2)$, differing by rotation by 2$\pi$, that are
homomorphically mapped on one element of $SO(3)$. Let us denote by
$\bar{\mathbf G}^+$ the universal covering group of ${\mathbf G}^+$.

Let us first give an intuitive motivation of why Galilean group acts on state
operators and POV measures. Given a quantum system ${\mathcal S}$ of type
$\tau$, classical apparatuses that are supposed to prepare and register
${\mathcal S}$ have well-defined Galilean transformations. Often, ${\mathcal
S}$ is subject to macroscopic influences external to ${\mathcal S}$ such as
classical external fields. These also possess non-trivial transformation laws
with respect to Galilean transformations.

Let us consider a measurement that consists of a preparation and a
registration of system $\mathcal S$ of type $\tau$ by apparatuses ${\mathcal
A}_p$ and ${\mathcal A}_r$, respectively, and let there be some external
fields $f$. Let ${\mathcal A}_p$ prepare object ${\mathcal S}$ in state
${\mathsf T}$ and let ${\mathcal A}_r$ register effect ${\mathsf E}(X)$. Next,
we can transport both ${\mathcal A}_p$ and ${\mathcal A}_r$ by $g \in {\mathbf
G}$ to $g({\mathcal A}_p)$ and $g({\mathcal A}_r)$. Let ${\mathsf T}_{g,f}$ be
the state prepared by $g({\mathcal A}_p)$ and ${\mathsf E}_{g,f}(X)$ the
effect registered by $g({\mathcal A}_r)$. If the external influences are also
transformed by $g : f \mapsto g(f)$, we obtain state ${\mathsf T}_{g,g(f)}$
and effect ${\mathsf E}_{g,g(f)}(X)$. The experiment has then been completely
transferred by $g$ and, therefore, Galilean relativity principle implies
\begin{prop}\label{propold9} State ${\mathsf T}_{g,g(f)}$ and effect ${\mathsf
E}_{g,g(f)}(X)$ satisfy
\begin{equation}\label{consprob} tr[{\mathsf T}{\mathsf E}(X)] = tr[{\mathsf
T}_{g,g(f)}{\mathsf E}_{g,g(f)}(X)]
\end{equation} for all $g \in {\mathbf G}$, ${\mathsf T} \in {\mathbf
T}({\mathbf H}_\tau)^+_1$ and ${\mathsf E}(X) \in {\mathbf L}_r({\mathbf
H}_\tau)^+_{\leq 1}$.
\end{prop}

Let us next compare measurable properties of the two states ${\mathsf T}$ and
${\mathsf T}_{g,g(f)}$. Proposition \ref{propold9} implies:
$$
tr[{\mathsf T}_{g,g(f)}{\mathsf E}(X)] = tr[{\mathsf T}{\mathsf
E}_{g^{-1},g^{-1}(f)}(X)]\ .
$$
Hence, values that can be registered on the state ${\mathsf T}_{g,g(f)}$ are
those on ${\mathsf T}$ transformed by $g^{-1}$.

\subsection{Closed systems} To keep the description of Galilean group action
simple, we restrict ourselves to systems for which all external macroscopic
influences can be assumed to be negligible and call such systems {\em
closed}. Then, $f = 0$, and we shall discard the second index at transformed
states and effects. The first basic assumption is
\begin{rl}\label{rlold5} Let ${\mathcal S}$ be an closed system of type
$\tau$. Then there is a unique linear map ${\mathsf U}(g) : {\mathbf H}_\tau
\mapsto {\mathbf H}_\tau$ for each element $g \in \bar{\mathbf G}^+$ so that
\begin{eqnarray}\label{actss1} {\mathsf T}_{g^{-1}} & = & {\mathsf
U}(g){\mathsf T}{\mathsf U}(g)^\dagger\ , \\ \label{actss2} {\mathsf
E}_{g^{-1}}(X) & = & {\mathsf U}(g){\mathsf E}(X){\mathsf U}(g)^\dagger
\end{eqnarray} for all ${\mathsf T} \in {\mathbf T}({\mathbf H}_\tau)^+_1$,
${\mathsf E}(X) \in {\mathbf L}_r({\mathbf H}_\tau)^+_{\leq 1}$ and $g \in
\bar{\mathbf G}^+ $\ .
\end{rl} For quantum mechanics, it is important that the universal covering group, rather than Galilean group itself,
acts on Hilbert spaces.

Then, Rules \ref{rlold4} and \ref{rlold5} imply
\begin{prop}\label{propnew1} $g \mapsto {\mathsf U}(g)$ is a unitary ray
representation of $\bar{\mathbf G}^+$ on ${\mathbf H}_\tau$
\end{prop} For proof and references see, e.g., \cite{KP}, pp.\ 285--292. The
'ray' representation means that
$$
{\mathsf U}(g_2g_1) = \exp[ i\omega(g_1,g_2)]{\mathsf U}(g_2){\mathsf U}(g_1)\
,
$$
where $\omega : \bar{\mathbf G}^+ \times \bar{\mathbf G}^+ \mapsto \mathbb R$
satisfies
$$
\omega(g_1,g_2) + \omega(g_1g_2,g_3) = \omega(g_1,g_2g_3) + \omega(g_2g_3)\ .
$$
Equations (\ref{actss1}) and (\ref{actss2}) and the fact that rotation by
$2\pi$ is always represented by 1 or $-$1 imply that it is indeed the proper
Galilean group and not only $\bar{\mathbf G}^+$ acting on ${\mathbf
T}({\mathbf H}_\tau)^+_1$ and ${\mathbf L}_r({\mathbf H}_\tau)^+_{\leq 1}$.

Clearly, the probability is preserved,
$$
tr[({\mathsf U}(g){\mathsf T}{\mathsf U}(g)^\dagger)({\mathsf U}(g){\mathsf
E}(X){\mathsf U}(g)^\dagger)] = tr[{\mathsf T}{\mathsf E}(X)]\ .
$$
We assume that Borel sets $X$ have a well-defined transformation with respect
to $g$, $\gamma(g) : {\mathbf F} \mapsto {\mathbf F}$ and that the two effects
${\mathsf E}(X)$ and ${\mathsf E}_g(X)$ belong to the same observable. Then
\begin{equation}\label{inversw} {\mathsf U}(g){\mathsf E}(X){\mathsf
U}(g)^\dagger) = {\mathsf E}(\gamma(g)^{-1}X)
\end{equation} (see Ref.\ \cite{DST} for generalisation of this relation).

Given a one-parameter group of unitary operators ${\mathsf U}(\lambda)$, then
according to Stone's theorem (see, e.g., \cite{BEH}), there exists a s.a.\
operator ${\mathsf G}$ satisfying
$$
{\mathsf U}(\lambda) = \exp(i{\mathsf G}\lambda)\ .
$$
It is called {\em generator} of group ${\mathsf U}(\lambda)$ and can be
calculated with the help of the formula
$$
i{\mathsf G} = {\mathsf U}(\lambda)^\dagger\frac{d{\mathsf
U}(\lambda)}{d\lambda}\ .
$$
If the parameter is not specified by any further convention, it is defined
only up to a real multiplier and so is the generator.

A small technical problem for application of Stone's theorem to our case is that
the set ${\mathsf U}(g(\lambda))$ for any one-parameter subgroup $g(\lambda)$
of $\bar{\mathbf G}^+$ does not necessarily form a group because map $U(g)$ is only a ray
representation. This is usually treated by working with the central extension
$\bar{\mathbf G}^+_c$ of group $\bar{\mathbf G}^+$ that has elements
$(g,\phi)$, $g \in \bar{\mathbf G}^+$ and $\phi \in ({\mathbb R}\ \text{mod}\
2\pi)$, multiplication law
$$
(g_1,\phi_1) \circ (g_2,\phi_2) = (g_1 \circ g_2, \phi_1 + \phi_2 +
\omega(g_1,g_2))
$$
and action on ${\mathbf H}_\tau$
$$
\tilde{\mathsf U}(g,\phi)|\psi\rangle = e^{i\phi}g|\psi\rangle\ .
$$
Then, $(g,\phi) \mapsto \tilde{\mathsf U}(g,\phi)$ is a unitary representation
of $\bar{\mathbf G}^+_c$ on ${\mathbf H}_\tau$.

Group $\bar{\mathbf G}^+_c$ has eleven parameters
$\lambda_1,\ldots,\lambda_{10}, \phi$ and each parameter has its generator,
which can be defined by the following conventions. Each one-parameter subgroup
of space translations can be represented on ${\mathbf H}_\tau$ by a unitary
maps of the form
$$
\exp\left(\frac{i}{\hbar}\vec{n}\cdot\vec{\mathsf P}\ a\right)\ ,
$$
where vector $\vec{n}$ is the unit vector in the direction, $a$ the distance,
of translation in a chosen system of units. The distance $a$ plays also the role of the parameter of the
subgroup.

Each one-parameter subgroup of $SU(2)$ can be represented by
$$
\exp\left(\frac{i}{\hbar}\vec{n}\cdot\vec{\mathsf J}\ \theta\right)\ ,
$$
where $\vec{n}$ is a unit vector along the rotation axis and the parameter is
an angle $\theta$ of rotation in the counterclockwise direction around the
axis. $\theta$ is an angle in radians, $\theta \in [0,4\pi$).

Each one-parameter boost subgroup can be represented by
$$
\exp\left(\frac{i}{\hbar}(\vec{n}\cdot \vec{\mathsf K}\ v\right)\ ,
$$
where $\vec{n}$ is the direction and $v$ the velocity of the boost in the
chosen system of units; $v$ is the parameter of the group.

Each time translation can be represented by
$$
\exp\left(-\frac{i}{\hbar}{\mathsf H}t\right)\ ,
$$
where the parameter $t$ is time in the chosen system of units and the
parameter of the group.

Finally, each phase transformation can be denoted by
$$
\exp\left(\frac{i}{\hbar}{\mathsf M}\phi\right)\ ,
$$
where the parameter $\phi$ has the dimension $1/[\text{mass}]$.
\begin{df}\label{dfold5} The three s.a.\ operators ${\mathsf P}^k$ are
components of {\em total momentum}, ${\mathsf J}^k$ are components of {\em
total angular momentum} and ${\mathsf Q}^k = (1/M){\mathsf K}^k$ are
components of {\em position} (centre of mass) of ${\mathcal S}$. $M$ is the
{\em total mass} of ${\mathcal S}$ and
$$
{\mathsf M} = M{\mathsf 1}
$$
is the operator of total mass of ${\mathcal S}$. Operator ${\mathsf H}$ is
{\em Hamiltonian} of ${\mathcal S}$.
\end{df} The self-adjoint operators defined by Definition \ref{dfold5} have a
common invariant domain (see, e.g.\ \cite{BR}), hence their sums and products
are well-defined. They are the most important operators for any system in the
sense that most quantum mechanical observables of the system are constructed
from them. For example, an internal angular momentum or spin ${\mathsf s}$ can
be defined in terms of these generators by
\begin{equation}\label{sJ} \vec{\mathsf s} = \vec{\mathsf J} - \vec{\mathsf
Q}\times \vec{\mathsf P}\ .
\end{equation}

However, not all unitary ray representations $\tilde{\mathsf U}(g,\phi)$ of
group $\tilde{\mathbf G}^+_c$ are physical. For example, the choice of phase
shift $\omega(g_1,g_2) = 0$ does not contradict any assumption above but has
wrong physical consequences. There is an additional rule restricting
$\omega(g_1,g_2)$:
\begin{rl}\label{rlbascom} The non-zero commutators of the generators are
\begin{equation}\label{Eucom} [{\mathsf J}^k,{\mathsf J}^l] =
i\sum_j\epsilon^{kjl}{\mathsf J}^j\ ,\quad [{\mathsf J}^k,{\mathsf P}^l] =
i\sum_j\epsilon^{kjl}{\mathsf P}^j\ ,
\end{equation}
\begin{equation}\label{boostcom} [{\mathsf J}^k,{\mathsf K}^l] =
i\sum_j\epsilon^{kjl}{\mathsf K}^j\ ,\quad [{\mathsf K}^k,{\mathsf H}] =
-i{\mathsf P}^k\ , \quad [{\mathsf K}^k,{\mathsf P}^l] = i\delta^{kl}M{\mathsf
1}\ .
\end{equation}
\end{rl} Thus, we can see why the mass is a generator in a Galilean invariant
theory.

We can also observe that the majority of observables dealt with in standard
textbooks of quantum mechanics are obtained from generators of Galilean group
and must, therefore, satisfy commutation relations dictated by the Lie algebra
of the group. But we can also observe that the situation in Newtonian
mechanics is analogous. Galilean group acts on the phase space as a group of
symplectic transformations and its generators are closely related to classical
observables. This explains the fact that the commutation relations of quantum
observables resemble Poisson brackets of classical observables.

Each Hilbert space ${\mathbf H}_\tau$ carries a definite representation of
$\bar{\mathbf G}^+_c$ depending only on $\tau$ ($\tau$ can contain more
information, e.g., about the electric charge, etc.). Construction of models can
start with the choice of a suitable representation. For example, the
representation is irreducible for particles. Irreducible representations are
classified by three numbers, $\mu, s$ and $E_0$, with the meaning of mass,
spin and ground state energy, respectively (see Ref.\ \cite{levy}, p.\
221). Let us describe this representation.

Consider the Schwartz space (see, e.g., Ref.\ \cite{RS}) ${\mathbf
S}^{2s+1}({\mathbb R}^{3})$ of rapidly decreasing $2s+1$-tuples of $C^\infty$
functions $\phi(\vec{x},m)$, where $m \in \{-s, -s+1,\cdots, s\}$ represents a
set of discrete parameters depending on the system spin. ${\mathbf
S}^{2s+1}({\mathbb R}^{3})$ is a common invariant domains of all generators.

Let ${\mathbf H}_\tau$ be the completion of ${\mathbf S}^{2s+1}({\mathbb
R}^{3})$ with respect to the inner product
$$
\langle\phi|\psi\rangle = \sum_m \int_{{\mathbb R}^3} d^3x \phi^*(\vec{x},m)
\psi(\vec{x},m)\ .
$$
The elements of ${\mathbf H}_\tau$ are called {\em wave functions}. To
describe the operators, it is sufficient to define the action of their
components on $C^\infty$ functions $\phi(\vec{x},m)$ because there is always
only one s.a.\ extension. The result is
\begin{equation}\label{posit} {\mathsf Q}^k\phi(\vec{x},m) =
x^k\phi(\vec{x},m)
\end{equation} and
\begin{equation}\label{moment} {\mathsf P}^k\phi(\vec{x},m) =
-i\hbar\frac{\partial}{\partial x^k}\phi(\vec{x},m)\ .
\end{equation}

The three components of spin are $(2s+1)\times(2s+1)$ Hermitian matrices $
s^k_{mn}$ that act on wave functions as follows
\begin{equation}\label{spin} {\mathsf s}^k\phi(\vec{x},m) = \sum_{n=-s}^s
s^k_{m\,n}\phi(\vec{x},n)\ .
\end{equation} For example, the matrices for $s = 0$ are all equal 1 and those
for $s = 1/2$ are
\begin{equation}\label{spinm} {\mathsf s}^1 = \hbar/2 \begin{pmatrix}0 & 1 \\
1 & 0
\end{pmatrix}\ ,\quad {\mathsf s}^2 = \hbar/2 \begin{pmatrix}0 & -i \\ i & 0
\end{pmatrix}\ ,\quad {\mathsf s}^3 = \hbar/2 \begin{pmatrix}1 & 0 \\ 0 & -1
\end{pmatrix}\ .
\end{equation} Then, the operator of angular momentum can be constructed from
equation (\ref{sJ}).

Finally, the Hamiltonian is
\begin{equation}\label{partham} {\mathsf H} = \frac{\vec{\mathsf
P}\cdot\vec{\mathsf P}}{2\mu} + E_0\ .
\end{equation}

This representation of $\bar{\mathsf G}^+_c$ is an example of structure that
is labelled by index $\tau$. The representation of Hilbert space and of
operators described above is called $Q$-{\em representation}\footnote{The
meaning of the term "representation" here is different from that of
group representation. It is called $Q$-representation because operators
$\vec{\mathsf Q}$ are diagonal in it. The same unitary representation of
$\bar{\mathsf G}^+_c$ that we are describing in $Q$-representation can also be
described in, say, $P$-representation, etc.}. All operators (not necessarily
group generators) on ${\mathbf H}_\tau$ in $Q$-representation can be written
in the form of integral operators with kernels
$$
{\mathsf A} = A(\vec{x},m;\vec{x}',m')\ ,
$$
the kernel $A(\vec{x},m;\vec{x}',m')$ being a generalised function of its
arguments acting on functions $\phi(\vec{x},m) \in {\mathbf S}^{2s+1}({\mathbb
R}^{3})$ as follows
$$
({\mathsf A}\phi)(\vec{x},m) = \sum_{m'} \int_{{\mathbb R}^3} d^3x'
A(\vec{x},m;\vec{x}',m') \phi(\vec{x}',m')\ ,
$$
where the right-hand side represents the action of the generalised function on
a test function. For example,
$$
({\mathsf Q}^k\phi)(\vec{x},m) = \sum_{m'} \int_{{\mathbb R}^3} d^3x' x^k
\delta(\vec{x} - \vec{x}')\delta_{mm'} \phi(\vec{x}',m')
$$
and
$$
({\mathsf P}^k\phi)(\vec{x},m) = \sum_{m'} \int_{{\mathbb R}^3} d^3x' i\hbar
\frac{\partial}{\partial x^k} \delta(\vec{x} - \vec{x}')\delta_{mm'}
\phi(\vec{x}',m')\ .
$$

The transformation from abstract notation to a representation can be
understood as expansion in an orthonormal basis. If $\{|n\rangle\}$ is a
basis, then any vector $|\psi\rangle$ can be represented by the function
$\psi(n) = \langle n|\psi\rangle$, and any operator ${\mathsf A}$ can be
represented by its matrix elements $A(n,m) = \langle n|{\mathsf
A}|m\rangle$. Only, for the $Q$-representation, we use a generalised basis
\cite{gelfand}.

The irreducible representation described above is a basic building block of
quantum mechanics. The models for all other systems can be constructed from
it. Many examples of such construction are given in textbooks of quantum
mechanics and we shall see some examples later.

\subsection{Time translations} In non-relativistic quantum mechanics, time is,
unlike position, just a parameter and time translation defines, unlike space
shift, the dynamics of the system. This is the asymmetry between time and
space in non-relativistic quantum mechanics. In this subsection, we drop the
assumption that systems are closed. For general systems, we assume
\begin{rl}\label{rlold8} Let ${\mathcal S}$ be a system of type $\tau$ and let
external fields $f$ be given. Then, time translation from $t_1$ to $t_2$ is
represented by unitary operator ${\mathsf U}(f,t_2,t_1)$ on ${\mathbf H}_\tau$
satisfying
$$
{\mathsf U}(f,t_3,t_1) = {\mathsf U}(f,t_3,t_2){\mathsf U}(f,t_2,t_1)
$$
and we have
$$
{\mathsf T}_{g(t_2 - t_1)^{-1},f} = {\mathsf U}(f,t_2,t_1){\mathsf T}{\mathsf
U}(f,t_2,t_1)^\dagger\ ,
$$
where $g(t_2 - t_1)$ is the group element for $\lambda_1 = \cdots = \lambda_9
= 0$, $\lambda_{10} = t_2 ? t_1$ and ${\mathsf T} \in {\mathbf T}({\mathbf
H}_\tau)^+_1$.
\end{rl} A given time translation element of ${\mathbf G}$ does not define a
unique map on ${\mathbf T}({\mathbf H}_\tau)^+_1$ in this case. ${\mathsf
U}(f,t_2,t_1)$ depends not only on $t_2 - t_1$ but also on the position of the
system with respect to the external fields (even if the fields are
stationary). Time translations ${\mathsf U}(f,t_2,t_1)$ do not form a
group. We shall let out the argument $f$ in ${\mathsf U}(f,t_2,t_1)$ in
agreement with the current practice.
\begin{df}\label{dfold6} The operator ${\mathsf H}(t)$ defined by
$$
{\mathsf H}(t) = i\hbar{\mathsf U}(t,t_0)^\dagger\frac{d{\mathsf
U}(t,t_0)}{dt}\ ,
$$
is the {\em Hamiltonian} of $\mathcal S$.
\end{df} Operators ${\mathsf H}(t)$ for different $t$ do not commute in
general.

An important observable in non-relativistic quantum mechanics is energy. The
corresponding operator for system $\mathcal S$ will be constructed from its
Hamiltonian in Section 2.2.3. On the other hand, any individually prepared
quantum system has a definite Hamiltonian with the form
$$
{\mathsf H} = {\mathsf H}(\text{operators, external fields})\ ,
$$
where function ${\mathsf H}$ symbolises the construction of the operator from
the other s.a.\ operators of $\mathcal S$. The form of Hamiltonian is a model
assumption of quantum mechanics. The choice of a Hamiltonian is usually the
most important step in model construction. This Hamiltonian is usually
different from the observable called energy of the system.

For closed systems, the choice of Hamiltonian is included in the choice of the
ray representation of group $\bar{\mathsf G}^+_c$. In fact, for each kind
$\tau$, there is also one system of this kind that is closed and the
corresponding representation can be viewed as a part of $\tau$. Then, for a
given system of kind $\tau$ that happens not to be closed, some generators of
the representation are not physical. Mostly, this is just the Hamiltonian, but
there are also cases when the physical momentum is different from that defined
by $\tau$. In general, the Hamiltonian of ${\mathcal S}$ defined by $\tau$ can
be decomposed in the free Hamiltonian of the mass-centre motion and the
internal Hamiltonian. The internal Hamiltonian is invariant with respect of
the Galilean group and is an objective (structural) property of ${\mathcal S}$
and some aspects of it, such as the spectrum, are in principle (indirectly)
observable even if it itself, as an operator, is not an observable.

Dynamics of quantum system $\mathcal S$ has to do with the time aspect of its
preparation and registration. Any preparation procedure finishes at some time
instant $t_p$ and any registration procedure starts at some time instant
$t_r$. The times $t_r$ and $t_p$ can serve as defining the time aspects
because the whole preparation or registration processes can themselves take
some time. The dynamics enables to calculate how the probabilities depend on
the times $t_r$ and $t_p$. The dependence can be obtained in two ways. We can
make either state ${\mathsf T}(t)$ to a function of $t$ by shifting ${\mathsf
T}(0)$ forwards or observable ${\mathsf E}(t)$ by shifting ${\mathsf E}(0)$
backwards by ${\mathsf U}(t,0)$. The probability $p_{\mathsf T}^{\mathsf
E}(X,t)$ corresponding to the time $t$ is then
$$
p_{\mathsf T}^{\mathsf E}(X,t) = tr[{\mathsf T}(t){\mathsf E}] = tr[{\mathsf
T}{\mathsf E}(t)]\ .
$$
The first method is called Schr\"{o}dinger picture, the second Heisenberg
picture.

Rule \ref{rlold8} implies:
\begin{prop}\label{propold10} Let quantum system $\mathcal S$ have a
Hamiltonian $\mathsf H(t)$. Then, the dynamical evolution of $\mathcal S$ in
Schr\"{o}dinger picture obeys von Neumann--Liouville equation of motion
\begin{equation}\label{dynamS} i\hbar\frac{d{\mathsf T}(t)}{dt} = [{\mathsf
H}(t),{\mathsf T}(t)]
\end{equation} and in Heisenberg picture Heisenberg equation of motion
\begin{equation}\label{dynamH} i\hbar\frac{d{\mathsf E}(X,t)}{dt} = [{\mathsf
E}(X,t),{\mathsf H}(t)]\ .
\end{equation}
\end{prop}

\begin{prop}\label{propold11} Let $\mathsf T$ be an extremal state
$|\phi\rangle\langle\phi|$. Then
$$
{\mathsf T}(t) = |\phi(t)\rangle\langle\phi(t)|\ ,
$$
where $\phi(t)$ obeys Schr\"{o}dinger equation
\begin{equation}\label{schrod} i\hbar\frac{d\phi(t)}{dt} = \mathsf H(t)\phi(t)
\end{equation} and
$$
\phi(t) = {\mathsf U}(t,0) \phi(0)\ .
$$
\end{prop} In particular, extremal states remain extremal states in unitary
evolution.

The unitary time evolution of a statistical decomposition must satisfy the
following rule
\begin{rl}\label{rlold9} Let $\mathsf T$ be the state with statistical
decomposition (\ref{physdec}) Then, its time evolution is a state operator
with statistical decomposition
\begin{equation}\label{evolmix} {\mathsf T}(t) = w{\mathsf T}_1(t)\ (+)_p\
(1-w){\mathsf T}_2(t)\ ,
\end{equation} where ${\mathsf T}_k(t)$ is determined by (\ref{dynamS}) for
each $k = 1,2$ and $w$ is time independent.
\end{rl} Of course, Eq.\ (\ref{evolmix}) is easily obtained from Eq.\
(\ref{physdec}) by multiplying both sides by $U(t,0)$ from the left and by
$U(t,0)^\dagger$ from the right. The non-trivial assumption is that the
statistical decomposition is conserved by unitary dynamics.

With the help of Hamiltonian, we can define the notion of symmetry of a closed
system.
\begin{df}\label{dfold7} Each unitary transformation ${\mathsf U} : {\mathbf
H}_\tau \mapsto {\mathbf H}_\tau$ that leaves the Hamiltonian ${\mathsf H}$ of
${\mathcal S}$ invariant,
$$
{\mathsf U}{\mathsf H}{\mathsf U}^\dagger = {\mathsf H}
$$
is called a {\em symmetry} of system ${\mathcal S}$.
\end{df} All symmetries of a system ${\mathcal S}$ form a unitary group that
is an objective (structural) property of ${\mathcal S}$. The generators of its
one-parameter subgroups are s.a.\ operators that commute with the Hamiltonian
and, as sharp observables, yield probability distributions that are
independent of time.

\chapter{Composition of quantum systems} This chapter studies composition of
two kinds of quantum systems: heterogeneous and identical. For the identical
systems, it introduces a number of new ideas.

Suppose that ${\mathcal S}_1$ and ${\mathcal S}_2$ are two quantum
systems. They can be particles or composites. Then, one can consider this pair
as one quantum system ${\mathcal S} = {\mathcal S}_1 +{\mathcal S}_2$. This is
called {\em composition} of systems. Clearly, composition is an effective tool
of model building so that any system can be constructed from particles. Also,
various kinds of interaction between quantum systems can be studied.

For the sake of simplicity, we assume that ${\mathcal S}_1$ and ${\mathcal
S}_2$ are {\em disjoint}, i.e., they have no subsystem in common. Composition
of systems that are not disjoint can clearly be reduced to composition of
disjoint ones. More important is a stronger condition, which we call {\em
heterogeneity}: systems ${\mathcal S}_1$ and ${\mathcal S}_2$ are
heterogeneous if there is no pair of particles ${\mathcal S}'_1$ and
${\mathcal S}'_2$ of the same type such that ${\mathcal S}'_1 \in {\mathcal
S}_1$ and ${\mathcal S}'_2 \in {\mathcal S}_2$.  The rules of composition are
different for systems that are or are not heterogeneous.

\section{Composition of heterogeneous systems} This section gives a brief
account of the composition of heterogeneous systems and the important
non-local effect of entanglement.

\subsection{Tensor product of Hilbert spaces} First, we describe the
mathematical apparatus. The tensor product ${\mathbf H} = {\mathbf H}_1\otimes
{\mathbf H}_2$ of ${\mathbf H}_1$ and ${\mathbf H}_2$ is the Cauchy completion
of the linear span of the set of products
$$
\phi\otimes\psi, \phi \in {\mathbf H}_1, \psi\in {\mathbf H}_2
$$
with respect to the inner product of ${\mathbf H}$, which is determined by
$$
\langle\phi\otimes\psi|\phi'\otimes\psi'\rangle = \langle\phi|\phi'\rangle
\langle\psi|\psi'\rangle\ .
$$
Tensor product operation "$\otimes$" is postulated to be associative
and distributive in both arguments. Hence, if $\{\phi_k\}$ and $\{\psi_k\}$
are bases of ${\mathbf H}_1$ and ${\mathbf H}_2$, then
$\{\phi_k\otimes\psi_l\}$ is a basis of ${\mathbf H}$. If the bases are
orthonormal then any $\Psi \in {\mathbf H}$ can be expressed as
$$
\Psi = \sum_{kl} \langle\phi_k\otimes\psi_l|\Psi\rangle\phi_k\otimes\psi_l\ .
$$
Thus, any vector $\Psi \in {\mathbf H}_1 \otimes {\mathbf H}_2$ can be
represented by the function of two arguments,
$$
\Psi(k_1,k_2) = \langle \phi_{k_1}| \otimes \langle \psi_{k_2}|\Psi\rangle\ .
$$
For instance, two-particle state $\Psi$ in $Q$-representation is described by
wave function $\Psi(\vec{x}_1,\vec{x_2})$.

If ${\mathsf A} \in {\mathbf L}({\mathbf H}_1)$ and ${\mathsf B} \in {\mathbf
L}({\mathbf H}_2)$, then their tensor product ${\mathsf A}\otimes {\mathsf B}$
on ${\mathbf H}_1\otimes{\mathbf H}_2$ is determined via the relation
$$
({\mathsf A}\otimes {\mathsf B})(\phi\otimes\psi) = {\mathsf A}\phi\otimes
{\mathsf B}\psi
$$
for all $\phi \in {\mathbf H}_1$ and $\psi \in {\mathbf H}_2$. It follows that
$$
tr[{\mathsf A} \otimes {\mathsf B}] = tr[{\mathsf A}]tr[{\mathsf B}]\ .
$$
In a basis $\{\phi_{k_1}\otimes\psi_{k_2}\}$, operator ${\mathsf A}\otimes
{\mathsf B}$ is described by its kernel
$$
K(k_1,k_2;k'_1,k'_2) = \langle\phi_{k_1}|{\mathsf A}|\phi_{k'_1}\rangle
\langle\psi_{k_2}|{\mathsf B}|\psi_{k'_2}\rangle\ .
$$
For instance, a kernel in the $Q$-representation is
$K(\vec{x}^k_1,\vec{x}^l_2;\vec{x}^{\prime k}_1,\vec{x}^{\prime l}_2)$.

An important example is tensor product of group representations. Let $(g,\phi)
\mapsto \tilde{\mathsf U}_1(g,\phi)$ be a unitary representation of
$\bar{\mathbf G}^+_c$ on ${\mathbf H}_1$ and $(g,\phi) \mapsto \tilde{\mathsf
U}_2(g,\phi)$ on ${\mathbf H}_2$. Then $(g,\phi) \mapsto \tilde{\mathsf
U}(g,\phi) = \tilde{\mathsf U}_1(g,\phi) \otimes \tilde{\mathsf U}_2(g,\phi)$
is a unitary representation of $\bar{\mathbf G}^+_c$ on ${\mathbf H}_1 \otimes
{\mathbf H}_2$.

Another example is the tensor product of states, ${\mathsf T}_1\otimes
{\mathsf T}_2$, of ${\mathsf T}_1 \in {\mathbf T}({\mathbf H}_1)^+_1$ and
${\mathsf T}_2 \in {\mathbf T}({\mathbf H}_2)^+_1$. It is a positive
trace-class operator with trace 1 and ${\mathsf T}_1\otimes {\mathsf T}_2 \in
{\mathbf T}({\mathbf H}_1 \otimes {\mathbf H}_2)^+_1$. However, ${\mathbf
T}({\mathbf H}_1\otimes{\mathbf H}_2)^+_1$ contains also convex combinations
of tensor products of elements from ${\mathbf T}({\mathbf H}_1)^+_1$ and
${\mathbf T}({\mathbf H}_2)^+_1$, which cannot themselves generally be written
as such tensor products.

The {\em partial trace} over the Hilbert space ${\mathbf H}_2$, say, is the
positive linear mapping
$$
\Pi_2:{\mathbf T}({\mathbf H}_1\otimes{\mathbf H}_2)^+_1 \mapsto {\mathbf
T}({\mathbf H}_1)^+_1
$$
defined via the relation
$$
tr[\Pi_2({\mathsf W}){\mathsf A}] = tr[{\mathsf W}({\mathsf A}\otimes {\mathsf
1}_2)]
$$
for all ${\mathsf A}\in {\mathbf L}_r({\mathbf H}_1)$, ${\mathsf W}\in
{\mathbf T}({\mathbf H}_1\otimes{\mathbf H}_2)^+_1$ and ${\mathsf 1}_2$ is the
identity operator on ${\mathbf H}_2$. State operator $\Pi_2({\mathsf W})$ is
uniquely defined because of Theorem \ref{propold6}.

If $\{\phi_k\}\subset {\mathbf H}_1$ and $\{\psi_k\}\subset {\mathbf H}_2$ are
orthonormal bases, then $\Pi_2({\mathsf W})$ can be written as
$$
\Pi_2({\mathsf W}) = \sum_{ijk}\langle\phi_i\otimes \psi_k|{\mathsf
W}(\phi_j\otimes \psi_k)\rangle |\phi_i\rangle \langle \phi_j|\ .
$$
Here $|\phi_i\rangle \langle \phi_j|$ is the bounded linear operator on
${\mathbf H}_1$ given by
$$
|\phi_i\rangle \langle \phi_j|(\phi) = \langle \phi_j|\phi\rangle \phi_i
$$
for all $\phi\in {\mathbf H}_1$. The partial trace over ${\mathbf H}_1$ is
defined similarly.

If ${\mathsf W} = {\mathsf T}_1\otimes {\mathsf T}_2$, then ${\mathsf T}_1 =
\Pi_2({\mathsf W})$ and ${\mathsf T}_2 = \Pi_1({\mathsf W})$ but, in general,
\begin{equation}\label{entang} {\mathsf W} \neq \Pi_2({\mathsf W})\otimes
\Pi_1({\mathsf W})\ .
\end{equation} In particular, if ${\mathsf W} = {\mathsf P}[\Psi]$, then
$$
{\mathsf P}[\Psi] = \Pi_2({\mathsf P}[\Psi])\otimes \Pi_1({\mathsf P}[\Psi])
$$
if and only if
$$
\Psi = \phi\otimes \psi
$$
for some $\phi\in {\mathbf H}_1$ and $\psi\in {\mathbf H}_2$. In that case
also
$$
\Pi_2 ({\mathsf P}[\Psi]) = {\mathsf P}[\phi]\ ,\quad \Pi_1 ({\mathsf
P}[\Psi]) = {\mathsf P}[\psi]\ .
$$
Thus, tensor products of extremal elements of the sets ${\mathbf T}({\mathbf
H}_1)^+_1$ and ${\mathbf T}({\mathbf H}_2)^+_1$ do not exhaust the set of
extremal elements of ${\mathbf T}({\mathbf H}_1\otimes{\mathbf H}_2)^+_1$.

Tensor product is also an operation for POV measures. Let ${\mathsf E}_1 :
{\mathbf F}_1 \mapsto {\mathbf L}_r({\mathbf H}_1)$ with dimension $n_1$ and
${\mathsf E}_2 : {\mathbf F}_2 \mapsto {\mathbf L}_r({\mathbf H}_2)$ with
dimension $n_2$ be two POV measures on Hilbert spaces ${\mathbf H}_1$ and
${\mathbf H}_2$ with value sets ${\mathbf \Omega}_1$ and ${\mathbf
\Omega}_2$. Then POV measure $({\mathsf E}_1\otimes {\mathsf E}_2) : ({\mathbf
F}_1\times{\mathbf F}_2) \mapsto {\mathbf L}_r({\mathbf H}_1\otimes{\mathbf
H}_2)$ on the tensor product ${\mathbf H}_1\otimes{\mathbf H}_2$ has dimension
$n_1 + n_2$, values set ${\mathbf \Omega}_1\times{\mathbf \Omega}_2$ and is
defined by
$$
({\mathsf E}_1\otimes {\mathsf E}_2)(X_1\times X_2) = {\mathsf
E}_1(X_1)\otimes {\mathsf E}_2(X_2)
$$
for all $X_1\subset {\mathbb R}^{n_1}$ and $X_2\subset {\mathbb
R}^{n_2}$. Tensor product of POV measures is associative but not commutative.

Tensor products of more Hilbert spaces and the corresponding notions and
relations can be obtained using the above axioms and relations.

Let us now turn to physical interpretation. Then, the Hilbert spaces are
associated with heterogeneous quantum systems and the indices distinguishing
the Hilbert spaces carry the information on the system types.
\begin{rl}\label{rlold10} Let ${\mathcal S}_1$ and ${\mathcal S}_2$ be two
heterogeneous quantum systems and their Hilbert spaces be ${\mathbf H}_1$ and
${\mathbf H}_2$, respectively. Then, the system ${\mathcal S}$ composed of
${\mathcal S}_1$ and ${\mathcal S}_2$ has the Hilbert space ${\mathbf H} =
{\mathbf H}_1\otimes{\mathbf H}_2$, its states are elements of ${\mathbf
T}({\mathbf H}_1\otimes{\mathbf H}_2)^+_1$ and its effects are elements of
${\mathbf L}_r({\mathbf H}_1\otimes{\mathbf H}_2)^+_{\leq 1}$.
\end{rl} An important assumption concerns the states and observables of
subsystems:
\begin{rl}\label{rlsubs} Let system ${\mathcal S}$ composed of heterogeneous
systems ${\mathcal S}_1$ and ${\mathcal S}_2$ be prepared in state ${\mathsf
T} \in {\mathbf T}({\mathbf H}_1\otimes{\mathbf H}_2)^+_1$. Then ${\mathcal
S}_1$ is simultaneously prepared in state $\Pi_2({\mathsf T})$ and ${\mathcal
S}_1$ in state $\Pi_2({\mathsf T})$. The observables of ${\mathcal S}_1$ and
${\mathcal S}_2$ can be identified with observables ${\mathsf E}_1 \otimes
{\mathsf 1}_2$ and ${\mathsf 1}_1 \otimes {\mathsf E}_2$ respectively, of the
composite. ${\mathcal S}_1$ and ${\mathcal S}_2$ are called {\em subsystems}
of ${\mathcal S}$.
\end{rl} Hence, in the case that a system is composed of heterogeneous
systems, these systems retain their individuality in the sense that they have
well-defined states and observables of their own.

The theory of composition would not be complete if we did not know how group
$\bar{\mathbf G}^+_c$ acts on Hilbert space ${\mathbf H}_1 \otimes {\mathbf
H}_2$ of the composite system. In this way, interaction between systems can be
defined.
\begin{df}\label{dfnoint} Let heterogeneous systems ${\mathcal S}_1$ and $
{\mathcal S}_2$ have Hilbert spaces ${\mathbf H}_1$ and ${\mathbf H}_2$. Let
the representative of $(g,\phi) \in \bar{\mathbf G}^+_c$ on ${\mathbf H}_k$ be
denoted by $\tilde{\mathsf U}_k(g,\phi)$ and that on ${\mathbf H}_1 \otimes
{\mathbf H}_2$ by $\tilde{\mathsf U}(g,\phi)$. If
\begin{equation}\label{groupcom} \tilde{\mathsf U}_(g,\phi) = \tilde{\mathsf
U}_1(g,\phi) \otimes \tilde{\mathsf U}_2(g,\phi)\ ,
\end{equation} we say that ${\mathcal S}_1$ and $ {\mathcal S}_2$ do not
{\em interact}.
\end{df} The consequence for the generators is that they are additive:
\begin{prop}\label{propgencom} Given a one parameter subgroup of $\bar{\mathbf
G}^+_c$ with generator ${\mathsf G}$ of its representation on ${\mathbf H}_1
\otimes {\mathbf H}_2$, ${\mathsf G}_1$ of its representations on ${\mathbf
H}_1$ and ${\mathsf G}_2$ of its representation on ${\mathbf H}_2$. Then, in
the case of non-interacting subsystems,
\begin{equation}\label{gencom} {\mathsf G} = {\mathsf G}_1 \otimes {\mathsf
1}_2 + {\mathsf 1}_1 \otimes {\mathsf G}_2\ .
\end{equation}
\end{prop} As an example, consider two particles ${\mathcal S}_1$ and $
{\mathcal S}_2$ of different types with Hilbert spaces ${\mathbf H}_1$ and
${\mathbf H}_2$ and let us composite them so that they do not interact. We can
then construct the representation of the group $\bar{\mathbf G}^+_c$ on
${\mathbf H}_1 \otimes {\mathbf H}_2$ from its representations on ${\mathbf
H}_1$ and ${\mathbf H}_2$ as follows.

Let the representations on ${\mathbf H}_i$ have the parameters
$\mu_i,s_i,E_{0i}$ and the corresponding group generators be $\vec{\mathsf
p}_i$, $\vec{\mathsf j}_i$, $\vec{\mathsf k}_i = \mu_i \vec{\mathsf x}_i$ and
${\mathsf h}_i = E_{0i} + \vec{\mathsf p}_i\cdot \vec{\mathsf p}_i/2\mu_i$
(see Section 1.3.1). Let the generators of the group on ${\mathbf H}_1 \otimes
{\mathbf H}_2$ that are to be determined be denoted by $\vec{\mathsf P}$,
$\vec{\mathsf J}$, $\vec{\mathsf K} = M \vec{\mathsf Q}$ and ${\mathsf H}$. We
use the $Q$-representation so that elements of ${\mathbf H}_1 \otimes {\mathbf
H}_2$ are constructed from rapidly decreasing $C^\infty$ functions
$\Psi(\vec{\mathsf x}_1,\vec{\mathsf x}_2)$, which also form the common
invariant domain of all group generators. Then, Proposition \ref{propgencom}
implies
\begin{eqnarray}\label{gencomp1} \vec{\mathsf P} = \vec{\mathsf p}_1 +
\vec{\mathsf p}_2\ & , & \quad \vec{Q} = \frac{\mu_1\vec{x}_1 +
\mu_2\vec{x}_2}{\mu_1 + \mu_2}\ , \\ \label{gencomp2} \quad M = \mu_1 + \mu_2\
& , & {\mathsf H} = E_{01} + E_{02} + \frac{\vec{\mathsf p}_1\cdot\vec{\mathsf
p}_1}{2\mu_1} + \frac{\vec{\mathsf p}_2\cdot\vec{\mathsf p}_2}{2\mu_2}\ ,
\end{eqnarray} where $\vec{\mathsf P}$, $\vec{\mathsf p}_1$ and $\vec{\mathsf
p}_2$ are the differential operators of the form (\ref{moment}).

It is advantageous to change variables $\vec{x}_1,\vec{x}_2$ to $Q,q$ so that
wave functions have the form $\Psi(Q,q)$ and
$$
[{\mathsf Q}^k,{\mathsf q}^l] = [{\mathsf p}^k,{\mathsf P}^l] = 0\ ,\quad
[{\mathsf Q}^k,{\mathsf P}^l] = [{\mathsf q}^k,{\mathsf p}^l] =
i\hbar\delta^{kl}\ .
$$
The transformation is uniquely determined by conditions:
\begin{eqnarray*} \vec{x}_1 = \vec{Q} - \frac{\mu_2}{\mu_1 + \mu_2}\vec{q}\
&,& \vec{x}_2 = \vec{Q} + \frac{\mu_1}{\mu_1 + \mu_2}\vec{q}\ , \\
\vec{\mathsf p}_1 = \frac{\mu_1}{mu_1 + \mu_2}\vec{\mathsf P} - \vec{\mathsf
p}\ &,& \vec{\mathsf p}_2 = \frac{\mu_2}{mu_1 + \mu_2}\vec{\mathsf P} +
\vec{\mathsf p}\ .
\end{eqnarray*} The transformed Hamiltonian and the angular momentum are
$$
{\mathsf H} = E_{01} + E_{02} + \frac{\vec{\mathsf P}\cdot\vec{\mathsf P}}{2M}
+ \frac{\vec{\mathsf p}\cdot\vec{\mathsf p}}{2\mu}\ ,\quad \vec{\mathsf J} =
\vec{Q} \times \vec{\mathsf P} + \vec{q} \times \vec{\mathsf p} + \vec{\mathsf
s}_1 + \vec{\mathsf s}_2\ ,
$$
where $\vec{\mathsf s}_1$ and $\vec{\mathsf s}_2$ are spin operators of
${\mathcal S}_1$ and ${\mathcal S}_2$ given by equation (\ref{spinm}) and
$$
\mu = \frac{\mu_1\mu_2}{\mu_1 + \mu_2}
$$
is the so-called reduced mass.

To construct models of {\em interacting} systems, we modify the above
generators in any way subject only to the condition that the commutation
relations (\ref{Eucom}) and (\ref{boostcom}) are preserved. For example, an
arbitrary term can be added to the Hamiltonian if that term commutes with all other
generators, i.e., it is invariant under Galilean group. For example,
$\vec{q}\cdot\vec{q}$ and any functions of it is an invariant. Such a
function, $V(\vec{q}\cdot\vec{q})$, is called {\em potential}.

The final assumption of this section concerns decomposable states.
\begin{rl}\label{rlold11} Suppose that the state of the system composed of
heterogeneous systems ${\mathcal S}_1 + {\mathcal S}_2$ is ${\mathsf T}$. The
necessary and sufficient condition for the statistical decomposition of the
state of ${\mathcal S}_1$ to be
$$
\Pi_2({\mathsf T}) = \left(\sum_k\right)_p w_k {\mathsf T}_{1k}
$$
is that ${\mathsf T}$ itself has statistical decomposition
$$
{\mathsf T} = \left(\sum_k\right)_p w_k {\mathsf T}_{1k} \otimes {\mathsf
T}_{2k}\ ,
$$
where ${\mathsf T}_{1k}$ are some states of ${\mathcal S}_1$ and ${\mathsf
T}_{2k}$ are some states of ${\mathcal S}_2$.
\end{rl} One can say that statistical decomposition is invariant with respect
to compositions. By registration of observables of ${\mathcal S}_1$ alone,
only the state operator $\Pi_2({\mathsf T})$ can be determined, not its
statistical decomposition. However, by registration of observables pertaining
to a composite object ${\mathcal S}_1 + {\mathcal S}_2$, some information
about the statistical decomposition of $\Pi_2({\mathsf T})$ can be obtained
from Rule \ref{rlold11}. Suppose, e.g., that ${\mathcal S}_1 + {\mathcal S}_2$
is in an extremal state ${\mathsf T} = {\mathsf P}[\Psi]$. Then state operator
$\Pi_2({\mathsf P}[\Psi])$ cannot have a non-trivial statistical
decomposition. This fact is at the root of the objectification problem in
quantum theory of measurement (cf.\ \cite{BLM}, our Sections 4.1 and 4.2).

Based on the mathematical properties of the tensor products, the description
of systems composed of arbitrary number of sub-systems can be obtained by
extension of the above methods. The only condition is that each two subsystems
are heterogeneous.

\subsection{Entanglement} Entanglement is a kind of mutual influence between
quantum systems that is very different from any effect known from classical
theories. It is not an interaction according to Definition \ref{dfnoint}.
\begin{df}\label{dfentang} If object ${\mathcal S}$ composed of two
heterogeneous quantum objects ${\mathcal S}_1$ and ${\mathcal S}_2$ is in an
indecomposable state ${\mathsf W}$ that satisfies condition (\ref{entang}), one
says that ${\mathcal S}_1$ and ${\mathcal S}_2$ are {\em entangled} or that state
${\mathsf W}$ is {\em entangled}.
\end{df} Consider a decomposable state
$$
{\mathsf W} = p{\mathsf T}_1 \otimes {\mathsf T}_2 \ (+)_p\ p'{\mathsf T}'_1
\otimes {\mathsf T}'_2\ ,
$$
where ${\mathsf T}_i$ and ${\mathsf T}'_i$, $i = 1,2$, are states of
${\mathcal S}_i$. Clearly, neither ${\mathsf T}_1 \otimes {\mathsf T}_2$ nor
${\mathsf T}'_1 \otimes {\mathsf T}'_2$ is entangled, and ${\mathsf W}$
cannot, therefore, be considered as entangled, either. But ${\mathsf W}$ does
satisfy condition (\ref{entang}). This is the reason for the condition that
${\mathsf W}$ be indecomposable in Definition \ref{dfentang}\footnote{There
are different definitions of entanglement, e.g., \cite{klyachko}. Our
definition seems to be intuitively clearer and mathematically much simpler. The
problem is that pure mathematics cannot distinguish decomposable state
operators from indecomposable ones.}.

Entanglement is a physical phenomenon that has measurable consequences. One of
these consequences is correlations between outcomes of registrations that are
performed simultaneously on the two (or more) entangled systems.

As an example, consider two heterogeneous objects ${\mathcal S}_1$ and
${\mathcal S}_2$ and two sharp observables ${\mathsf A}_i : {\mathbf H}_i
\mapsto {\mathbf H}_i$, $i=1,2$. Let $|a_i\rangle \in {\mathbf H}_i$ and
$|b_i\rangle \in {\mathbf H}_i$ be four eigenstates,
$$
{\mathsf A}_i|a_i\rangle = a_i|a_i\rangle\ ,\quad {\mathsf A}_i|b_i\rangle =
b_i|b_i\rangle\ ,
$$
and let $b_i > a_i$, $i=1,2$. The state ${\mathsf P}[\Psi]$ of the composite
object ${\mathcal S}_1 + {\mathcal S}_2$, where
$$
|\Psi\rangle = \frac{1}{\sqrt{2}}(|a_1\rangle\otimes |b_2\rangle
+|b_1\rangle\otimes |a_2\rangle)
$$
is entangled. Indeed,
\begin{equation}\label{parti} \Pi_2({\mathsf P}[\Psi]) =
\frac{1}{2}(|a_1\rangle\otimes\langle a_1| + |b_1\rangle\otimes\langle b_1|)\
,\quad \Pi_1({\mathsf P}[\Psi]) = \frac{1}{2}(|a_2\rangle\otimes\langle a_2| +
|b_2\rangle\otimes\langle b_2|)
\end{equation} and ${\mathsf P}[\Psi]$ is not a tensor product of these two
states.

Let us calculate the correlations of two sharp observables, ${\mathsf
A}_1\otimes{\mathsf 1}$ and ${\mathsf 1}\otimes{\mathsf A}_2$ on ${\mathbf
H}_1\otimes{\mathbf H}_2$. The observables commute hence Definition
\ref{dfold10} is applicable. We need the first two moments of observables
${\mathsf A}_1\otimes{\mathsf 1}$ and ${\mathsf 1}\otimes{\mathsf A}_2$ in
state ${\mathsf P}[\Psi]$:
$$
\langle\Psi|{\mathsf A}_1\otimes{\mathsf 1}|\Psi\rangle = \frac{1}{2}(a_1 +
b_1)\ ,\quad\langle\Psi|{\mathsf 1}\otimes{\mathsf A}_2|\Psi\rangle =
\frac{1}{2}(a_2 + b_2)\
$$
and
$$
\langle\Psi|({\mathsf A}_1\otimes{\mathsf 1})^2|\Psi\rangle =
\frac{1}{2}(a_1^2 + b_1^2)\ ,\quad\langle\Psi|({\mathsf 1}\otimes{\mathsf
A}_2)^2|\Psi\rangle = \frac{1}{2}(a_2^2 + b_2^2)\ .
$$
The variances (Definition \ref{dfold9}) are
$$
\Delta({\mathsf A}_1\otimes{\mathsf 1}) = \left(\frac{b_1-a_1}{2}\right)^2\
,\quad \Delta({\mathsf 1}\otimes{\mathsf A}_2) =
\left(\frac{b_2-a_2}{2}\right)^2\ .
$$
The average of the product $({\mathsf A}_1\otimes{\mathsf 1})({\mathsf
1}\otimes{\mathsf A}_2) = {\mathsf A}_1\otimes{\mathsf A}_2$ is
$$
\langle\Phi|({\mathsf A}_1\otimes{\mathsf A}_2)\Phi\rangle =
\frac{a_1b_2+b_1a_2}{2}
$$
so that, finally
$$
C({\mathsf A}_1\otimes{\mathsf 1},{\mathsf 1}\otimes{\mathsf A}_2,{\mathsf
P}[\Psi])= -1\ .
$$
The result is that the observables are strongly anticorrelated in state
${\mathsf P}[\Psi]$. What this means can be seen from the probability
distributions for different possible outcomes by measuring the two
observables. The corresponding PV measures define the four projections
$$
{\mathsf P}_{aa} = {\mathsf E}_1(\{a_1\})\otimes{\mathsf E}_2(\{a_2\})\ ,\quad
{\mathsf P}_{ab} = {\mathsf E}_1(\{a_1\})\otimes{\mathsf E}_2(\{b_2\})\ ,
$$
$$
{\mathsf P}_{ba} = {\mathsf E}_1(\{b_1\})\otimes{\mathsf E}_2(\{a_2\})\ ,\quad
{\mathsf P}_{bb} = {\mathsf E}_1(\{b_1\})\otimes{\mathsf E}_2(\{b_2\})\ ,
$$
and we obtain
$$
\langle\Psi|{\mathsf P}_{aa}|\Psi\rangle = 0\ ,\quad \langle\Psi|{\mathsf
P}_{ab}|\Psi\rangle = \frac{1}{2}\ ,
$$
$$
\langle\Psi|{\mathsf P}_{ba}|\Psi\rangle = \frac{1}{2}\ ,\quad
\langle\Psi|{\mathsf P}_{bb}|\Psi\rangle = 0\ .
$$
It follows: if the registration of ${\mathsf A}_1\otimes {\mathsf 1}$ gives
$a_1$ ($b_1$) then the registration of ${\mathsf 1}\otimes {\mathsf A}_1$
gives $b_2$ ($a_2$) with certainty, and vice versa, the correlation being
symmetric with respect to ${\mathcal S}_1$ and ${\mathcal S}_2$.

Existence of such correlations has surprising consequences. For example, it
allows to register observable ${\mathsf A}_1$ by an apparatus that interacts
only with ${\mathsf A}_2$: If the apparatus register observable ${\mathsf
A}_2$ and gives value $b_2$ for it then it simultaneously gives value $a_1$
for ${\mathsf A}_1$. This is an example of indirect registration.

State $\Pi_2({\mathsf P}[\Psi])$ in equation (\ref{parti}) is an example of a
non-extremal state that is indecomposable. This follows from Rule
\ref{rlold11} and the fact that $\Psi$ is extremal.

Another aspect of entanglement is the amount of information that the entangled
state and the states of the subsystems carry. This can be studied with the
help of von Neumann entropy (\ref{vNentropy}). Von Neumann entropy of
composite objects satisfies a number of interesting inequalities, see, e.g.,
Ref.\ \cite{wehrl}. One is the so-called sub-additivity \cite{AL}, which is
described by the following:
\begin{prop}\label{propold14} Let $\mathcal S$ be composed of heterogeneous
systems ${\mathcal S}_1$ and ${\mathcal S}_2$ of different types. Let
${\mathsf T}$ be the state of $\mathcal S$ and let
$$
{\mathsf T}_1 = \Pi_2({\mathsf T})\ ,\quad {\mathsf T}_2 = \Pi_1({\mathsf T})\
.
$$
Then
$$
S({\mathsf T}) \leq S({\mathsf T}_1) + S({\mathsf T}_2)
$$
and the equality sign is valid only if
$$
{\mathsf T} = {\mathsf T}_1 \otimes {\mathsf T}_2\ .
$$
\end{prop}

In the example above,
$$
S({\mathsf P}[\Psi]) = 0\ ,\quad S(\Pi_2({\mathsf P}[\Psi])) =
S(\Pi_2({\mathsf P}[\Psi])) = \ln 2\ ,
$$
confirming the inequality and showing that an entangled state of two systems
can contain more information than the sum of the information in both
subsystems. This possibility is utilised by quantum computers.

Suppose next that ${\mathcal S}_1$ and ${\mathcal S}_2$ are far from each
other, ${\mathcal S}_1$ near point $\vec{x}_1$ and ${\mathcal S}_2$ near point
$\vec{x}_2$. State $\Psi$ of the composite is independent of the distance
between $\vec{x}_1$ and $\vec{x}_2$. Could one use the strong anticorrelation
to send signals from $\vec{x}_1$ to $\vec{x}_2$, say? There are two reasons
why one cannot. First, one has no choice of the value, $a_1$ or $b_1$, that is
obtained at $\vec{x}_1$. As is easily seen from the state $\Pi_2({\mathsf
P}[\Psi])$ of ${\mathcal S}_1$, Eq.\ (\ref{parti}), the probability of each
outcome is 1/2. One has, therefore, no control about what signal will be
sent. Second, suppose that the state ${\mathsf P}[\Psi]$ is prepared many
times and let the observer at
$\vec{x}_1$ register ${\mathsf A}_1$ every time in the first ensemble of experiments and do noting in the second ensemble. Is there any difference between the two cases that could be
recognized at $\vec{x}_2$? In the first case, the state of ${\mathcal S}_2$
given by Eq.\ (\ref{parti}) is the statistical decomposition of the state,
whereas, in the second case, the state operator is the same but the state is
indecomposable because the composite system is in an extremal state. However,
by direct registrations of any observable pertinent to ${\mathcal S}_2$, the
observer at $\vec{x}_2$ cannot distinguish different statistical decomposition
of the same state operator from each other.

The next interesting question is, how the influence of a registered value at
$\vec{x}_1$ on that at $\vec{x}_2$ is to be understood. Even in classical
mechanics, one can arrange strong correlations. For example, if a body with
zero angular momentum with respect to its centre of mass decays into two
bodies flying away from each other, the angular momentum of the first is
exactly the opposite to that of the second. This strong anticorrelation cannot
be used to send signals either. It is moreover clear that a measurement of
the first angular momentum giving the value $\vec{L}_1$ is not a cause of the
second angular momentum having the value $-\vec{L}_1$. Rather, the decay is
the common cause of the two values being opposite\footnote{The condition of
common cause can be formulated rigorously \cite{reichenbach}, see also the
discussion in Ref.\ \cite{fraass}, pp.\ 83--94.}. The process of creating the
two opposite values at distant points is also completely local: the decay is a
local phenomenon and the movement of each of the debris is governed by a local
equation of motion. Moreover, the values $\vec{L}_1$ and $-\vec{L}_1$ are
objective, that is, they exist on the debris independently of any measurement
and can be, in this way, transported from the decay point to the measurements
in a local way.

In quantum mechanics, such an explanation of the correlations is not
possible. A value of an observable is created only during its registration. It
does not exist in any form before the registration, except in the special case
of state which is an eigenstate. However, in our example, $\Psi$ is not an
eigenstate either of ${\mathsf A}_1\otimes {\mathsf 1}$ or of ${\mathsf
1}\otimes {\mathsf A}_2$. The registrations performed simultaneously at
$\vec{x}_1$ and at $\vec{x}_2$, which can be very far from each other, are
connected by a relation that is utterly non-local. How can the apparatuses
together with parts of the quantum system at two distant points $\vec{x}_1$
and at $\vec{x}_2$ "know" what values they are to create so
that the correlations result? Note that the rejection of objectivity of
observables leads more directly to non-locality than assumption of any sort of
realism.

The nature of simultaneity with which the correlations take place can be
described in more details. In Ref.\ \cite{KS}, the entangled state is created
by a decay of one particle that defines a preferred frame---its rest frame,
which, in turn, defines the simultaneity. Lead by this example, we adopt the
following general assumption about entanglement. First, the entangled state on
which the correlations between remote registrations can be observed must be
prepared and the apparatus that prepares it (or some of its parts) defines a
unique preferred frame similarly as in the case considered in Ref.\
\cite{KS}. Second, the apparatus location and the time interval of the
preparation process defines a spacetime region $D$ that must lie inside the
common past of the events at which the simultaneous correlations are
registered. Region $D$ can be arbitrarily large (it does not have to be just a small
neighbourhood of the particle-decay event as in Ref.\ \cite{KS}). This
assumption is weaker than Reichenbach's condition of common cause because it
holds, e.g., for EPR effect. The point is that it does not require the causal
independence of the registrations that are correlated (allowing for
non-locality).

The non-local correlation between the registration outcomes can be ascertained
only if both values are known and can be compared. It may therefore be more
precise if we say that the non-local correlations can only be seen if the
non-local observable ${\mathsf A}_1\otimes {\mathsf A}_2$ is registered. This
non-locality of quantum correlations in entangled states does not lead to any
internal contradictions in quantum mechanics and is compatible with other
successful theories (such as special relativity) as well as with existing
experimental data. Nonetheless it is very surprising and it has been very
thoroughly studied. In this way, various conditions (e.g., Bell inequality)
have been found that had to be satisfied by values of observables if the values were
objective and locality were satisfied. Experiments show that such conditions
are violated, and, moreover, that their violation can even be exploited in
quantum communication techniques \cite{brukner}.

Finally, let us repeat here that non-local correlations and non-objectivity of
observables is accepted by our interpretation of quantum mechanics in the full
extent and that this does not lead to any contradiction with Basic Ontological
Hypothesis of Quantum Mechanics and with the Realist Model Approach (see
Section 0.1.3).

\section{Composition of identical systems} This section gives account of
composition of non-heterogeneous systems, explains why the standard theory is
inadequate and introduces all necessary corrections.

An efficient mathematical tool for dealing with such systems are Fock-space
methods (see, e.g., \cite{peres}, p.\ 137 and \cite{KP}). However, the
Fock-space methods are not advantageous for the presentation of our ideas on
general identical systems and we shall not use them in this paper. This does
not mean that they cannot be utilised and give an efficient help to solve
various mathematical problems that occur in specific cases.

\subsection{Identical subsystems} If a prepared object has more than one
subsystem of the same type (identical subsystems), then these subsystems are
indistinguishable. This idea can be mathematically expressed as invariance
with respect to permutations.

Let ${\mathbf S}_N$ be the permutation group of $N$ objects, that is, each
element $g$ of ${\mathbf S}_N$ is a bijective map $g : \{1,\cdots,N\} \mapsto
\{1,\cdots,N\}$, the inverse element to $g$ is the inverse map $g^{-1}$ and
the group product of $g_1$ and $g_2$ is defined by $(g_1 g_2)(k) =
g_1(g_2(k))$, $k \in \{1,\cdots,N\}$.

Given a Hilbert space $\mathbf H$, let us denote by ${\mathbf H}^N$ the tensor
product of $N$ copies of $\mathbf H$,
$$
{\mathbf H}^N = {\mathbf H} \otimes {\mathbf H} \otimes \cdots \otimes
{\mathbf H}\ .
$$
On ${\mathbf H}^N$, the permutation group ${\mathbf S}_N$ acts as follows. Let
$\psi_k \in {\mathbf H}$, $k=1,\cdots,N$, then
$$
\psi_1 \otimes \cdots \otimes \psi_N \in {\mathbf H}^N
$$
and
\begin{equation}\label{perm} {\mathsf g} (\psi_1 \otimes \cdots \otimes
\psi_N) = \psi_{g(1)} \otimes \cdots \otimes \psi_{g(N)}\ .
\end{equation} $\mathsf g$ preserves the inner product of ${\mathbf H}^N$ and
is, therefore, bounded and continuous. Hence, it can be extended by linearity
and continuity to the whole of ${\mathbf H}^N$. The resulting operator on
${\mathbf H}^N$ is denoted by the same symbol $\mathsf g$ and is a unitary
operator by construction. The action (\ref{perm}) thus defines a unitary
representation of the group ${\mathbf S}_N$ on ${\mathbf H}^N$.

All vectors of ${\mathbf H}^N$ that transform according to a fixed unitary
representation ${\mathcal R}$ of ${\mathbf S}_N$ form a closed linear subspace
of ${\mathbf H}^N$ that will be denoted by ${\mathbf H}^N_{\mathcal R}$. The
representations being unitary, the subspaces ${\mathbf H}^N_{\mathcal R}$ are
orthogonal to each other. Let us denote by ${\mathsf P}^{(N)}_{\mathcal R}$
the orthogonal projection operator,
$$
{\mathsf P}^{(N)}_{\mathcal R} : {\mathbf H}^N \mapsto {\mathbf H}^N_{\mathcal
R}\ .
$$
An important property of the subspaces is their invariance with respect to
tensor products of unitary operators. Let ${\mathsf U}$ be a unitary
transformation on ${\mathbf H}$, then ${\mathsf U} \otimes {\mathsf U} \otimes
\ldots \otimes {\mathsf U}$ is a unitary transformation on ${\mathbf H}
\otimes {\mathbf H} \otimes \ldots \otimes {\mathbf H}$ and each subspace
${\mathbf H}^N_{\mathcal R}$ is invariant with respect to it. Hence, ${\mathsf
U} \otimes {\mathsf U} \otimes \ldots \otimes {\mathsf U}$ acts as a unitary
transformation on ${\mathbf H}^N_{\mathcal R}$ for each ${\mathcal R}$.

The location order of a given state in a tensor product can be considered as
information about the identity of the corresponding system. Such
information has no physical meaning and a change of the ordering is just a
kind of gauge transformation\footnote{A different and independent part
(ignored here) of the theory of identical particles is that states of two
identical systems can also be swapped in a physical process of continuous
evolution, and can so entail a non-trivial phase factor at the total state
(anyons, see, e.g.\ Ref.\ \cite{wilczek}).}. Motivated by this idea, we look
for one-dimensional unitary representations of ${\mathbf S}_N$ because only
these transform vectors by a phase factor multiplication. ${\mathbf S}_N$ has
exactly two one-dimensional unitary representations: the symmetric (trivial)
one, $g \mapsto {\mathsf 1}$, and the alternating one $g \mapsto
\eta(g){\mathsf 1}$ for each $g \in {\mathbf S}_N$, where $\eta(g) = 1$ for
even and $\eta(g) = -1$ for odd permutations $g$. If ${\mathcal R}$ is the
symmetric (alternating) representation we use symbol ${\mathbf H}^N_s$
(${\mathbf H}^N_a$) for ${\mathbf H}^N_{\mathcal R}$. Let ${\mathsf
P}_s^{(N)}$ (${\mathsf P}_a^{(N)}$) be the orthogonal projection on ${\mathbf
H}^N_s$ (${\mathbf H}^N_a$). Note that the usual operation of symmetrisation
or antisymmetrisation on a vector $\Psi \in{\mathbf H}^N$, such as
$$
\psi \otimes \phi \mapsto (1/2)(\psi \otimes \phi \pm \phi \otimes \psi)
$$
in ${\mathbf H}^2$, is nothing but ${\mathsf P}^{(N)}_s\Psi$ or ${\mathsf
P}^{(N)}_a\Psi$, respectively.

Now, we are ready to formulate the basic assumption concerning identical
subsystems. From relativistic quantum field theory \cite{Weinberg}, we take
over the following result.
\begin{rl}\label{rlold12} Let ${\mathcal S}^N$ be a quantum system composed
of $N$ subsystems ${\mathcal S}$, each of type $\tau$ with Hilbert space
${\mathbf H}_\tau$. Then, the Hilbert space of ${\mathcal S}^N$ is ${\mathbf
H}_{\tau s}^N$ for subsystems with integer spin and ${\mathbf H}^N_{\tau a}$
for those with half-integer spin. If systems ${\mathcal S}$ are closed and do
not interact, then the representation of group $\bar{\mathbf G}^+_c$ on
${\mathbf H}_{\tau s}^N$ or ${\mathbf H}_{\tau a}^N$ is the tensor product of
its $N$ representations on ${\mathbf H}_\tau$.
\end{rl}
\begin{df}\label{dfold11} Systems with integer spin are called {\em bosons}
and those with half-integer spin are called {\em fermions}. The symmetry
properties of states lead to Bose--Einstein statistics for bosons and
Fermi--Dirac one for fermions.
\end{df} We can, therefore, introduce a useful notation, a common symbol
${\mathbf H}^N_{{\mathcal R}(\tau)}$ for the subspaces of the symmetric and
anti-symmetric representations and ${\mathsf P}^{(N)}_\tau$ for the
corresponding projections because the representation is determined by the
system type $\tau$.

For the Galilean group, we have:
\begin{prop}\label{propsymG} Let ${\mathsf G}$ be the generator of subgroup
$g(t)$ of $\bar{\mathbf G}^+_c$ on ${\mathbf H}_\tau$. Then, the generator
$\tilde{\mathsf G}$ of $g(t)$ on ${\mathbf H}^N_{{\mathcal R}(\tau)}$ is given
by
\begin{equation}\label{symG} \tilde{\mathsf G} = {\mathsf G} \otimes {\mathsf
1}_2 \otimes \ldots \otimes {\mathsf 1}_N + {\mathsf 1}_1 \otimes {\mathsf G}
\otimes {\mathsf 1}_3 \otimes \ldots \otimes {\mathsf 1}_N + \ldots + {\mathsf
1}_1 \otimes \ldots \otimes {\mathsf 1}_{N-1} \otimes {\mathsf G}\ .
\end{equation}
\end{prop} Observe that the form of the generator is independent on whether
the space is symmetric or anti-symmetric. Proposition \ref{propsymG} also
suggests a way of constructing an operator on ${\mathbf H}^N_{{\mathcal
R}(\tau)}$ from that on ${\mathbf H}_\tau$ that will be important later. Let
us study the properties of the construction with the help of simple example.

Suppose that ${\mathcal S}$ is a bosonic particle with Hilbert space ${\mathbf
H}_\tau$ and let ${\mathcal S}^2$ be composed of two such bosons with Hilbert
space ${\mathbf H}^2_{\tau s}$. Let ${\mathsf A}$ be a bounded s.a.\ operator
on ${\mathbf H}_\tau$ with discrete non-degenerate value set $\{a_k\}$ and let
the projection on the eigenspace of $a_k$ be ${\mathsf P}_k = |\psi_k\rangle
\langle \psi_k|$ for all $k$, where $ \psi_k$ is an eigenvector of ${\mathsf
A}$, ${\mathsf A |\psi_k\rangle} = a_k |\psi_k\rangle$. Consider the operator
${\mathsf A}\otimes{\mathsf 1} + {\mathsf 1}\otimes{\mathsf A}$ on ${\mathbf
H}^2_{\tau s}$. We obtain
\begin{equation}\label{twobos} {\mathsf A}\otimes {\mathsf 1} + {\mathsf
1}\otimes{\mathsf A} = \sum_{k=1}^\infty a_k({\mathsf P}_k \otimes {\mathsf 1}
+ {\mathsf 1}\otimes{\mathsf P}_k)\ .
\end{equation} However, ${\mathsf P}_k \otimes {\mathsf 1} + {\mathsf
1}\otimes{\mathsf P}_k$ is not a projection. Indeed,
$$
({\mathsf P}_k \otimes {\mathsf 1} + {\mathsf 1}\otimes{\mathsf P}_k)^2 =
{\mathsf P}_k \otimes {\mathsf 1} + {\mathsf 1}\otimes{\mathsf P}_k + 2
{\mathsf P}_k \otimes {\mathsf P}_k\ .
$$
The eigenvectors of ${\mathsf A}\otimes {\mathsf 1} + {\mathsf 1}\otimes
{\mathsf A}$ are
\begin{equation}\label{eigenvtb} \frac{1}{\sqrt{2}}(|\psi_k\rangle \otimes
|\psi_l\rangle + |\psi_l\rangle \otimes |\psi_k\rangle)
\end{equation} with eigenvalue $a_k + a_l$, for all $k<l$, and
$$
|\psi_k\rangle \otimes |\psi_k\rangle
$$
with eigenvalue $2a_k$ for all $k$. There can be a degeneracy if there are
values $k<l$ and $k'<l'$ such that $a_k + a_l = a_{k'} + a_{l'}$. The
projection on eigenvector (\ref{eigenvtb}) is
$$
\frac{1}{2}{\mathsf P}_k \otimes {\mathsf P}_l + \frac{1}{2} {\mathsf P}_l
\otimes {\mathsf P}_k +\frac{1}{2}(|\psi_k\rangle \langle \psi_l|\otimes
|\psi_l\rangle \langle \psi_k| + |\psi_l\rangle \langle \psi_k|\otimes
|\psi_k\rangle \langle \psi_l|)\ ,
$$
which cannot be easily expressed in terms of ${\mathsf P}_k$ and ${\mathsf
P}_l$. We can see from this example that it is no straightforward business to
construct a POV measure for a system composed of identical particles from
that of its subsystems. But we can still use expression (\ref{twobos}) and its
analogues in calculations. We shall generalise this construction later.

For states and observables, we have:
\begin{prop}\label{propold12} Possible states of system ${\mathcal S}^N$
composed of $N$ systems of type $\tau$ are elements of ${\mathbf T}({\mathbf
H}^N_{{\mathcal R}(\tau)})^+_1$ and the effects of ${\mathcal S}^N$ are
elements of ${\mathbf L}_r({\mathbf H}^N_{{\mathcal R}(\tau)})^+_{\leq 1}$.
\end{prop} Proposition \ref{propold12} follows directly from the definition of
a system with a given Hilbert space. To see its significance, let us make a
comparison with the case of subsystems of different types. Let system
${\mathcal S}'$ be composed of two particles ${\mathcal S}_1$ and ${\mathcal
S}_2$ of different types. Let us prepare ${\mathcal S}'$ in a state ${\mathsf
P}[\Psi(\vec{x}_1,\vec{x}_2)] \in {\mathbf T}({\mathbf H}_1 \otimes {\mathbf
H}_2)^+_1$. Then, ${\mathcal S}_1$ is in state ${\mathsf T} \in {\mathbf
T}({\mathbf H}_1)^+_1$ given by kernel
$$
T(\vec{x}_1;\vec{x}'_1) = \int d^3x_2
\Psi^*(\vec{x}_1,\vec{x}_2)\Psi(\vec{x}'_1,\vec{x}_2)
$$
and sharp observable of ${\mathcal S}_1$ with kernel ${\mathsf A}$ can be
measured in such a way that we measure sharp observable
\begin{equation}\label{Ahet} {\mathsf A} \otimes {\mathsf 1}_2
\end{equation} of ${\mathcal S}'$. This is an expression of the obvious fact
that observing anything on a subsystem is tantamount to observing something on
the whole system. Thus, composing heterogeneous systems does not disturb their
individuality and rules valid for each of them separately (even in the cases
they are entangled).

Looking at the composition of identical systems we observe a very different
picture. A simple example of two spin-zero particles will give a sufficient
illustration of the phenomena. Let us consider two experiments.
\par \vspace*{.4cm} \noindent {\bf Experiment I}: State ${\mathsf P}[\psi]$ of
particle ${\mathcal S}_1$ of type $\tau$ is prepared in our laboratory.
\par \vspace*{.4cm} \noindent {\bf Experiment II}: State ${\mathsf P}[\psi]$
is prepared as in Experiment I and state ${\mathsf P}[\phi]$ of particle
${\mathcal S}_2$ of the same type $\tau$ is prepared simultaneously in a
remote laboratory.
\par \vspace*{.4cm} \noindent If our laboratory does not know about the second
one, it believes that the state of ${\mathcal S}_1$ is ${\mathsf P}[\psi] \in
{\mathbf T}({\mathbf H}_\tau)^+_1$. If it does then it believes that the state
of the composite system ${\mathcal S}'$ is, according to Proposition
\ref{propold12}, ${\mathsf P}[\Psi] \in {\mathbf T}({\mathbf H}^2_{{\mathcal
R}(\tau)})^+_1$ given by
\begin{equation}\label{symstate} \Psi(\vec{x}_1,\vec{x}_2) =
\nu\bigl(\psi(\vec{x}_1)\phi(\vec{x}_2) +
\phi(\vec{x}_1)\psi(\vec{x}_2)\bigr)\ ,
\end{equation} where $\nu = [2(1 + |c|^2)]^{-1/2}$ is a normalisation factor,
$c = \langle \psi |\phi \rangle$. This is true even if states $\psi$ and
$\phi$ are localized (the wave functions have supports) within the respective
laboratories.

Localisation of the states $\psi$ and $\phi$ removes at least the following
difficulty. Suppose that a fermion is prepared in a remote laboratory in a
state $\phi$. Then a fermion of the same type cannot be prepared in the same
state $\psi = \phi$ in our laboratory: the composite state of the two fermions
had then to be zero (Pauli exclusion principle). However, if the wave function
of the fermion prepared in our laboratory falls off rapidly outside our
laboratory and that in the remote laboratory does the same outside the remote
one, then the two wave functions would be different, $\psi(x) \neq \phi(x)$,
and their antisymmetric combination would not be zero even if $\psi$ was just
a Euclidean group transform of $\phi$. The requirement of the fall-off is very
plausible indeed. It would be technically impossible to prepare a state with
the same wave function $\psi$ in a laboratory located in Prague, and
simultaneously in a laboratory in Bern, say, even for bosons.

Returning to our original experiments, it seems that the intuitive notions of
preparation reaches its limits and becomes ambiguous. Has this ambiguity any
observable consequences? To answer this question, let us first consider
Experiment I supplemented by a registration corresponding to the sharp
observable ${\mathsf A}$ of ${\mathcal S}_1$ and let the registration be made
in our laboratory. Measurements of this kind lead to average value $\langle
\psi|{\mathsf A}|\psi\rangle$.  Second, perform Experiment II supplemented by
the registration by the same apparatus in our laboratory as above. Because the
apparatus cannot distinguish between the contributions by the two particles,
the correct observable corresponding to this registration now is:
\begin{equation}\label{symobs} {\mathsf A} \otimes {\mathsf 1}_2 + {\mathsf
1}_1 \otimes {\mathsf A}\ .
\end{equation} We can see a physical motivation for the construction suggested
above by the considerations about the Galilean group: here, it is not the
additivity as in the case of heterogeneous systems but the way identical
particles contribute to averages. Operator (\ref{symobs}) acts on ${\mathbf
H}^2_{{\mathcal R}(\tau)}$ so that it satisfies Proposition \ref{propold12}
while (\ref{Ahet}) does not. The proposed measurements lead to the average
value defined by Eqs.\ (\ref{symstate}) and (\ref{symobs}):
\begin{equation}\label{aver2} \frac{\langle \psi | {\mathsf A}\psi \rangle +
\langle \phi | {\mathsf A}\phi \rangle + c \langle \phi | {\mathsf A}\psi
\rangle + c^*\langle \psi | {\mathsf A}\phi \rangle}{1 + |c|^2}\ .
\end{equation} Average (\ref{aver2}) appreciably differs from $\langle
\psi|{\mathsf A}|\psi\rangle$ for many choices of ${\mathsf A}$, such as all
generators of group $\bar{\mathbf G}^+_c$ and most operators constructed from
them such as position, momentum, spin, angular momentum, energy, etc. (the majority
of operators dealt with in any textbook). In particular, if the states are
localised inside each laboratory that prepares them, then $c = 0$ and average
(\ref{aver2}) is $\langle \psi | {\mathsf A}\psi \rangle + \langle \phi |
{\mathsf A}\phi \rangle$. Consider the position operator ${\mathsf Q}$. In
$Q$-representation, the difference between the two averages is
$$
\int d^3x \vec{x} \phi^*(\vec{x}) \phi(\vec{x})\ .
$$
This can be made arbitrarily large by choosing the remote laboratory far
enough.

We conclude that system ${\mathcal S}_1$ does not seem to be in state
${\mathsf P}[\psi]$ prepared in our laboratory and registration of its
observable ${\mathsf A}$ in our laboratory seems to give values influenced by
external circumstances that are not under our control. Thus, Proposition
\ref{propold12} would lead to violation of Rules stated in Chapter 1 if it
were not supplemented by some further assumptions.

\subsection{Cluster separability} There was a proposal (Ref.\ \cite{peres},
p. 128) that the problem described in the previous section can be avoided by
adding some kind of locality to properties of observables. Such locality
assumptions are quite popular in various branches of quantum theory. Let us
briefly look at them.

The full relativistic theory starts with the requirement that space-time
symmetries of a closed system be realised by unitary representations of
Poincar\'{e} group on the Hilbert space of states, see Refs.\ \cite{Weinberg}
and \cite{haag}. Then, the {\em cluster decomposition principle}, a locality
assumption, states that if multi-particle scattering experiments are studied
in distant laboratories, then the $S$-matrix element for the overall process
factorizes into those concerning only the experiments in the single
laboratories. This ensures a factorisation of the corresponding transition
probabilities, so that an experiment in one laboratory cannot influence the
results obtained in another one. Cluster decomposition principle implies
non-trivial local properties of the theory underlying the $S$-matrix, in
particular it plays a crucial part in suggesting that local field theory is
inevitable (cf. Ref.\ \cite{Weinberg}, Chap. 4).

In the phenomenological theory of relativistic or non-relativistic many-body
systems, Hilbert space of a closed system must also carry a unitary
representation of Poincar\'{e} or Galilei group. Then, the so-called cluster
separability is a locality assumption, see, e.g., Refs.\ \cite{KP} or
\cite{coester} and references therein. It is a condition on interaction terms
in the generators of the space-time symmetry group saying: if the system is
divided into disjoint subsystems (i.e.,clusters) by a sufficiently large spacelike
separation, then each subsystem behaves as a closed system with a suitable
representation of space-time symmetries on its Hilbert space, see Ref.\
\cite{KP}, Section 6.1. Let us call this principle {\em Cluster Separability
I}.

Peres' proposal contains another special case of locality assumption. Let us
reformulate it as follows:
\par \vspace{.5cm}\noindent {\bf Cluster Separability II} No quantum
experiment with a system in a local laboratory is affected by the mere
presence of an identical system in remote parts of the universe.
\par \vspace{.5cm} \noindent It is well known (see, e.g., Ref.\ \cite{peres},
p. 136) that this principle leads to restrictions on possible statistics
(fermions, bosons). What is less well known is that it also motivates
non-trivial locality conditions on observables.

Some locality condition is already formulated in Ref.\ \cite{peres}, p. 128:
\begin{quote} \mbox{...} a state $\mathsf w$ is called remote if $\|{\mathsf
A}{\mathsf w}\|$ is vanishingly small, for any operator ${\mathsf A}$ which
corresponds to a quantum test in a nearby location. ... We can now show that
the entanglement of a local quantum system with another system in a remote
state (as defined above) has no observable effect.
\end{quote} This is a condition on ${\mathsf A}$ inasmuch as there has to be
at least one remote state for ${\mathsf A}$.

However, Peres does not warn that most operators of quantum theory do not
satisfy his condition on ${\mathsf A}$ \cite{hajicek2}. Indeed, observables of
quantum field theory and many-body theories that are constructed from
generators of Poincar\'{e} or Galilei groups do not satisfy the locality
condition. It follows that Cluster Separability II is logically independent
from the cluster decomposition or from Cluster Separability I.

The present section reformulates and extends Peres' ideas. Let us explain
everything on an example of single spin-zero particle, working in
$Q$-representation of Hilbert space ${\mathbf H}_\tau$ and of operators on
it. A more general theory will be developed in the next subsection.

We introduce an important locality property of observables \cite{hajicek2}. A
similar local condition on observables has been introduced in
\cite{wan,wanb}: Let $D \subset {\mathbb R}^{3}$ be open. Operator with
kernel $a(\vec{x}_1;\vec{x}'_1)$ is $D$-{\em local} if
$$
\int d^3x'_1\, a(\vec{x}_1;\vec{x}'_1) f(\vec{x}'_1) = \int d^3x_1\,
a(\vec{x}_1;\vec{x}'_1) f(\vec{x}_1) = 0\ ,
$$
for any test function $f$ that vanishes in $D$.

Now assume for Experiment II that our laboratory is inside open set $D \subset
{\mathbb R}^{3}$ and that $\text{supp}\,\phi \cap D = \emptyset$. Then, the
second term in (\ref{aver2}) vanishes for all $D$-local observables and
averages $\langle\psi|{\mathsf A}|\psi\rangle$ and (\ref{aver2}) agree in this
case. Hence, in this case, the two approaches are compatible: First, system
${\mathcal S}_1$ is in state $\psi$ and the observable that is measured on it
is ${\mathsf A}$. Second, system ${\mathcal S}_1 + {\mathcal S}_2$ is in state
(\ref{symstate}) and the observable being measured is (\ref{symobs}).

This suggests the following idea. Let ${\mathcal S}$ be prepared in a state
$\psi(\vec{x})$ such that supp\,$\psi \cap D \neq \emptyset$ and that
registration of any $D$-local observable ${\mathsf A}$ of ${\mathcal S}$ lead
to average $\langle \psi(\vec{x})|{\mathsf A}\psi(\vec{x})\rangle$. In such a
case, we say that ${\mathcal S}$ has {\em separation status} $D$.

This includes the following condition that must be fulfilled by the apparatus
${\mathcal A}$ registering observable ${\mathsf A}$: The states of those
subsystems of ${\mathcal A}$ that are identical with ${\mathcal S}$ do not
disturb the registration of ${\mathsf A}$ by ${\mathcal A}$ (but may disturb
other registrations). A model of ${\mathcal A}$ satisfying the condition is
given in Section 4.3.

In the above example, the reason why the registration of ${\mathsf A}$ is not
disturbed by other identical systems is that the wave functions of these
systems vanish in $D$. This could also be weakened to
$$
\int_D d^3x \phi^*(\vec{x})\phi(\vec{x}) \approx 0\ ,
$$
so that the registration apparatuses are not sensitive enough to react to
$\phi$. In any case, the vanishing of the wave functions of all external
identical systems provides a good mathematical model of separation status and
we shall assume that the mathematical results obtained for such models are
approximately valid for more realistic situations.

Now, we can introduce a correction to our intuitive notion of preparation. The
separation status must be understood as new condition on preparations and
observables that could be formulated as follows:
\begin{rl}\label{rlprep} Let ${\mathcal S}$ be a quantum system of type $\tau$
with Hilbert space ${\mathbf H}_\tau$. Any preparation of ${\mathcal S}$ must
give it separation status $D$ satisfying $D \neq \emptyset$. Then the prepared
state of ${\mathcal S}$ is an element of ${\mathbf T}({\mathbf H}_\tau)^+_1$
and $D$-local effects of ${\mathbf L}_r({\mathbf H}_\tau)^+_{\leq 1}$ are
individually registrable on ${\mathcal S}$ but {\em only these are}.
\end{rl} Thus, any preparation must also provide sufficient isolation of the
prepared system ${\mathcal S}$ from influence of particles of the environment
that are identical to constituents of the prepared system. In this way,
control about what is prepared and what is registered might be regained and
Cluster Separability II would hold.

What is the nature and aim of the limitation of observables to $D$-local ones?
A current opinion is that, on the one hand, the states of each quantum system
of type $\tau$ with Hilbert space ${\mathbf H}_\tau$ are all positive
normalised s.a.\ operators on ${\mathbf H}_\tau$ and its observables are all
s.a.\ operators on ${\mathbf H}_\tau$. On the other, in the everyday mess of
practical conditions of each particular experiment, only some of these states
occur and only some of these effects are registrable. That is all that the
current opinion can say.

In this sense, the current language of quantum mechanics is too abstract and
too far away from real experiments. So far away that it cannot give a detailed
account of experiments without getting entangled in contradictions. This is
one of the reasons why it has problems with realist interpretation, classical
properties and theory of measurement. Our aim is to refine the language so
that it becomes more suitable for description of real experiments.

For example, the concept "quantum system of type $\tau$" is an ideal
theoretical one that has no real counterpart. It has been abstracted by
generalisation from many particular experiments with real objects. Only
objects that have been prepared are real and we have to study these real
objects first to arrive at useful abstract concepts. We must view each real
physical quantum object ${\mathcal S}$ as a system with a {\em modified} set
of observables. The modification depends on the bounded open set $D$ that is
determined by the preparation of $\mathcal S$. Hence, in general, any
registrable observable must be $D$-local for some bounded open set $D$. The
price is that, strictly speaking, no quantum system is proper (see Section
1.2.2).

Simple examples of separation status are $D = \emptyset$ and $D = {\mathbb
R}^3$. The first, the so-called {\em trivial} separation status, is the case
when ${\mathcal S}$ is not separated from a system composed of $N$ particles
of the same kind. Then, ${\mathcal S}$ has no individual states and
observables. An example of the second is the case that ${\mathcal S}$ is
isolated: the whole space ${\mathbb R}^3$ is empty except for ${\mathcal
S}$. The standard quantum mechanics works just with these two separation
statuses, assuming tacitly that systems are approximately isolated. Our
definition of general separation status $D$ explains what may be meant by
"approximately" and why quantum mechanics works at all in practical
applications.

\subsection{Mathematical theory of $D$-local observables} In the two foregoing
sections, some raw physical ideas on composition of identical systems have
been put forward. Now, we start to give a general mathematical formulation of
these ideas. In the present section, we extend the definition of $D$-local
operators, to composite systems containing more than just one kind of
particles and to non-vector states, and we study the question of how $D$-local POV
measures can be constructed from general ones. We limit ourselves to
one-dimensional POV's, that is, ${\mathbf F}$ is the set of subsets of
${\mathbb R}$.

Let ${\mathcal S}$ be a general $N$-particle system. We shall work in $Q$
representation throughout, suppress spin indices and consider only those
operators $\mathsf A$ on ${\mathbf H}$ that fulfil the following
condition. The kernel
$A(\vec{x}_1,\cdots,\vec{x}_N;\vec{x}'_1,\cdots,\vec{x}'_N)$ of $\mathsf A$ is
a distribution on Schwartz space (its elements are rapidly decreasing
$C^\infty$ functions, see, e.g., Ref.\ \cite{RS}) ${\mathbf S}({\mathbb
R}^{3N})$ in variables $\vec{x}'_1,\cdots,\vec{x}'_N$ for any fixed value of
$\vec{x}_1,\cdots,\vec{x}_N$ and that
$$
g(\vec{x}_1,\cdots,\vec{x}_N) = \int d^{3N}x'
A(\vec{x}_1,\cdots,\vec{x}_N;\vec{x}'_1,\cdots,\vec{x}'_N)
f(\vec{x}'_1,\cdots,\vec{x}'_N) \in {\mathbf S}({\mathbb R}^{3N})
$$
for any $f(\vec{x}_1,\cdots,\vec{x}_N) \in {\mathbf S}({\mathbb
R}^{3N})$. This is usually satisfied, see Section 1.3.1.

The general and formal definition of $D$-local operators is the following:
\begin{df}\label{dfold13} Let $D \subset {\mathbb R}^{3}$ be open, let
$\mathsf A$ be an operator on ${\mathbf H}$ and let the following conditions
hold:
\begin{enumerate}
\item $A(\vec{x}_1,\cdots,\vec{x}_N;\vec{x}'_1,\cdots,\vec{x}'_N)$ is the zero
distribution for any $\vec{x}_1,\cdots,\vec{x}_N \in {\mathbb R}^{3N}
\setminus D^N$.
\item
$$
\int d^{3N}x' A(\vec{x}_1,\cdots,\vec{x}_N;\vec{x}'_1,\cdots,\vec{x}'_N)
f(\vec{x}'_1,\cdots,\vec{x}'_N) = 0
$$
for any test function $f$ such that
$$
\text{supp}f(\vec{x}'_1,\cdots,\vec{x}'_N) \subset {\mathbb R}^{3N} \setminus
D^N\ .
$$
\end{enumerate} Then $\mathsf A$ is called $D${\em -local}.
\end{df} Let $\mathsf A$ be $D$-local for some open set $D$, let $D'$ be open
and $D \subset D'$. Then, $\mathsf A$ is $D'$-local. Let $\mathsf A$ be
$D$-local and also $D'$-local for two open sets $D$ and $D'$. Then, $\mathsf
A$ is ($D \cap D'$)-local. Thus, all open sets $D$ such that $\mathsf A$ is
$D$-local form a filter in the Boolean lattice of open subsets of ${\mathbb
R}^3$.
\par \noindent {\bf Example 1} $N = 1$, $A(\vec{x};\vec{x}') =
x^1\delta({\vec{x} - \vec{x}'})$. $\mathsf A$ is ${\mathbb R}^3$-local and the
filter is ${\mathbb R}^3$.
\par \noindent {\bf Example 2} $N = 1$, $\psi(\vec{x})$ is a wave function
with support $D$. Then $|\psi(\vec{x})\rangle \langle \psi(\vec{x})|$ is
$D'$-local, where $D'$ is the interior of $D$, and the filter is the family of
all open sets containing $D'$.

Definition \ref{dfold13} can be applied both to state operators and to effects
(of a POV measure, see Section 1.2.1). However, the definition of a $D$-local
POV measure is more complicated than just the condition that its effects be
$D$-local. Let us denote by ${\mathbf H}(D)$ the Hilbert space obtained by
completion of the linear space of rapidly decreasing $C^\infty$-functions with
support in $D^N$ with respect to the inner product of ${\mathbf H}$. ${\mathbf
H}(D)$ is a closed linear subspace of ${\mathbf H}$ as is the set ${\mathbf
H}^\perp(D)$ of all vectors in ${\mathbf H}$ orthogonal to ${\mathbf
H}(D)$. Let ${\mathsf P}[{\mathbf H}(D)]$ be the orthogonal projection from
${\mathbf H}$ onto ${\mathbf H}(D)$. Then ${\mathsf 1} - {\mathsf P}[{\mathbf
H}(D)]$ is the projection onto ${\mathbf H}^\perp(D)$.

\begin{df}\label{dfold14} A POV measure ${\mathsf E}(X)$ of dimension 1 is
called $D${\em -local} if
\begin{enumerate}
\item effect ${\mathsf E}(X)$ is $D$-local for all $X \in {\mathbf F}$ such
that $0 \notin X$ and
\item
$$
{\mathsf E}(X) = {\mathsf 1} - {\mathsf P}[{\mathbf H}(D)] + {\mathsf E}'(X)
$$
for all $X \in {\mathbf F}$ such that $0 \in X$, where ${\mathsf E}'(X)$ is a
$D$-local effect.
\end{enumerate}
\end{df} In principle, real registrations can register only $D$-local POV
measures for some bounded $D$ because each registration apparatus takes only a
limited region $D$ of space and is not sensitive to any system localised
outside $D$. But there is no a priory bound on $D$.

Next, we shall prove that each one-dimensional POV measure ${\mathsf E}$ of a
general system ${\mathcal S}$ is associated, for a given open set $D$, with a
unique one-dimensional $D$-local POV measure $\Lambda_D({\mathsf E})$, the
so-called $D$-localisation of ${\mathsf E}$, such that the condition
\begin{equation}\label{trLambda} tr[{\mathsf E}(X) {\mathsf T}] =
tr[\Lambda_D({\mathsf E})(X)) {\mathsf T}]
\end{equation} is satisfied for all $X \in {\mathbf F}$ and for all $D$-local
states ${\mathsf T}$. Thus, any standard observable ${\mathsf E}$, such as
position, momentum, etc., can be registered under the condition that the system
is within $D$ by registering $\Lambda_D({\mathsf E})$ and the probability
distribution would be the same as if we had registered ${\mathsf E}$. We are
going to propose that exactly this has been done if somebody claims he has
measured position, momentum, etc.

Let ${\mathsf A} \in {\mathbf L}_r({\mathbf H})$. Then ${\mathsf P}[{\mathbf
H}(D)]{\mathsf A}{\mathsf P}[{\mathbf H}(D)] \in {\mathbf L}_r({\mathbf
H})$. Moreover, if ${\mathsf A}$ is positive, so is ${\mathsf P}[{\mathbf
H}(D)]{\mathsf A}{\mathsf P}[{\mathbf H}(D)]$, and we have also
$$
\|{\mathsf P}[{\mathbf H}(D)]{\mathsf A}{\mathsf P}[{\mathbf H}(D)]\| \leq \|
{\mathsf A}\|\ .
$$
Hence, if ${\mathsf A} \in {\mathbf L}_r({\mathbf H})^+_{\leq 1}$ then
${\mathsf P}[{\mathbf H}(D)]{\mathsf A}{\mathsf P}[{\mathbf H}(D)] \in
{\mathbf L}_r({\mathbf H})^+_{\leq 1}$. Let us call
$$
{\mathsf P}[{\mathbf H}(D)]{\mathsf A}{\mathsf P}[{\mathbf H}(D)]
$$
$D$-projection of ${\mathsf A}$. Of course, the $D$-projection is not a
unitary map and it changes the spectral measure of the operator. For example,
the spectral measure of
$$
{\mathsf P}[{\mathbf H}(D)]{\mathsf A}{\mathsf P}[{\mathbf H}(D)]
$$
must contain ${\mathsf P}(\{0\}) = {\mathsf 1} - {\mathsf P}[{\mathbf H}(D)]$
even if that of ${\mathsf A}$ does not because all vectors in ${\mathbf
H}^\perp_\tau(D)$ are eigenvectors of ${\mathsf P}[{\mathbf H}(D)]{\mathsf
A}{\mathsf P}[{\mathbf H}(D)]$ to eigenvalue 0.

Now, we are ready to give the following definition:
\begin{df}\label{dfold15} Let ${\mathsf E}$ be an one-dimensional POV measure
on ${\mathbf H}$. Then,\newline $D${\em -localisation}
$\Lambda_D({\mathsf E})$ of ${\mathsf E}$ is defined by
\begin{enumerate}
\item
$$
\Lambda_D({\mathsf E})(X) = {\mathsf P}[{\mathbf H}(D)]{\mathsf E}(X){\mathsf
P}[{\mathbf H}(D)]
$$
for all $X \in {\mathbf F}$ such that $0 \not\in X$ and
\item
$$
\Lambda_D({\mathsf E})(X) = {\mathsf P}[{\mathbf H}(D)]{\mathsf E}(X){\mathsf
P}[{\mathbf H}(D)] + {\mathsf 1} - {\mathsf P}[{\mathbf H}(D)]
$$
for all $X \in {\mathbf F}$ such that $0 \in X$.
\end{enumerate}
\end{df} Clearly,
$$
\Lambda_D({\mathsf E})({\mathbb R}) = {\mathsf P}[{\mathbf H}(D)]{\mathsf
1}{\mathsf P}[{\mathbf H}(D)] + {\mathsf 1} - {\mathsf P}[{\mathbf H}(D)] =
{\mathsf 1}\ ,
$$
and the normalisation condition holds. Then, all other conditions on POV
measures are also satisfied by $\Lambda_D({\mathsf E})$ and equation
(\ref{trLambda}) is true.

Let us study this definition for a simple example. Let ${\mathsf A}$ be a
bounded s.a.\ operator on ${\mathbf H}$ with discrete value set $\{a_k\}$
including $a_0 = 0$ and let the projection on the eigenspace of $a_k$ be
${\mathsf P}_k$ for all $k$. Then
\begin{equation}\label{specdA} {\mathsf A} = \sum_{k=1}^\infty a_k {\mathsf
P}_k\ .
\end{equation} The corresponding PV measure is
$$
{\mathsf E}^{\mathsf A}(X) = \sum_{k\in {\mathbf K}(X)} {\mathsf P}_k\ ,
$$
where $k\in {\mathbf K}(X)$ if $a_k \in X$. Let $D$ be open, ${\mathbf H}(D)$
and ${\mathsf P}[{\mathbf H}(D)]$ be defined as above. Equation (\ref{specdA})
and Definition \ref{dfold15} imply that $D$-projection of ${\mathsf A}$ can
then be written as
\begin{equation}\label{DprojA} {\mathsf P}[{\mathbf H}(D)] {\mathsf A}
{\mathsf P}[{\mathbf H}(D)] = \sum_{k=0}^\infty a_k \Lambda_D({\mathsf P}_k)\
,
\end{equation} where
$$
\Lambda_D({\mathsf P}_k) = {\mathsf P}[{\mathbf H}(D)] {\mathsf P}_k {\mathsf
P}[{\mathbf H}(D)] + \delta_{0k}({\mathsf 1} - {\mathsf P}[{\mathbf H}(D)])
$$
and
$$
\Lambda_D({\mathsf E}^{\mathsf A})(X) = \sum_{k\in {\mathbf K}(X)}
\Lambda_D({\mathsf P}_k)\ .
$$
Clearly, ${\mathsf P}[{\mathbf H}(D)] {\mathsf P}_k {\mathsf P}[{\mathbf
H}(D)]$ is a projection only if ${\mathsf P}$ and ${\mathsf P}[{\mathbf
H}(D)]$ commute. Projections onto subspaces ${\mathbf H}(D)$ and ${\mathbf
H}_k$ of ${\mathbf H}$ commute only if either ${\mathbf H}(D) \cap {\mathbf
H}_k = 0$, where $0$ is the zero vector of ${\mathbf H}$, or ${\mathbf H}(D)
\subset {\mathbf H}_k$ or ${\mathbf H}_k \subset {\mathbf H}(D)$.

Of course, the $k=0$ element of the right-hand side of sum (\ref{DprojA})
vanishes and it is not necessary for the validity of the equation to care how
$\Lambda_D({\mathsf P}_0)$ is to be defined. With our definition, however, the
normalisation condition on the POV measures is satisfied.

As ${\mathsf A}$ is bounded and s.a., so is ${\mathsf P}[{\mathbf H}(D)]
{\mathsf A} {\mathsf P}[{\mathbf H}(D)]$, hence it also possesses its own
spectral decomposition, which is in general different from equation
(\ref{DprojA}). Its value set will be different from $\{a_k\}$ and its
projections different from ${\mathsf P}[{\mathbf H}(D)] {\mathsf P}_k {\mathsf
P}[{\mathbf H}(D)]$ (there does not seem to be any general way calculate
${\mathsf E}^{{\mathsf P}[{\mathbf H}(D)] {\mathsf A} {\mathsf P}[{\mathbf
H}(D)]}$ from ${\mathsf E}^{\mathsf A}$). Hence, our definition of the
$D$-localisation of an observable is different from $D$-projection of the
corresponding operator. In particular, our definition preserves the value set
(adding possibly 0 to it) but does not preserve the sharpness. Moreover, it
makes sense for unbounded operators, too.

The above definitions of $D$-local observables and of $D$-localisation could
be generalised as follows. As yet, $D$ has been an open subset in the spectrum
of the position $\vec{\mathsf Q}$ of ${\mathcal S}$. We can take any sharp
observable instead of position and all definitions and their consequences
remain valid. A very important example is the following. As we shall see in
Chapter 4, every registration apparatus ${\mathcal A}$ must include
detectors. It seems that any detector can work only if the total energy of the
detected system is higher than certain threshold $E({\mathcal A}) > 0$. Hence,
the observable that is registered by such an apparatus must be $D$-local where
$D$ is the subset $E > E({\mathcal A})$ of the energy spectrum. If the energy
of all systems identical with ${\mathcal S}$ that are in the environment is
smaller than $E({\mathcal A})$ then the apparatus cannot be influenced by them
similarly as it is not influenced by wave functions that vanish in a
neighbourhood of ${\mathcal A}$. Under such conditions, ${\mathcal S}$
prepared so that its energy lies in $D$ has the generalised separation status
$D$. Here, the term "local" loses its purely space-like character.

\subsection{Separation status} The generalisation of the notion of separation
status introduced in Section 2.2.2 to non-vector states of composite systems
and to one-dimensional POV measures is straightforward:
\begin{df} Let $D \subset {\mathbb R}^3$ be an open set and system ${\mathcal
S}$ be prepared in state ${\mathsf T}$ that satisfies the following
conditions:
\begin{enumerate}
\item There is at least one $D$-local POV measure ${\mathsf E}'$ such that its
average on ${\mathsf T}$ does not vanish,
$$
\int_{\mathbb R} \iota tr[{\mathsf T}d{\mathsf E}'] \neq 0\ .
$$
\item The average of any $D$-local one-dimensional POV measure ${\mathsf E}$
as registered on ${\mathsf T}$ is given by
$$
\int_{\mathbb R} \iota tr[{\mathsf T}d{\mathsf E}]\ .
$$
\end{enumerate} Then, open set $D$ is called {\em separation status} of
${\mathcal S}$.
\end{df} The second condition means that the registration of ${\mathsf E}$ on
${\mathsf T}$ is not disturbed by any other existing identical system.

Rule \ref{rlprep} then guarantees that control is regained, external
influences are removed and possible ambiguity is harmless. An example of such
an ambiguity can be constructed from Experiments I and II if the preparations are
viewed as hierarchically nested. More generally, let system $\bar{\mathcal S}$ be
prepared in state $\bar{\mathsf T}$ with separation status $\bar{D}$ and have
a subsystem ${\mathcal S}$ that is simultaneously prepared in state ${\mathsf
T}$ with separation status $D \subset \bar{D}$. Suppose further that
$\bar{\mathcal S}$ contains at least two particles ${\mathcal S}_1$ and
${\mathcal S}_2$ of the same type such that ${\mathcal S}_1 \subset {\mathcal
S}$ and ${\mathcal S}_2 \not\subset {\mathcal S}$. Then $\bar{\mathcal S}$ has
more subsystems that are different from, but contain particles of the same
type as, ${\mathcal S}$. Thus, ${\mathcal S}$ would in general be only a
mathematical entity because it could not be physically distinguished from some
other subsystems of $\bar{\mathcal S}$. However, in our special case, one of
these subsystems, viz.\ ${\mathcal S}$, has a separation status $D$ and is,
therefore, recognisable and has state ${\mathsf T}$, on which $D$-local
observables of ${\mathcal S}$ can be registered.

Let us set up a mathematical model of such an ambiguity. It concerns
composition of general non-heterogeneous systems $\mathcal S$ and ${\mathcal
S}'$ and the construction of observables of $\bar{\mathcal S} = {\mathcal S} +
{\mathcal S}'$ from those of $\mathcal S$ analogous to (\ref{symobs}).

We assume that there is, generally, a fixed number $K$ of fermion particle
types and a and fixed number $L$ of boson particle types in quantum
mechanics. Let ${\mathcal S}$ be a general $N$-particle system composed of
numbers $F_1,\ldots,F_K$ of the fermions and $B_1,\ldots,B_L$ of bosons, where
$F_k$ and $B_l$ are non-negative integers so that
$$
\sum_{k=1}^K F_k + \sum_{l=1}^L B_l = N
$$
is the total particle number in ${\mathcal S}$. We call $F_k$ and $B_l$
occupation numbers. Let the Hilbert space of $F_k$ fermions be ${\mathbf
H}_a^{F_k}$, $k=1,\dots,K$ and that of $B_l$ bosons be ${\mathbf H}_s^{B_l}$,
$l=1,\dots,L$ according to Rule \ref{rlold12} so that the Hilbert space of
${\mathcal S}$ is
\begin{equation}\label{compg} {\mathbf H} = \prod_{k=1}^K \otimes {\mathbf
H}_a^{F_k} \otimes \prod_{l=1}^L \otimes {\mathbf H}_s^{B_l}
\end{equation} according to Rule \ref{rlold10}. The tensor product on the
right hand side is to be understood so that the factor ${\mathbf H}_a^{F_k}$
is left out if $F_k = 0$ and similarly for the bosons. Let the occupation
numbers of ${\mathcal S}'$ be $F'_1,\ldots,F'_K$ and $B'_1,\ldots,B'_L$, and
those of $\bar{\mathcal S}$ be $\bar{F}_1,\ldots,\bar{F}_K$ and
$\bar{B}_1,\ldots,\bar{B}_L$. We have
$$
N' = \sum_{k=1}^KF'_k + \sum_{l=1}^LB'_l\ ,\quad F_k + F'_k =\bar{F}_k\ ,\quad
B_k + B'_k = \bar{B}_k
$$
for all $k$. The Hilbert spaces ${\mathbf H}_\tau$, ${\mathbf H}_{\tau'}$ and
${\mathbf H}_{\tau\tau'}$ of $\mathcal S$, ${\mathcal S}'$ and $\bar{\mathcal
S}$, respectively, are given by equations analogous to (\ref{compg}). Let us
write the Hilbert space ${\mathbf H}_\tau \otimes {\mathbf H}_{\tau'}$ on
which ${\mathsf T} \otimes {\mathsf T}'$ is a state operator, as follows:
$$
{\mathbf H}_\tau \otimes {\mathbf H}_{\tau'} = \prod_{k=1}^K \otimes
\big({\mathbf H}_a^{F_k} \otimes {\mathbf H}_a^{F'_k}\big) \otimes
\prod_{l=1}^L \otimes \big({\mathbf H}_s^{B_l} \otimes {\mathbf
H}_s^{B'_l}\big)\ .
$$
For any factor ${\mathbf H}_\rho^k \otimes {\mathbf H}_\rho^{k'}$ in the
expression for ${\mathbf H}_\tau \otimes {\mathbf H}_{\tau'}$, there is a
factor ${\mathbf H}_\rho^{k+k'}$ in the expression for ${\mathbf
H}_{\tau\tau'}$, as $\bar{k} = k + k'$. Here, $\rho$ is either $a$ or $s$ and
$k$ is either $F_k$ or $B_l$, etc. Let ${\mathsf P}^{k+k'}_\rho$ be the
orthogonal projection,
$$
{\mathsf P}^{k+k'}_\rho : {\mathbf H}_\rho^k \otimes {\mathbf H}_\rho^{k'}
\mapsto {\mathbf H}_\rho^{k+k'}
$$
consisting of total symmetrisation or anti-symmetrisation depending on
$\rho$. We define
$$
{\mathsf P}_{\tau\tau'} : {\mathbf H}_\tau \otimes {\mathbf H}_{\tau'} \mapsto
{\mathbf H}_{\tau\tau'}
$$
by
\begin{equation}\label{projTT'} {\mathsf P}_{\tau\tau'} = \prod_{k=1}^K
\otimes {\mathsf P}_a^{F_k + F'_k} \otimes \prod_{l=1}^L \otimes {\mathsf
P}_s^{B_l + B'_l}\ .
\end{equation} From the definition, it follows immediately that ${\mathsf
P}_{\tau\tau'}$ is an orthogonal projection. Let us define
$$
{\mathsf J} : {\mathbf T}({\mathbf H}_\tau)^+_1 \times {\mathbf T}({\mathbf
H}_{\tau'})^+_1 \mapsto {\mathbf T}({\mathbf H}_{\tau\tau'})^+_1
$$
by
\begin{equation}\label{operj} {\mathsf J}({\mathbf T},{\mathbf T}') =
\frac{{\mathsf P}_{\tau\tau'}({\mathsf T}\otimes {\mathsf T}'){\mathsf
P}_{\tau\tau'}}{tr[{\mathsf P}_{\tau\tau'}({\mathsf T}\otimes {\mathsf
T}'){\mathsf P}_{\tau\tau'}]}\ .
\end{equation}

The symmetry properties of states such as ${\mathsf J}\big({\mathsf
T},{\mathsf T}'\big)$ are stronger than the symmetry properties of operators that
are necessary to ensure that their action does not change the symmetry
properties of the states. For example, the kernel of operator (\ref{symobs})
or (\ref{symG}) are not of the form ${\mathsf P}{\mathsf A}{\mathsf P}$, where
${\mathsf P}$ is a (anti-)symmetrising projection such as ${\mathsf
P}_{\tau\tau'}$. Rather, they are symmetrised in whole variable pairs
$(x_i;x'_i)$. Formulas (\ref{symobs}) and (\ref{symG}) are examples of a
construction of an operator $\bar{\mathsf A}$ of a system $\bar{\mathcal S}$
from an operator ${\mathsf A}$ of its subsystem ${\mathcal S}$. The general
form of such an operator is
\begin{equation}\label{extenO} \bar{\mathsf A} = \sum_{\tilde{\mathcal
S}+{\mathcal S}'=\bar{\mathcal S}} {\mathsf A}_{\tilde{\mathcal S}} \otimes
{\mathsf 1}_{{\mathcal S}'}\ ,
\end{equation} where the sum is over all different pairs of subsystems
$\tilde{\mathcal S}$ and ${\mathcal S}'$ such that $\tilde{\mathcal S}$ is of
the same kind as ${\mathcal S}$. We call $\bar{\mathsf A}$ {\em extension} of
${\mathsf A}$ to the composite. It is clear that our definition of extension
is also valid for the composition of two heterogeneous systems. In this case,
the extension would be $\bar{\mathsf A} = {\mathsf A} \otimes {\mathsf 1}$.

Now, we can tackle the composition of general systems. The composition of
systems of the same type is determined by Rule \ref{rlold12} and Proposition
\ref{propold12} together with Rule \ref{rlprep} so that we can choose a state
for the system that is prepared from the corresponding state space. The
generalisation to arbitrary systems can be formulated as follows.
\begin{df}\label{dfsepst} Let system $\mathcal S$ be prepared in state
${\mathsf T}$ with separation status $D \neq \emptyset$ and system ${\mathcal
S}'$ in state ${\mathsf T}'$ with separation status $D' \neq \emptyset$ such
that $D \cap D = \emptyset$. Then, $\mathcal S$ and $\mathcal S'$ are called
{\em separated}.
\end{df} This is a generalisation of the situation occurring in Experiments I
and II.
\begin{rl}\label{rlold14} Let $\mathcal S$ and $\mathcal S'$ be
separated. Then system ${\mathcal S} + {\mathcal S}'$ can be considered as
prepared in state $\bar{\mathsf T} = {\mathsf J}({\mathsf T},{\mathsf T}')$
with separation status $D \cup D'$. The operators of the form (\ref{extenO})
for ${\mathsf A} \in {\mathbf A}_{\mathcal S}(D)$ are observables of $\mathcal
S + {\mathcal S}'$. Alternatively, $\mathcal S + {\mathcal S}'$ can be
considered as prepared in state ${\mathsf T} \otimes {\mathsf T}'$ and
operators of the form ${\mathsf A} \otimes {\mathsf 1}'$ are observables of
$\mathcal S + {\mathcal S}'$, where ${\mathsf A} \in {\mathbf A}_{\mathcal
S}(D)$ and ${\mathsf 1}' \in {\mathbf A}_{{\mathcal S}'}(D')$.
\end{rl} Now, it also ought to be clear why we do not employ Fock-space method
to deal with identical systems: it automatically (anti-)symmetrises over all
systems of the same type.

To show that the ambiguity is innocuous, it may be helpful to consider a
simple example giving the Euclidean space just one dimension. Let ${\mathcal
S}$ has occupation numbers $F_1 = 2$, $B_1 = 1$ and ${\mathcal S}'$ has $F'_1
= 1$ and $B'_1 = 1$, and let the prepared state of ${\mathcal S}$ be
$\psi_1(x_1,x_2)\phi_1(x_3)$ and that of ${\mathcal S}'$ be
$\psi_2(x_4)\phi_2(x_5)$. The wave function $\psi_1(x_1,x_2) =
-\psi_1(x_2,x_1)$ is antisymmetric in its arguments and the normalisation is
\begin{eqnarray*} \int dx_1dx_2\psi^*_1(x_1,x_2)\psi_1(x_1,x_2) &=& \quad \int
dx_3\phi^*_1(x_3)\phi_1(x_3) \\ = \quad \int dx_4\psi^*_2(x_4)\phi_1(x_4) &=&
\quad \int dx_5\phi^*_2(x_5)\phi_2(x_5) = 1\ .
\end{eqnarray*} The state of $\bar{\mathcal S}$ then is
\begin{multline}\label{symst'} \bar{\psi}(x_1,x_2,x_3,x_4,x_5) =
\nu\Big[\psi_1(x_1,x_2)\psi_2(x_4) + \psi_1(x_4,x_1)\psi_2(x_2) +
\psi_1(x_2,x_4)\psi_2(x_1)\Big] \\ \times \Big[\phi_1(x_3)\phi_2(x_5) +
\phi_1(x_5)\phi_2(x_3)\Big]\ ,
\end{multline} where $\nu$ is a normalisation factor. Observe that the kernel
$$
\bar{\psi}(x_1,x_2,x_3,x_4,x_5) \bar{\psi}^*(x'_1,x'_2,x'_3,x'_4,x'_5)
$$
is antisymmetric in three variables $x_1,x_2,x_4$ and independently so in
further three variables $x'_1,x'_2,x'_4$ and similarly symmetric in two
variables $x_3,x_5$ and independently so in further two variables $x'_3,x'_5$.

Kernels of observables have less symmetry. For example, let
$A(x_1,x_2,x_3;x'_1,x'_2,x'_3)$ be the kernel of an operator for ${\mathcal
S}$. It must be an operator on Hilbert space ${\mathbf H}^2_a\otimes {\mathbf
H}$ of ${\mathcal S}$. To satisfy this requirement, it is sufficient that
$$
A(x_1,x_2,x_3;x'_1,x'_2,x'_3) = A(x_2,x_1,x_3;x'_2,x'_1,x'_3)\ ,
$$
as a simple calculation easily shows. The extension of ${\mathsf A}$ to
$\bar{\mathcal S}$ is
\begin{multline}\label{extobs}
\bar{A}(x_1,x_2,x_3,x_4,x_5;x'_1,x'_2,x'_3,x'_4,x'_5) = \\
A(x_1,x_2,x_3;x'_1,x'_2,x'_3)\delta(x_4,x'_4)\delta(x_5,x'_5) \\ +
A(x_2,x_4,x_3;x'_2,x'_4,x'_3)\delta(x_1,x'_1)\delta(x_5,x'_5) \\ +
A(x_4,x_1,x_3;x'_4,x'_1,x'_3)\delta(x_2,x'_2)\delta(x_5,x'_5) \\ +
A(x_1,x_2,x_5;x'_1,x'_2,x'_5)\delta(x_4,x'_4)\delta(x_3,x'_3) \\ +
A(x_2,x_4,x_5;x'_2,x'_4,x'_5)\delta(x_1,x'_1)\delta(x_3,x'_3) \\ +
A(x_4,x_1,x_5;x'_4,x'_1,x'_5)\delta(x_2,x'_2)\delta(x_3,x'_3)\ .
\end{multline} The six terms are obtained by exchanging identical particles
only between the different subsystems.

Suppose next that state $\psi_1(x_1,x_2)\phi_1(x_3)$ and operator
$A(x_1,x_2,x_3;x'_1,x'_2,x'_3)$ are both $D$-local while state
$\psi_2(x_4)\phi_2(x_5)$ is $D'$-local so that $D \cap D' = \emptyset$. In
such a case, calculations of traces simplifies considerably. To see how it
comes about, consider $tr[|\bar{\psi}\rangle \langle \bar{\psi}|]$ or
$$
\int dx_1dx_2dx_3dx_4dx_5\bar{\psi}^*(x_1,x_2,x_3,x_4,x_5)
\bar{\psi}(x_1,x_2,x_3,x_4,x_5)\ .
$$
Taking the product of two arbitrary terms of (\ref{symst'}), e.g.,
$$
\int dx_1dx_2dx_3dx_4dx_5\psi^*_1(x_1,x_2)\psi^*_2(x_4)
\phi^*_1(x_5)\phi^*_2(x_3) \psi_1(x_2,x_4)\psi_2(x_1) \phi_1(x_3)\phi_2(x_5)\
,
$$
we observe that e.g.\ the integral over $x_1$ must vanish because it connects
the functions $\psi^*_1(x_1,x_2)$ and $\psi_2(x_1)$ that are non-zero in two
different non-overlapping domains $D$ and $D'$ of $x_1$. It is clear that only
terms that are obtained by the same permutation of the original variables in
both factors, such as
\begin{multline*} \int dx_1dx_2dx_3dx_4dx_5\psi^*_1(x_1,x_2)\psi^*_2(x_4)
\phi^*_1(x_5)\phi^*_2(x_3) \psi_1(x_1,x_2)\psi_2(x_4) \phi_1(x_5)\phi_2(x_3)
\\ = \langle \psi_1|\psi_1\rangle \langle \phi_1|\phi_1\rangle \langle
\psi_2|\psi_2\rangle \langle \phi_2|\phi_2\rangle = 1\ ,
\end{multline*} can give a non-zero result. Hence,
$$
\int dx_1dx_2dx_3dx_4dx_5\bar{\psi}^*(x_1,x_2,x_3,x_4,x_5)
\bar{\psi}(x_1,x_2,x_3,x_4,x_5) = 6
$$
and $\nu = 1/\sqrt{6}$. The same observation holds for the traces containing
$D$-local observable. For example,
$$
tr[\bar{\mathsf A}|\bar{\psi}\rangle \langle \bar{\psi}|]
$$
contains $6^3$ terms but only six survive, namely those in which the same
permutation of variables $x_1,x_2,x_3,x_4,x_5$ meet each other in
$\bar{\mathsf A}$ and $|\bar{\psi}\rangle \langle \bar{\psi}|$ and the same
permutation of variables $x'_1,x'_2,x'_3,x'_4,x'_5$ meet each other in
$\bar{\mathsf A}$ and $|\bar{\psi}\rangle \langle \bar{\psi}|$. For instance,
\begin{multline*} \int dx_1dx_2dx_3dx_4dx_5 dx'_1dx'_2dx'_3dx'_4dx'_5
A(x_2,x_4,x_5;x'_2,x'_4,x'_5) \\ \times \delta(x_1,x'_1) \delta(x_3,x'_3)
\psi^*_1(x_2,x_4)\psi^*_2(x_1) \phi^*_1(x_5)\phi^*_2(x_3)
\psi_1(x'_2,x'_4)\psi_2(x'_1) \phi_1(x'_5)\phi_2(x'_3) \\ = \int dx_2dx_4dx_5
dx'_2dx'_4dx'_5A(x_2,x_4,x_5;x'_2,x'_4,x'_5) \psi^*_1(x_2,x_4)
\phi^*_1(x_5)\psi_1(x'_2,x'_4) \phi_1(x'_5)\ .
\end{multline*} Hence,
$$
tr[\bar{\mathsf A}\bar{\mathsf T}] = tr[{\mathsf A}{\mathsf T}]
$$
where
$$
\bar{\mathsf T} = |\bar{\psi}\rangle \langle \bar{\psi}|
$$
and
$$
{\mathsf T} = |\psi_1\phi_1\rangle \langle \psi_1\phi_1|\ .
$$

We can state the following general key property of composition of
non-heterogeneous systems:
\begin{thm}\label{propold13} Let systems ${\mathcal S}$ and ${\mathcal S}'$ be
separated. Then, system ${\mathcal S} + {\mathcal S}'$ is prepared in state
$\bar{\mathsf T} = {\mathsf J}({\mathsf T},{\mathsf T}')$ with separation
status $D \cup D'$. Let further ${\mathsf A}$ be a $D$-local s.a.\ operator
for ${\mathcal S}$ and $\bar{\mathsf A}$ its extension to ${\mathcal S} +
{\mathcal S}'$. Then,
\begin{equation}\label{examp6} tr[\bar{\mathsf A}\bar{\mathsf T}] =
tr[{\mathsf A}{\mathsf T}]\ .
\end{equation}
\end{thm} In fact, this "Theorem" is only a conjecture because we
have not proved it for the general case stated in it.

Thus, the ambiguity of preparation has no observable consequences. The
resulting methods that use tensor products instead of full (anti-)symmetrised
tensor products are in agreement with the common practice in quantum
mechanics. In fact, they make quantum mechanics viable because the state of
the whole environment is never known. For example, in the theory of the
experiment described in Section 0.1.2, the state is prepared as a state of an
individual electron and its entanglement with all other electrons, which
exist, in fact, everywhere in huge amounts, is serenely ignored. Due to
Theorem \ref{propold13}, such method cannot lead to any problems.

\subsection{Change of separation status} In classical mechanics, the possible
states of system ${\mathcal S}$ are points of the phase space ${\mathbf
\Gamma}$ and possible observables are real function on ${\mathbf
\Gamma}$. Clearly, all such observables have definite values on ${\mathcal S}$
in a fixed state independently of external circumstances. ${\mathbf \Gamma}$
is uniquely associated with the system alone and forms the basis of its
kinematic description. Alternatively, we can always consider ${\mathcal S}$ as
a subsystem of a larger system $\bar{\mathcal S}$ with bigger phase space
$\bar{\mathbf \Gamma}$. In Newtonian mechanics, ${\mathbf \Gamma}$ is then a
subspace of $\bar{\mathbf \Gamma}$ and observables of ${\mathcal S}$ can be
extended to $\bar{\mathcal S}$ by defining them to vanish outside of ${\mathbf
\Gamma}$. Hence, there is an analogous ambiguity in the choice of space of
states in Newtonian mechanics as in quantum mechanics. However, no additional
conditions, such as suitable separation statuses, are needed there. Thus, the
quantum theory of observables is much more complicated than the Newtonian one:
not only their values cannot be ascribed to microsystem ${\mathcal S}$ alone
but some of them are not even registrable in principle due to the environment
of ${\mathcal S}$.

We assume that the quantum kinematics of a microsystem is defined
mathematically by possible states represented by all positive normalised
(trace one) operators, and possible observables represented by some POV
measures, on the Hilbert space associated with the system. Then the transition
from state ${\mathsf T} \otimes {\mathsf T}'$ to ${\mathsf J}({\mathsf T},
{\mathsf T}')$ as it occurs in Rule \ref{rlold14} is a change of kinematic
description.

Let us study this transition in more detail. We observe that ${\mathsf
P}_{\tau\tau'} : {\mathbf H}_\tau \otimes {\mathbf H}_{\tau'} \mapsto {\mathbf
H}_{\tau\tau'}$ is a linear but in general non-invertible and non-unitary
operator and that the normalisation is an even non-linear operation on the two
states. We can however show that the maps can be invertible in a special case
of separation statuses $D$ and $D'$ of ${\mathsf T}$ and ${\mathsf T}'$.

Let system ${\mathcal S}$ consist of $N$ and ${\mathcal S}'$ of $N'$
particles. Consider first vector states $\phi$ of ${\mathcal S}$ and
$\phi'$ of ${\mathcal S}'$. Let
$$
\Phi_{as} = {\mathsf P}_{\tau\tau'} (\phi \otimes \phi')\ ,\quad \Phi_{asn} =
{\mathsf J}(\phi, \phi') = \frac{{\mathsf P}_{\tau\tau'} (\phi \otimes
\phi')}{\sqrt{\langle{\mathsf P}_{\tau\tau'} (\phi \otimes \phi')|{\mathsf
P}_{\tau\tau'} (\phi \otimes \phi')\rangle}}\ .
$$
If ${\mathcal S}$ and ${\mathcal S}'$ are separated ($D \cap D' = \emptyset$),
then $\phi$ and $\phi'$ satisfy:
$$
\int d^3 x_i f'(\vec{x}_i)\phi(\vec{x}_1,\ldots,\vec{x}_N) = 0
$$
for any $i = 1,\ldots,N$ and for any test function $f'$ with $\text{supp} f'
\subset D'$, and
$$
\int d^3 x_i f(\vec{x}_i)\phi'(\vec{x}_1,\ldots,\vec{x}_N') = 0
$$
for any $i = 1,\ldots,N'$ and for any test function $f$ with $\text{supp} f
\subset D$.

Let $f'$ be a test function such that $f' \in {\mathbf H}_{\tau'}$ with
$\text{supp} f \subset (D'\times)^{N'}$, where $(D'\times)^{N'}$ is an
abbreviation for the Cartesian product of $N'$ factors $D'$. Let us define map
$R[f',D'] : {\mathbf H}_{\tau\tau'} \mapsto {\mathbf H}_\tau$ by
\begin{multline*} (R[f',D'] \Phi_{as})(\vec{x}_1,\ldots,\vec{x}_N) \\ = \int
d^3 x_{N+1}\ldots d^3 x_{N+N'}
f'(\vec{x}_{N+1},\ldots,\vec{x}_{N+N'})\Phi_{as}(\vec{x}_1,\ldots,\vec{x}_N,\vec{x}_{N+1},\ldots,\vec{x}_{N+N'})\
,
\end{multline*} and similarly, for test function $f \in {\mathbf H}_\tau$ and
$\text{supp} f \subset (D\times)^N$, $R[f,D] : {\mathbf H}_{\tau\tau'} \mapsto
{\mathbf H}_{\tau'}$ by
\begin{multline*} (R[f,D] \Phi_{as})(\vec{x}_{N+1},\ldots,\vec{x}_{N+N'}) \\ =
\int d^3 x_1\ldots d^3 x_N
f(\vec{x}_1,\ldots,\vec{x}_N)\Phi_{as}(\vec{x}_1,\ldots,\vec{x}_N,\vec{x}_{N+1},\ldots,\vec{x}_{N+N'})\
.
\end{multline*} Then, we obtain easily:
$$
R[f',D']\Phi_{as} = \nu_f' \phi(\vec{x}_1,\ldots,\vec{x}_N)\ ,
$$
where
$$
\nu_f' = \nu_{\tau\tau'} \int d^3 x_{N+1}\ldots d^3
x_{N+N'}f'(\vec{x}_{N+1},\ldots,\vec{x}_{N+N'})\phi'(\vec{x}_{N+1},\ldots,\vec{x}_{N+N'})\
,
$$
and $\nu_{\tau\tau'}$ is the normalisation factor defined by
$P_{\tau\tau'}$. $\nu_f'$ is non-zero for at least some $f'$. Similarly,
$$
R[f,D]\Phi_{as} = \nu_f \phi'(\vec{x}_{N+1},\ldots,\vec{x}_{N+N'})\ ,
$$
where
$$
\nu_f = \nu_{\tau\tau'} \int d^3 x_1\ldots d^3 x_N
f(\vec{x}_1,\ldots,\vec{x}_N) \phi(\vec{x}_1,\ldots,\vec{x}_N) \ .
$$
Thus, we obtain both functions $\phi(\vec{x}_1,\ldots,\vec{x}_N)$ and
$\phi'(\vec{x}_{N+1},\ldots,\vec{x}_{N+N'})$ up to normalisation. As the
functions are normalised, they can be reconstructed. Analogous steps work for
$\Phi_{asn}$.

For the generalisation of these ideas to state operators, we shall need
adjoints of operators $R[f',D']$ and $R[f,D]$. The definition of
$R[f,D']^\dagger : {\mathbf H}_\tau \mapsto {\mathbf H}_{\tau\tau'}$ is
$$
(R[f',D']^\dagger \phi,\Phi) = (\phi,R[f',D']\Phi)
$$
for all $\phi \in {\mathbf H}_\tau$ and $\Phi \in {\mathbf H}_{\tau\tau'}$. A
simple calculation yields
$$
R[f',D']^\dagger \phi = {\mathsf P}_{\tau\tau'}(\phi \otimes f^{\prime *})\ .
$$
Similarly,
$$
R[f,D]^\dagger \phi' = {\mathsf P}_{\tau\tau'}(f^* \otimes \phi')\ .
$$

Map ${\mathsf T} \otimes {\mathsf T}' \mapsto {\mathsf P}_{\tau\tau'}({\mathsf
T} \otimes {\mathsf T}'){\mathsf P}_{\tau\tau'}$ is linear in both ${\mathsf
T}$ and ${\mathsf T}'$ and its result is an operator on ${\mathbf H}_\tau
\otimes {\mathbf H}_{\tau'}$ that leaves ${\mathbf H}_{\tau\tau'}$
invariant. Operator ${\mathsf P}_{\tau\tau'}({\mathsf T} \otimes {\mathsf
T}'){\mathsf P}_{\tau\tau'} : {\mathbf H}_{\tau\tau'} \mapsto {\mathbf
H}_{\tau\tau'}$ is self-adjoint and positive if ${\mathsf T}$ and ${\mathsf
T}'$ are state operators but it is not normalised. Let $\{\psi_n\}$ be a basis
${\mathbf H}_\tau$ and $\{\psi'_\alpha\}$ that of ${\mathbf H}_\tau$. We can
write
$$
{\mathsf T} = \sum_{mn} T_{mn}|\psi_m\rangle\langle\psi_n|\ ,\quad {\mathsf
T}' = \sum_{\alpha\beta} T'_{\alpha\beta}
|\psi'_\alpha\rangle\langle\psi'_\beta|\ .
$$
Then
$$
{\mathsf P}_{\tau\tau'}({\mathsf T} \otimes {\mathsf T}'){\mathsf
P}_{\tau\tau'} = \sum_{mn}\sum_{\alpha\beta}T_{mn}T'_{\alpha\beta} |{\mathsf
P}_{\tau\tau'}(\psi_m\otimes \psi'_\alpha)\rangle \langle{\mathsf
P}_{\tau\tau'}(\psi_n \otimes \psi'_\beta)|\ .
$$
Now, the above proof that vector states $\phi$ and $\phi'$ can be
reconstructed from ${\mathsf J}(\phi, \phi')$ can be easily extended to
general state operators ${\mathsf T}$ and ${\mathsf T}'$ by expanding the
state operators into the bases and acting by $R$'s from the left and
$R^\dagger$'s from the right on them.

Moreover, for separated systems, the "individual" observables from
${\mathbf A}[{\mathcal S}]_D$ and ${\mathbf A}[{\mathcal S}']_{D'}$ can be
recovered from operators on ${\mathbf H}_{\tau\tau'}$ that are extensions of
operators either of ${\mathbf A}[{\mathcal S}]_D$ or of ${\mathbf A}[{\mathcal
S}']_{D'}$. Let us show it for a simple example.
\par \vspace*{.4cm} \noindent {\bf Example} Let ${\mathcal S}$ be a fermion
particle and ${\mathcal S}'$ a composite of one fermion of the same type as
${\mathcal S}$ and some particle of a different type. Let $\phi(\vec{x}_1)$ be
an arbitrary element of ${\mathbf H}$ and $\phi'(\vec{x}_2,\vec{x}_3)$ that of
${\mathbf H}_{\tau'}$, $\vec{x}_2$ being the coordinate of the fermion. Then
$$
\Psi(\vec{x}_1,\vec{x}_2,\vec{x}_3) = {\mathsf
P}_{\tau\tau'}\big(\phi(\vec{x}_1)\phi'(\vec{x}_2,\vec{x}_3)\big) =
\frac{1}{2}\big(\phi(\vec{x}_1)\phi'(\vec{x}_2,\vec{x}_3) -
\phi(\vec{x}_2)\phi'(\vec{x}_1,\vec{x}_3)\big)\ .
$$
\par \vspace*{.4cm} \noindent Let ${\mathsf a} \in {\mathbf A}[{\mathcal
S}]_D$. Then its extension ${\mathsf A}$ is an operator on ${\mathbf
H}_{\tau\tau'}$ defined by its kernel
$$
a(\vec{x}_1;\vec{x}'_1) \delta(\vec{x}_2 - \vec{x}'_2) \delta(\vec{x}_3 -
\vec{x}'_3) + a(\vec{x}_2;\vec{x}'_2) \delta(\vec{x}_1 - \vec{x}'_1)
\delta(\vec{x}_3 - \vec{x}'_3)
$$
so that
$$
({\mathsf A}\Psi)(\vec{x}_1,\vec{x}_2,\vec{x}_3) = \frac{1}{2}\big(({\mathsf
a}\phi)(\vec{x}_1)\phi'(\vec{x}_2,\vec{x}_3) - ({\mathsf
a}\phi)(\vec{x}_2)\phi'(\vec{x}_1,\vec{x}_3)\big)\ .
$$
Then,
$$
R[f',D']({\mathsf A}\Psi) = \nu_f' ({\mathsf a}\phi)(\vec{x}_1)\ ,
$$
where
$$
\nu_f' = \frac{1}{2} \int d^3 x_2d^3 x_3 f'(\vec{x}_2,\vec{x}_3)
\phi'(\vec{x}_2,\vec{x}_3)\ .
$$
But $\phi(\vec{x}_1)$, $\phi'(\vec{x}_2,\vec{x}_3)$ and
$f'(\vec{x}_2,\vec{x}_3)$ are known, hence, as $\phi$ is arbitrary, ${\mathsf
a}$ is well-defined.

To summarise: for separated systems ${\mathcal S}$ and ${\mathcal S}'$, there
are two equivalent descriptions: the {\em standard QM description} of
${\mathcal S} + {\mathcal S}'$ on the Hilbert space ${\mathbf H}_{\tau\tau'}$
and the {\em untangled QM description} on ${\mathbf H}_\tau \otimes {\mathbf
H}_{\tau'}$ explained above.

As yet, the considerations apply to situations at a fixed instant of time. The
new aspect that time evolution can introduce is that separation status of a
system in a state can change in time. Let us define mathematically what this means.

First, we come to the notion of formal evolution.
\begin{df} Let system ${\mathcal S}$ be initially ($t = t_1$) prepared in
state ${\mathsf T}$, another quantum system ${\mathcal S}'$ in state ${\mathsf
T}'$ and let them be separated at $t_1$. Let the composite have a
time-independent Hamiltonian defining a unitary group ${\mathsf U}(t-t_1)$ of
evolution operators on ${\mathbf H}_{\tau\tau'}$. Then, the standard quantum
mechanical evolution of ${\mathcal S} + {\mathcal S}'$,
\begin{equation}\label{sqme} \bar{\mathsf T}(t) = {\mathsf U}(t-t_1){\mathsf
J}\Big({\mathsf T} \otimes {\mathsf T}'\Big){\mathsf U}(t-t_1)^\dagger\ ,
\end{equation} is called {\em formal evolution} of two interacting systems
${\mathcal S}$ and ${\mathcal S}'$.
\end{df} The idea is analogous to the well-know time-dependent Hartree--Fock
method in the theory of nuclear fusion \cite{SAL}. Thus, the formal evolution
uses the standard QM description. It is called "formal" because the
character of the separation statuses can change during the evolution and it is
not clear whether the standard quantum mechanics is then still
applicable. Indeed, this evolution does not agree with observation of
separation status changes that occur during registrations. However, the formal
evolution is our first step in the mathematical analysis of separation status
changes. With its help, we can decide whether a change of separation status
has taken place in a given theoretical model. Let us study an example in some
detail.

Let ${\mathcal S}$ and ${\mathcal S}'$ be two quantum systems, ${\mathcal S}$
containing $N$ particles and ${\mathcal S}'$ containing $N'$ particles. Let
the systems be prepared, at time $t_1$, in states ${\mathsf T}$ and ${\mathsf
T}'$ with non-trivial separation statuses $D_1$ and $D'$, respectively, and
$D_1 \cap D' = \emptyset$. Thus, ${\mathcal S}$ and ${\mathcal S}'$ are
separated at $t_1$. Let the formal evolution of the composite ${\mathcal S} +
{\mathcal S}'$ for the initial state $\bar{\mathsf T}(t_1) = {\mathsf
J}({\mathsf T}, {\mathsf T}')$ be described by its kernel in
$Q$-representation:
$$
\bar{T}(t)(\vec{x}_1,\ldots,\vec{x}_N, \vec{x}_{N+1},\ldots,\vec{x}_{N+N'};
\vec{x}'_1,\ldots,\vec{x}'_N,\vec{x}'_{N+1},\ldots,\vec{x}'_{N+N'})\ .
$$
\begin{enumerate}
\item Suppose that, for some $t_2 > t_1$, $\text{supp}\ \bar{T}(t_2) =
(D'\times)^{2(N+N')}$\footnote{This can easily be generalised to a more
realistic condition, e.g., $\int_{(D'\times)^{N+N'}}d^3x_1\ldots
d^3x_{N+N'}\bar{T}(t_2)(\vec{x}_1,\ldots,\vec{x}_{N+N'};\vec{x}_1,\ldots,\vec{x}_{N+N'})
\approx 1$.}. Then we can say: at time $t_2$, the separation status of
${\mathcal S}$ is $\emptyset$, that of ${\mathcal S}'$ is $D'$ and that of the
composite ${\mathcal S} + {\mathcal S}'$ is also $D'$ or, that ${\mathcal S}$
is {\em swallowed} by ${\mathcal S}'$.
\item Suppose that, for some $t_3 > t_2$, there is an open set $D_3 \subset
{\mathbb R}^3$, $D_3 \cap D' = \emptyset$, such that the kernel $T_{\mathsf
J}(t_3)$ has the properties:
\begin{enumerate}
\item For any test function $f' \in {\mathbf H}_{\tau'}$ and
$$
\text{supp}\,f' = (D'\times)^{N'}\ ,\quad R[f',D']T_{\mathsf
J}(t_3)R[f',D']^\dagger \neq 0\ ,
$$
$\nu R[f',D']\bar{T}(t_3)R[f',D']^\dagger$ is a state operator of ${\mathcal
S}$ independent of $f'$, where $\nu$ is the normalisation factor.
\item For any test function $f \in {\mathbf H}_\tau$ and
$$
\text{supp}\,f = (D_3 \times)^N\ ,\quad R[f,D_3]T_{\mathsf
J}(t_3)R[f,D_3]^\dagger \neq 0\ ,
$$
$\nu R[f,D_3]\bar{T}(t_3)R[f,D_3]^\dagger$ is a state operator of ${\mathcal
S}'$ independent of $f$, where $\nu$ is the normalisation factor.
\item For any test function $g \in {\mathbf H}_\tau$ and $\text{supp}\,g =
(D_3 \times)^N$, we have
$$
R[f',D']T_{\mathsf J}(t_3)R[f',D']^\dagger|g\rangle = 0\ .
$$
\item For any test function $g' \in {\mathbf H}_{\tau'}$ and $\text{supp}\, g'
= (D' \times)^{N'}$, we have
$$
R[f,D_3]T_{\mathsf J}(t_3)R[f,D_3]^\dagger|g'\rangle = 0\ .
$$
\end{enumerate} Then we can say: the systems become separated again at time
$t_3 > t_2$, system ${\mathcal S}$ being in state $\nu R[f',D']T_{\mathsf
J}(t_3)R[f',D']^\dagger$ with separation status $D_3$ and system ${\mathcal
S}'$ in state $\nu R[f,D_3]T_{\mathsf J}(t_3)R[f,D_3]^\dagger$ with separation
status $D'$.
\end{enumerate}

Thus, we judge on separation statuses of the two systems by studying the
supports of the kernels of the $Q$-representation of their state operators
during the formal (i.e., ordinary unitary) evolution of their composite. The
change from $D_1$ at $t_1$ to $\emptyset$ at $t_2$ and to $D_3$ at $t_3$ is a
complicated function of the evolution of the whole composite system. As for
the observables, their unitary evolution is, in fact, irrelevant to what can
be registered. Consider, e.g., position of ${\mathcal S}$. The standard
position operator ${\mathsf Q}$ can never serve as observable "position
of ${\mathcal S}$". The $D_1$ localisation of ${\mathsf Q}$ is
registrable and its meaning is "position of ${\mathcal S}$" at
$t_1$. But it is not "position of ${\mathcal S}$" at $t_2$ or $t_3$
because at $t_2$, ${\mathcal S}$ does not possess any observable of its own,
including position. At $t_3$, "position of ${\mathcal S}$" is
$D_3$-localisation of ${\mathsf Q}$. At any time, one can construct the
extensions of the corresponding localisations to the whole composite, but the
registrable meaning of these extensions changes with time. Thus, the
observables change with time even if we are working in Schr\"{o}dinger
representation.

Although we have based the time process on a unitary evolution (the formal
evolution), the time evolution of genuinely registrable properties of
${\mathcal S}$ does not look like a unitary evolution. And, although we can
find the separation statuses of ${\mathcal S}$ and ${\mathcal S}'$ by studying
the formal evolution of ${\mathcal S} +{\mathcal S}'$, we cannot claim that
the formal evolution gives the physical state of the composite. The question
even seems natural, whether the formal evolution ought to be further corrected
in the case that it leads to separation-status changes.

What has been said up to now shows that standard quantum mechanics is
incomplete in the following sense:
\begin{enumerate}
\item It accepts and knows only two separation statuses:
\begin{enumerate}
\item that of isolated systems, $D = {\mathbb R}^3$, with the standard
operators (position, momentum, energy, spin, etc.) as observables, and
\item that of a member of a system of identical particles, $D = \emptyset$,
with no observables of its own.
\end{enumerate}
\item It disregards the fact that separation status can change during time
evolution. In particular, it does change during preparations and
registrations, and that makes the measurement a process physically
different from most other processes considered by quantum mechanics. The
question naturally arises, whether the unitary evolution law provides an
adequate description to such changes.
\end{enumerate} This suggests that quantum mechanics can be supplemented by a
theory of general separation status and by new rules that govern processes in
which separation status changes. The new rules must not contradict the rest of
quantum mechanics and ought to agree with, and to explain, observational
facts.

\section{State reduction} The standard quantum theory of indistinguishable
particles as explained in the foregoing sections leads to an important but as
yet insufficiently studied or even ignored phenomenon: A prepared state of a
quantum system will often be mangled and degraded during an interaction with a
large system such as a macroscopic body. We consider this to be an objective
change of the state similarly as worn boots are objectively different from new
boots. Let us give a simple example.

In many optical experiments, such as \cite{RSH}, polarisers, such as
Glan--Thompson ones, are employed. A polariser is a macroscopic body that
decomposes the coming light into two orthogonal-polarisation parts. One part
disappears inside an absorber and the other is left through practically
unchanged. Similarly, in most quantum experiments, one or more screens are
used. A screen is a macroscopic body that decomposes the incoming state into
one part that disappears inside the body and the other that evolves further.

Disappearance of a quantum system ${\mathcal S}$ in a macroscopic body
${\mathcal B}$ is the following process. The body is assumed to be a perfect
absorber. First, ${\mathcal S}$ enters ${\mathcal B}$ and ditches most of its
kinetic energy somewhere inside ${\mathcal B}$. Second, the energy passed to
${\mathcal B}$ is dissipated and distributed homogeneously through ${\mathcal
B}$ in a process aiming at thermodynamic equilibrium. In this way, ${\mathcal
S}$ ceases to be separated from the other systems of the same type within
${\mathcal B}$ by its energy. Moreover, a photon might be annihilated and a
massive particle becomes entangled with all other particles of its type inside
${\mathcal B}$ and its separation status by position becomes trivial. In this
way, ${\mathcal S}$ ceases to exist as an individual object and no more
registrations can be done on it. This can be viewed as a complete or
partial loss of the system because it becomes undistinguishable from all
subsystems of ${\mathcal B}$ that are of the same type as ${\mathcal S}$.

Mathematical description of the initial and the final state of the composite
${\mathcal S} + {\mathcal B}$ can be easily given. Let ${\mathcal S}$ be a
particle of type $\tau$ and $\psi(\vec{x})$ the wave function of its initial
state prepared with a separation status $D$. Let screen ${\mathcal B}$ be a
macroscopic quantum system of type $\tau'$ with separation status $D'$ having
sufficiently large common boundary with $D$. Let
\begin{equation}\label{decomp1} \psi(\vec{x}) = c_{\text{thr}}
\psi_{\text{thr}}(\vec{x}) + c_{\text{sw}} \psi_{\text{sw}}(\vec{x})
\end{equation} be the decomposition of the initials state, where
$\psi_{\text{thr}}(\vec{x})$ is a normalised wave function of the part that
will be left through and $\psi_{\text{sw}}(\vec{x})$ that that will be
swallowed by ${\mathcal B}$. This decomposition is determined by the nature of
${\mathcal B}$: for a polariser, these are the two orthogonal polarisation
states, and for a simple screen, these can be calculated from the geometry of
${\mathcal B}$ and the incoming beam as it is usually done e.g.\ in accounts
of a double-slit experiment.

The initial state of ${\mathcal B}$ is a classical state, which is a high
entropy one (see Chapter 3). It is, therefore, described by a state operator
${\mathsf T}$. The initial state of the composite is then
\begin{equation}\label{screeni} \bar{\mathsf T}_i = |\psi\rangle \langle\psi|
\otimes {\mathsf T}\ .
\end{equation} Now, the initial state for the formal evolution of the
composite is
$$
\bar{\mathsf T}_{\text{fei}} = \nu {\mathsf P}_{\tau\tau'}(|\psi\rangle
\langle\psi| \otimes {\mathsf T}){\mathsf P}_{\tau\tau'}\ ,
$$
where $\nu = tr[{\mathsf P}_{\tau\tau'}(|\psi\rangle \langle\psi| \otimes
{\mathsf T}){\mathsf P}_{\tau\tau'}]$ and ${\mathsf P}_{\tau\tau'}$ is defined
by Equation (\ref{projTT'}). Using decomposition (\ref{decomp1}), we can write
\begin{multline}\label{formscreeni} {\mathsf T}_{\text{fei}} = \nu
\Big(c^*_{\text{thr}}c_{\text{thr}}{\mathsf
P}_{\tau\tau'}(|\psi_{\text{thr}}\rangle \langle\psi_{\text{thr}}| \otimes
{\mathsf T}){\mathsf P}_{\tau\tau'} + c_{\text{thr}}c^*_{\text{sw}}{\mathsf
P}_{\tau\tau'} (|\psi_{\text{thr}}\rangle \langle\psi_{\text{sw}}| \otimes
{\mathsf T}){\mathsf P}_{\tau\tau'} \\ + c_{\text{sw}}c^*_{\text{thr}}{\mathsf
P}_{\tau\tau'} (|\psi_{\text{sw}}\rangle \langle\psi_{\text{thr}}| \otimes
{\mathsf T}){\mathsf P}_{\tau\tau'} + c^*_{\text{sw}}c_{\text{sw}} {\mathsf
P}_{\tau\tau'} (|\psi_{\text{sw}}\rangle \langle\psi_{\text{sw}}| \otimes
{\mathsf T}){\mathsf P}_{\tau\tau'}\Big)\ .
\end{multline}

Let ${\mathsf U}$ be the unitary operator that describes the formal evolution
on the Hilbert space ${\mathbf H}_{\tau\tau'}$ of the composite. After the
process is finished, we obtain
\begin{multline}\label{formscreenf} \bar{\mathsf T}_{\text{fef}} = \nu
\Big(c^*_{\text{thr}}c_{\text{thr}}{\mathsf
P}_{\tau\tau'}(|\psi'_{\text{thr}}\rangle \langle\psi'_{\text{thr}}| \otimes
{\mathsf T}_{\text{thr}}){\mathsf P}_{\tau\tau'} +
c_{\text{thr}}c^*_{\text{sw}} {\mathsf U}{\mathsf P}_{\tau\tau'}
(|\psi_{\text{thr}}\rangle \langle\psi_{\text{sw}}| \otimes {\mathsf
T}){\mathsf P}_{\tau\tau'}{\mathsf U}^\dagger \\ +
c_{\text{sw}}c^*_{\text{thr}} {\mathsf U}{\mathsf P}_{\tau\tau'}
(|\psi_{\text{sw}}\rangle \langle \psi_{\text{thr}}| \otimes {\mathsf
T}){\mathsf P}_{\tau\tau'}{\mathsf U}^\dagger \Big) +
c^*_{\text{sw}}c_{\text{sw}}\bar{\mathsf T}'\ ,
\end{multline} where
$$
\bar{\mathsf T}' = \nu{\mathsf U}{\mathsf P}_{\tau\tau'}
(|\psi_{\text{sw}}\rangle \langle\psi_{\text{sw}}| \otimes {\mathsf
T}){\mathsf P}_{\tau\tau'}{\mathsf U}^\dagger
$$
is the end state of the screen with the swallowed part of ${\mathcal S}$ and
we have assumed that
$$
{\mathsf U}{\mathsf P}_{\tau\tau'} (|\psi_{\text{thr}}\rangle
\langle\psi_{\text{thr}}| \otimes {\mathsf T}){\mathsf P}_{\tau\tau'}{\mathsf
U}^\dagger = {\mathsf P}_{\tau\tau'}(|\psi'_{\text{thr}}\rangle
\langle\psi'_{\text{thr}}| \otimes {\mathsf T}_{\text{thr}}){\mathsf
P}_{\tau\tau'}\ ,
$$
where $\psi'_{\text{thr}}$ is the wave function of $\mathcal S$ with
separation status $D_{\text{thr}}$ describing the part that went through,
$D_{\text{thr}} \cap D = \emptyset$, $D_{\text{thr}} \cap D' = \emptyset$, and
${\mathsf T}_{\text{thr}}$ is the corresponding state of the screen with
separation status $D'$.

The crucial step now is that the two terms containing products of the
left-through and the swallowed parts of $\mathcal S$ are discarded so that the
physical final state of the composite is
\begin{equation}\label{screenf} \bar{\mathsf T}_f = p_{\text{thr}}
\frac{|\psi'_{\text{thr}}\rangle \langle\psi'_{\text{thr}}|}{p_{\text{thr}}}
\otimes {\mathsf T}_{\text{thr}}\ (+)_p\ p_{\text{sw}} \bar{\mathsf T}'\ ,
\end{equation} where
$$
p_{\text{thr}} = c^*_{\text{thr}}c_{\text{thr}}\ ,\quad p_{\text{sw}} =
c^*_{\text{sw}}c_{\text{sw}}\ ,
$$

The change from (\ref{formscreenf}) to (\ref{screenf}) is called {\em state
reduction}. It is not a unitary transformation: the non-diagonal terms in
(\ref{formscreenf}) have been erased. The sign "$(+)_p$" suggests
that the convex combination is a statistical decomposition (see Section
1.1.2). Thus, not only some terms have been erased but also the state of the
composite has been further changed.

What part of the state operator is to be erased must be judged both with the
help of an assessment of the experimental arrangement resulting in a
theoretical model thereof and calculation of the unitary evolution of an
initial state, which had to lead to decomposition (\ref{decomp1}). Thus,
equation (\ref{screenf}) is a result of a judicious decision that may but does
not necessarily work because it may but does not necessarily express the reality with sufficient
accuracy. It is analogous to the decision of which state has been prepared by a
given experimental setup of a preparation apparatus.

In our previous work \cite{hajicek2,survey,hajicek4}, we have assumed that
transformation analogous to that from (\ref{decomp1}) to (\ref{screenf})
result from application of some general alternative dynamical law and that
such a law must be postulated. An extended study of all empirical cases that
came to mind as yet has shown that a number of specific details must in each
case be taken into account so as to make an adequate theoretical description
of what happens. Schr\"{o}dinger equation gives a physical
evolution only in ideal cases of isolated systems. Under more general
conditions, Schr\"{o}dinger equation gives only a formal evolution. Then,
the results of Schr\"{o}dinger evolution must be suitably corrected to
express the resulting state degradation. This is the content of the following
rule.

\begin{rl}\label{rldegrad} Let ${\mathcal S}$ be a microscopic quantum system
and ${\mathcal A}$ a macroscopic one in a classical (high entropy) state. Let
there be process with an interaction between them such that the resulting
change in the state of ${\mathcal A}$ includes a dissipation of a portion of
the state within a macroscopic part of the degree of freedom of ${\mathcal
A}$. Then the end state of the formal evolution of ${\mathcal S} + {\mathcal
A}$ must be corrected by discarding all terms that express correlation between
macroscopically different end states of the composite, and the resulting
convex combination of states is a statistical decomposition.
\end{rl}

The state reduction above is formally similar to what is often called the
collapse of wave function or the state reduction or the dynamical state
reduction. State reduction was declared to be a basic new kind of dynamics
that sometimes replaces the unitary dynamics \cite{JvN} or always corrects a
unitary evolution \cite{GRW,pearle} but no cause for the state reduction has
been given (for an extended discussion see Sections 4.1 and 4.2). As explained
above, the state reduction runs parallel to, and the reason for it is provided
by, the unitary evolution together with relevant specific empirical data for
each case. Its basic feature is a complete or partial disappearance of the
system due to the dissipation and this is the cause of a state degradation, the
result of which is a state reduction.

Rule \ref{rldegrad} formulates only few general features and leaves some
freedom in the choice of exact mathematical details that must be assumed
separately for each particular object studied so that the resulting model
corresponds well to its observed properties. This is, in fact, similar to
Schr\"{o}dinger equation, which is generally restricted only in its
overall general features and its details must also be assumed for each model
separately with the aim to yield a good model of the object.

The described action of the screen on incoming individual system ${\mathcal
S}$ is not a registration: by itself, it does not deliver the value of any
observable of ${\mathcal S}$. Hence, it is either a preparation or a part of a
registration. For example, in experiment \cite{RSH}, the position of photons
leaving the polariser is measured by a photodiode. Thus, a detector is needed
to accomplish a registration. We shall study various models of registration in
Chapter 4 and many examples will show how Rule \ref{rldegrad} is applied.

\part{The models} The treasure of successful models is the primary part of any
physical theory. Textbooks of quantum mechanics dedicate most of their text to
models of atoms and molecules, to scattering theory of particles on atoms and
molecules, to solid bodies etc. A general method of such constructions has
been described in Section 1.3.1. The only part of the textbooks that has to be
changed concerns the operators that are used in the construction and referred
by the textbooks as "observables". This name is not correct because
we have seen in Section 2.2.1 that the measurement of all such operators would
be disturbed by the environment.

For particles, a $D$-localisation (see Definition \ref{dfold15}) of these
operators are already genuine observables. For composite systems, such as
atoms, the construction of position and momentum observables of their mass
centre is analogous, but other observables may lead to more
complications. Real experiments must be carefully studied and the
corresponding observable must be constructed accordingly. For example, the
energy spectrum of hydrogen atom is usually measured indirectly via the energy
of photons scattered off the atoms. The corresponding observable will not be
just a $D$-localisation of the Hamiltonian operator. How observables
describing indirect registrations are constructed is well known (see, e.g.,
\cite{peres}, p.\ 282). We do not expect any contradictions to our new rules
or difficult mathematical problems that would hinder such
constructions. Hence, we shall skip the whole menagerie of models of
microscopic systems and restrict ourselves only to those models that are
immediately important for our main aim: to deal with the problems of classical
properties (Chapter 3) and of quantum measurement (Chapter 4).

To construct models of classical world will require, in addition to the
already described changes of language, some further new ideas, which are
specific to the particular objects to be modelled.

\chapter{Quantum models of classical properties} There are many classical
aspects of real objects that have been successfully modelled by quantum
mechanics, such as electrical conductivity or specific heats. These are
typically phenomena that occur in systems with very many degrees of freedom so
that statistical methods can be used. The statistical methods were invented
already before quantum mechanics was born and introduced some elements that
could be understood only later by quantum mechanics. For example, the
microcanonical or canonical ensemble is, in fact, methods of preparation of
thermodynamic systems. Or, theoretical results are given in the form of
averages and variances. The modern condensed matter theory around room
temperature can, therefore, be included into our theory of classical
properties without much change. We would just utilise the objectivity of
averages and variances in our interpretation of quantum mechanical results.

However, the Galilean invariance of quantum theory leads to separation of the
overall motion from all other degrees of freedom. The motion of mass centre
and of the total angular momentum with respect to the mass centre comprises
only six degrees of freedom that do not seem to allow statistical
methods. Exactly this kind of motion is studied by Newtonian mechanics. Thus,
the situation is that there are quantum models of classical thermodynamic
properties but none of mechanical properties that would be really
satisfactory.

Quantum modelling of non-thermodynamic properties of classical systems
encounters two main problems. First, a key feature of Newtonian mechanics (and
any other classical theory as well) is that each system objectively has a
sharp trajectory. Any fuzziness is just due to incomplete knowledge. In
particular, the state of a Newtonian system is described by a point of its
phase space, and the system is always in a definite state, i.e., it cannot be
at two points of the phase space simultaneously (see also the discussion at
the end of Section 1.2.2). Second, the system is robust so that measurements
can be done on it without changing its properties. For example, the state of a
system can be determined or confirmed by a suitable set of measurements on the
system.

Thus, any quantum model of a classical system must satisfy the first two
conditions of what Leggett has called Principle of Macroscopic Realism
\cite{leggett}:
\begin{enumerate}
\item A macroscopic system that has available to it two or more distinct
macroscopic states is at any given time in a definite one of those states.
\item It is possible in principle to determine which of these states the
system is in without any effect on the state itself or on the subsequent
system dynamics.
\end{enumerate}

In trying to model the sharpness of classical states and trajectories, one may
be mislead to overestimate the importance of quantum states of minimum
uncertainty, which is of coherent states. However, such states are always
extremal states, which can be linearly superposed, and quantum mechanics
requires that linear superpositions of available states are also an available
state. Moreover, measurements of the classical parameters of a coherent state
necessarily disturb the value of the parameters.

To solve this problem, one could e.g.\ assume that some as yet unknown
phenomena exist at the macroscopic level that are not compatible with
standard quantum mechanics. For example, they may prevent linear
superpositions (see, e.g., \cite{leggett} and the references
therein). However, no such phenomena have been observed.

Another strategy is to assume that the macroscopic realism is only apparent in the
sense that there {\em are} linear superpositions of macroscopic states but the
corresponding interference phenomena are difficult or impossible to
observe. For example, the quantum decoherence theory \cite{Zurek, Zeh} works
only if certain observables concerning both the environment and the quantum
system cannot be measured (see the analysis in \cite{d'Espagnat,bub}). Another
example is the theories based on coarse-grained operators
\cite{peres,poulin,Kofler} being measurable but fine grained being not. The
third example is the Coleman--Hepp theory \cite{Hepp,Bell3,Bona} and its
modifications \cite{Sewell,Primas,wanb}: they are based on some particular
theorems that hold only for infinite systems (see the analysis in
\cite{Bell3}) or for asymptotic regions \cite{wanb}.

However, if we turn from theory to experiment, we may notice that any
well-founded scientific observation of classical properties always has a
statistical form. A measurement or observation is only viewed as well
understood if it is given as an average with a variance. This fact does not
by itself contradict the sharp character of the corresponding theory. The
usual excuse is that the observation methods are beset with inaccuracy but
that improvement of techniques can lead to better and better results
approaching the "objective sharp" values arbitrarily closely. In any
case, however, the measured classical parameters of real objects are much
fuzzier than the minimal quantum uncertainty requires.

Moreover, that popular excuse is clearly incompatible with the assumption that
the classical world is only an aspect of a deeper quantum world and that each
classical model is nothing but a kind of incomplete description of the
underlying quantum system. If we assume such universality of quantum theory,
then the statistical character of classical observational results must not
only be due to inaccuracy of observational methods but also to genuine
uncertainty of quantum origin. This point of view is due to Exner
\cite{Exner}, p.~669, and Born \cite{Born} and will be adopted here as a
starting point of our theory of classical properties.

We can formulate this idea in terms of the Realist Model Approach as
follows. The language part of classical theories contains the notions of sharp
state and trajectory. These are idealised notions that do not possess any
counterpart in the real world, but they are useful for model construction.

In this chapter, we first formulate some general hypotheses that can be
applied to both thermodynamic and mechanical properties, introducing thus a
unified theory of classical properties: they turn out to be selected objective
properties of high-entropy quantum states of macroscopic systems. Next, we
show in detail how these ideas are to be applied to Newtonian mechanics,
introducing states called ME packets. Then, we construct a quantum model of a
classical rigid body. Finally, we modify the well-known model of a
simultaneous measurement of position and momentum of a Gaussian wave packet to
that of position and momentum of a ME packet.

Thus, our project to construct quantum models of observed classical systems
seems to work nicely. What remains open is the question of what is the origin
of all the high-entropy states that are observed in such a great abundance
around us.

\section{Modified correspondence principle} The Born--Exner assumption has
quite radical consequences, which is only seldom realised. First, the exactly
sharp states and trajectories of classical theories are not objective. They do
not exist in reality but are only idealisations. What really exist are
fuzzy states and trajectories. The objectivity of fuzzy states of classical
models is a difficult point to accept and understand. Let us explain it in
more detail.

In quantum mechanics, the basis of objectivity of dynamical properties is the
objectivity of the conditions that define preparation procedures. In other
words, if a property is uniquely determined by a preparation, then it is an
objective property. If we look closely, one hindrance to try the same idea in
classical theories is the custom always to speak about initial data instead of
preparations. An initial datum can be and mostly is a sharp state. The
question on how an exactly sharp state can come into being is ignored. This in
turn seems justified by the hypothesis that sharp states are objective, that
is, they just exist by themselves.

To come away from this self-deception, we accept that preparation procedures
play the same basic role in classical as in quantum physics. Then, the
nature and form of necessary preparation procedures must be specified and the
corresponding states described. In this way, the Exner--Born idea leads to a
rather radical change of interpretation of classical theories and this will
enable us to construct quantum models of all classical aspects of real
objects.

An obvious starting point of such constructions is that all classical systems
are also quantum systems. Let us now make this more precise. Consider a real
physical object (so to speak, independent of any theory). If there is an
adequate classical model ${\mathcal S}_c$ of some aspects of this system the
system is called classical. Adequate means that some properties of ${\mathcal
S}_c$ approximately represent important properties of the real system. In
addition, the same real object must also be understood in terms of a quantum
model ${\mathcal S}_q$. ${\mathcal S}_q$ could be richer than ${\mathcal S}_c$
so that classical properties of ${\mathcal S}_c$ can be identified with some
quantum properties of ${\mathcal S}_q$. In particular, these quantum
properties ought to be objective. This follows from the fact that all
classical properties can be assumed to be objective without any danger of
contradictions.

The construction of quantum model ${\mathcal S}_q$ consists of the following
points. 1) The composition of ${\mathcal S}_q$ must be defined. 2) The
observables that can be measured on ${\mathcal S}_q$ are to be determined. On
the one hand, this is a non-trivial problem because there are relatively
strong restrictions on what observables of macroscopic systems can be measured
(see Section 2.2.2). On the other, as any quantum observable is measurable
only by a classical apparatus, the existence of such apparatuses is tacitly
assumed from the very beginning. Quantum model ${\mathcal S}_q$ will thus
always depend on some classical elements. This does not mean that classicality
has been smuggled in because, in our approach, classical properties are
specific quantum ones. 3) A Hamiltonian operator of the system must be set
up. 4) Suitable quantum states must be chosen. Finally, the known classical
properties of ${\mathcal S}_c$ must be listed and each derived as an objective
property of ${\mathcal S}_q$ from the four sets of assumptions above. This is
a self-consistent framework for a non-trivial problem.

It follows that there must be some at least approximate relation between
classical observables of ${\mathcal S}_c$ and quantum observables of
${\mathcal S}_q$ as well as between the classical states of ${\mathcal S}_c$
and the quantum states of ${\mathcal S}_q$. The following model assumption on such
a relation might be viewed as a version of Correspondence Principle, let us
call it Modified Correspondence Principle.
\begin{assump}\label{assold4}
\begin{enumerate}
\item The state of classical model ${\mathcal S}_c$ in a given classical
theory is described by a set of $n$ numbers $\{a_1,\cdots,a_n\}$ that
represent values of some classical observables. The set is not uniquely
determined. Let us call any such set \underline{state coordinates}.
\item We assume that state coordinates $\{a_1,\cdots,a_n\}$ can be chosen so
that there is a subset $\{{\mathsf a}_1,\cdots,{\mathsf a}_n\}$ of sharp
observables of quantum model ${\mathcal S}_q$ and a state ${\mathsf T}$ of
${\mathcal S}_q$ such that
\begin{equation}\label{MCP} tr[{\mathsf T} {\mathbf a}_k] = a_k\ .
\end{equation}
\item All such states form a subset of ${\mathbf T}({\mathbf H}_{{\mathcal
S}_q})^+_1$. Some of these states satisfy the condition that all properties of
${\mathcal S}_c$ can be (at least approximately) obtained from ${\mathcal
S}_q$ if it is in such states. They are called \underline{classicality states}
of quantum system ${\mathcal S}_q$.
\end{enumerate}
\end{assump}

Clearly, Modified Correspondence Principle does not need the assumption that
all observables from $\{{\mathsf a}_1,\cdots,{\mathsf a}_n\}$ commute with
each other. Then, even if they themselves are not jointly measurable, their
fuzzy values can be (see Section 3.7). Further, it does not follow that each
classical property is an average of a quantum operator. That would be
false. We assume only that the classical state coordinates ${a_1,...,a_n}$ can
be chosen in such a way.

It is important to realise that Modified Correspondence Principle suggests how
Principle of Macroscopic Realism is to be understood. For example, macroscopic
systems also have extremal states that satisfy equation (\ref{MCP}). These
seem to be macroscopic states available to the system. However, extremal
states are readily linearly superposed and any quantum registration that ought
to find the parameters of a coherent state (a generalized measurement:
positive operator valued measure) would strongly change the state (for a
general argument, see Ref.\ \cite{BLM}, p. 32). We assume that the validity of
Principle of Macroscopic Realism can be achieved if the words "distinct
macroscopic states" are replaced with "distinct classicality
states". Let us try to motivate a proposal of what such classicality
states might be.

An interesting subset of classical properties of macroscopic system is the
thermodynamic ones. They are important for us because quantum models of these
properties are available. Existing models based on statistical physics need
one non-trivial assumption: the states of sufficiently small macroscopic
systems that we observe around us are approximately states of maximum
entropy. As it has been discussed in Section 0.1.4, entropy is an objective
property of quantum systems because it is defined by their preparation. Thus,
the validity of thermodynamics depends on the preparation conditions, or the
origin, of observed macroscopic systems. The averages and variances that
result from the models based on the maximum-entropy assumption agree with
observations. In particular, they explain why classical states and properties
are relatively sharp. Moreover, high entropy states are very far from extremal
and linear superposition does not make any sense for them.

The physical foundations of thermodynamics are not yet well-understood but
there are many ideas around about the origin of high-entropy states. Their
existence might follow partially from logic (Bayesian approach, \cite{Jaynes})
and partially from quantum mechanics (thermodynamic limit, \cite{thirring},
Vol.\ 4). Some very interesting models of how maximum entropy quantum states
come into being are based on entanglement \cite{GMM,
short1,short2,goldstein}).  However, just in order to construct quantum models
of classical properties, they can be used as one of the main assumptions
without really understanding their origin.

We generalise the statistical methods as follows \cite{hajicek}.
\begin{assump}\label{assold5} All classicality states are states of high
entropy.
\end{assump} Assumption \ref{assold5} is a heuristic one and it is therefore
formulated a little vaguely. It will be made clearer after some examples of
its use will be studied in this chapter. But some brief discussion can be
given already now.

Consider first states of macroscopic systems that are at or near absolute zero
of temperature. These are approximately or exactly extremal and maximize
entropy at the same time. Thus, they are not classicality states but the
entropy, though maximal, is not high, either. Second, consider states of
macroscopic systems at room temperature that are not at their thermodynamic
equilibrium but are close to it. There are many such states, and they and the
systems can be described by classical physics to a good approximation. They
are not in maximum- but in high-entropy states.

\section{Maximum entropy assumption \\ in classical mechanics} In this
Section, we follow loosely Ref.\ \cite{hajicek}. As explained at the start of
this chapter, the basic notions of the language part of Newtonian mechanics
are that of a sharp state---a point of the phase space---and of a sharp
trajectory---a curve in the phase space of an isolated system. We accept this
language without assuming that the sharp trajectories have any real
counterpart in the world because this does no prevent us from building fuzzy
models that have a more direct relation to reality.

However, most physicists take the existence of sharp trajectories seriously
and try to obtain them from quantum mechanics as exactly as possible. Hence,
they focus at quantum states the phase-space picture of which is as sharp as
possible. That are states with minimum uncertainty allowed by quantum
mechanics. For one degree of freedom, described by coordinate ${\mathsf q}$
and momentum ${\mathsf p}$, the uncertainty is given by the quantity
\begin{equation}\label{uncertainty} \nu = \frac{2\Delta {\mathsf q}\Delta
{\mathsf p}}{\hbar},
\end{equation} where $\Delta {\mathsf a}$ denotes the variance of quantity
${\mathsf a}$, as defined by Eq.\ (\ref{variance}).

The states with minimum uncertainty $\nu = 1$ are, however, very special
extremal states. Such states do exist for macroscopic systems but are very
difficult to prepare unlike the usual states of macroscopic systems described
by classical mechanics. As we explained in Section 1.1.2, they have a number
of properties that are very strange from the point of view of classical
theories and they are therefore not what we have called classicality states.

We feel that there is no point in attempts to derive the language part of
Newtonian mechanics from that of quantum mechanics. Instead, we propose that
the classical limit is to be considered at the level of models. That is,
properties of successful Newtonian models are to be obtained from some quantum
models under suitable conditions. We assume further that a good model of
Newtonian mechanics is necessarily fuzzy and that the fuzziness is determined
by the preparation of the system similarly as in quantum mechanics. Let us
give some examples.

Consider a gun in a position that is fixed in a reproducible way and that
shoots bullets using cartridges of a given provenance. All shots made under
these conditions form an ensemble with average trajectory
$(Q^k_{\text{gun}}(t),P^k_{\text{gun}}(t))$ and the trajectory variance
$(\Delta Q^k_{\text{gun}}(t),\Delta P^k_{\text{gun}}(t))$ that describes
objective properties of the ensemble. The Newtonian model of this ensemble is
the evolution $\rho_{\text{gun}}(Q^k,P^k;t)$ of a suitable distribution
function on the phase space. According to Newtonian mechanics, each individual
shot has a sharp trajectory $(Q^k(t),P^k(t))$. Each individual shot is also an
element of the ensemble and this is a property of the individual that can be
considered also as objective, even in Newtonian mechanics.

The existence of this fuzzy property of an individual shot does not contradict
the fact that some more precise observations (optical, say) of this one shot
can give a different fuzzy structure. Indeed, such an optical measurement
method ought to have been studied on other ensembles and already well
established itself, which will allow to estimate its error (variance) and
hence to understand the result of the measurement as saying that a given,
fixed trajectory is an element of a thought ensemble with an average
$(Q^k_{\text{opt}}(t),P^k_{\text{opt}}(t))$ and variance $(\Delta
Q^k_{\text{opt}}(t),\Delta P^k_{\text{opt}}(t))$, where
$$
Q^k_{\text{opt}}(t) \approx Q^k_{\text{gun}}(t)\ ,\quad \Delta
Q^k_{\text{opt}}(t) \ll \Delta Q^k_{\text{gun}}(t)\ ,
$$
and similarly for the momentum part. Still, $\Delta Q^k_{\text{opt}}(t)\cdot
\Delta P^k_{\text{opt}}(t)$ must be much larger than the minimum quantum
uncertainty $\hbar/2$.

The simplest way to construct a fuzzy model is to fix initial averages and
variances of coordinates and momenta, $Q^k$, $\Delta Q^k$, $P^k$, $\Delta
P^k$, and leave everything else as fuzzy as possible. To calculate the
corresponding probability distributions in classical mechanics and the state operators
in quantum mechanics, we shall, therefore, apply the maximum entropy
principle. The resulting states will be called {\em maximum-entropy packets},
abbreviated as ME packets. The averages of coordinates and momenta take over
the role of coordinate and momenta in classical mechanics. In any case these
averages represent measurable aspects of these variables. Quantities $Q^k$,
$\Delta Q^k$, $P^k$, $\Delta P^k$ will also play the role of classical state
coordinates defined by Assumption \ref{assold4}. To limit ourselves just to
given averages and variances of coordinates and momenta is a great
simplification that enables us to obtain interesting results easily. One can
imagine, however, more complicated models, where further moments are fixed, or
moments of different observables (e.g., mass centre, total momentum, angles
and total angular momentum) are fixed.

The variances are not assumed small. How large they are depends on the
accuracy of a preparation or of a measurement, as the gun example shows.

In fact, the dynamical evolution of variances is an important indicator of the
applicability of the model one is working with. It determines the time
intervals within which reasonable predictions are possible. Consider a
three-body system that is to model the Sun, Earth and Jupiter in Newtonian
mechanics. It turns out that generic trajectories starting as near to each
other as, say, the dimension of the irregularities of the Earth surface will
diverge from each other by dimensions of the Earth--Sun distance after the time
of only about $10^7$ years. This seems to contradict the $4\times 10^{12}$
years of relatively stable Earth motion around the Sun that is born out by
observations. The only way out is the existence of a few special trajectories
that are much more stable than the generic ones and the fact that bodies following
an unstable trajectory have long ago fallen into the Sun or have been ejected
from the solar system. By the way, this spontaneous evolution can be
considered as a preparation procedure of solar system.

\section{Classical ME packets} Let us first consider a system ${\mathcal S}$
with one degree of freedom and then generalise it to any number of
degrees. Let the coordinate be $q$ and the momentum $p$. A state is a
distribution function $\rho(q,p)$ on the phase space spanned by $q$ and
$p$. The function $\rho(q,p)$ is dimensionless and normalized by
$$
\int\frac{dq\,dp}{v}\,\rho = 1\ ,
$$
where $v$ is an auxiliary phase-space volume to make $\rho$ dimensionless. The
entropy of $\rho(q,p)$ can be defined by
$$
S := -\int\frac{dq\,dp}{v}\,\rho \ln\rho\ .
$$
The value of entropy will depend on $v$ but most other results will
not. Classical mechanics does not offer any idea of how to fix $v$. We shall
get its value from quantum mechanics.

\subsection{Definition and properties}
\begin{df}\label{dfold21} {\em ME packet} is the distribution function $\rho$ that
maximizes the entropy subject to the conditions:
\begin{equation}\label{21.4} \langle q\rangle = Q\ ,\quad \langle q^2\rangle =
\Delta Q^2 + Q^2\ ,
\end{equation} and
\begin{equation}\label{21.5} \langle p\rangle = P\ ,\quad \langle p^2\rangle =
\Delta P^2 + P^2\ ,
\end{equation} where $Q$, $P$, $\Delta Q$ and $\Delta P$ are given values.
\end{df} We have used the abbreviation
$$
\langle x\rangle = \int\frac{dq\,dp}{v}\,x\rho\ .
$$

The explicit form of $\rho$ can be found using the partition-function method
as described in Ref.\ \cite{Jaynes}. The variational principle yields
\begin{equation}\label{H5} \rho =
\frac{1}{Z(\lambda_1,\lambda_2,\lambda_3,\lambda_4)} \exp(-\lambda_1q -
\lambda_2p - \lambda_3q^2 - \lambda_4p^2)\ ,
\end{equation} where
$$
Z = \int \frac{dq\,dp}{v}\exp(-\lambda_1q - \lambda_2p - \lambda_3q^2 -
\lambda_4p^2)\ ,
$$
and $\lambda_1$, $\lambda_3$, $\lambda_2$ and $\lambda_4$ are the four
Lagrange multipliers corresponding to the four conditions (\ref{21.4}) and
(\ref{21.5}). Hence, the partition function for classical ME packet is
\begin{equation}\label{22.1} Z=
\frac{\pi}{v}\frac{1}{\sqrt{\lambda_3\lambda_4}}
\exp\left(\frac{\lambda_1^2}{4\lambda_3} +
\frac{\lambda_2^2}{4\lambda_4}\right)\ .
\end{equation} The expressions for $\lambda_1$, $\lambda_2$, $\lambda_3$ and
$\lambda_4$ in terms of $Q$, $P$, $\Delta Q$ and $\Delta P$ can be obtained by
solving the equations
$$
\frac{\partial\ln Z}{\partial\lambda_1} = -Q\ , \quad \frac{\partial\ln
Z}{\partial\lambda_3} = -\Delta Q^2 - Q^2\ ,
$$
and
$$
\frac{\partial\ln Z}{\partial\lambda_2} = -P\ , \quad \frac{\partial\ln
Z}{\partial\lambda_4} = -\Delta P^2 - P^2\ .
$$
The result is:
\begin{equation}\label{lagrancl1} \lambda_1 = -\frac{Q}{\Delta Q^2}\ ,\quad
\lambda_3 = \frac{1}{2\Delta Q^2}\ ,
\end{equation} and
\begin{equation}\label{lagrancl2} \lambda_2 = -\frac{P}{\Delta P^2}\ ,\quad
\lambda_4 = \frac{1}{2\Delta P^2}\ .
\end{equation} Substituting this into Eq.\ (\ref{H5}), we obtain the
distribution function of a one-dimensional ME packet. The generalization to
any number of dimensions is:
\begin{thm}\label{propold19} The distribution function of the ME packet for a
system with given averages and variances $Q_1,\cdots,Q_n$, $\Delta
Q_1,\cdots,\Delta Q_n$ of coordinates and $P_1,\cdots,P_n$, $\Delta
P_1,\cdots,\Delta P_n$ of momenta, is
\begin{equation}\label{23.1} \rho =
\left(\frac{v}{2\pi}\right)^n\prod_{k=1}^n\left(\frac{1}{\Delta Q_k\Delta
P_k}\exp\left[-\frac{(q_k-Q_k)^2}{2\Delta Q_k^2} -\frac{(p_k-P_k)^2}{2\Delta
P_k^2}\right]\right)\ .
\end{equation}
\end{thm}

We observe that all averages obtained from $\rho$ are independent of $v$ and
that the right-hand side of equation (\ref{23.1}) is a Gaussian distribution
in agreement with Jaynes' conjecture that the maximum entropy principle gives
the Gaussian distribution if the only conditions are fixed values of the first
two moments.

As $\Delta Q$ and $\Delta P$ approach zero, $\rho$ becomes a $\delta$-function
and the state becomes sharp. For some quantities this limit is sensible, for
others it is not. In particular, the entropy, which can easily be calculated,
$$
S = 1 + \ln\frac{2\pi\Delta Q\Delta P}{v}\ ,
$$
diverges to $-\infty$. This is due to a general difficulty in giving a
definition of entropy for a continuous system that would be satisfactory in
every respect. What one could do is to divide the phase space into cells of
volume $v$ so that $\Delta Q\Delta P$ could not be chosen smaller than
$v$. Then, the limit $\Delta Q\Delta P \rightarrow v$ of entropy would make
more sense.

The average of any monomial of the form $q^k p^l q^{2m} p^{2n}$ can be
calculated with the help of partition-function method as follows:
\begin{equation}\label{43.1} \langle q^k p^l q^{2m} p^{2n}\rangle =
\frac{(-1)^{\mathbf N}}{Z}\ \frac{\partial^{\mathbf N}
Z}{\partial\lambda_1^k \partial\lambda_2^l \partial\lambda_3^m \partial\lambda_4^n
}\ ,
\end{equation} where ${\mathbf N} = k+l+2m+2n$, $Z$ is given by
Eq. (\ref{22.1}) and the values (\ref{lagrancl1}) and (\ref{lagrancl2}) must
be substituted for the Lagrange multipliers after the derivatives are taken.

Observe that this enables to calculate the average of a monomial in several
different ways. Each of these ways, however, leads to the same result due the
identities
$$
\frac{\partial^2 Z}{\partial \lambda_1^2} = -\frac{\partial Z}{\partial
\lambda_3}\ ,\quad \frac{\partial^2 Z}{\partial \lambda_2^2} = -\frac{\partial
Z}{\partial \lambda_4}\ ,
$$
which are satisfied by the partition function.

\begin{assump}\label{assMEcl} ME packet (\ref{23.1}) is a part of a
satisfactory model for many systems in Newtonian mechanics.
\end{assump}

\subsection{Classical equations of motion} Let us assume that the Hamiltonian
of ${\mathcal S}$ has the form
\begin{equation}\label{35.1} H = \frac{p^2}{2\mu} + V(q)\ ,
\end{equation} where $\mu$ is the mass and $V(q)$ the potential function. The
equations of motion are
$$
\dot{q} = \{q,H\}\ ,\quad\dot{p} = \{p,H\}\ .
$$
Inserting (\ref{35.1}) for $H$, we obtain
\begin{equation}\label{36.6} \dot{q} = \frac{p}{\mu}\ ,\quad\dot{p} =
-\frac{dV}{dq}\ .
\end{equation} The general solution to these equations can be written in the
form
\begin{equation}\label{36.65} q(t) = q(t;q,p)\ ,\quad p(t) = p(t;q,p)\ ,
\end{equation} where
\begin{equation}\label{36.3} q(0;q,p) = q\ ,\quad p(0;q,p) = p\ ,
\end{equation} $q$ and $p$ being arbitrary initial values. This implies for
the time dependence of the averages and variances, if the initial state is an
ME packet:
\begin{equation}\label{36.7} Q(t) = \langle q(t; q,p)\rangle\ ,\quad \Delta
Q(t) = \sqrt{\langle (q(t;q,p)- Q(t))^2\rangle}
\end{equation} and
\begin{equation}\label{36.8} P(t) = \langle p(t; q,p)\rangle\ ,\quad \Delta
P(t) = \sqrt{\langle (p(t;q,p)- P(t))^2}\rangle\ .
\end{equation} In general, $Q(t)$ and $P(t)$ will depend not only on initial
$Q$ and $P$, but also on $\Delta Q$ and $\Delta P$.

Let us consider the special case of at most quadratic potential:
\begin{equation}\label{36.1} V(q) = V_0 + V_1 q + \frac{1}{2} V_2 q^2\ ,
\end{equation} where $V_k$ are constants with suitable dimensions. If $V_1 =
V_2 =0$, we have a free particle, if $V_2 = 0$, it is a particle in a
homogeneous force field and if $V_2 \neq 0$, it is a harmonic or
anti-harmonic oscillator.

In this case, general solution (\ref{36.65}) has the form
\begin{eqnarray}\label{37.1} q(t) &=& f_0(t) + q f_1(t) + p f_2(t)\ , \\
\label{37.2} p(t) &=& g_0(t) + q g_1(t) + p g_2(t)\ ,
\end{eqnarray} where $f_0(0) = f_2(0) = g_0(0) = g_1(0) = 0$ and $f_1(0) =
g_2(0) = 1$. If $V_2 \neq 0$, the functions are
\begin{equation}\label{37.4} f_0(t) = -\frac{V_1}{V_2}(1-\cos\omega t)\ ,\quad
f_1(t) = \cos \omega t\ ,\quad f_2(t) = \frac{1}{\xi}\sin\omega t\ ,
\end{equation}
\begin{equation}\label{37.5} g_0(t) = -\xi\frac{V_1}{V_2}\sin\omega t\ ,\quad
g_1(t) = -\xi\sin \omega t\ ,\quad g_2(t) = \cos\omega t\ ,
\end{equation} where
$$
\xi = \sqrt{\mu V_2}\ ,\quad \omega  = \sqrt{\frac{V_2}{\mu}}\ .
$$
Only for $V_2 > 0$, the functions remain bounded. If $V_2 = 0$, we obtain
\begin{equation}\label{37.9a} f_0(t) = -\frac{V_1}{2\mu}t^2\ ,\quad f_1(t) =
1\ ,\quad f_2(t) = \frac{t}{\mu}\ ,
\end{equation}
\begin{equation}\label{37.9b} g_0(t) = -V_1t\ ,\quad g_1(t) = 0\ ,\quad g_2(t)
= 1\ .
\end{equation}

The resulting time dependence of averages and variances resulting from Eqs.\
(\ref{36.65}), (\ref{21.4}) and (\ref{21.5}) are \cite{hajicek}
\begin{equation}\label{38.3} Q(t) = f_0(t) + Q f_1(t) + P f_2(t)
\end{equation} and
\begin{multline}\label{38.4} \Delta Q^2(t) + Q^2(t) = f_0^2(t) + (\Delta Q^2 +
Q^2) f_1^2(t) + (\Delta P^2 + P^2) f_2^2(t)\\ + 2Qf_0(t)f_1(t) +
2Pf_0(t)f_2(t) + 2\langle qp\rangle f_1(t)f_2(t)\ .
\end{multline} For the last term, we have from Eq.\ (\ref{43.1})
$$
\langle qp\rangle = \frac{1}{Z}\frac{\partial^2
Z}{\partial\lambda_1\partial\lambda_2}\ .
$$
Using Eqs.\ (\ref{22.1}), (\ref{lagrancl1}) and (\ref{lagrancl2}), we obtain
from Eq.\ (\ref{38.4})
\begin{equation}\label{39.1} \Delta Q(t) = \sqrt{f_1^2(t)\Delta Q^2 +
f_2^2(t)\Delta P^2}\ .
\end{equation} Similarly,
\begin{eqnarray}\label{39.2} P(t) &=& g_0(t) + Q g_1(t) + P g_2(t)\ ,\\
\label{39.3} \Delta P(t) &=& \sqrt{f_g^2(t)\Delta Q^2 + g_2^2(t)\Delta P^2}\ .
\end{eqnarray} We observe that, if functions $f_1(t)$, $f_2(t)$, $g_1(t)$ and
$g_2(t)$ remain bounded, the variances also remain bounded and the predictions
are possible in arbitrary long intervals of time. Otherwise, there will always
be only limited time intervals in which the theory can make predictions.

In the case of general potential, the functions (\ref{36.65}) can be expanded
in products of powers of $q$ and $p$, and the averages of these products will
contain powers of the variances. However, as one easily sees form formula
(\ref{43.1}) and (\ref{22.1}),
$$
\langle q^kp^l\rangle = Q^kP^l +X\Delta Q +Y\Delta P\ ,
$$
where $X$ and $Y$ are bounded functions. It follows that the dynamical
equations for averages coincide, in the limit $\Delta Q \rightarrow 0, \Delta
P \rightarrow 0$, with the exact dynamical equations for $q$ and $p$. It is an
idealisation that we consider as not realistic, even in principle, but it may
still be useful for calculations.

Let us expand a general potential function in powers of $q$,
\begin{equation}\label{50.2} V(q) = \sum_{k=0}^\infty \frac{1}{k!}V_k q^k\ ,
\end{equation} where $V_k$ are constants of appropriate dimensions. The
Hamilton equations can be used to calculate all time derivatives at
$t=0$. First, we have
$$
\frac{dq}{dt} = \{q,H\} = \frac{p}{\mu}\ .
$$
This equation can be used to calculate all derivatives of $q$ in terms of
those of $p$:
\begin{equation}\label{50.4} \frac{d^nq}{dt^n} =
\frac{1}{\mu}\frac{d^{n-1}p}{dt^{n-1}}\ .
\end{equation}

A simple iterative procedure gives:
\begin{equation}\label{50.5} \frac{dp}{dt} =
-V_1-V_2q-\frac{V_3}{2}q^2-\frac{V_4}{6}q^3 + r_5\ ,
\end{equation}
\begin{equation}\label{50.6} \frac{d^2p}{dt^2} =
-\frac{V_2}{\mu}p-\frac{V_3}{\mu}qp-\frac{V_4}{2\mu}q^2p + r_5\ ,
\end{equation}
\begin{multline}\label{50.7} \frac{d^3p}{dt^3} =
-\frac{V_3}{\mu^2}p^2-\frac{V_4}{\mu^2}qp^2 + \frac{V_1V_2}{\mu} +
\frac{V_1V_3+V_2^2}{\mu}q + \frac{3V_2V_3+V_1V_4}{2\mu}q^2 \\ +
\frac{4V_2V_4+3V_3^2}{6\mu}q^3 + \frac{5V_3V_4}{12\mu}q^4 +
\frac{V_4^2}{12\mu}q^5 + r_5\ ,
\end{multline} and
\begin{multline}\label{50.8} \frac{d^4p}{dt^4} = -\frac{V_4}{\mu^3}p^3 +
\frac{3V_1V_3+V_2^2}{\mu^2}p + \frac{3V_1V_4+5V_2V_3}{\mu^2}qp +
\frac{5V_3^2+8V_2V_4}{2\mu^2}q^2p \\ + 3\frac{V_3V_4}{\mu^2}q^3p +
\frac{3V_4^2}{4\mu^2}q^4p + r_5\ ,
\end{multline} where $r_k$ is the rest term that is due to all powers in
(\ref{50.2}) that are not smaller than $k$ (the rests symbolize different
expressions in different equations). The purpose of having time derivatives up
to the fourth order is to see better the difference to quantum corrections
that will be calculated in Section 3.4.3.

Taking the average of both sides of Eqs.\ (\ref{50.5})--(\ref{50.8}), and
using Eq.\ (\ref{43.1}), (\ref{22.1})--(\ref{lagrancl2}), we obtain
\begin{equation}\label{51.1} \frac{dP}{dt} =
-V_1-V_2Q-\frac{V_3}{2}Q^2-\frac{V_4}{6}Q^3-\frac{V_3+V_4Q}{2}\Delta Q^2 +
r_5\ ,
\end{equation}
\begin{equation}\label{51.2} \frac{d^2P}{dt^2} = -\frac{V_2}{\mu}P +
\frac{V_3}{\mu}QP + \frac{V_4}{2\mu}Q^2P + \frac{V_4}{2\mu}P\Delta Q^2 + r_5\
,
\end{equation}
\begin{multline}\label{51.3} \frac{d^3P}{dt^3} =
-\frac{V_3}{\mu^2}P^2-\frac{V_4}{\mu^2}QP^2 + \frac{V_1V_2}{\mu} +
\frac{V_1V_3+V_2^2}{\mu}Q + \frac{3V_2V_3+V_1V_4}{2\mu}Q^2 \\+
\frac{4V_2V_4+3V_3^2}{6\mu}Q^3 + \frac{5V_3V_4}{12\mu}Q^4 +
\frac{V_4^2}{12\mu}Q^5-\left(\frac{V_3}{\mu^2}+\frac{V_4}{\mu^2}Q\right)\Delta
P^2 \\+ \left(\frac{3V_2V_3+V_1V_4}{2\mu} + \frac{4V_2V_4+3V_3^2}{2\mu}Q +
\frac{5V_3V_4}{2\mu}Q^2 + \frac{5V_3V_4}{4\mu}\Delta Q^2 \right. \\+
\left. \frac{5V_4^2}{6\mu}Q^3 + \frac{5V_4^2}{4\mu}Q\Delta Q^2\right)\Delta
Q^2 + r_5\ ,
\end{multline} and
\begin{multline}\label{51.4} \frac{d^4P}{dt^4} = -\frac{V_4}{\mu^3}P^3 +
\frac{3V_1V_3+V_2^2}{\mu^2}P + \frac{3V_1V_4+5V_2V_3}{\mu^2}QP \\+
\frac{5V_3^2+8V_2V_4}{2\mu^2}Q^2P + 3\frac{V_3V_4}{\mu^2}Q^3P +
\frac{3V_4^2}{4\mu^2}Q^4P-\frac{3V_4}{\mu^3}P\Delta P^2
\\+\left(\frac{5V_3^2+8V_2V_4}{2\mu^2}P + \frac{9V_3V_4}{\mu^2}QP +
\frac{9V_4^2}{2\mu^2}Q^2P + \frac{9V_4^2}{4\mu^2}P\Delta Q^2\right)\Delta Q^2
+ r_5\ .
\end{multline}

We can see, that the limit $\Delta Q\rightarrow 0,\Delta P \rightarrow 0$ in
Eqs.\ (\ref{51.1})--(\ref{51.4}) lead to equations that coincide with Eqs.\
(\ref{50.5})--(\ref{50.8}) if $Q\rightarrow q,P\rightarrow p$ as promised.

\section{Quantum ME packets} Let us now turn to quantum mechanics and try to
solve an analogous problem.
\begin{df}\label{dfold22} Let the quantum model ${\mathcal S}_q$ of system
${\mathcal S}$ has spin 0, position ${\mathsf q}$ and momentum ${\mathsf
p}$. State ${\mathsf T}$ that maximizes von Neumann entropy (see Section
1.2.2)
\begin{equation}\label{VNE} S = -tr({\mathsf T}\ln{\mathsf T})
\end{equation} under the conditions
\begin{equation}\label{12.1} tr[{\mathsf T}{\mathsf q}] = Q\ ,\quad
tr[{\mathsf T} {\mathsf q}^2] = Q^2 + \Delta Q^2\ ,
\end{equation}
\begin{equation}\label{12.2} tr[{\mathsf T}{\mathsf p}] = P\ ,\quad
tr[{\mathsf T} {\mathsf p}^2] = P^2 + \Delta P^2\ ,
\end{equation} where $Q$, $P$, $\Delta Q$ and $\Delta P$ are given numbers, is
called {\em quantum ME packet}.
\end{df}

\subsection{Calculation of the state operator} To solve the mathematical
problem, we use the method of Lagrange multipliers as in the classical
case. Thus, the following equation results:
\begin{equation}\label{12.5}( dS -\lambda_0 d\ tr[{\mathsf T}]-\lambda_1
d\,tr[{\mathsf T} {\mathsf q}]-\lambda_2 d\,tr[{\mathsf T} {\mathsf p}]
-\lambda_3 d\,tr[{\mathsf T} {\mathsf q}^2]-\lambda_4 d\,tr[{\mathsf T}
{\mathsf p}^2] = 0\ .
\end{equation} The differentials of the terms that are linear in $\rho$ are
simple to calculate:
$$
d\,tr[{\mathsf T} {\mathsf x}] = \sum_{mn}x_{nm}dT_{mn}.
$$
Although not all elements of the matrix $dT_{mn}$ are independent (it is a
Hermitian matrix), we can proceed as if they were because the matrix $x_{nm}$
is to be also Hermitian. The only problem is to calculate $dS$. We have the
following
\begin{lem}
\begin{equation}\label{10.1} dS = -\sum_{mn}[\delta_{mn} + (\ln
T)_{mn}]dT_{mn}\ .
\end{equation}
\end{lem} {\bf Proof} Let ${\mathsf M}$ be a unitary matrix that diagonalizes
${\mathsf T}$,
$$
{\mathsf M}^\dagger {\mathsf T} {\mathsf M} = {\mathsf R}\ ,
$$
where ${\mathsf R}$ is a diagonal matrix with elements $R_n$. Then $S =
-\sum_n R_n\ln R_n$. Correction to $R_n$ if ${\mathsf T} \mapsto {\mathsf T} +
d{\mathsf T}$ can be calculated by the first-order formula of the stationary
perturbation theory. This theory is usually applied to Hamiltonians but it
holds for any perturbed Hermitian operator. Moreover, the formula is exact for
infinitesimal perturbations. Thus,
$$
R_n \mapsto R_n + \sum_{kl}M^\dagger_{kn}M_{ln}dT_{kl}\ .
$$
In this way, we obtain
\begin{multline*} dS = -\sum_n\left(R_n +
\sum_{kl}M^\dagger_{kn}M_{ln}dT_{kl}\right)\\ \times \ln\left[R_n\left(1 +
\frac{1}{R_n}\sum_{rs}M^\dagger_{rn}M_{sn}dT_{rs}\right)\right] -\sum_n R_n\ln
R_n\\ = -\sum_n\left[\ln R_n\sum_{kl}M^\dagger_{kn}M_{ln}dT_{kl}
+\sum_{kl}M^\dagger_{kn}M_{ln}d\rho_{kl}\right] \\
=-\sum_{kl}\left[\delta_{kl} + (\ln{\mathsf T})_{kl}]\right]dT_{kl}\ ,
\end{multline*} QED.

With the help of Lemma 1, Eq.\ (\ref{12.5}) becomes
$$
tr\left[(1 + \ln{\mathsf T} -\lambda_0-\lambda_1 {\mathsf q} -\lambda_2
{\mathsf p}-\lambda_3 {\mathsf q}^2-\lambda_4{\mathsf p}^2)d{\mathsf T}\right]
= 0
$$
so that we have
\begin{equation}\label{12.6} {\mathsf T} = \exp(-\lambda_0-1-\lambda_1
{\mathsf q}-\lambda_2 {\mathsf p}-\lambda_3 {\mathsf q}^2-\lambda_4 {\mathsf
p}^2)\ .
\end{equation} The first two terms in the exponent determine the normalization
constant
$$
e^{-\lambda_0-1}
$$
because they commute with the rest of the exponent and are independent of the
dynamical variables. Taking the trace of Eq.\ (\ref{12.6}), we obtain
$$
e^{-\lambda_0-1} = \frac{1}{Z(\lambda_1,\lambda_2,\lambda_3,\lambda_4)}\ ,
$$
where $Z$ is the partition function,
\begin{equation}\label{13.1} Z(\lambda_1,\lambda_2,\lambda_3,\lambda_4) = tr[
\exp(-\lambda_1 {\mathsf q}-\lambda_2 {\mathsf p}-\lambda_3 {\mathsf
q}^2-\lambda_4 {\mathsf p}^2)]\ .
\end{equation} Thus, the state operator has the form
\begin{equation}\label{13.2} {\mathsf T} =
\frac{1}{Z(\lambda_1,\lambda_2,\lambda_3,\lambda_4)} \exp(-\lambda_1 {\mathsf
q}-\lambda_2 {\mathsf p}-\lambda_3 {\mathsf q}^2-\lambda_4 {\mathsf p}^2)\ .
\end{equation}

At this stage, the quantum theory begins to differ from the classical one. It
turns out that, for the case of non-commuting operators in the exponent of the
partition function, formula (\ref{43.1}) is not valid in general. We can only
show that it holds for the first derivatives. To this aim, we prove the
following
\begin{lem} Let ${\mathsf A}$ and ${\mathsf B}$ be Hermitian matrices. Then
\begin{equation}\label{14.1} \frac{d}{d\lambda}tr[\exp({\mathsf A}+{\mathsf
B}\lambda)] = tr[{\mathsf B}\exp({\mathsf A}+{\mathsf B}\lambda)]\ .
\end{equation}
\end{lem} {\bf Proof} We express the exponential function as a series and then
use the invariance of trace with respect to any cyclic permutation of its
argument.
\begin{multline*} d\,tr[\exp({\mathsf A}+{\mathsf B}\lambda)] =
\sum_{n=0}^\infty\frac{1}{n!}tr[d({\mathsf A}+{\mathsf B}\lambda)^n] \\ =
\sum_{n=0}^\infty\frac{1}{n!}tr\left[\sum_{k=1}^n({\mathsf A}+{\mathsf
B}\lambda)^{k-1}{\mathsf B}({\mathsf A}+{\mathsf
B}\lambda)^{n-k}\right]d\lambda \\
=\sum_{n=0}^\infty\frac{1}{n!}\sum_{k=1}^ntr\left[{\mathsf B}({\mathsf
A}+{\mathsf B}\lambda)^{n-1}\right]d\lambda = tr[{\mathsf B}\exp({\mathsf
A}+{\mathsf B}\lambda)]d\lambda\ ,
\end{multline*} QED.

The proof of Lemma 2 shows why formula (\ref{43.1}) is not valid for higher
derivatives than the first in the quantum case: the operator $B$ does not
commute with ${\mathsf A}+{\mathsf B}\lambda$ and cannot be shifted from its
position to the first position in product
$$
({\mathsf A}+{\mathsf B}\lambda)^k{\mathsf B}({\mathsf A}+{\mathsf
B}\lambda)^l\ .
$$
For the first derivative, it can be brought there by a suitable cyclic
permutation. However, each commutator $[{\mathsf B},({\mathsf A}+{\mathsf
B}\lambda)]$ is proportional to $\hbar$. Hence, formula (\ref{43.1}) with
higher derivatives is the leading term in the expansion of averages in powers
of $\hbar$.

Together with Eq.\ (\ref{13.1}), Lemma 2 implies the formulae:
\begin{equation}\label{13.3} \frac{\partial \ln Z}{\partial\lambda_1} = -Q\
,\quad \frac{\partial \ln Z}{\partial\lambda_3} = -Q^2-\Delta Q^2
\end{equation} and
\begin{equation}\label{13.4} \frac{\partial \ln Z}{\partial\lambda_2} = -P\
,\quad \frac{\partial \ln Z}{\partial\lambda_4} = -P^2-\Delta P^2\ .
\end{equation} The values of the multipliers can be calculated from Eqs.\
(\ref{13.3}) and (\ref{13.4}), if the form of the partition function is known.

Variational methods can find locally extremal values that are not necessarily
maxima. We can however prove that our state operator maximizes entropy. The
proof is based on the generalized Gibbs' inequality,
$$
tr[{\mathsf T}\ln{\mathsf T} - {\mathsf T}\ln{\mathsf S}] \geq 0
$$
for all pairs $\{{\mathsf T},{\mathsf S}\}$ of state operators (for proof of
the inequality, see \cite{peres}, p.\ 264). The proof of maximality is then
analogous to the "classical" proof (see, e.g., \cite{Jaynes}, p.\
357). The first proof of maximality in the quantum case was given by von
Neumann \cite{JvN}.

The state operator (\ref{13.2}) can be inserted in the formula (\ref{VNE}) to
give the value of the maximum entropy,
\begin{equation}\label{15.1} S = \ln Z + \lambda_1\langle {\mathsf q}\rangle +
\lambda_2\langle {\mathsf p}\rangle + \lambda_3\langle {\mathsf q}^2\rangle +
\lambda_4\langle {\mathsf p}^2\rangle\ .
\end{equation} This, together with Eqs. (\ref{13.3}) and(\ref{13.4}), can be
considered as the Legendre transformation from the function $\ln
Z(\lambda_1,\lambda_2,\lambda_3,\lambda_4)$ to the function $S(\langle
{\mathsf q}\rangle,\langle {\mathsf p}\rangle,\langle {\mathsf
q}^2\rangle,\langle {\mathsf p}^2\rangle )$.

\subsection{Diagonal representation} The exponent in Eq. (\ref{13.2}) can be
written in the form
\begin{equation}\label{19.1} \frac{\lambda_1^2}{4\lambda_3} +
\frac{\lambda_2^2}{4\lambda_4} -2\sqrt{\lambda_3\lambda_4}{\mathsf K}\ ,
\end{equation} where
\begin{equation}\label{19.2} {\mathsf K} =
\frac{1}{2}\sqrt{\frac{\lambda_3}{\lambda_4}}\left({\mathsf q} +
\frac{\lambda_1}{2\lambda_3}\right)^2 +
\frac{1}{2}\sqrt{\frac{\lambda_4}{\lambda_3}}\left({\mathsf p} +
\frac{\lambda_2}{2\lambda_4}\right)^2\ .
\end{equation} This is an operator acting on the Hilbert space of our
system. ${\mathsf K}$ has the form of the Hamiltonian\footnote{The operator
${\mathsf K}$ must not be confused with the Hamiltonian ${\mathsf H}$ of our
system, which can be arbitrary.} of a harmonic oscillator with the coordinate
${\mathsf u}$ and momentum ${\mathsf w}$
\begin{equation}\label{16.1} {\mathsf u} = {\mathsf q} +
\frac{\lambda_1}{2\lambda_3}\ ,\quad {\mathsf w} = {\mathsf p} +
\frac{\lambda_2}{2\lambda_4}\ ,
\end{equation} that satisfy the commutation relation $[{\mathsf u},{\mathsf
w}] = i\hbar$. The oscillator has mass $M$ and frequency $\Omega$,
\begin{equation}\label{16.2} M = \sqrt{\frac{\lambda_3}{\lambda_4}}\ ,\quad
\Omega = 1\ .
\end{equation} The normalized eigenstates $|k\rangle$ of the operator form a
basis in the Hilbert space of our system defining the so-called {\em diagonal
representation} and its eigenvalues are $\hbar/2 + \hbar k$. As usual, we
introduce operator ${\mathsf A}$ such that
\begin{eqnarray}\label{ua} {\mathsf u} &=& \sqrt{\frac{\hbar}{2M}}({\mathsf
A}+{\mathsf A}^\dagger)\ , \\
\label{qa} {\mathsf w} &=& -i\sqrt{\frac{\hbar M}{2}}({\mathsf A}-{\mathsf
A}^\dagger)\ , \\
\label{Hamilt} {\mathsf K} &=& \frac{\hbar}{2}({\mathsf A}^\dagger {\mathsf A}
+ {\mathsf A}{\mathsf A}^\dagger))\ , \\
\label{aact1} {\mathsf A}|k\rangle &=& \sqrt{k}|k-1\rangle\ , \\
\label{aact2} {\mathsf A}^\dagger|k\rangle &=& \sqrt{k+1}|k+1\rangle\ .
\end{eqnarray}

To calculate $Z$ in the diagonal representation is easy:
\begin{multline*} Z = tr\left[\exp\left(\frac{\lambda_1^2}{4\lambda_3} +
\frac{\lambda_2^2}{4\lambda_4} -2\sqrt{\lambda_3\lambda_4}{\mathsf
K}\right)\right] \\ = \sum_{k=0}^\infty\langle
k|\exp\left(\frac{\lambda_1^2}{4\lambda_3} + \frac{\lambda_2^2}{4\lambda_4}
-2\sqrt{\lambda_3\lambda_4}{\mathsf K}\right)|k\rangle \\ =
\exp\left(\frac{\lambda_1^2}{4\lambda_3} + \frac{\lambda_2^2}{4\lambda_4}
-\hbar\sqrt{\lambda_3\lambda_4}\right)\sum_{k=0}^\infty\exp(-2\hbar\sqrt{\lambda_3\lambda_4}k)\
.
\end{multline*} Hence, the partition function for the quantum ME-packets is
\begin{equation}\label{17.3} Z =
\frac{\exp\left(\frac{\lambda_1^2}{4\lambda_3} +
\frac{\lambda_2^2}{4\lambda_4}\right)}{2\sinh(\hbar\sqrt{\lambda_3\lambda_4})}\
.
\end{equation}

Now, we can express the Lagrange multipliers in terms of the averages and
variances. Eqs.\ (\ref{13.3}) and (\ref{13.4}) yield
\begin{equation}\label{lagranqu1} \lambda_1 = -\frac{Q}{\Delta
Q^2}\frac{\nu}{2}\ln\frac{\nu+1}{\nu-1}\ ,\quad \lambda_2 = -\frac{P}{\Delta
P^2}\frac{\nu}{2}\ln\frac{\nu+1}{\nu-1}\ ,
\end{equation} and
\begin{equation}\label{lagranqu2} \lambda_3 = \frac{1}{2\Delta
Q^2}\frac{\nu}{2}\ln\frac{\nu+1}{\nu-1}\ ,\quad \lambda_4 = \frac{1}{2\Delta
P^2}\frac{\nu}{2}\ln\frac{\nu+1}{\nu-1}\ ,
\end{equation} where $\nu$ is defined by Eq.\ (\ref{uncertainty}).

From Eqs.\ (\ref{15.1}), (\ref{lagranqu1}) and (\ref{lagranqu2}), we obtain
the entropy:
\begin{equation}\label{18.5} S = -\ln 2 + \frac{\nu+1}{2}\ln(\nu+1)
-\frac{\nu-1}{2}\ln(\nu-1)\ .
\end{equation} Thus, $S$ depends on $Q$, $P$, $\Delta Q$, $\Delta P$ only via
$\nu$. We have
$$
\frac{dS}{d\nu} = \frac{1}{2}\ln\frac{\nu+1}{\nu-1} > 0\ ,
$$
so that $S$ is an increasing function of $\nu$. Near $\nu = 1$,
$$
S \approx -\frac{\nu-1}{2}\ln(\nu-1)\ .
$$
Asymptotically ($\nu \rightarrow\infty$),
$$
S \approx \ln\nu+1-\ln 2\ .
$$
In the classical region, $\nu \gg 1$, $S\approx\ln\nu$.

It is clear that the choice of Q and P cannot influence the entropy. The
independence of $S$ from $Q$ and $P$ does not contradict the Legendre
transformation properties. Indeed, usually, one would have
$$
\frac{\partial S}{\partial Q} = \lambda_1\ ,
$$
but here
$$
\frac{\partial S}{\partial Q} = \lambda_1 + 2\lambda_3 Q\ ,
$$
which is zero.

The resulting state operator, generalised to $n$ degrees of freedom, is
described by the following
\begin{thm}\label{propold20} The state operator of the ME packet of a system
with given averages and variances $Q_1,\cdots,Q_n$, $\Delta Q_1,\cdots,\Delta
Q_n$ of coordinates and $P_1,\cdots,P_n$, $\Delta P_1,\cdots,$ $\Delta P_n$ of
momenta, is
\begin{equation}\label{32.1} {\mathsf T}=
\prod_{k=1}^n\left[\frac{2}{\nu_k^2-1}\exp\left(-\frac{1}{\hbar}\ln\frac{\nu_k+1}{\nu_k-1}{\mathsf
K}_k\right)\right]\ ,
\end{equation} where
\begin{equation}\label{20.3} {\mathsf K}_k = \frac{1}{2}\frac{\Delta
P_k}{\Delta Q_k}({\mathsf q}_k-Q_k)^2 + \frac{1}{2}\frac{\Delta Q_k}{\Delta
P_k}({\mathsf p}_k-P_k)^2
\end{equation} and
\begin{equation}\label{20.3b} \nu_k = \frac{2\Delta P_k\Delta Q_k}{\hbar}\ .
\end{equation}
\end{thm}

Strictly speaking, the state operator (\ref{32.1}) is not a Gaussian
distribution. Thus, it seems to be either a counterexample to, or a
generalization of, Jaynes' hypothesis.
\begin{assump}\label{assMEqu} The quantum model ${\mathcal S}_q$ corresponding
to the classical model ${\mathcal S}_c$ described by Assumption \ref{assMEcl}
is the ME packet (\ref{32.1}).
\end{assump}

Let us study the properties of quantum ME packets. In the diagonal
representation, we have for $n=1$:
\begin{equation}\label{20.2a} {\mathsf K} = \sum_{k=0}^\infty
R_m|m\rangle\langle m|\ .
\end{equation} We easily obtain for $R_m$ that
\begin{equation}\label{20.2} R_m = 2\frac{(\nu-1)^m}{(\nu+1)^{m+1}} .
\end{equation} Hence,
$$
\lim_{\nu\rightarrow 1}R_m = \delta_{m0}\ ,
$$
and the state ${\mathsf T}$ becomes $|0\rangle\langle 0|$. In general, states
$|m\rangle$ depend on $\nu$. The state vector $|0\rangle$ expressed as a
function of $Q$, $P$, $\Delta Q$ and $\nu$ is given, for any $\nu$, by
\begin{equation}\label{20.1} \psi(q) = \left(\frac{1}{\pi} \frac{\nu}{2\Delta
Q^2}\right)^{1/4} \exp\left[-\frac{\nu}{4\Delta Q^2}(q-Q)^2 +
\frac{iPq}{\hbar}\right]\ .
\end{equation} This is a Gaussian wave packet that corresponds to other values
of variances than the original ME packet but has the minimum uncertainty. For
$\nu\rightarrow 1$, it remains regular and the projection $|0\rangle\langle
0|$ becomes the state operator of the original ME packet. Hence, Gaussian wave
packets are special cases of quantum ME packets.

The diagonal representation offers a method for calculating averages of
coordinates and momenta products that replaces the partition function way. Let
us denote such a product X. We have
\begin{equation}\label{avX} \langle {\mathsf X}\rangle = \sum_{k=0}^\infty
R_k\langle k|{\mathsf X}|k\rangle\ .
\end{equation} To calculate $\langle k|{\mathsf X}|k\rangle$, we use Eqs.\
(\ref{ua}), (\ref{qa}), (\ref{16.1}), (\ref{16.2}), (\ref{lagranqu1}) and
(\ref{lagranqu2}) to obtain
$$
{\mathsf q} = Q +\frac{\Delta Q}{\sqrt{\nu}}({\mathsf A}+{\mathsf A}^\dagger)\
,\quad {\mathsf p} = P-i\frac{\Delta P}{\sqrt{\nu}} ({\mathsf A}-{\mathsf
A}^\dagger)\ .
$$
By substituting these relations into ${\mathsf X}$ and using the commutation
relations $[{\mathsf A},{\mathsf A}^\dagger] = {\mathsf 1}$, we obtain
$$
{\mathsf X} = {\mathcal P}({\mathsf N}) + {\mathcal Q}({\mathsf A},{\mathsf
A}^\dagger)\ ,
$$
where ${\mathsf N} = {\mathsf A}^\dagger {\mathsf A}$ and where, in each
monomial of the polynomial ${\mathcal Q}$, the number of ${\mathsf A}$-factors
is different from the number of ${\mathsf A}^{\dagger}$-factors. Thus,
$$
\langle k|{\mathsf X}|k\rangle = {\mathcal P}(k)\ .
$$
In Eq.\ (\ref{avX}), there are, therefore, sums
$$
\sum_{k=0}^\infty k^nR_k\ .
$$
With Eq.\ (\ref{20.2}), this becomes
$$
\sum_{k=0}^\infty k^nR_k = \frac{2}{\nu+1}I_n\ ,
$$
where
$$
I_n(\nu) = \sum_{k=0}^\infty k^n\left(\frac{\nu-1}{\nu+1}\right)^k\ .
$$
We easily obtain
$$
I_n = \left(\frac{\nu^2-1}{2}\frac{d}{d\nu}\right)^n\frac{\nu+1}{2}\ .
$$
The desired average value is then given by
\begin{equation}\label{average} \langle {\mathsf X} \rangle =
\frac{2}{\nu+1}{\mathcal
P}\left(\frac{\nu^2-1}{2}\frac{d}{d\nu}\right)\frac{\nu+1}{2}\ .
\end{equation} The calculation of the polynomial ${\mathcal P}$ for a given
${\mathsf X}$ and the evaluation of the right-hand side of Eq.\
(\ref{average}) are the two steps of the promised method.

\subsection{Quantum equations of motion} Let the Hamiltonian of ${\mathcal
S}_q$ be ${\mathsf H}$ and the unitary evolution group be ${\mathsf
U}(t)$. The dynamics in the Schr\"{o}dinger picture leads to the time
dependence of ${\mathsf T}$:
$$
{\mathsf T}(t) = {\mathsf U}(t){\mathsf T} {\mathsf U}(t)^\dagger\ .
$$
Substituting for ${\mathsf T}$ from Eq.\ (\ref{32.1}) and using a well-known
property of exponential functions, we obtain
\begin{equation}\label{32.2} {\mathsf T}(t) =
\frac{2}{\nu^2-1}\exp\left(-\frac{1}{\hbar}\ln\frac{\nu+1}{\nu-1}{\mathsf
U}(t){\mathsf K}{\mathsf U}(t)^\dagger\right)\ .
\end{equation}

In the Heisenberg picture, ${\mathsf T}$ remains constant, while ${\mathsf q}$
and ${\mathsf p}$ are time dependent and satisfy the equations
\begin{equation}\label{33.1} i\hbar\frac{d{\mathsf q}}{dt} = [{\mathsf
q},{\mathsf H}]\ ,\quad i\hbar\frac{d{\mathsf p}}{dt} = [{\mathsf p},{\mathsf
H}]\ .
\end{equation} They are solved by
$$
{\mathsf q}(t) = {\mathsf U}(t)^\dagger{\mathsf q}{\mathsf U}(t)\ ,\quad
{\mathsf p}(t) = {\mathsf U}(t)^\dagger{\mathsf p}{\mathsf U}(t)\ ,
$$
where ${\mathsf q}$ and ${\mathsf p}$ are the initial operators, ${\mathsf
q}={\mathsf q}(0)$ and ${\mathsf p}={\mathsf p}(0)$. The resulting operators
can be written in the form of operator functions analogous to classical
expressions (\ref{36.65}) so that Eqs.\ (\ref{36.7}) and (\ref{36.8}) can
again be used.

The example with potential function (\ref{36.1}) is solvable in quantum
theory, too, and we can use it for comparison with the classical dynamics as
well as for a better understanding of the ME packet dynamics. Eqs.\
(\ref{33.1}) have then the solutions given by (\ref{37.1}) and (\ref{37.2})
with functions $f_n(t)$ and $g_n(t)$ given by (\ref{37.4}) and (\ref{37.5}) or
(\ref{37.9a}) and (\ref{37.9b}). The calculation of the averages and variances
is analogous to the classical one and we obtain Eqs.\ (\ref{38.3}) and Eq.\
(\ref{39.1}) again with the difference that the term $2\langle {\mathsf
q}{\mathsf p}\rangle$ on the right hand side of (\ref{38.4}) is now replaced
by $\langle {\mathsf q}{\mathsf p}+{\mathsf p}{\mathsf q}\rangle$.

To calculate $\langle {\mathsf q}{\mathsf p}+{\mathsf p}{\mathsf q}\rangle$,
we use the method introduced in the previous section. We have
$$
{\mathsf q}{\mathsf p}+{\mathsf p}{\mathsf q} = 2QP + 2\frac{P\Delta
Q}{\sqrt{\nu}}({\mathsf A}+{\mathsf A}^\dagger)-2i\frac{Q\Delta
P}{\sqrt{\nu}}({\mathsf A}-{\mathsf A}^\dagger)-2i\frac{\Delta Q\Delta P}{\nu}
({\mathsf A}^2-{\mathsf A}^{\dagger 2})\ .
$$
hence, ${\mathcal P} = 2QP$, and
$$
\langle{\mathsf q}{\mathsf p}+{\mathsf p}{\mathsf q}\rangle = 2QP\ .
$$
The result is again Eq.\ (\ref{39.1}). Similarly for ${\mathsf p}$, the
results are given by Eqs.\ (\ref{39.2}) and (\ref{39.3}).

We have shown that the averages and variances of quantum ME packets have
exactly the same time evolution as those of classical ME packets in the
special case of at-most-quadratic potentials. From Eqs.\ (\ref{39.1}) and
(\ref{39.3}) we can also see an interesting fact. On the one hand, both
variances must increase near $t=0$. On the other hand, the entropy must stay
constant because the evolution of the quantum state is unitary. As the
relation between entropy and $\nu$ is fixed for ME packets, the ME packet form
is not preserved by the evolution (the entropy ceases to be maximal). This is
similar for Gaussian-packet form or for coherent-state form.

For general potentials, there are two types of corrections to the dynamics of
the averages: terms containing the variances and terms containing $\hbar$. To
obtain these corrections, let us calculate time derivatives for the quantum
analogue of Hamiltonian (\ref{35.1}) with potential (\ref{50.2}). The
Heisenberg-picture equations of motion give again
$$
\frac{d{\mathsf q}}{dt} = \frac{1}{\mu}{\mathsf p}\ ,
$$
so that Eq.\ (\ref{50.4}) is valid. The other equation,
$$
i\hbar\frac{d{\mathsf p}}{dt} = [{\mathsf p},{\mathsf H}]\ ,
$$
can be applied iteratively as in the classical case so that all time
derivatives of ${\mathsf p}$ can be obtained. Thus,
\begin{equation}\label{52.1} \frac{d{\mathsf p}}{dt} = -V_1-V_2{\mathsf
q}-\frac{V_3}{2}{\mathsf q}^2-\frac{V_4}{6}{\mathsf q}^3 + r_5\ ,
\end{equation} and
$$
\frac{d^2{\mathsf p}}{dt^2} = -\frac{V_2}{\mu}{\mathsf p} -
\frac{V_3}{2\mu}({\mathsf q}{\mathsf p}+{\mathsf p}{\mathsf q}) -
\frac{V_4}{6\mu}({\mathsf q}^2{\mathsf p}+{\mathsf q}{\mathsf p}{\mathsf
q}+{\mathsf p}{\mathsf q}^2) + r_5\ .
$$
This differs from the classical equation only by factor ordering. We can use
the commutator $[{\mathsf q},{\mathsf p}]=i\hbar$ to simplify the last term,
\begin{equation}\label{52.2} \frac{d^2{\mathsf p}}{dt^2} =
-\frac{V_2}{\mu}{\mathsf p}-\frac{V_3}{2\mu}({\mathsf q}{\mathsf p}+{\mathsf
p}{\mathsf q})-\frac{V_4}{2\mu}{\mathsf q}{\mathsf p}{\mathsf q} + r_5\ .
\end{equation} Similarly,
\begin{multline}\label{52.3} \frac{d^3{\mathsf p}}{dt^3} =
-\frac{V_3}{\mu^2}{\mathsf p}^2-\frac{V_4}{\mu^2}{\mathsf p}{\mathsf
q}{\mathsf p} + \frac{V_1V_2}{\mu} + \frac{V_1V_3+V_2^2}{\mu}{\mathsf q} +
\frac{3V_2V_3+V_1V_4}{2\mu}{\mathsf q}^2 \\ +
\frac{4V_2V_4+3V_3^2}{6\mu}{\mathsf q}^3 + \frac{5V_3V_4}{12\mu}{\mathsf q}^4
+ \frac{V_4^2}{12\mu}{\mathsf q}^5 + r_5\ ,
\end{multline} and
\begin{multline}\label{52.4} \frac{d^4{\mathsf p}}{dt^4} =
-\frac{V_4}{\mu^3}{\mathsf p}^3 + \frac{3V_1V_3+V_2^2}{\mu^2}{\mathsf p} +
\frac{3V_1V_4+5V_2V_3}{2\mu^2}({\mathsf q}{\mathsf p}+{\mathsf p}{\mathsf q})
+ \frac{5V_3^2+8V_2V_4}{2\mu^2}{\mathsf q}{\mathsf p}{\mathsf q} \\ +
\frac{3V_3V_4}{2\mu^2}({\mathsf q}^3{\mathsf p}+{\mathsf p}{\mathsf q}^3) +
\frac{3V_4^2}{4\mu^2}{\mathsf q}^2{\mathsf p}{\mathsf q}^2 + r_5\ .
\end{multline}

Next, we calculate quantum averages with the help of Eq.\
(\ref{average}). The quantum averages of the monomials that are linear in one
of the variables ${\mathsf q}$ or ${\mathsf p}$ can differ from their classical
counterparts only by terms that are of the first order in $1/\nu$ and purely
imaginary. For example,
$$
\langle {\mathsf q}{\mathsf p}\rangle = QP + \frac{i\hbar}{2}\ ,
$$
or
$$
\langle {\mathsf q}^3{\mathsf p}\rangle = Q^3P + 3QP\Delta Q^2 +
3i\frac{Q^2\Delta Q\Delta P}{\nu} + 3i\frac{\Delta Q^3\Delta P}{\nu}\ .
$$
These corrections clearly cancel for all symmetric factor orderings. The first
term in which a second-order correction occurs is ${\mathsf q}^2{\mathsf p}^2$
and we obtain for it:
$$
\langle {\mathsf p}{\mathsf q}^2{\mathsf p}\rangle = \langle
q^2p^2\rangle_{\text{class}} + 2\frac{\Delta Q^2\Delta P^2}{\nu^2}\ .
$$

The equations (\ref{52.1})--(\ref{52.4}) do not contain any such terms and so
their averages coincide exactly with the classical equations
(\ref{51.1})--(\ref{51.4}). The terms ${\mathsf q}^2{\mathsf p}^2$ with
different factor orderings occur in the fifth time derivative of ${\mathsf p}$
and have the form
$$
\frac{3V_3V_4}{2\mu^2}\left[{\mathsf q}^3{\mathsf p}+{\mathsf p}{\mathsf
q}^3,\frac{{\mathsf p}^2}{2\mu}\right] +
\frac{V_3V_4}{2\mu^3}\left[\frac{1}{3}{\mathsf q}^3,{\mathsf p}^3\right]
=i\hbar\frac{V_3V_4}{2\mu^3}(21{\mathsf p}{\mathsf q}^2{\mathsf p}-11\hbar^2)\
.
$$
The average of the resulting term in the fifth time derivative of ${\mathsf
p}$ is
$$
\frac{V_3V_4}{2\mu^3}\left(21Q^2P^2 + 21P^2 \Delta Q^2 + 21 Q^2\Delta P^2 +
21\Delta Q^2\Delta P^2-\frac{\hbar^2}{2}\right)\ .
$$
If we express $\hbar$ as $2\Delta Q\Delta P/\nu$, we can write the last two
terms in the parentheses as
$$
\Delta Q^2\Delta P^2\left(21-\frac{2}{\nu^2}\right)\ .
$$
A similar term appears in the third time derivative of ${\mathsf p}$, if we
allow $V_5\neq 0$ in the expansion (\ref{50.2}):
$$
\left[-\frac{V_5}{12\mu}({\mathsf q}^3{\mathsf p}+{\mathsf p}{\mathsf
q}^3),\frac{{\mathsf p}^2}{2\mu}\right] =
i\hbar\left[-\frac{V_5}{4\mu^2}(2{\mathsf p}{\mathsf q}^2{\mathsf
p}+\hbar^2)\right]\ ,
$$
which contributes to $d^3P/dt^3$ by
$$
-\frac{V_5}{2\mu^2}\left(\langle q^2p^2\rangle_{\text{class}} + \frac{4\Delta
Q^2\Delta P^2}{\nu^2}\right)\ .
$$
Again, the correction is of the second order in $\nu^{-1}$.

We can conclude. The quantum equations begin to differ from the classical
ones only for the higher order terms in $V$ or in the higher time derivatives
and the correction is of the second order in $1/\nu$. This seems to be very
satisfactory: our quantum model reproduces the classical dynamic very
well. Moreover, Eq.\ (\ref{20.1}) shows that Gaussian wave packets are special
cases of ME packets with $\nu=1$. Thus, they approximate classical
trajectories less accurately than ME packets with large $\nu$. Of course,
these results have as yet been shown only for the first four time
derivatives. It would be nice if a general theorem could be proved.

\section{Classical limit} Let us now look to see if our equations give some
support to the statement that $\nu\gg 1$ is the classical regime.

The quantum partition function (\ref{17.3}) differs from its classical
counterpart (\ref{22.1}) by the denominator
$\sinh(\hbar\sqrt{\lambda_3\lambda_4})$. If
\begin{equation}\label{ll} \hbar\sqrt{\lambda_3\lambda_4} \ll 1\ ,
\end{equation} we can write
$$
\sinh(\hbar\sqrt{\lambda_3\lambda_4}) = \hbar\sqrt{\lambda_3\lambda_4}[1 +
O((\hbar\sqrt{\lambda_3\lambda_4})^2)]
$$
The leading term in the partition function then is
$$
Z = \frac{\pi}{h}\frac{1}{\sqrt{\lambda_3\lambda_4}}
\exp\left(\frac{\lambda_1^2}{4\lambda_3} +
\frac{\lambda_2^2}{4\lambda_4}\right)\ ,
$$
where $h = 2\pi\hbar$. Comparing this with Eq.\ (\ref{22.1}) shows that the
two expressions are identical if we set
$$
v = h\ .
$$
We can interpret this by saying that quantum mechanics gives us the value of
$v$. Next, we have to express condition (\ref{ll}) in terms of the averages
and variances. Equations (\ref{lagranqu1}) and (\ref{lagranqu2}) imply
$$
\hbar\sqrt{\lambda_3\lambda_4} = \frac{1}{2}\ln\frac{\nu+1}{\nu-1}\ .
$$
Hence, condition (\ref{ll}) is equivalent to
\begin{equation}\label{gg} \nu \gg 1\ .
\end{equation}

The result can be formulated as follows. Classical mechanics allows not only
sharp, but also fuzzy trajectories and the comparison of some classical and
quantum fuzzy trajectories shows a very good match. The fuzzy states chosen
here are the so-called ME packets. Their fuzziness is described by the
quantity $\nu = 2\Delta Q\Delta P/\hbar$. The entropy of an ME packet depends
only on $\nu$ and is an increasing function of it. The time evolution of
classical and quantum ME packets with the same initial values of averages and
variances defines the averages as time functions. The larger $\nu$ is, the
better the quantum and the classical evolutions of average values have been
shown to agree for the first four terms in the expansion in powers of
time. Thus, the classical regime is neither $\Delta Q = \Delta P = 0$
(absolutely sharp trajectory) nor $\nu=1$ (minimum quantum uncertainty). This
is the most important result of Ref.\ \cite{hajicek}. The time functions
coincide for the two theories in the limit $\nu \rightarrow \infty$. Hence, in
our approach, this is the classical limit. This is just the opposite to the
usual assumption that the classical limit must yield the variances as small as
possible. Of course, $\nu$ can be very large and still compatible with
classically negligible variances.

One also often requires that commutators of observables vanish in classical
limit. This is however only motivated by the assumption that all basic quantum
properties are single values of observables. Within our interpretation, this
assumption is replaced by the following claim: If classical observables are
related to quantum operators then only in such a way that they are average values of the
operators in classicality states. Then, first, all such averages are defined
by a preparation and do exist simultaneously, independently of whether the
operators commute or not. For example, $Q$ and $P$ are such simultaneously
existing variables for ME packets. Second, a joint measurement of fuzzy values
of non-commuting observables is possible. This will be explained in Section
3.7.

It might be helpful to emphasise that construction of models of Newtonian
mechanics and the so-called semi-classical or WKB approximation to quantum
mechanics are two different things. Indeed, the semi-classical approximation
is a mathematical method, usually defined as the expansion in powers of $h$ in
some quantum expressions \cite{peres}, to calculate approximately correct
values of quantum expressions in suitable applications. Equations resulting
from $h \rightarrow 0$ may be similar to the corresponding classical
equations. In fact, limit $\nu \rightarrow \infty$ also results from $h
\rightarrow 0$ if the variances are kept constant. The suitable applications
can be more general than the above construction of models in that they, e.g.,
do not necessarily concern fuzzy trajectories and macroscopic systems.

\section{A model of classical rigid body} To show how the above theory of
classical properties works, we construct a one dimensional model of a free
solid body. The restrictions to one dimension and absence of external forces
enable us to calculate everything explicitly---the model is completely
solvable. The real object is a thin solid rod of mass $M$ and length $L$. Its
classical model ${\mathcal S}_c$ is a one-dimensional continuum of the same
mass and length, with mass density $M/L$, internal energy $E$, centre of mass
$X$ and total momentum $P$. The classical state coordinates (see Section 3.1)
are $M$, $L$, $X$, $P$ and $E$.

The construction of its quantum model ${\mathcal S}_q$ entails that, first,
the structural properties of the system must be defined, second, some
assumptions on the state of the system must be done; third, the objective
properties must be found that correspond to the classical properties $M$, $L$, $E$,
$X$ and $P$. Large parts of this section follow reference \cite{PHJT}.

\subsection{Composition, Hamiltonian and spectrum}
\begin{assump}\label{assrodham} ${\mathcal S}_q$ is an isolated linear chain
of $N$ identical particles of mass $\mu$ distributed along the $x$-axis with
the quantum Hamiltonian
\begin{equation}\label{Ham} {\mathsf H} = \frac{1}{2\mu}\sum_{n=1}^N {\mathsf
p}_n^2 + \frac{\kappa^2}{2}\sum_{n=2}^N ({\mathsf x}_n - {\mathsf x}_{n-1} -
\xi)^2,
\end{equation} involving only nearest-neighbour elastic forces. Here operator
${\mathsf x}_n$ is the position, operator ${\mathsf p}_n$ the momentum of the
$n$-th particle, $\kappa$ the oscillator strength and $\xi$ the equilibrium
interparticle distance.
\end{assump} The parameters $N$, $\mu$, $\kappa$ and $\xi$ are structural
properties (determining the Hamiltonian of a closed system, see Section
1.3.2).

This kind of chain seems to be different from most chains that are studied in
literature: the positions of the chain particles are dynamical variables so
that the chain can move as a whole and the invariance with respect to Galilean
group is achieved. However, the chain can still be solved by methods that are
described in \cite{Kittel,Rutherford}.

First, we find the variables ${\mathsf u}_n$ and ${\mathsf q}_n$ that
diagonalize the Hamiltonian and thus define the so-called normal modes. The
transformation is
\begin{equation}\label{xu} {\mathsf x}_n = \sum_{m=0}^{N-1}Y^m_n{\mathsf u}_m
+ \left(n - \frac{N+1}{2}\right)\xi,
\end{equation} and
\begin{equation}\label{pu} {\mathsf p}_n = \sum_{m=0}^{N-1}Y^m_n{\mathsf q}_m,
\end{equation} where the mode index $m$ runs through $0,1,\cdots,N-1$ and
$Y^m_n$ is an orthogonal matrix; for even $m$,
\begin{equation}\label{evenm} Y^m_n = A(m)\cos\left[\frac{\pi
m}{N}\left(n-\frac{N+1}{2}\right)\right],
\end{equation} while for odd $m$,
\begin{equation}\label{oddm} Y^m_n = A(m)\sin\left[\frac{\pi
m}{N}\left(n-\frac{N+1}{2}\right)\right]\ ,
\end{equation} and the normalization factors are given by
\begin{equation}\label{factor} A(0) = \frac{1}{\sqrt{N}},\quad A(m) =
\sqrt{\frac{2}{N}},\quad m>0.
\end{equation}

To show that ${\mathsf u}_n$ and ${\mathsf q}_n$ do represent normal modes, we
substitute Eqs.\ (\ref{xu}) and (\ref{pu}) into (\ref{Ham}) and obtain, after
some calculation,
$$
{\mathsf H} = \frac{1}{2\mu}\sum_{m=0}^{N-1}{\mathsf q}_m^2 +
\frac{\mu}{2}\sum_{m=0}^{N- 1}\omega_m^2{\mathsf u}_m^2,
$$
which is indeed diagonal. The mode frequencies are
\begin{equation}\label{spectr} \omega_m =
\frac{2\kappa}{\sqrt{\mu}}\sin\frac{m}{N}\frac{\pi}{2}.
\end{equation}

Consider the terms with $m=0$. We have $\omega_0=0$, and
$Y^0_n=1/\sqrt{N}$. Hence,
$$
{\mathsf u}_0 = \sum_{n=1}^N\frac{1}{\sqrt{N}}{\mathsf x}_n, \quad {\mathsf
q}_0 = \sum_{n=1}^N\frac{1}{\sqrt{N}}{\mathsf p}_n,
$$
so that
$$
{\mathsf u}_0 = \sqrt{N}{\mathsf X},\quad {\mathsf q}_0 =
\frac{1}{\sqrt{N}}{\mathsf P},
$$
where ${\mathsf X}$ is the centre-of-mass coordinate of the chain and
${\mathsf P}$ is its total momentum. The "zero" terms in the
Hamiltonian then reduce to
\begin{equation}\label{bulkham} \frac{1}{2M_0}{\mathsf P}^2
\end{equation} with $M_0 = N\mu$. Thus, the "zero mode" describes a
straight, uniform motion of the chain as a whole. The fact that the centre of
mass degrees of freedom decouple from other (internal) ones is a consequence
of Galilean invariance.

The other modes are harmonic oscillators called "phonons" with
eigenfrequencies $\omega_m$, $m = 1,2,\dots,N-1$. The energy of the phonons,
\begin{equation}\label{intenop} {\mathsf E} = {\mathsf H} -
\frac{1}{2M_0}{\mathsf P}^2\ ,
\end{equation} is the internal energy of our system and its spectrum is built
from the mode frequencies by the formula
\begin{equation}\label{phonons} E = \sum_{m=1}^{N-1}\nu_m \hbar\omega_m,
\end{equation} where $\{\nu_m\}$ is an $(N-1)$-tuple of non-negative
integers---phonon occupation numbers.

Let us define the operator describing the mass by
$$
{\mathsf M} = M_0 +\frac{\mathsf E}{c^2}
$$
and the length of the body by
\begin{equation}\label{length} {\mathsf L} = {\mathsf x}_N - {\mathsf x}_1.
\end{equation} We assume that the second term in the expression for the mass
can be safely neglected in the non-relativistic regime in which we are
working.  The length can be expressed in terms of modes ${\mathsf u}_m$ using
Eq.~(\ref{xu}),
$$
{\mathsf L} = (N-1)\xi + \sum_{m=0}^{N-1}(Y^m_N-Y^m_1){\mathsf u}_m.
$$
The differences on the right-hand side are non-zero only for odd values of
$m$, and equal then to $-2Y^m_1$. We easily find, using Eqs.~(\ref{oddm}) and
(\ref{factor}):
\begin{equation}\label{L} {\mathsf L} = (N-1)\xi - \sqrt{\frac{8}{N}}\
\sum_{m=1}^{[N/2]}(-
1)^m\cos\left(\frac{2m-1}{N}\frac{\pi}{2}\right)\,{\mathsf u}_{2m-1}.
\end{equation}

\subsection{Maximum-entropy assumption} The next point is the choice of
classicality states. We write the Hilbert space of ${\mathcal S}_q$ as
$$
{\mathbf H} = {\mathbf H}_{\text{CM}} \otimes {\mathbf H}_{\text{int}}\ ,
$$
where ${\mathbf H}_{\text{CM}}$ is constructed from the wave functions
$\Psi(X)$ (see Section 1.3.1) and ${\mathbf H}_{\text{int}}$ has the phonon
eigenstates as a basis.
\begin{assump}\label{assrodstate} The classicality states have the form
$$
{\mathsf T}_{\text{CM}} \otimes {\mathsf T}_{\text{int}}\ .
$$
Internal state ${\mathsf T}_{\text{int}}$ maximises the entropy under the
condition of fixed average of the internal energy,
$$
\text{Tr}\left[{\mathsf T}_{\text{int}}\left({\mathsf H} -
\frac{1}{2M_0}{\mathsf P}^2\right)\right] = E\ .
$$
The external state ${\mathsf T}_{\text{CM}}$ is the ME packet for given
averages $\langle X\rangle$, $\langle P\rangle$, $\langle \Delta X\rangle$ and
$\langle \Delta P\rangle$.
\end{assump} Let us first focus on ${\mathsf T}_{\text{int}}$. It is the state
of thermodynamic equilibrium or the Gibbs state, which we denote by ${\mathsf
T}_{E}$ (see, e.g., \cite{Jaynes}).

The maximum of entropy does not represent an additional condition but rather
the absence of any, see Section 0.1.4. This is, of course, also a condition,
and its validity in overwhelming number of real cases is an interesting
problem. For ${\mathsf T}_{\text{int}}$, it must have to do with the
preparation (not by physicists but by nature). Physically, the thermodynamic
equilibrium can settle down spontaneously starting from an arbitrary state
only if some weak but non-zero interaction exists both between the phonons and
between the rod and the environment. We assume that this can be arranged so
that the interaction can be neglected in the calculations of the present
section.

The internal energy has itself a very small relative variance in the Gibbs
state if $N$ is large. This explains why it appears to be sharp. All other
classical internal properties will turn out to be functions of the classical
internal energy. Hence, for the internal degrees of freedom, $E$ forms itself
a complete set of state coordinates introduced in Assumption
\ref{assold4}. The properties of internal energy are well known and we shall
not repeat the calculations here.

\subsection{The length of the body} The mathematics associated with the
maximum entropy principle is variational calculus. The condition of fixed
average energy is included with the help of Lagrange multiplier denoted by
$\lambda$. It becomes a function $\lambda(E)$ for the resulting state. As it
is well known, $\lambda(E)$ has to do with temperature.

The phonons of one species are excitation levels of a harmonic oscillator, so
we have
$$
{\mathsf u}_m = \sqrt{\frac{\hbar}{2\mu\omega_m}}({\mathsf a}_m + {\mathsf
a}^\dagger_m),
$$
where ${\mathsf a}_m$ is the annihilation operator for the $m$-th species. The
diagonal matrix elements between the energy eigenstates $\mid\nu_m\rangle$
that we shall need then are
\begin{equation}\label{averu} \langle\nu_m|{\mathsf u}_m|\nu_m\rangle =
0,\quad \langle\nu_m| {\mathsf u}^2_m|\nu_m\rangle =
\frac{\hbar}{2\mu\omega_m}(2\nu_m + 1).
\end{equation}

For our system, the phonons of each species form statistically independent
subsystems, hence the average of an operator concerning only one species in
the Gibbs state ${\mathsf T}_E$ of the total system equals the average in the
Gibbs state for the one species. Such a Gibbs state operator for the $m$-th
species has the form
$$
{\mathsf T}_m = \sum_{\nu_m=0}^{\infty}|\nu_m\rangle
p_{\nu_m}^{(m)}\langle\nu_m|,
$$
where
$$
p_{\nu_m}^{(m)} = Z^{-1}_{m}\exp\left(-\hbar\lambda\omega_m\nu_m\right)
$$
and $Z_m$ is the partition function for the $m$-th species
\begin{equation}\label{partf} Z_{m}(\lambda) = \sum_{\nu_{m}=0}^\infty
e^{-\lambda\hbar\omega_m\nu_m} = \frac{1}{1-e^{-\lambda\hbar\omega_m}}.
\end{equation} The average length is obtained using Eq.\ (\ref{L}),
\begin{equation}\label{averL} \langle {\mathsf L}\rangle_E = (N-1)\xi.
\end{equation} It is a function of objective properties $N$, $\xi$ and $E$.

Eq.\ (\ref{L}) is an important result. It shows that contributions to the
length are more or less evenly distributed over all odd modes. Such a
distribution leads to a very small variance of ${\mathsf L}$ in Gibbs
states. A lengthy calculation \cite{PHJT} gives for large $N$
\begin{equation} \frac{\Delta {\mathsf L}}{\langle {\mathsf L}\rangle_E}
\approx \frac{2\sqrt{3}}{\pi\kappa\xi\sqrt{\lambda}}\frac{1}{\sqrt{N}}.
\end{equation}

Thus, the small relative variance for large $N$ does not need to be assumed from the
start. The only assumptions are values of some structural properties and that
an average value of energy is fixed. We have obtained even more information,
viz.\ the internal-energy dependence of the length (in this model, the
dependence is trivial). This is an objective relation that can be in principle
tested by measurements.

Similar results can be obtained for further thermodynamic properties such as
specific heat, elasticity coefficient\footnote{If we extend the classical
model so that it contains the elasticity coefficient, we could calculate the
coefficient for an extended quantum model, in which the rod would be placed
into a non-homogeneous "gravitational" field described by, say, a
quadratic potential. This would again give a solvable model.} etc. All these
quantities are well known to have small variances in Gibbs states. The reason
is that the contributions to these quantities are evenly distributed over the
normal modes and the modes are mechanically and statistically independent.

\subsection{The bulk motion} The mechanical properties of the system are the
centre of mass and the total momentum. The contributions to them are evenly
distributed over all atoms, not modes: the bulk motion is mechanically and
statistically independent of all other modes and so its variances will not be
small in Gibbs states defined by a fixed average of the total energy. Still,
generalized statistical methods of Sections 3.2--3.4 can be applied to
it. This is done in the present subsection.

First, we assume that the real rod we are modelling cannot possess a sharp
trajectory. Thus, satisfactory models of it can be ME packets in both
Newtonian and quantum mechanics. Then, according to Assumption
\ref{assrodstate} and Theorem \ref{propold19}, the external state of the
classical model can be chosen as
\begin{equation}\label{MErodc} \rho = \frac{v}{2\pi}\frac{1}{\langle \Delta
X\rangle\langle \Delta P\rangle}\exp\left[-\frac{(X-\langle
X\rangle)^2}{2\langle \Delta X\rangle^2} -\frac{(P-\langle
P\rangle)^2}{2\langle \Delta P\rangle^2}\right]\ .
\end{equation} (For the definition of $v$, see Section 3.3.) Similarly,
Theorem \ref{propold20} implies that the external state of the quantum model
can be chosen as
\begin{equation}\label{MErodq} {\mathsf T}_{\text{CM}}=
\frac{2}{\nu^2-1}\exp\left(-\frac{1}{\hbar}\ln\frac{\nu+1}{\nu-1}{\mathsf
K}\right)\ ,
\end{equation} where
$$
{\mathsf K} = \frac{1}{2}\frac{\langle \Delta P\rangle}{\langle \Delta
X\rangle}({\mathsf X}-\langle X\rangle)^2 + \frac{1}{2}\frac{\langle \Delta
X\rangle}{\langle \Delta P\rangle}({\mathsf P}-\langle P\rangle)^2
$$
and
$$
\nu = \frac{2\langle \Delta P\rangle\langle \Delta X\rangle}{\hbar}\ .
$$

The Hamiltonian for the bulk motion of both models is given by equation
(\ref{bulkham}). Thus, as explained in Section 3.4.3, the quantum trajectory
coincides with the classical one exactly. (Recall that trajectory has been
defined as the time dependence of averages and variances.)

Hopefully, this simple rod example has sufficiently illustrated how our idea
of model construction works in the case of classical properties and we can
finish the comparison of classical and quantum models here.

\section{Joint measurement of position \\ and momentum} The existence of an
observable that represents a joint measurement of position and momentum plays
some role in the theory of classical properties. To show it, we generalise the
construction of such an observable for a simplified model that was first
proposed in Ref.\ \cite{AK}. We follow Ref.\ \cite{survey}. The model system
${\mathcal S}$ is a free one-dimensional spin-zero particle with position
${\mathsf q}$ and momentum ${\mathsf p}$. The Hilbert space is $L^2({\mathbb
R})$ and the operators are defined by equations analogous to (\ref{posit}) and
(\ref{moment}).

Operators ${\mathsf q}$ and ${\mathsf p}$ have an invariant common domain and
their commutator is easily calculated to be
$$
[{\mathsf q},{\mathsf p}] = i\hbar\ .
$$
Hence, the joint measurement may be a problem.

The general construction of a non-trivial POV measure for system ${\mathcal
S}$ introduces another system, ancilla, that forms a composite system with
${\mathcal S}$. Let our ancilla ${\mathcal A}$ be a similar particle with
position ${\mathsf Q}$ and momentum ${\mathsf P}$. We work in
$Q$-representation so that the Hilbert space of the composite system
${\mathcal S} + {\mathcal A}$ is $L^2({\mathbb R})\otimes L^2({\mathbb R})$,
which can be identified with $L^2({\mathbb R}^2)$. Then, we have wave
functions $\Psi(q,Q)$ and integral operators with kernels of the form
${K}(q,Q;q',Q')$.

The dynamical variables ${\mathsf A} = {\mathsf q}-{\mathsf Q}$ and ${\mathsf
B} = {\mathsf p}+{\mathsf P}$ of the composite system ${\mathcal S} +
{\mathcal A}$ commute and can therefore be measured jointly. The value space
of PV observable ${\mathsf E}^{{\mathsf A}\wedge {\mathsf B}}$ is ${\mathbb
R}^2$ with coordinates $a$ and $b$ (see end of Section 1.2.3).

The next step is to smear ${\mathsf E}^{{\mathsf A}\wedge {\mathsf B}}$ to
obtain a realistic POV measure ${\mathsf E}_{kl}$, where $k$ and $l$ are
integers. Let us divide the $ab$ plane into disjoint rectangular cells
$X_{kl}= [a_{k}, a_{k+1}]\times [b_{l}, b_{l+1}]$ covering the entire
plane. Each cell is centred at $(a_k,b_l)$, $a_k = (a_{k+1} + a_{k})/2$, $b_l
= (b_{l+1} + b_{l})/2$ and $S_{kl} = (a_{k+1} - a_{k})(b_{l+1} - b_{l})$ is
its area. Then,
\begin{equation}\label{cell} {\mathsf E}_{kl} = {\mathsf E}^{{\mathsf A}\wedge
{\mathsf B}}(X_{kl}) = {\mathsf E}^{\mathsf A}([a_{k},a_{k+1}]){\mathsf
E}^{\mathsf B}([b_{l}, b_{l+1}])\ .
\end{equation} The cells can be arbitrarily small.

The probability to obtain the outcome $\{k,l\}$ in state $\mathsf T$ of the
composite system is
\begin{equation}\label{149} tr[{\mathsf E}_{kl}{\mathsf T}] = \int
dq\,dQ\,dq'\,dQ'\, {E}_{kl}(q,Q;q',Q'){T}(q',Q';q,Q)\ .
\end{equation} We assume that the composite system ${\mathcal S} + {\mathcal
A}$ is prepared in a factorised state
\begin{equation}\label{factors} {\mathsf T} ={\mathsf T}_{\mathcal
S}\otimes{\mathsf T}_{\mathcal A}
\end{equation} and express the probability (\ref{149}) in terms of the state
${\mathsf T}_{\mathcal S}$. The action of the projections ${\mathsf
E}^{\mathsf A}([a_{k}, a_{k+1}])$ and ${\mathsf E}^{\mathsf B}([b_{l},
b_{l+1}])$ on vector states of the form $\Psi(q,Q)$ is
$$
{\mathsf E}^{\mathsf A}([a_{k},a_{k+1}])\Psi(q,Q) =
\chi[a_{k},a_{k+1}](q-Q)\Psi(q,Q)\ ,
$$
where
\begin{eqnarray*} \chi[a_{k},a_{k+1}](x) &=& 1\quad \forall\, x \in
[a_{k},a_{k+1}]\ ,\\ \chi[a_{k},a_{k+1}](x) &=& 0\quad \forall\, x\,\not\in
[a_{k},a_{k+1}]
\end{eqnarray*} and
\begin{multline*} {\mathsf E}^{\mathsf B}([b_{l},b_{l+1}])\Psi(q,Q) =
(2\pi\hbar)^{-2}\int dp\,dP\,dq'\,dQ' \\ \exp\frac{i}{\hbar}[p(q-q') +
P(Q-Q')]\chi[b_{l},b_{l+1}](p + P)\Psi(q',Q')\ ,
\end{multline*} where
\begin{eqnarray*} \chi[b_{l},b_{l+1}](x) &=& 1\quad \forall\, x \in
[b_{l},b_{l+1}]\ ,\\ \chi[b_{l},b_{l+1}](x) &=& 0\quad \forall\, x\ \not\in
[b_{l},b_{l+1}]\ .
\end{eqnarray*}

The trace (\ref{149}) can be calculated in several steps as follows. First,
\begin{multline*} tr[{\mathsf E}_{kl}{\mathsf T}] = tr[{\mathsf E}^{\mathsf
A}([a_{k},a_{k+1}]){\mathsf E}^{\mathsf B}([b_{l},b_{l+1}]) {\mathsf T}] =
tr[{\mathsf E}^{\mathsf B}([b_{l},b_{l+1}]) {\mathsf T}{\mathsf E}^{\mathsf
A}([a_{k},a_{k+1}])] \\ = (2\pi\hbar)^{-2}\int
dq\,dQ\,dp\,dP\,dq'\,dQ'\,\exp\frac{i}{\hbar}[p(q-q') + P(Q-Q')] \\
\chi[b_{l},b_{l+1}](p + P) \chi[a_{k},a_{k+1}](q - Q) {T}(q',Q';q,Q)\ .
\end{multline*} Second, introduce new integration variables $q$, $a$, $p$,
$b$, $q'$, $a'$,
\begin{multline*} tr[{\mathsf E}_{kl}{\mathsf T}] = (2\pi\hbar)^{-2}\int
dq\,da\,dp\,db\,dq'\,da'\, \exp\frac{i}{\hbar}[p(q-q') + (b - p)(q - a-q' +
a')] \\ \times \chi[b_{l},b_{l+1}](b) \chi[a_{k},a_{k+1}](a) \ {T}(q',q' -
a';q,q - a)\ .
\end{multline*} Third, if the cells are sufficiently small, the integrands do
not change appreciably inside the integration intervals of $a$ and $b$ so that
they can be approximated as follows:
$$
\int da\,db\,\chi[a_{k},a_{k+1}](a)\chi[b_{l},b_{l+1}](b)f(a,b) \approx
S_{kl}f(a_k,b_l)\ .
$$
In this way, we obtain
\begin{multline*} tr[{\mathsf E}_{kl}{\mathsf T}] \approx
\frac{S_{kl}}{(2\pi\hbar)^2}\int dq\,dp\,dq'\,da' \\
\exp\frac{i}{\hbar}[(p(a_k-a') + b_l(q - q' - a_k + a')] \ T(q',q' - a';q,q -
a_k)\ .
\end{multline*} But, fourth, the factor containing the integral over $p$ is a
$\delta$-function,
$$
\int dp\,\exp\frac{i}{\hbar}p(a_k-a') = 2\pi\hbar\delta(a' -a_k)\ .
$$
Thus, we obtain,
$$
tr[{\mathsf E}_{kl}{\mathsf T}] \approx \frac{S_{kl}}{(2\pi\hbar)^2}\int
dq\,dq'\,\exp\frac{i}{\hbar}b_l(q - q') \ T(q',q' - a_k;q,q - a_k)\ .
$$
Fifth, we use Eq.\ (\ref{factors}),
$$
{\mathsf T}(q',Q';q,Q) = {\mathsf T}_{\mathcal S}(q',q) {\mathsf T}_{\mathcal
A}(Q',Q)\ ,
$$
with the result:
\begin{prop}\label{propold18} The probability of the outcome $\{k,l\}$ to be
found in the factorised state (\ref{factors}) is given approximately by
\begin{equation}\label{factorT} tr[{\mathsf E}_{kl}{\mathsf T}] \approx
tr\left[{\mathsf T}_{\mathcal S}\frac{S_{kl}}{(2\pi\hbar)^2}{\mathsf
T}_{\mathcal A}[a_k,b_l]\right]\ ,
\end{equation} where
$$
{\mathsf T}_{\mathcal A}[a_k,b_l] = \exp\frac{i}{\hbar}b_l(q - q') \ T(q',q' -
a_k;q,q - a_k)
$$
is the state ${\mathsf T}_{\mathcal A}$ first shifted by $a_k$ and then
boosted by $-b_l$. The approximation improves if the cells are smaller.
\end{prop}

If the cells are sufficiently small, we have $q \approx Q + a_k$ and $p
\approx b_l - P$. In this way, the coordinate of $\mathcal S$ is shifted by
$a_k$ while the inverted momentum is shifted by $-b_l$ with respect to those
of $\mathcal A$.

Eq.\ (\ref{factorT}) shows that there is an 'effective' POV measure ${\mathsf
E}_{{\mathcal S}kl}$ for system $\mathcal S$ defined by
$$
{\mathsf E}_{{\mathcal S}kl} = \frac{S_{kl}}{(2\pi\hbar)^2}{\mathsf
T}_{\mathcal A}[a_k,b_l]
$$
that yields the probability of the outcome $k$ of the above registration
considered as a registration performed on $\mathcal S$.

The state ${\mathsf T}_{\mathcal A}$ is completely arbitrary. To construct a
useful quantity, one usually chooses a vector state in the form of a Gaussian
wave packet (see, e.g., Ref.\ \cite{peres}, p. 418),
$$
\Psi_\sigma = (\pi\sigma^2)^{-1/4}\exp\left(-\frac{Q^2}{2\sigma^2}\right)\ .
$$
Easy calculation yields
$$
\langle Q\rangle = \langle P\rangle = 0\ ,\quad \Delta Q =
\frac{\sigma}{\sqrt{2}}\ ,\quad \Delta P = \frac{\hbar}{\sigma\sqrt{2}}
$$
so that $\Psi_\sigma$ is a state of minimum uncertainty. If we shift
$\Psi_\sigma$ by $a_k$ and then boost it by $-b_l$, we obtain
$$
\Psi_\sigma[a_k,b_l] =
(\pi\sigma^2)^{-1/4}\exp\left(-\frac{(Q-a_k)^2}{2\sigma^2}
-\frac{i}{\hbar}b_lQ\right)\ ,
$$
which is the Gaussian packet concentrated at $(a_k,-b_l)$,
\begin{equation}\label{avvar1} \langle Q\rangle = a_k\ ,\quad \langle P\rangle
= -b_l\ ,
\end{equation} and
\begin{equation}\label{avvar2} \quad \Delta Q = \frac{\sigma}{\sqrt{2}}\
,\quad \Delta P = \frac{\hbar}{\sigma\sqrt{2}}\ .
\end{equation} As the Gaussian wave packet is uniquely determined by its
averages and variances, we can interpret the observable ${\mathsf E}_{kl}$ as
giving the probability that the corresponding registration applied to state
${\mathsf T}_{\mathcal S}$ detects the state of $\mathcal S$ with the averages
and variances given by Eqs.\ (\ref{avvar1}) and (\ref{avvar2}).

Equation (\ref{factorT}) is a general formula valid for an arbitrary state
${\mathsf T}_{\mathcal A}$ of the ancilla. We have chosen ${\mathsf
T}_{\mathcal A}$ to be a Gaussian wave packet, which has $\nu = 1$. However,
${\mathsf T}_{\mathcal A}$ can also be chosen as a quantum ME packet with $Q =
0$, $P = 0$ and $\Delta Q$ and $\Delta P$ allowing arbitrary large $\nu$. The
resulting observable represents the detection of a shifted and boosted ME
packet.

\chapter{Quantum models of preparation and registration} Discussions about the
nature of quantum measurement were started already by founding fathers of
quantum theory, persisted throughout and seem even to amplify at the present
time.

This chapter begins by a short survey of the so-called old theory of quantum
measurement and its criticisms and then describes the theory that has evolved
from papers \cite{hajicek2,hajicek4}.

\section{Old theory of measurement} Let us first briefly describe the quantum
theory of measurement that is generally found in textbooks, such as \cite{BLM}
or \cite{WM}. We call this theory "old", in spite of the fact that it
is in frequent use today \cite{WM}, in order to distinguish it from the theory
introduced in \cite{hajicek2} and to show what is the difference between the
two.

The theory considers two quantum systems: {\em object system} ${\mathcal S}$
with Hilbert space ${\mathbf H}$ on which the measurement is to be done, and
{\em meter} or {\em apparatus} ${\mathcal A}$ with Hilbert space ${\mathbf
H}_{\mathcal A}$ that performs the registration and can be microscopic or
macroscopic. In quantum mechanics, the fact that an arbitrary large number of
identical copies of ${\mathcal S}$ is available enables to carry out a whole
ensemble of equivalent individual measurements as explained in the
Introduction. The equivalence of the individual measurements is defined by the
classical devices that are used to prepare the initial states, to manipulate
the states and to read the meter.

The first step of an individual measurement is to prepare ${\mathcal S}$ in
state ${\mathsf T}$ and ${\mathcal A}$ independently in state ${\mathsf
T}_{\mathcal A}$ so that the initial state of ${\mathcal S} + {\mathcal A}$ is
${\mathsf T} \otimes {\mathsf T}_{\mathcal A}$. This is the preparation stage
of the measurement.

The second step is to bring both systems into interaction that leads to their
entanglement. The interaction is mathematically represented by unitary
transformation ${\mathsf U}$ on ${\mathbf H} \otimes {\mathbf H}_{\mathcal
A}$. At the end of this step, the state of the composite ${\mathcal S} +
{\mathcal A}$ is ${\mathsf U} {\mathsf T} \otimes {\mathsf T}_{\mathcal A}
{\mathsf U}^\dagger$. This step is sometimes called {\em premeassurement}.

The third step is {\em reading} the meter. The requirement that each
individual reading gives a definite result $r$ from value set $\Omega$ is
called {\em objectification requirement} \cite{BLM}. For the sake of
simplicity, we assume $\Omega$ to be a discrete subset of ${\mathbb
R}$. Generalisations to a vector of numbers and to a continuous value set
$\Omega$ are easy and do not lead to any conceptually new problems.

The reading is a rather mysterious procedure in the old theory. To find the
state of the meter, another apparatus is needed, and this apparatus is also a
quantum system; hence, to find what is the result of its reading, another
apparatus would be necessary, etc. This is the so-called von Neumann chain
\cite{JvN}. At some stage before reaching the mind of the observer, the chain
must be cut by applying the so-called {\em projection postulate}. At this cut,
known as {Heisenberg cut} \cite{heisenberg}, the result $r$ becomes
well-defined and the reading is considered as having been made.

After performing many equivalent individual measurements, we obtain the
statistic of the results that can be described by probability distribution
$p_r$, normalised by $\sum_r p_r = 1$.

The old theory avoids discussion of von Neumann series and concentrates on
quantities that describe the measurement more or less phenomenologically
(operational approach). This is done by two important quantities. The first is
POV measure ${\mathsf E}_r$ representing the observable of ${\mathcal S}$ that
has been measured, so that
$$
p_r = tr[{\mathsf T}{\mathsf E}_r]\ .
$$
The second is an operation ${\mathcal O}_r$ that maps initial state ${\mathsf
T}$ of ${\mathcal S}$ to its final state ${\mathsf T}_r$ after value $r$ has
been read i.e., after an individual registration has been made:
$$
{\mathsf T}_r = \frac{{\mathcal O}_r({\mathsf T})}{tr[{\mathsf T}{\mathsf
E}_r]}\ .
$$
An operation is a map ${\mathcal O} : {\mathbf L}_r({\mathbf H}) \mapsto
{\mathbf L}_r({\mathbf H})$ that brings positive operators to positive
operators, and must satisfy some further mathematical conditions which will
not be important for us (for details, see \cite{WM,BLM}). State ${\mathsf
T}_r$ of the object system is called {\em conditional} state or state
conditioned on the readout, or also {\em selective} state. The state of the
object system obtained by averaging the conditional states after the
registrations,
$$
\sum_r p_r {\mathsf T}_r\ ,
$$
is called {\em unconditional} or {\em non-selective}. An important property of
the conditional or unconditional states is that they are not determined by
unitary transformation ${\mathsf U}$ so that we have in general, for an
unconditional state,
$$
\sum_r p_r {\mathsf T}_r \neq \Pi_{\mathcal A}[{\mathsf U} {\mathsf T} \otimes
{\mathsf T}_{\mathcal A} {\mathsf U}^\dagger]
$$
or, for a conditional state,
$$
{\mathsf T}_r \neq \Pi_{\mathcal A}[{\mathsf U} {\mathsf T} \otimes {\mathsf
T}_{\mathcal A} {\mathsf U}^\dagger]\ ,
$$
where $\Pi_{\mathcal A}$ is the partial trace over states of ${\mathcal A}$
(Section 2.1.1). The additional change of the state of the object system in
comparison with state $\Pi_{\mathcal A}[{\mathsf U} {\mathsf T} \otimes {\mathsf
T}_{\mathcal A} {\mathsf U}^\dagger]$ is called {\em state reduction} or {\em
collapse of wave function}.

Experiments can be classified by mathematical properties of ${\mathsf E}$ and
${\mathcal O}$ (see, e.g., \cite{WM}). All the beautiful modern experiments
such as weak measurements \cite{svensson}, non-demolition measurements
\cite{bragin}, experiments with photons such as "seing a photon without
destroying it" \cite{NROBRH} can be analysed in this way
\cite{WM,svensson}.

\subsection{Example: Beltrametti--Cassinelli--Lahti model} The general
discussion above can be explained in terms of a simple example such as
Beltrametti--Cassinelli--Lahti (BCL) model \cite{belt} described in Ref.\
\cite{BLM}, p.\ 38. Let a discrete observable ${\mathsf O}$ of object system
${\mathcal S}$ of type $\tau$ with Hilbert space ${\mathbf H}_\tau$ be
measured. Let $o_k$ be eigenvalues and $\{\phi_{kj}\}$ be the complete
orthonormal set of eigenvectors of ${\mathsf O}$,
$$
{\mathsf O}\phi_{kj} = o_k \phi_{kj}\ .
$$
We assume that $ k = 1,\cdots,N$ so that there is only a finite number of
different eigenvalues $o_k$. This is justified either by the nature of the
whole measurement if only few initial states of the systems are prepared at
all and one can work with a finite Hilbert space, or by the fact that no real
registration apparatus can distinguish all elements of an infinite set from
each other. It can therefore measure only a function of an observable that
maps its spectrum onto a finite set of real numbers. Our observable ${\mathsf
O}$ is then such a function. The projection ${\mathsf E}^{\mathsf O}_k$ on the
eigenspace of $o_k$ is then ${\mathsf E}^{\mathsf O}_k = \sum_j
|\phi_{kj}\rangle\langle\phi_{kj}|$. Variable $o_k$ will play the role of $r$.

Let the meter be a quantum system ${\mathcal A}$ with Hilbert space ${\mathbf
H}_{\mathcal A}$ and an observable ${\mathsf A}$. Let ${\mathsf A}$ be a
non-degenerate, discrete observable with the same eigenvalues $o_k$ and with
orthonormal set of eigenvectors $\psi_k$,
$$
{\mathsf A}\psi_k = o_k \psi_k
$$
with possible further eigenstates (such as $\psi$ of Proposition
\ref{propold21}) and eigenvalues. The projection on an eigenspace is ${\mathsf
E}^{\mathsf A}_k = |\psi_k\rangle\langle\psi_k|$. ${\mathsf A}$ is sometimes
called {\em pointer observable} \cite{BLM}.

The measurement starts with the preparation of ${\mathcal S}$ in state
${\mathsf T}$ and the independent preparation of ${\mathcal A}$ in state
${\mathsf T}_{\mathcal A}$. Let the measurement coupling be ${\mathsf U}$, a
unitary transformation on ${\mathbf H}_\tau \otimes {\mathbf H}_{\mathcal
A}$. Thus, the state of the meter after the premeasurement is $\Pi_{\mathcal
S}\bigl[{\mathsf U}({\mathsf T}\otimes {\mathsf T}_{\mathcal A}){\mathsf
U}^\dagger\bigr]$.

Let us require that state $\Pi_{\mathcal S}\bigl[{\mathsf U}({\mathsf
T}\otimes {\mathsf T}_{\mathcal A}){\mathsf U}^\dagger\bigr]$ gives the same
probability measure for the pointer observable as the initial state ${\mathsf
T}$ predicted for the observable ${\mathsf O}$:
$$
tr[{\mathsf T}{\mathsf E}^{\mathsf O}_k] = tr\bigl[\Pi_{\mathcal S}[{\mathsf
U}({\mathsf T}\otimes {\mathsf T}_{\mathcal A}){\mathsf U}^\dagger]{\mathsf
E}^{\mathsf A}_k\bigr]\ .
$$
This is called {\em probability reproducibility condition}. Now, there is a
theorem \cite{belt}:
\begin{prop}\label{propold21} Let a measurement fulfil all assumptions and
conditions listed above, in particular the probability reproducibility. Then,
for any initial vector state $\psi$ of ${\mathcal A}$, there is a set
$\{\varphi_{kl}\}$ of unit vectors in ${\mathbf H}_\tau$ satisfying the
orthogonality conditions
\begin{equation}\label{orth} \langle \varphi_{kl}|\varphi_{kj}\rangle =
\delta_{lj}
\end{equation} for all $k$ such that ${\mathsf U}$ is a unitary extension of
the map
\begin{equation}\label{unitar} \phi_{kl}\otimes \psi \mapsto
\varphi_{kl}\otimes \psi_k
\end{equation} for all $k,l$.
\end{prop}

Suppose that the initial state of ${\mathcal S}$ is an eigenstate of ${\mathsf
O}$, ${\mathsf T} =|\phi_{kl}\rangle\langle\phi_{kl}|$, with the eigenvalue
$o_k$. Then, Eq.\ (\ref{unitar}) implies that the final state of meter
${\mathcal A}$ is $|\psi_k\rangle\langle\psi_k|$. In this case, the
registration of the pointer observable ${\mathsf A}$ on the meter yields a
definite result.

Suppose next that the initial state is an arbitrary vector state, ${\mathsf T}
=|\phi\rangle\langle\phi|$. Decomposing $\phi$ into the eigenstates,
\begin{equation}\label{decomp} \phi = \sum_{kl} c_{kl}\phi_{kl}\ ,
\end{equation} we obtain from Eq.\ (\ref{unitar})
\begin{equation}\label{finalSA} \Phi_{\text{end}} = {\mathsf U} (\phi \otimes
\psi) = \sum_k \sqrt{p_k}\varphi^1_k\otimes \psi_k\ ,
\end{equation} where
\begin{equation}\label{Phik} \varphi^1_k = \frac{\sum_l
c_{kl}\varphi_{kl}}{\sqrt{\langle \sum_l c_{kl}\varphi_{kl}|\sum_j
c_{kj}\varphi_{kj}\rangle}}
\end{equation} and
$$
p_k = \left\langle \sum_l c_{kl}\varphi_{kl}\Biggm|\sum_j
c_{kj}\varphi_{kj}\right\rangle
$$
is the probability that a registration of ${\mathsf O}$ performed on vector
state $\phi$ gives the value $o_k$. The final state of apparatus ${\mathcal
A}$ then is
\begin{equation}\label{finalA} \Pi_{\mathcal S}[{\mathsf U}({\mathsf T}\otimes
{\mathsf T}_{\mathcal A}){\mathsf U}^\dagger] = \sum_{kl}
\sqrt{p_k}\sqrt{p_l}\langle\varphi^1_k|\varphi^1_l\rangle
|\psi_k\rangle\langle\psi_l|\ .
\end{equation}

Because of the orthonormality of $|\psi_k\rangle$'s, the probability that the
result is $o_k$ if ${\mathsf A}$ is registered on ${\mathcal A}$ in this final
state is $p_k$, which is what the probability reproducibility requires.

Independently of how long the von Neumann series of registration apparatuses
is and of where the Heisenberg cut is performed, we can assume that the
reading of value $o_k$ causes the conditional state of the composite to be
\begin{equation}\label{condSA} \varphi^1_k\otimes \psi_k
\end{equation} instead of the right-hand side of Eq.\ (\ref{finalSA}). In
this way, the old theory leads to definitive and useful results.

From Eq.\ (\ref{condSA}), we find that the operation ${\mathsf O}_k$
describing this registration is defined by
$$
{\mathsf O}_k(|\phi\rangle \langle \phi|) = p_k |\varphi_k\rangle \langle
\varphi_k|
$$
and that the observable being registered is ${\mathsf E}^{\mathsf O}$.

The state of ${\mathcal S}$ conditional on reading $o_k$ is then
$$
|\varphi_k\rangle \langle \varphi_k|
$$
and the non-conditional state of ${\mathcal S}$ after the measurement is the
convex combination
$$
\sum_k p_k |\varphi_k\rangle \langle \varphi_k|\ .
$$
Clearly, each individual registration can be considered as a preparation of
the object system in the corresponding conditional state. Hence, the
non-conditional state is obtained by a statistical mixture of preparations and
the state average represents its statistical decomposition according to
Section 1.1.2. Thus, we can write it as
\begin{equation}\label{noncondSA} \left(\sum_k\right)_p p_k |\varphi_k\rangle
\langle \varphi_k|\ .
\end{equation}

Similarly, the conditional state of the meter after the reading of value $o_k$
is $|\psi_k\rangle \langle \psi_k|$ and its non-conditional state is
$\left(\sum_k\right)_p p_k |\psi_k\rangle \langle \psi_k|$, which is different
from the state (\ref{finalA}) of the meter before reading in two respects:
first, the non-conditional state contains only the diagonal terms of the state
operator defined by vector (\ref{finalA}) and second, it is a statistical
decomposition.

The model can be varied in a great number of ways \cite{WM}. For example, one
can give up the requirement of probability reproducibility. One can also
abandon any direct relation between pointer observable ${\mathsf A}$ and
observable ${\mathsf O}$ as well as the condition that the states of the meter
obtained by the evolution defined by the measurement coupling that occur in formula
(\ref{unitar}) are eigenstates of the pointer observable ${\mathsf A}$, or
that they are orthogonal to each other (this is vital for weak measurements,
see \cite{svensson}).

\subsection{Attempts at improvement of the old theory} The old theory is
conceptually not satisfactory even if it is sufficient for all practical
purposes (abbreviation FAPP introduced by John Bell). The problem is the
objectification requirement. Indeed, it is impossible to evolve right-hand
side of Eq.\ (\ref{finalSA}) to that of Eq.\ (\ref{noncondSA}) in terms
of a unitary transformation. Some attempts to solve this difficulty start from
the assumption that the difference between (\ref{finalSA}) and
(\ref{noncondSA}) is not observable because the registration of the
observables that would reveal the difference is either very difficult or that
such observables do not exist. One can then deny that the transition from
(\ref{finalSA}) to (\ref{noncondSA}) really takes place and so assume that the
objectification is only apparent (no-collapse scenario). Other attempts do
assume that the reduction is a real process and postulate a new dynamics that
leads directly to (\ref{noncondSA}) taking into account that some measurement
could disprove this assumption. There are three most important no-collapse
approaches:
\begin{enumerate}
\item Quantum decoherence theory \cite{Zeh,Zurek,schloss}. The idea is that
system ${\mathcal S} + {\mathcal A}$ composed of a quantum object and an
apparatus cannot be isolated from environment ${\mathcal E}$. Then the unitary
evolution of ${\mathcal S} + {\mathcal A} + {\mathcal E}$ leads to a
non-unitary evolution of ${\mathcal S} + {\mathcal A}$ that can erase all
correlations and interferences from ${\mathcal S} + {\mathcal A}$ that hinder
the objectification \cite{Zeh,Zurek,schloss}. This leads to the necessary
convex combination of the end states but not to the statistical decomposition
needed for objectification (see discussion in Refs.\
\cite{d'Espagnat,Ghirardi,bub,BLM,PHJT,hajicek}).
\item Superselection sectors approach \cite{Hepp,Primas,wanb}. Here, classical
properties are described by superselection observables of ${\mathcal A}$ that
commute with each other and with all other observables of ${\mathcal A}$ (see
Section 1.2.5). Then, the state of ${\mathcal A}$ after the measurement is
equivalent to a suitable convex combination but again it is not the required
statistical decomposition.
\item Modal interpretation \cite{bub}. One assumes that there is a subset of
projections that, first, can have determinate values in the state of
${\mathcal S} + {\mathcal A}$ before the registration in the sense that the
assumption does not violate contextuality (see Section 1.2.4), and second, that
one can reproduce all important results of ordinary quantum mechanics with the
help of these limited set of observables. Thus, one must postulate that other
observables are not registered.
\end{enumerate}

The second kind of approach is known as dynamical reduction program
\cite{GRW,pearle}. It postulates new universal dynamics that is non-linear and
stochastic. The main idea is that of spontaneous localisation. That is, linear
superpositions of different positions spontaneously decay, either by jumps
\cite{GRW} or by continuous transitions \cite{pearle}. The form of this decay
is chosen judiciously to take very long time for microsystems so that the
unitary evolution is a good approximation, and take very short time for
macrosystems so that the evolution between Eqs.\ (\ref{finalSA}) and
(\ref{noncondSA}) is granted. In this way, a simple explanation of the
definite positions of macroscopic systems and of the pointers of registration
apparatuses is achieved. Moreover, it is a realist approach to quantum
mechanics.

\section{New theory of measurement} What we have called "new theory"
assumes that the state reduction satisfying Rule \ref{rldegrad} really takes
place, that it happens at a well-defined time and place, namely in a detector
(see below), and that it seems to contradict quantum mechanics only because
quantum mechanics and registration process are not well understood. Indeed,
Section 2.2 has shown that quantum mechanics is not well understood because
the disturbance of registration by identical particles and the necessity of
changes of separation status is disregarded, as in the old theory. As we are
going to explain in the present section, registration process is not well
understood because the old-theory notion of "meter" does not
distinguish between microscopic ancilla and macroscopic detector.

To explain the second point, let us describe what a detector is (for more
details see e.g.\ \cite{leo,stefan}). The first important part of a detector
is the so-called {\em sensitive matter} forming a macroscopic {\em active
volume}. The second important part is a kind of signal collector. For example,
the sensitive matter of an ionisation detector is a gas. When detected
microscopic system ${\mathcal S}$ interacts with the molecules of the gas, it
turns them into electron-ion pairs if its energy is higher than the energy of
one ionisation. If it is much higher than this threshold, many pairs are
produced. Then, the signal collector is the electrodes with sufficient voltage
difference that attract and collect the resulting negative and positive
charges.

In the so-called cryogenic detectors \cite{stefan}, ${\mathcal S}$ interacts,
e.g., with superheated superconducting granules by scattering off a nucleus in
a granule and the resulting phonons induce the phase transition from the
superconducting into the normally conducting phase. An active volume can
contain very many granules (typically $10^9$) in order to enhance the
probability of such scattering if the interaction between ${\mathcal S}$ and
the nuclei is very weak (weakly interacting massive particles,
neutrino). Then, there is signal collector: a solenoid around the active
volume and an independent strong magnetic field. The phase transition of only
one granule leads to a change in magnetic current through the solenoid giving
a perceptible electronic signal.

Modern detectors are constructed so that their signal is electronic. For
example, to a scintillating matter, a photomultiplier is attached, see Section
0.1.2. We assume that there is a signal collected immediately after the
sensitive matter changes its phase, which we call {\em primary}
signal. Primary signal may still be amplified and filtered by other electronic
apparatuses to transform it into the final signal of the detector. For
example, the light signal of a scintillation film in Tonomura experiment
described in Section 0.1.2 is a primary signal. It is then transformed into an
electronic signal by a photocathode and the resulting electronic signal is
further amplified by a photomultiplier, but this is already a transformation
of the primary signal.

In order to make a detector respond ${\mathcal S}$ must lose some of its
energy to the sensitive matter of the detector. The larger the loss, the
better the signal. Thus, most detectors are built in such a way that
${\mathcal S}$ loses all its kinetic energy and is absorbed by the detector
(in this way, also its total momentum can be measured). Let us call such
detectors {\em absorbing}. If the bulk of the sensitive matter is not large
enough, ${\mathcal S}$ can leave the detector after the interaction with it,
in which case we call the detector {\em non-absorbing}.

Observe that a detector is absorbing even if most copies of ${\mathcal S}$
leave the detector without causing a response but cannot leave if there is a
response (some neutrino detectors). Suppose that ${\mathcal S}$ is prepared in
such a way that it must cross a detector. Then, the probability of the
detector response is generally $\eta < 1$. We call a detector {\em ideal}, if
$\eta = 1$.

The main idea of detectors seems to be that the sensitive matter together with
the signal collector is in a classical state of metastable equilibrium (in
thermodynamic sense) and that the interaction with the detected system
disturbs the equilibrium. The detector then relaxes to a new equilibrium and
this leads to the signal on the one hand and to the dissipation of the state
on the other. Clearly, such high-entropy states cannot be described by wave
functions unlike the states of an ancilla.

An important observation is that a real detector gives a definite signal
(remaining silent is also a kind of signal) in each individual
registration. The main assumption of the new theory is that the stage of
registration at which the state reduction occurs is the interaction of a
detected system with a detector:
\begin{assump}\label{assold6} Any registration apparatus for microsystems must
contain at least one detector. Every von Neumann series contains an
interaction of the detected microsystem with sensitive matter of the detector
and the Heisenberg cut is this interaction. Every "reading of a pointer
value" is a signal from the detector.
\end{assump} We call Assumption \ref{assold6} {\em Pointer Hypothesis}.

We also propose a reason why unitary evolution may be distorted by the
detection process. The sensitive matter contains many particles of the same
type as, or that are subsystems of, the detected system and this lead to a
change of separation status of the system. The change can be a complete
loss---the status becomes trivial---or a partial loss of the status. This is
why the states of the system composed of the registered system and the detector become degraded (see Section
2.3).

Now, we can specify how the old theory of measurement is corrected by the new
theory. First, the Heisenberg cut is a definite stage of the von Neumann
series, viz.\ an interaction with a detector or a screen (see Section
2.3). Second, the meter of the old theory is a microscopic quantum system that
can itself interact with a detector, and which we prefer to be called {\em
ancilla}. An ancilla can be described by wave functions and the equations of
the old theory are applicable to the interaction of an object system with an
ancilla. After this interaction, one can use detectors either to detect the
ancilla, or the object system, or both. In fact, all scattering experiments
and most modern quantum experiments (such as non-demolition or weak
measurements) are {\em indirect} measurements in the sense that the object
system first interact with an ancilla and the ancilla is then registered {\em
directly}, being manipulated by fields and screens and finally exciting a
detector. Thus, the old theory does not become obsolete: it gives a valid
account of the interaction of the object system with the ancilla and of its
conditional state resulting after a registration of the ancilla by a detector.

However, the old theory is not applicable to processes running in detectors
because it completely disregards changes of separation status. In particular,
individual states of the object system after an interaction with a detector
such as $\varphi^1_k$ do not necessarily exist and quantities such as ${\mathcal
O}_k$ are not necessarily well-defined. That is the reason why the meter cannot be a
detector. We shall study many examples of direct registration in the next
section.

\section{Models of direct registrations} As explained in previous section, the
old theory of measurement is not applicable to direct registrations. It is the
purpose of the present section to supply the missing part of the theory. We
shall modify the BCL model of Section 4.1.1 by introducing detectors,
describing the formal evolution and specifying Rule \ref{rldegrad} by model
assumptions valid for several different cases of registration. These will be
our model assumptions for each detection process. We shall also observe that
the overall design of the detector determines what is sometimes (see, e.g.,
\cite{schloss}) called {\em preferred basis}: the basis in which the reduced
state becomes diagonal. The exposition follows, more or less, Ref.\
\cite{hajicek4}.

\subsection{Ideal detectors} First, we simplify things by assumption that the
detectors are ideal. For an ideal detector, the number of events registered by
the detector equals the number of events impinging on it ({\em intrinsic
efficiency} equal to 1). We also restrict ourselves to the active
volume\footnote{An active volume is a quantum system, not just a volume of
space.} of the detector, denote it by ${\mathcal D}$ and speak of it as of the
detector. It is a macroscopic quantum system (not just a space volume) with
Hilbert space ${\mathbf H}_{\tau'}$. Object system ${\mathcal S}$ (which can
be the ancilla of some registration) has Hilbert space ${\mathbf
H}_\tau$. Initially, ${\mathcal S}$ and ${\mathcal D}$ are separated. We can,
therefore, speak of initial states $\phi_{mk}$ of ${\mathcal S}$ as in Section
4.1.1 and ${\mathsf T}$ of ${\mathcal D}$, where ${\mathsf T}$ is assumed to
be a stationary, high-entropy state.

Eq.\ (\ref{unitar}) has now to be replaced by the formal evolution of
${\mathcal S} + {\mathcal D}$ on ${\mathsf H}_{\tau\tau'} = {\mathsf
P}_{\tau\tau'}({\mathsf H}_\tau \otimes {\mathsf H}_{\tau'})$, where ${\mathsf
P}_{\tau\tau'}$ is defined by Eq.\ (\ref{projTT'}). Let us write a
suitable initial state as follows:
\begin{equation}\label{init} {\mathsf T}_{\text{init}}({\mathrm c}) = {\mathsf
J}\left(\sum_{kl} c_k c^*_l |\phi_{mk}\rangle\langle \phi_{ml}| \otimes
{\mathsf T}\right) = \nu \left(\sum_{kl} c_k c^*_l{\mathsf
P}_{\tau\tau'}(|\phi_{mk}\rangle\langle \phi_{ml}| \otimes {\mathsf
T}){\mathsf P}_{\tau\tau'}\right)\ ,
\end{equation} where $c_k$ are components of a unit complex vector ${\mathrm
c}$, ${\mathsf J}$ is defined by Eq.\ (\ref{operj}) and $\nu$ is a
normalisation factor. For any unitary map ${\mathsf U}$ holds
\begin{equation}\label{NUP} {\mathsf U}{\mathsf T}_{\text{init}}({\mathrm
c}){\mathsf U}^\dagger = \nu \left(\sum_{kl} c_k c^*_l{\mathsf U}{\mathsf
P}_{\tau\tau'}(|\phi_{mk}\rangle\langle \phi_{ml}| \otimes {\mathsf
T}){\mathsf P}_{\tau\tau'}{\mathsf U}^\dagger\right)\ .
\end{equation} It is, therefore, sufficient to consider operators ${\mathsf
P}_{\tau\tau'}(|\phi_{mk}\rangle \langle\phi_{ml}| \otimes {\mathsf
T}){\mathsf P}_{\tau\tau'}$ and their evolution for different possible values
of $m$, $k$ and $l$.

Let the formal evolution on ${\mathsf H}_{\tau\tau'}$ between the initial and
an end state be given by unitary map ${\mathsf U}$. Now, we use the strategy
of Ref.\ \cite{belt} and describe ${\mathsf U}$ by suitable initial and final
states so that probability reproducibility is satisfied. Thus, ${\mathsf U}$
defines operators ${\mathsf T}'_{mkl}$ on ${\mathsf H}_{\tau\tau'}$:
\begin{equation}\label{fulev} {\mathsf U}{\mathsf
P}_{\tau\tau'}(|\phi_{mk}\rangle \langle\phi_{ml}| \otimes {\mathsf
T}){\mathsf P}_{\tau\tau'} {\mathsf U}^\dagger = N{\mathsf T}'_{mkl}
\end{equation} where $N$ is a normalisation constant due to map ${\mathsf
P}_{\tau\tau'}$ not preserving norms. It is chosen so that
$$
tr[{\mathsf T}'_{mkl}] =\delta_{kl}\ .
$$
Let us formulate our model assumptions in terms of operators ${\mathsf
T}'_{mkl}$.
\begin{description}
\item[A] For any complex unit vector ${\mathrm c}$, state $\sum_{kl} c_k c^*_l
{\mathsf T}'_{mkl}$ includes a direct signal of the detector.
\item[B] For any pair of complex unit vectors ${\mathrm c}$ and ${\mathrm
c}'$, the states $\sum_{kl} c_k c^*_l {\mathsf T}'_{mkl}$ and $\sum_{kl} c'_k
c'^*_l {\mathsf T}'_{mkl}$ are not macroscopically different. That is, the
signal of the detector depends only on $m$ so that the detector registers
${\mathsf O}$.
\item[C] For any complex unit vector ${\mathrm c}$, state $\sum_{kl} c_k c^*_l
{\mathsf T}'_{mkl}$ describes system ${\mathcal S}$ being swallowed by
${\mathcal D}$, that is, the separation status of ${\mathcal S}$
changes. Hence, we cannot reproduce any particular state operator on ${\mathbf
H}_\tau$ as an end state of ${\mathcal S}$ and on ${\mathbf H}_{\tau'}$ as an
end state of ${\mathcal D}$. In general, it is not true that ${\mathcal S}$
and ${\mathcal D}$ are each in a well-defined state at the end.
\end{description}

If the formal evolution were applied to general initial state $\phi$ of
${\mathcal S}$ with decomposition (\ref{decomp}) then the end state of the
composite ${\mathcal S} + {\mathcal D}$ would contain linear superposition of
different detector signals and the objectification requirement would be
violated. We shall therefore apply Rule \ref{rldegrad} next to weaken the
assumption of unitarity.

\subsubsection{Flexible-signal detectors} Detectors can be divided in {\em
fixed-signal} and {\em flexible-signal} ones. For a fixed-signal detector, the
amplification erases differences of states $\sum_{kl} c_k c^*_l {\mathsf
T}'_{mkl}$ so that the signal is independent not only of ${\mathrm c}$ but
also of $m$. An example is a Geiger--Mueller counter. A flexible-signal
detector, such as a proportional counter, gives different signals for
different $m$.

The minimal change of the unitarity assumption results from the consequence of
assumptions A and B that the formal evolution of initial states $\phi$
constructed from all eigenstates of ${\mathcal S}$ with one fixed eigenvalue,
$$
\phi = \sum_k c_k \phi_{mk}\ ,
$$
does not lead to violation of objectification requirement.

Let us call this part of formal evolution a {\em channel} or {\em $m$-th
channel}. For a general initial state $\phi$, decomposition (\ref{decomp}) can
be written as
$$
\phi = \sum_{mk} \sqrt{p_m} \frac{c_{mk}}{\sqrt{p_m}}\ \phi_{mk}
$$
and $(\sqrt{p_m})^{-1}c_{mk}$ is a complex unit vector for each $m$. Thus,
$\phi$ is now a linear superposition of different channels and we have to put
all channels together so that the result agrees with the objectification
requirement. The unique possibility is:
\begin{equation}\label{endflex} {\mathsf T}_{flex} =
\left(\sum_{m=1}^N\right)_p p_m \sum_{kl} \frac{c_{mk}c^*_{ml}}{p_m} {\mathsf
T}'_{mkl}\ .
\end{equation} End state (\ref{endflex}) has the form of a convex combination
of states of the composite ${\mathcal S} + {\mathcal D}$, each of which
includes only one detector signal, and the combination is the statistical
decomposition of the end state. In general, such an additional {\em reduction}
of the end state to a non-trivial statistical decomposition cannot be the
result of a unitary evolution. The formal evolution defines the channels but
remains valid only within each channel. Observe that this is sufficient to
recognise whether the separation status has changed or not. Moreover, we can
accept the validity of Eq.\ (\ref{fulev}) and all properties A--C of operators
${\mathsf T}'_{mkl}$ as model assumptions without requiring full unitarity.

\subsubsection{Fixed-signal detectors} This is the case considered in Ref.\
\cite{hajicek2}. Let state $\phi$ of particle ${\mathcal S}$ be prepared with
separation status $D$. Let ${\mathcal S}$ be manipulated by fields and screens
in $D$ so that beams corresponding to different eigenvalues of ${\mathsf O}$
become spatially separated.

Let the detector ${\mathcal D}$ be an array of $N$ fixed-signal sub-detectors
${\mathcal D}^{(m)}$ prepared in initial states ${\mathsf T}^{(m)}$ with
separation statuses $D^{(m)}$ where $D^{(n)} \cap D^{(m)} = \emptyset$ for all
$n \neq m$ and $ D^{(m)} \cap D = \emptyset$ for all $m$. We assume further
that the sub-detectors are placed at the boundary of $D$ in such a way that
the beam corresponding to eigenvalue $o_m$ will impinge on sub-detector
${\mathcal D}^{(m)}$ for each $m$. Each sub-detector ${\mathcal D}^{(m)}$
interacts with ${\mathcal S}$ as a whole and processes running in different
sub-detectors do not influence each other.

It has been shown that every observable can in principle be registered by this
kind of measurement (see \cite{wanb}, Section 3.6). The definition feature of
it is that different eigenvalues of the observable are associated with
disjoint regions of space and its registration can then be reduced to that of
position. However, even if the objectification problem could be solved for
such registrations, it still remains unsolved for other kinds of registration
(such as that described in the previous section), which undoubtedly exist and
exhibit the objectification effect.

In general, ${\mathcal S}$ hits all sub-detectors simultaneously because it is
present in all beams simultaneously. However, ${\mathcal S}$ in initial state
$\sum_{kl}c_kc^*_l |\phi_{mk}\rangle\langle\phi_{ml}|$ for any complex unit
vector ${\mathrm c}$ interacts only with sub-detector ${\mathcal
D}^{(m)}$. The formal evolution of ${\mathcal S} + {\mathcal D}^{(m)}$ can
then be decomposed into
$$
{\mathsf U}{\mathsf P}_{\tau\tau'}(|\phi_{mk}\rangle\langle\phi_{ml}| \otimes
{\mathsf T}^{(m)}){\mathsf P}_{\tau\tau'}{\mathsf U}^\dagger = N {\mathsf
T}^{(m)\prime}_{kl}
$$
and we adopt assumptions A--C for operators ${\mathsf T}^{(m)\prime}_{kl}$.

Again, we have to put all channels together in the correct way. The end state
of ${\mathcal S} + {\mathcal D}$ for any initial state $\phi$ of ${\mathcal
S}$ then is
\begin{equation}\label{endfix} {\mathsf T}_{\text{fix}} =
\left(\sum_{m=1}^N\right)_p p_m \sum_{kl} \frac{c_{mk}c^*_{ml}}{p_m} {\mathsf
T}^{(m)\prime}_{kl} \otimes \prod_{r=1}^{N\setminus m}\otimes {\mathsf
T}^{(r)}\ ,
\end{equation} where $\prod_{r=1}^{N\setminus m}$ denotes the product of all
terms except for that with $r=m$ and coefficients $c_{mk}$ are defined by Eq.\
(\ref{decomp}). Again, Eq.\ (\ref{endfix}) represents a non-trivial
reduction, where only the channels evolve unitarily.

\subsubsection{Some comments and generalisations} To discuss Eqs.\
(\ref{endflex}) and (\ref{endfix}), let us distinguish absorbing and
non-absorbing detectors \cite{hajicek2}. An absorbing detector never
releases a particle that it detects, whereas a non-absorbing detector always releases
it. We can consider only Eq.\ (\ref{endfix}), which will be needed later, the
other case is similar. If the detectors are absorbing, then state ${\mathsf
T}_{\text{fix}}$ evolves with ${\mathcal S}$ staying inside ${\mathcal
D}$. ${\mathcal S}$ is not manipulable and can be considered as lost in the
detector.

The case of non-absorbing detectors is more interesting. Extension of the
formal evolution in each channel then leads to separation of the two systems
at some later time (see Section 2.2.5). Further evolution of operator
${\mathsf P}_{\tau\tau'}(|\phi_{mk}\rangle \langle\phi_{ml}| \otimes {\mathsf
T}){\mathsf P}_{\tau\tau'}$ depends on the Hamiltonian. The simplest
imaginable end result is ${\mathsf P}_{\tau\tau'}(|\varphi_{mk}\rangle
\langle\varphi_{ml}| \otimes {\mathsf T}^{(m) \prime\prime}){\mathsf
P}_{\tau\tau'}$, where $\varphi_{mk}$ is a state of a system identical to
${\mathcal S}$ with separation status $D_m$, $D_m \cap D^{(n)} = \emptyset$
for all $m$ and $n$, and ${\mathsf T}^{(m) \prime\prime}$ is a state of
${\mathcal D}^{(m)}$ with separation status $D^{(m)}$. Thus, end state
${\mathsf T}_{\text{release}}$ that can be reconstructed from the formal
evolution is
\begin{equation}\label{release} {\mathsf T}_{\text{release}} =
\left(\sum_{m=1}^N\right)_p p_m \sum_{kl} \frac{c_{mk}c^*_{ml}}{p_m}
|\varphi_{mk}\rangle \langle\varphi_{ml}| \otimes {\mathsf T}^{(m)
\prime\prime} \otimes \prod_{r=1}^{N\setminus m}\otimes {\mathsf T}^{(r)}\ .
\end{equation} As a result, there is a random mixture of states
$$
\sum_{kl} \frac{c_{mk}c^*_{ml}}{p_m} |\varphi_{mk}\rangle \langle\varphi_{ml}|
$$
of a system ${\mathcal S}'$ of the same type as ${\mathcal S}$ and each of
these states is correlated with a macroscopic state ${\mathsf T}^{(m)
\prime\prime}$ of detector ${\mathcal D}^{(m)}$ including a macroscopic
signal. System ${\mathcal S}'$ has a non-trivial separation status again
($D_m$) so that the release in each channel can be understood as an instance
of preparation for ${\mathcal S}'$ and the whole evolution as a statistical
mixture (Definition \ref{dfstatprep}) of these single preparations. The
different individual preparations are distinguished by the different
macroscopic states ${\mathcal D}^{(m)}$ of different sub-detectors and
accompanied by partial dissipation of the original state within the
detector. The formula (\ref{release}) preserves the state reduction, which is
due to the loss of separation status of the original system ${\mathcal S}$
inside detector ${\mathcal D}$. One could even define the conditional state
${\mathsf T}_r$ and operation ${\mathcal O}_r$ (see Section 4.1) for this
special case of direct registration.

The new rules that have been proposed as yet always correct the unitary formal
evolution determined by standard quantum mechanics by a reduction of the state
operator. The reduced state occurs in the formulas as the so-called "end
state". We assume that the time instant at which each end state formula
is valid is the time at which the detector gives its macroscopic signal. No
detail of the time evolution to this end state is given. The end state itself
as well as any time evolution to it cannot be derived from quantum mechanics
but must simply be guessed and subjected to experimental checks. The question
of detailed time evolution is left open. An example of such a more detailed
evolution may be given by some scenario of the dynamical reduction program. Of
course, the motivation for the reduction would then be very different.

An interesting case, which has some relevance to the end-time question and
which is a hybrid of the registration by non-absorbing and absorbing detector,
is the Einstein, Podolsky and Rosen (EPR) experiment \cite{EPR}. We consider
Bohm's form of it \cite{bohm}. A spin-zero particle decays into two spin-1/2
ones, ${\mathcal S}_1$ and ${\mathcal S}_2$, that run in two opposite
directions. The state of composite ${\mathcal S}_1 \otimes {\mathcal S}_2$ is
then
\begin{equation}\label{EPRin} \frac{1}{\sqrt{2}}(|1_+\rangle\otimes
|2_-\rangle - |1_-\rangle\otimes |2_+\rangle)\ ,
\end{equation} where $|1_+\rangle$ is the spin-up state of ${\mathcal S}_1$,
etc. Finally, the spin of ${\mathcal S}_1$ is registered after some time at
which the particles ${\mathcal S}_1$ and ${\mathcal S}_2$ may be far away from
each other. Let the detector be a special case of fixed-signal one, as
described in Section 4.3.1. Hence, there are two sub-detectors, ${\mathcal
D}_1^{(+)}$ and ${\mathcal D}_1^{(-)}$ so that spin up of ${\mathcal S}_1$ is
associated with a signal from ${\mathcal D}_1^{(+)}$ and spin down with that
from ${\mathcal D}_1^{(-)}$. Let the state of ${\mathcal S}_1 + {\mathcal
D}_1^{(+)}$ containing the signal be ${\mathsf T}_1^{(+)\prime}$ and that of
${\mathcal S}_1 + {\mathcal D}_1^{(-)}$ be ${\mathsf
T}_1^{(-)\prime}$. Although ${\mathcal S}_1$ will be swallowed by the detector
(see Section 4.3.1), the left particle may remain accessible to
registration. Thus, our new rule is analogous to Eq.\ (\ref{release}):
\begin{equation}\label{EPR} {\mathsf T}_{\text{EPR}} = \frac{1}{2}
|2_+\rangle\langle 2_+| \otimes {\mathsf T}_1^{(+)} \otimes {\mathsf
T}_1^{(-)\prime}\ (+)_p\ \frac{1}{2} |2_-\rangle\langle 2_-| \otimes {\mathsf
T}_1^{(+)\prime} \otimes {\mathsf T}_1^{(-)}\ ,
\end{equation} where ${\mathsf T}_1^{(+)}$ and ${\mathsf T}_1^{(-)}$ are the
non-excited states of the corresponding sub-detectors. The state reduction
takes place at the time of the detector signal and has a {\em non-local}
character. We do not see any paradox in it. The only problem comes with the
generalisation to a relativistic theory: what is the correct simultaneity
plane? This problem has been solved by Keyser and Stodolsky \cite{KS}, see
also the discussion in Section 2.1.2.

Eqs.\ (\ref{endflex}), (\ref{endfix}) and (\ref{release}) can readily be
generalised to registration on a non-extremal state ${\mathsf S}$ of
${\mathcal S}$. First, we have to decompose ${\mathsf S}$ into eigenstates of
${\mathsf O}$,
\begin{equation}\label{nonvec} {\mathsf S} = \sum_{nkml} S_{nkml}
|\phi_{nk}\rangle\langle\phi_{ml}|\ ;
\end{equation} the probability to register eigenvalue $o_k$ on ${\mathcal S}$
is
$$
p_m = \sum_k S_{mkmk}\ .
$$
Second, because of the linearity of ${\mathsf U}$, everything we must do is to
replace the expressions in Eqs.\ (\ref{endflex}), (\ref{endfix}) and
(\ref{release}) as follows:
\begin{equation}\label{repl} \frac{c_{mk}c^*_{ml}}{p_m} \mapsto
\frac{S_{mkml}}{p_m}\ .
\end{equation}

The last case of direct registration to consider in this section is that the
registered particle can miss the detectors and enter into environment. We can
use Eq.\ (\ref{endfix}) again by modelling the part of the environment that
the particle must join if it misses the detector by one of the sub-detectors,
${\mathcal D}^{(N)}$, say.

This also explains the fact that the Schr\"{o}dinger's cat is never observed in
the linear superposition of life and death states. Indeed, in the case of
Schr\"{o}dinger's cat, there is a radioactive substance releasing
alpha-particles and a detector of alpha-particles, the signal of which leads
to the death of the cat. Then, we can decompose the state of an alpha-particle
into that of it being in the nucleus or of being released and missing the
detector and that of hitting the detector, so that the above analysis is
applicable.

\subsubsection{Registration of composite systems} Eqs.\
(\ref{endflex})--(\ref{release}) were obtained for registrations of
one-particle systems. This section will generalise them to many-particle
ones. Composite systems can be classified into {\em bound} and {\em
unbound}. Bound systems such as atoms and molecules can be dealt with in an
analogous way as particles. The only change is that map ${\mathsf
P}_{\tau\tau'}$ is more complicated. Then, Eqs.\
(\ref{endflex})--(\ref{release}) are valid for bounded composite
systems. Unbounded composite systems are different. A system ${\mathcal S}$
that contains $K$ unbound particles can excite more detectors simultaneously,
at most $K$ detectors.

Generalisation to such systems is not completely straightforward because it
must achieve, on the one hand, that there can be some non-trivial correlations
between the signals from different detectors and, on the other, that the
detectors are never in a linear superposition of their different signals,
which in turn erases some correlations between different detectors. Of course,
for one-particle systems, signals of different detectors are always
anti-correlated in a trivial way. Non-trivial correlations that can emerge for
unbounded many-particle systems are, e.g.,\ Hanbury--Brown--Twiss (HBT) ones
\cite{HBT} or Eistein--Podolski--Rosen (EPR) ones. Let us start with HBT effect.

In the original experiment by Hanbury Brown and Twiss, two photomultiplier
tubes separated by about 6 m distance, were aimed at the star Sirius. An
interference effect was observed between the two intensities, revealing a
positive correlation between the two signals. Hanbury Brown and Twiss used the
interference signal to determine the angular size of Sirius. The theory of the
effect \cite{fano} studies a model in which the signal consists of two photons
that impinge simultaneously on two detectors. Our strategy will be to
construct a non-relativistic model of Hanbury Brown and Twiss effect following
closely Fano's ideas \cite{fano} and try then to modify it similarly as the
BCL model has been modified for the case of one-particle systems in Section
4.3.1.

Let us limit ourselves to ${\mathcal S} = {\mathcal S}_1 + {\mathcal S}_2$
consisting of two bosons, $K = 2$, with Hilbert spaces ${\mathbf H}_1$ and
${\mathbf H}_2$. To simplify further, let the registered observable be
${\mathsf O}_1 + {\mathsf O}_2$, ${\mathsf O}_k$ having only two eigenvalues
$+1$ and $-1$ and eigenvectors $|k_+\rangle$ and $|k_-\rangle$, $k = 1,2$
satisfying
$$
{\mathsf O}_k |k_+\rangle = +|k_+\rangle\ ,\quad {\mathsf O}_k |k_-\rangle =
-|k_-\rangle\ .
$$
Let, moreover, the one-particle Hilbert spaces be two-dimensional,
i.e., vectors $|k_+\rangle$ and $|k_-\rangle$ form a basis of ${\mathbf
H}_k$. Let the projections onto these states be denoted by ${\mathsf P}_{k+}$
and ${\mathsf P}_{k-}$ so that we have:
\begin{equation}\label{proj1} {\mathsf P}_{k+}{\mathsf P}_{k+} = {\mathsf
P}_{k+}\ ,\quad {\mathsf P}_{k-}{\mathsf P}_{k-} = {\mathsf P}_{k-}\ ,\quad
{\mathsf P}_{k+}{\mathsf P}_{k-} = 0\ .
\end{equation} The generalisation to more particles of arbitrary kinds,
general observables and general Hilbert spaces is straightforward.

The Hilbert space ${\mathbf H}$ of the composite system has then basis
$\{|++\rangle,|--\rangle,|+-\rangle\}$, where
\begin{eqnarray*} |++\rangle &=& |1_+\rangle |2_+\rangle\ ,\\ |--\rangle &=&
|1_-\rangle |2_-\rangle\ ,\\ |+-\rangle &=& \frac{1}{\sqrt{2}}(|1_+\rangle
|2_-\rangle + |1_-\rangle |2_+\rangle)\ .
\end{eqnarray*} It is the basis formed by eigenvectors of ${\mathsf O}_1 +
{\mathsf O}_2$ with eigenvalues $2$, $-2$ and $0$, respectively. The
corresponding projections are
\begin{eqnarray*}\label{proj2} {\mathsf P}_{++} &=& {\mathsf P}_{1+}{\mathsf
P}_{2+}\ ,\\ {\mathsf P}_{--} &=& {\mathsf P}_{1-}{\mathsf P}_{2-}\ ,\\
{\mathsf P}_{+-} &=& {\mathsf P}_{1+}{\mathsf P}_{2-} + {\mathsf
P}_{1-}{\mathsf P}_{2+}\ .
\end{eqnarray*} It follows from Eq.\ (\ref{proj1}) that these are indeed
projections.

To calculate the correlation in a state ${\mathsf S}$ of system ${\mathcal S}$
between the values $\pm 1$ of any subsystem ${\mathcal S}_1$ or ${\mathcal
S}_2$, which is intended to model the correlation measured by Hanbury Brown
and Twiss, we need probability $p_+$ that eigenvalue $+1$ will be registered
at least on one subsystem and similarly $p_-$ for $-1$. These are given by
\begin{eqnarray*} p_+ &=& tr[{\mathsf S}({\mathsf P}_{++} + {\mathsf
P}_{+-})]\ ,\\ p_- &=& tr[{\mathsf S}({\mathsf P}_{--} + {\mathsf P}_{+-})]\ ,
\end{eqnarray*} respectively. If we define
$$
{\mathsf P}_+ = {\mathsf P}_{++} + {\mathsf P}_{+-}\ ,\quad {\mathsf P}_- =
{\mathsf P}_{--} + {\mathsf P}_{+-}\ ,
$$
we have
$$
{\mathsf P}_{+-} = {\mathsf P}_+ {\mathsf P}_-\ .
$$
The normalised correlation (see Section 1.2.2) is then given by
\begin{equation}\label{corr} C({\mathsf S}) = \frac{tr[{\mathsf S}{\mathsf
P}_+ {\mathsf P}_-] - tr[{\mathsf S}{\mathsf P}_+]tr[{\mathsf S}{\mathsf
P}_-]}{\sqrt{tr[{\mathsf S}{\mathsf P}_+] - (tr[{\mathsf S}{\mathsf
P}_+])^2}\sqrt{tr[{\mathsf S}{\mathsf P}_-] - (tr[{\mathsf S}{\mathsf
P}_-])^2}}\ .
\end{equation}

For example, let $|\Phi\rangle$ be a general vector state in $\mathbf H$:
$$
|\Phi\rangle = a|++\rangle + b|--\rangle + c|+-\rangle\ ,
$$
where $a$, $b$ and $c$ are complex numbers satisfying
$$
|a|^2 + |b|^2 + |c|^2 = 1\ .
$$
Then,
$$
C(\Phi) = -\frac{|a|^2 |b|^2}{\sqrt{(|a|^2 - |a|^4)(|b|^2 - |b|^4)}}\ .
$$
The correlation lies, in general, between $0$ and $-1$. The value $-1$ occurs
for $c= 0$, means the strong anti-correlation and is the standard (trivial)
case for one-particle systems.

Next, we construct a suitable detector. System ${\mathcal S}$ can be prepared
in vector state $|\Phi\rangle$ with separation status $D$ where then fields
and screens split the beam $B$ of single particles corresponding to
$|\Phi\rangle$ into two beams, $B_+$ and $B_-$, each associated with an
eigenvalue $\pm 1$ of observable ${\mathsf O}_1$ or ${\mathsf O}_2$. Let
detector ${\mathcal D}$ consist of two sub-detectors, ${\mathcal D}^{(+)}$
placed in the way of the beam $B_+$ and ${\mathcal D}^{(-)}$ placed in the way
of $B_-$ so that the signal of ${\mathcal D}^{(+)}$ registers eigenvalue $+1$
and that of ${\mathcal D}^{(+)}$ eigenvalue $-1$ on the registered particle
similarly as in our model of fixed signal detector (Eq.\
(\ref{endfix})). Let the Hilbert spaces of the sub-detectors be ${\mathbf
H}_+$ and ${\mathbf H}_-$.

Let the sub-detectors be prepared in initial states $|{\mathcal
D}^{(+)}0\rangle$ and $|{\mathcal D}^{(-)}0\rangle$ with separation statuses
$D^{(+)}$ and $D^{(-)}$, $D^{(+)} \cap D^{(-)} = \emptyset$, $D \cap D^{(\pm)}
= \emptyset$. After the interaction between ${\mathcal S}$ and ${\mathcal D}$,
the following states are relevant: $|{\mathcal D}^{(+)}1\rangle \in {\mathsf
P}_{1+}({\mathbf H}_1 \otimes {\mathbf H}_+)$, $|{\mathcal D}^{(-)}1\rangle
\in {\mathsf P}_{1-}({\mathbf H}_1 \otimes {\mathbf H}_-)$, $|{\mathcal
D}^{(+)}2\rangle \in {\mathsf P}_{2+}({\mathbf H}_2 \otimes {\mathbf H}_+)$,
$|{\mathcal D}^{(-)}2\rangle \in {\mathsf P}_{2-}({\mathbf H}_2 \otimes
{\mathbf H}_-)$, $|{\mathcal D}^{(+)}12\rangle \in {\mathsf P}_{12+}({\mathbf
H}_1 \otimes {\mathbf H}_2 \otimes {\mathbf H}_+)$ and $|{\mathcal
D}^{(-)}12\rangle \in {\mathsf P}_{12-}({\mathbf H}_1 \otimes {\mathbf H}_2
\otimes {\mathbf H}_-)$. These states describe one or two of the particles
being swallowed by one of the sub-detectors and they are associated with
changes of their separation status and include detector signals. Here,
(symmetrising) projection ${\mathsf P}_{1+}$ is defined by Eq.\
(\ref{projTT'}) for Hilbert spaces ${\mathbf H}_1$ and ${\mathbf H}_+$ and
analogously for the other projections.

Finally, to register ${\mathsf O}_1 + {\mathsf O}_2$, the measurement coupling
${\mathsf U}$ must satisfy
\begin{eqnarray}\label{U++} {\mathsf U}{\mathsf P}_{12+-}(|++\rangle \otimes
|{\mathcal D}^{(+)}0\rangle \otimes |{\mathcal D}^{(-)}0\rangle) &=& {\mathsf
P}_{12+-}(|{\mathcal D}^{(+)}12\rangle \otimes |{\mathcal D}^{(-)}0\rangle)\ ,
\\ \label{U--} {\mathsf U}{\mathsf P}_{12+-}(|--\rangle \otimes |{\mathcal
D}^{(+)}0\rangle \otimes |{\mathcal D}^{(-)}0\rangle) &=& {\mathsf
P}_{12+-}(|{\mathcal D}^{(+)}0\rangle \otimes |{\mathcal D}^{(-)}12\rangle)\ ,
\\ \label{U+-} {\mathsf U}{\mathsf P}_{12+-}(|+-\rangle \otimes |{\mathcal
D}^{(+)}0\rangle \otimes |{\mathcal D}^{(-)}0\rangle) &=& {\mathsf
P}_{12+-}(|{\mathcal D}^{(+)}1\rangle \otimes |{\mathcal D}^{(-)}2\rangle)\ .
\end{eqnarray} Observe that operator ${\mathsf P}_{12+-}$ also exchanges
particles 1 and 2, which is a non-trivial operation on the right-hand side of
Eq.\ (\ref{U+-}).

Eqs.\ (\ref{U++}), (\ref{U--}) and (\ref{U+-}) describe the formal evolution
defining the three channels of the measurement. Each channel leads to the
composite signal due to a registration of one copy of two-particle system
${\mathcal S}$. Thus, it can include signals of two detectors (Eq.\
(\ref{U+-})).

The formal evolution of state $\Phi$ would yield for the end state of the
system ${\mathcal S} + {\mathcal D}$:
\begin{multline}\label{end0a} {\mathsf U}{\mathsf J}(|\Phi\rangle \otimes
|{\mathcal D}^{(+)}0\rangle \otimes |{\mathcal D}^{(-)}0\rangle) = a{\mathsf
J}(|{\mathcal D}^{(+)}12\rangle \otimes |{\mathcal D}^{(-)}0\rangle) \\ +
b{\mathsf J}(|{\mathcal D}^{(+)}0\rangle \otimes |{\mathcal D}^{(-)}12\rangle)
+ c{\mathsf J}(|{\mathcal D}^{(+)}1\rangle \otimes |{\mathcal
D}^{(-)}2\rangle)\ ,
\end{multline} where map $\mathsf J$ is defined by Eq.\
(\ref{operj}). According to our theory, this state must be reduced to a
decomposable state with component states, each of them corresponding to a
single channel. Thus, the correct end state ${\mathsf T}_{\text{comp}}$ of the
whole system ${\mathcal S} + {\mathcal D}$ after the measurement process
described above is
\begin{multline}\label{end1a} {\mathsf T}_{\text{comp}} = |a|^2{\mathsf
J}(|{\mathcal D}^{(+)}12\rangle\langle{\mathcal D}^{(+)}12|) \otimes
|{\mathcal D}^{(-)}0\rangle\langle{\mathcal D}^{(-)}0| \\ (+)_p |b|^2
|{\mathcal D}^{(+)}0\rangle\langle{\mathcal D}^{(+)}0| \otimes {\mathsf
J}(|{\mathcal D}^{(-)}12\rangle\langle{\mathcal D}^{(-)}12|) \\ (+)_p |c|^2
{\mathsf J}\left(|{\mathcal D}^{(+)}1\rangle\langle{\mathcal D}^{(+)}1|
\otimes |{\mathcal D}^{(-)}2\rangle\langle {\mathcal D}^{(-)}2|\right)\ .
\end{multline} We assume that Eq.\ (\ref{end1a}) describes a special case
of the registration of many-particle systems by many detectors and that it
illustrates a method that can be used for more general cases. State ${\mathsf
T}_{\text{comp}}$ is an operator on ${\mathbf H} \otimes {\mathbf H}_+ \otimes
{\mathbf H}_-$ and it is a convex combination of three states each on a
different subspace of it. These three states are obtained by reconstruction
from the corresponding results of formal evolution in accordance with the
separation statuses. For example, the formal evolution gives for the first
state
$$
{\mathsf J}(|{\mathcal D}^{(+)}12\rangle\langle{\mathcal D}^{(+)}12| \otimes
|{\mathcal D}^{(-)}0\rangle\langle{\mathcal D}^{(-)}0|)\ ,
$$
but both particles are inside ${\mathcal D}^{(+)}$ and are, together with
${\mathcal D}^{(+)}$, separated from ${\mathcal D}^{(-)}$.

One can see that the pair of sub-detectors is in a well-defined signal state
after each individual registration on ${\mathcal S}$ and, at the same time,
the correlation contained in state $|\Phi\rangle$ that models the HTB
correlation is preserved and can be read off the signals of the
sub-detectors. This is of course due to the fact that HTB correlation is a
function of the absolute values $|a|$, $|b|$ and $|c|$, none of which is
erased by reduction of Eq.\ (\ref{end0a}) to Eq.\ (\ref{end1a}), while the
extra correlations due to the linear superposition depend on mixed products
such as $ab^*$, etc.

A different but analogous case is the EPR experiment. The composite system of
two fermions ${\mathcal S}_1$ and ${\mathcal S}_2$ is in initial state
(\ref{EPRin}). The detector consists of four sub-detectors, ${\mathcal
D}^{(+)}_1$, ${\mathcal D}^{(-)}_1$, ${\mathcal D}^{(+)}_2$ and ${\mathcal
D}^{(-)}_2$, where the first pair interacts only with ${\mathcal S}_1$ and the
second only with ${\mathcal S}_2$. The initial states of the sub-detectors are
${\mathsf T}^{(\pm)}_k$. The symbol ${\mathsf T}^{(\pm)\prime}_k$ denotes the
state of system ${\mathcal D}^{(\pm)}_k + {\mathcal S}_k$ in which the
sub-detector ${\mathcal D}^{(\pm)}_k$ swallows particle ${\mathcal S}_k$ and
sends its signal. Procedure analogous to that leading to formula (\ref{end1a})
will now give for the end state
\begin{equation}\label{end1b} \frac{1}{2}{\mathsf T}^{(+)}_1 \otimes {\mathsf
T}^{(-)\prime}_1 \otimes {\mathsf T}^{(-)\prime}_2 \otimes {\mathsf
T}^{(+)\prime}_2 (+)_p \frac{1}{2}{\mathsf T}^{(+)\prime}_1 \otimes {\mathsf
T}^{(-)\prime}_1 \otimes {\mathsf T}^{(-)}_2 \otimes {\mathsf
T}^{(+)\prime}_2\ .
\end{equation} Again, EPR anti-correlation of the sub-detector signals is
preserved even if the quadruple of the sub-detectors is always in a
well-defined signal state at the end.

\subsection{Non-ideal detectors} Non-ideal detectors may be the natural and
dominating case, from the experimental point of view. If a non-ideal detector
${\mathcal D}$ is hit by a system ${\mathcal S}$, there is only probability $0
< \eta < 1$, the intrinsic efficiency, that it will give a signal. From the
theoretical point of view, they are important because our simple method of
channels does not work for them.

We restrict ourselves to flexible-signal detectors with possible signals
enumerated by $m = 1,\ldots,N$ and suppose that, in general, $\eta_m$ depends
on $m$. The other cases can be dealt with in an analogous way. Let the
separation status of ${\mathcal D}$ be $D_{\mathcal D}$. If ${\mathcal S}$ is
prepared in an eigenstate of ${\mathsf O}$ with eigenvalue $o_m$ which
formally evolves to ${\mathcal S}$ being inside $D_{\mathcal D}$ with
certainty, then the probability that ${\mathcal D}$ signals is $\eta_m$ and
not 1. Thus, the condition of probability reproducibility is not satisfied in
this case. Instead, we introduce the notion of {\em approximate probability
reproducibility}. Its meaning is that the detector does register eigenvalue
$o_m$ on ${\mathcal S}$ if it gives $m$-th signal, but we do not know anything,
if it remains silent.

To construct a model of this situation, we must first modify Eq.\
(\ref{unitar}) that expresses the idea of probability reproducibility into
what expresses the approximate probability reproducibility (within standard
quantum mechanics):
\begin{equation}\label{unitar'} {\mathsf U}(\phi_{mk} \otimes \psi) = C_m^1
\varphi_{mk} \otimes \psi^1_m + C_m^0\phi'_{mk} \otimes \psi^0_m\ ,
\end{equation} where $\phi'_{mk}$ is a suitable time evolution of $\phi_{mk}$
into $D_{\mathcal D}$ and $\varphi_{mk}$ are states of ${\mathcal S}$ while
$\psi$ is the initial state, $\psi^1_m$ the signal state and $\psi^0_m$ a no-signal state
of ${\mathcal D}$. These states satisfy orthogonality relations
$$
\langle \psi |\psi^1_m \rangle = 0\ ,\quad \langle \psi^1_m |\psi^1_n \rangle
= \delta_{mn}\ ,\quad \langle \psi^0_m |\psi^1_n \rangle = 0\ ,\quad \langle
\varphi_{mk} |\varphi_{ml}\rangle = \langle \phi'_{mk} |\phi'_{ml}\rangle =
\delta_{kl}\ .
$$
The coefficients $C_m^1$ and $C^0_m$ are related by
$$
|C_m^1|^2 + |C_m^0|^2 = 1\ ,\quad |C_m^1|^2 = \eta_m\ .
$$

Measurement coupling ${\mathsf U}$ commutes with ${\mathsf P}_{\tau\tau'}$
because the Hamiltonian leaves ${\mathbf H}_{\tau\tau'}$ invariant and with
normalisation because it is a unitary map (see Section 2.2.5). We can,
therefore, replace Eq.\ (\ref{unitar'}) by the corresponding formal evolution:
\begin{equation}\label{formev1} {\mathsf U}{\mathsf J}(\phi_{mk} \otimes \psi)
= C_m^1 {\mathsf J}(\varphi_{mk} \otimes \psi^1_m) + C_m^0 {\mathsf
J}(\phi'_{mk} \otimes \psi^0_m)\ .
\end{equation} This is not a channel because it is not the formal evolution of
an initial state into an end state with a single detector signal. Indeed, no
signal is also a macroscopically discernible detector state. We have to return
to the formal evolution that starts with general state $\phi$ of ${\mathcal
S}$:
\begin{multline}\label{formev2} {\mathsf U}{\mathsf
J}(|\phi\rangle\langle\phi| \otimes |\psi\rangle\langle\psi|){\mathsf
U}^\dagger = \sum_{mn}\sum_{kl}c_{mk}c^*_{nl}\Big(C_m^1 C^{1*}_n |{\mathsf
J}(\varphi_{mk} \otimes \psi^1_m)\rangle \langle{\mathsf J}(\varphi_{nl}
\otimes \psi^1_n)| \\ + C_m^1 C^{0*}_n |{\mathsf J}(\varphi_{mk} \otimes
\psi^1_m)\rangle \langle{\mathsf J}(\phi'_{nl} \otimes \psi^0_n)| + C_m^0
C^{1*}_n |{\mathsf J}(\phi'_{mk} \otimes \psi^0_m)\rangle \langle{\mathsf
J}(\varphi_{nl} \otimes \psi'_n)| \\ + C_m^0 C^{0*}_n |{\mathsf J}(\phi'_{mk}
\otimes \psi^0_m)\rangle \langle{\mathsf J}(\phi'_{nl} \otimes
\psi^0_n)|\Big)\ ,
\end{multline}

To obtain a correct end state of a non-ideal detector, we have to discard the
cross-terms between $\psi^1_m$ and $\psi^1_n$ and between $\psi^1_m$ and
$\psi^0_n$. This is a general method that works also in the case that there
are channels. The result is
\begin{multline}\label{nonidvec} {\mathsf T}_{\text{nonid1}} =
\left(\sum_{m=1}^N\right)_p p_m \eta_m \sum_{kl} \frac{c_{mk} c^*_{ml}}{p_m}
|{\mathsf J}(\varphi_{mk} \otimes \psi^1_m)\rangle \langle{\mathsf
J}(\varphi_{ml} \otimes \psi^1_m)| \\ (+)_p \sum_{mn}\sum_{kl} c_{mk}
c^*_{nl}C_m^0 C^{0*}_n |{\mathsf J}(\phi'_{mk} \otimes \psi^0_m)\rangle
\langle{\mathsf J}(\phi'_{nl} \otimes \psi^0_n)|\ .
\end{multline} This is not yet a practical formula because the detector is
always in a state with high entropy, which is not a vector state. Hence, the
initial state is $|\phi\rangle\langle\phi| \otimes {\mathsf T}$, and the end
state is
\begin{equation}\label{nonid} {\mathsf T}_{\text{nonid2}} =
\left(\sum_{m=1}^N\right)_p p_m \eta_m \sum_{kl} \frac{c_{mk} c^*_{ml}}{p_m}
{\mathsf T}^1_{mkl} (+)_p \sum_{mn}\sum_{kl} c_{mk} c^*_{nl}{\mathsf
T}^0_{mnkl}\ ,
\end{equation} where we have made the replacements
$$
|{\mathsf J}(\varphi_{mk} \otimes \psi^1_m)\rangle \langle{\mathsf
J}(\varphi_{ml} \otimes \psi^1_m)| \mapsto {\mathsf T}^1_{mkl}
$$
and
$$
C_m^0 C^{0*}_n|{\mathsf J}(\phi'_{mk} \otimes \psi^0_m)\rangle \langle{\mathsf
J}(\phi'_{nl} \otimes \psi^0_n)| \mapsto {\mathsf T}^0_{mnkl}\ .
$$
Operators ${\mathsf T}^1_{mkl}$ and ${\mathsf T}^0_{mnkl}$ are determined by
the initial state and the formal evolution and satisfy the conditions:
\begin{description}
\item[A']
$$
tr[{\mathsf T}^1_{mkl}] = \delta_{kl}\ ,\quad tr[{\mathsf T}^0_{mnkl}] =
(1-\eta_m)\delta_{mn}\delta_{kl}\ .
$$
\item[B'] For any unit complex vector with components $c_k$,
$$
\sum_{kl} c_kc^*_l {\mathsf T}^1_{mkl}
$$
is a state operator on ${\mathbf H}_{\tau\tau'}$ and the state includes direct
$m$-th signal from the detector.
\item[C'] For any unit complex vector with components $c_{mk}$ (for all $m$
and $k$)
$$
\left(\sum_m p_m(1-\eta_m)\right)^{-1}\sum_{mn}\sum_{kl} c_{mk}
c^*_{nl}{\mathsf T}^0_{mnkl}
$$
is a state operator on ${\mathbf H}_{\tau\tau'}$ and the state includes no
detector signal from the detector.
\end{description}

\subsection{Particle tracks in detectors} Particle tracks in a Wilson chamber
look suspiciously similar to classical trajectories and have been an
interesting problem for quantum mechanics since the end of 1920s. There is
the classical paper by Mott \cite{mott} (see also \cite{heisenberg}), which
shows by applying Schr\"{o}dinger equation that there is an overwhelming
probability of getting a second scattering event very close to the ray
pointing away from the decay centre through the location of the first scattering
event. A more rigorous calculation is given in \cite{DFT}, which uses the same
idea for a one-dimensional model. The initial situation in Ref.\ \cite{mott}
is spherically symmetric and the interaction between the alpha particle and
the detector also is. Thus, the resulting state must also be spherically
symmetric and not just one radial track. A consequence of the linearity of
Schr\"{o}dinger equation then is that the end state is a linear
superposition of all possible radial tracks. A way to save one single radial
track is the state reduction at least for the first ionisation, which is
apparently assumed tacitly. This separation of state reduction and unitary
evolution does not exactly correspond to what is going on because we have in
fact a chain of state reductions with a unitary evolution in between.

In this section, we apply our theory to the problem, but we simplify it by
assuming, instead of the spherical symmetry, that the particle momentum has a
large average value $\langle \vec{p} \rangle$ and the detector has the plane
symmetry with the plane being perpendicular to $\langle \vec{p} \rangle$.

The registration model obeying formula (\ref{release}) can be characterised as
a single transversal layer of detectors: each beam is registered once. What we
now have can be viewed as an arrangement of many transversal detector layers:
one beam passes through all layers successively causing a multiple
registration. Examples of such arrangements are cloud chambers or MWPC
telescopes for particle tracking \cite{leo}. The latter is a stack of the
so-called multiwire proportional chambers (MWPC) so that the resulting system
of electronic signals contains the information about a particle track. Here,
we restrict ourselves to cloud chambers, but the generalisation needed to
describe MWPC telescopes does not seem difficult.

Then, a model of a Wilson chamber is a system of sub-detectors ${\mathcal
D}^{(nk)}$, where $n$ distinguishes different transversal layers and $k$
different sub-detectors in each such layer. Let the space occupied by
${\mathcal D}^{(nk)}$ be $D^{(nk)}$ and let it be at the same time its
separation status. We shall assume that $D^{(nk)}$ are small cubes with edge
$d$ that is approximately equal to the diameter of the resulting clouds in the
Wilson chamber. We denote the $n$-th layer by ${\mathcal D}^{(n)}$ so that
${\mathcal D}^{(n)} = \cup_{k=1}^N{\mathcal D}^{(nk)}$. To simplify the
subsequent analysis, we assume that coordinates can be chosen in a
neighbourhood of ${\mathcal D}^{(n)}$ so that each $D^{(nk)}$ in the
neighbourhood can be described by
$$
x^1 \in (u_k^1,u_k^1+d)\ ,\quad x^2 \in (u_k^2,u_k^2 +d)\ ,\quad x^3 \in
(u_n^3,u_n^3+d)\ .
$$

The observable ${\mathsf O}^{(n)}$ that is registered by each layer ${\mathcal
D}^{(n)}$ is equivalent to the position within the cubes. The eigenfunctions
and eigenvalues are
$$
{\mathsf O}^{(n)}\phi^{(nk)}_{l_1l_2l_3}(\vec{x}) = k
\phi^{(nk)}_{l_1l_2l_3}(\vec{x})\ ,
$$
where $\{l_1,l_2,l_3\}$ is a triple of integers that replaces the degeneration
index $l$,
$$
\phi^{(nk)}_{l_1l_2l_3}(\vec{x}) = d^{-3/2} \exp\left(\frac{2\pi
l_1i}{d}(x^1-u^1_k) + \frac{2\pi l_2i}{d}(x^2-u^2_k) + \frac{2\pi
l_3i}{d}(x^3-u^3_n)\right)
$$
for $\vec{x} \in D^{(nk)}$ and $\phi^{(nk)}_{l_1l_2l_3}(\vec{x}) = 0$
elsewhere.

The state ${\mathsf S}_n$ of ${\mathcal S}$ impinging on ${\mathcal D}^{(n)}$
can be defined as the state ${\mathcal S}$ would have after being released by
the layer ${\mathcal D}^{(n-1)}$. The interaction of ${\mathcal S}$ with
${\mathcal D}^{(n)}$ can then be described by formula (\ref{release}) with
replacement (\ref{repl}). The decomposition (\ref{nonvec}) must, of course,
use functions $\phi^{(nk)}_{l_1l_2l_3}$ instead of $\phi^{(n)}_k$ and the
support of $\varphi^{(nk)}_{l_1l_2l_3}$ is $D^{(nk)}$. The procedure can be
repeated for all $n$.

The first layer "chooses" one particular
$\varphi^{(1k)}_{l_1l_2l_3}$ with the support $D^{(1k)}$ in each individual
act of registration even in the case that the state arriving at it is a plane
wave. Hence, the "choice" in the next layer is already strongly
limited. In this way, a straight particle track of width $d$ results during
each individual multiple registration. Formally, of course, the resulting
state of ${\mathcal S}$ is a state decomposable in such straight tracks, which
would have the plane symmetry if the original wave arriving at the detector
stack were a plane wave.

\subsection{General assumption for models of direct registration} We have
discussed different models of direct registration. A case by case analysis
trying to take into account the idiosyncrasy of each experiment and to isolate
the relevant features of its results has lead to Eqs.\ (\ref{endflex}),
(\ref{endfix}), (\ref{end1a}), (\ref{end1b}) and (\ref{nonid}). We can now try
to formulate a general model assumption specifying Rule \ref{rldegrad} for
direct registrations:
\par \vspace{.5cm} \noindent {\bf State Reduction in Direct Registration} {\it
Let microscopic system ${\mathcal S}$ be prepared in state ${\mathsf
T}_{\mathcal S}$ with separation status $D_{\mathcal S}$ and detector
${\mathcal D}$ in state ${\mathsf T}_{\mathcal D}$ with separation status
$D_{\mathcal D}$, where $D_{\mathcal S} \cap D_{\mathcal D} = \emptyset$. The
initial state is then ${\mathsf T}_{\mathcal S} \otimes {\mathsf T}_{\mathcal
D}$ according to Rule \ref{rlold14}. Let the formal evolution describing the
interaction between ${\mathcal S}$ and ${\mathcal D}$ lead to separation
status change of ${\mathcal S}$. Then the state of the composite ${\mathcal S}
+ {\mathcal D}$ given by the formal evolution must be corrected by state
reduction to decomposable state
\begin{equation}\label{genrule} {\mathsf T}_{\text{end}} =
\left(\sum_m\right)_p p_m {\mathsf T}'_m\ ,
\end{equation} where each state ${\mathsf T}'_m$ includes only one (possibly
composite) direct signal from the whole detector. States ${\mathsf T}'_m$ of
${\mathcal S} + {\mathcal D}$ are determined by the formal evolution. The
state ${\mathsf T}_{\text{end}}$ refers then to any time after the signals.}
\par \vspace{.5cm} \noindent Hence, the evolution during a separation status
change brings three changes: first, the change of kinematic description
${\mathsf T}_{\mathcal S} \otimes {\mathsf T}_{\mathcal D} \mapsto {\mathsf
J}({\mathsf T}_{\mathcal S} \otimes {\mathsf T}_{\mathcal D})$, second, the
standard unitary evolution of state ${\mathsf J}({\mathsf T}_{\mathcal S}
\otimes {\mathsf T}_{\mathcal D})$, and third, the state reduction of the
evolved state into (\ref{genrule}). Afterwards, the state evolves unitarily
with a possible change of kinematics if ${\mathcal S}$ and ${\mathcal D}$
become separated again. Its form (the statistical decomposition) is then
uniquely determined by detector signals. It is interesting to observe that the
signals result in a process of relaxation, in which the sensitive matter of
the detector approach its thermal equilibrium. This seems to be in accordance with
our theory of classical states in Chapter 3 and grants the dissipation of the
state obtained by the formal evolution.

A tenet adopted for the search of the assumption State Reduction in Direct
Registration has been that corrections to standard quantum mechanics ought to
be the smallest possible changes required just by the experiments. The
assumption is of course guessed and not derived and could yet be falsified in
confrontation with further observational evidence concerning different cases
of direct registration. It could also be further extended, e.g., to describe
how the postulated end states evolved in more detail (for example, in analogy
to a scenario of the dynamical reduction program). However, for such an
evolution, there does not seem to exist as yet any experimental evidence to
lead us. Let us emphasise that the clean decomposition of a separation status
change into three steps, viz.\ change of kinematics, unitary evolution and
state reduction, is just a method enabling a mathematically well defined
application of State Reduction in Direct Registration, but it is definitely
not a description of the time dependence of the real process.

\section{Comparison with other changes of separation status} It is the
existence of separation-status change that allows us to choose the statistical
decomposition, such as Eq.\ (\ref{endflex}), of the end states so that the
theory agrees with the observational fact of objectification. However,
separation status changes can also occur in processes that have nothing to do
with registrations. Must there be any reduction to decomposable states in such
processes, too? We have already studied the case of screen in Section
2.3. Here, we analyse scattering on a macroscopic body and the question of
linear superposition of large quantum systems.

\subsection{Scattering on macroscopic bodies} Let us restrict ourselves to a
scattering of a microsystem by a macroscopic target and observe that there can
then be separation status changes, one when the system enters the target and
other when it is released. First, let us consider no-entanglement processes
such as the scattering of electrons on a crystal of graphite with a resulting
interference pattern \cite{DG} or the splitting of a laser beam by a
down-conversion process in a crystal of KNbO$_3$ (see, e.g., Ref.\
\cite{MW}). No-entanglement processes can be described by the following
model. Let the initial state of the target ${\mathcal D}$ be $\mathsf T$ with
separation status $D_{\mathcal D}$ and that of the microsystem ${\mathcal S}$
be $\phi$ with separation status $D_1$, $D_1 \cap D_{\mathcal D} =
\emptyset$. Let the formal evolution lead to two subsequent changes of
separation status of ${\mathcal S}$: first, it is swallowed by ${\mathcal D}$
in $D_{\mathcal D}$ and, second, it is released by ${\mathcal D}$ in state
$\varphi$ with separation status $D_2$, $D_2 \cap D_{\mathcal D} =
\emptyset$. We assume that the end state of the target, ${\mathsf T}'$, is
independent of $\phi$ and that we have a unitary evolution:
$$
|\phi\rangle\langle\phi| \otimes {\mathsf T}
\mapsto|\varphi\rangle\langle\varphi| \otimes {\mathsf T}'\ ,
$$
which can be reconstructed from the formal evolution because the systems are
separated initially and finally. The two systems are not entangled by their
interaction, hence there is no necessity to divide the resulting correlations
between ${\mathcal S}$ and ${\mathcal D}$ in what survives and what is
erased. The dissipation at any stage of the process can be neglected. The end
state is in fact of the form (\ref{release}): it is a trivial statistical
decomposition.

Another example of this situation is a particle prepared in a cavity $D$ with
imperfect vacuum. We can model this situation in the above way and so in
effect suppose that the particle has separation status $D$.

A more interesting case is an {\em entanglement scattering} during which two
subsequent changes of separation status of the scattered particle also
occur. Examples include the scattering of neutrons on spin waves in ferromagnets or ionising an
atom of an ideal gas in a vessel. Let microsystem ${\mathcal S}$
in initial state $\phi$ with separation status $D$ be scattered by a
macrosystem ${\mathcal A}$ in initial state ${\mathsf T}$ with separation
status $D_{\mathcal A}$, $D \cap D_{\mathcal A} = \emptyset$. For simplicity,
we assume that the formal evolution leads to supp$\phi \subset D_{\mathcal A}$
at some time $t_{\text{scatt}}$. Such $t_{\text{scatt}}$ does not need to be uniquely
determined but the subsequent calculations are valid for any possible choice
of it. A more general situation can be dealt with by the method applied in the
case of a microsystem that can miss a detector (see the discussion after Eq.\
(\ref{repl})).

The experimental setup determines two Hilbert spaces ${\mathbf H}$ and
${\mathbf H}_{\mathcal A}$ and unitary map
\begin{equation}\label{scattU} {\mathsf U} : {\mathbf H} \otimes {\mathbf
H}_{\mathcal A} \mapsto {\mathbf H} \otimes {\mathbf H}_{\mathcal A}
\end{equation} describing the interaction according to standard quantum
mechanics.

The experimental setup studied in the previous section also determined a
basis $\{\phi_{mk}\}$ of ${\mathbf H}$, namely the eigenvectors of registered
observable ${\mathsf O}$ as well as sets of states $\{\varphi_{m k}\}$ in
${\mathbf H}$ and $\{\psi_m\}$ of ${\mathbf H}_{\mathcal A}$. This together
with the assumption that ${\mathcal A}$ measures ${\mathsf O}$ (with exact or
approximate probability reproducibility) restricted the possible ${\mathsf
U}$. These particular properties enabled us to choose a unique statistical
decomposition for the end state. The question is how any statistical
decomposition of the end result can be even formally well-defined for
processes described by Eq.\ (\ref{scattU}), where the physical situation does
not determine any such special sets of states. This is analogous to the
well-known Problem of Preferred Basis \cite{schloss}.

To be able to give an account of the situation, let us first introduce the
formal evolution ${\mathsf U}_f$ on ${\mathsf P}_{\tau\tau'}({\mathbf H}
\otimes {\mathbf H}_{\mathcal A})$, from which ${\mathsf U}$ can be
reconstructed. Second, we decompose map ${\mathsf U}_f$ into two steps,
${\mathsf U}_f = {\mathsf U}_{f2} \circ {\mathsf U}_{f1}$, where ${\mathsf
U}_{f1}$ develops up to $t_{\text{scatt}}$ and ${\mathsf U}_{f2}$ further from
$t_{\text{scatt}}$.

Then, the correct intermediate state ${\mathsf T}_{\text{interm}}$ at
$t_{\text{scatt}}$ is
$$
{\mathsf T}_{\text{interm}} = {\mathsf N}({\mathsf U}_{f1}{\mathsf
P}_{\tau\tau'}(|\phi\rangle\langle \phi| \otimes {\mathsf T}){\mathsf
P}_{\tau\tau'}{\mathsf U}_{f1})\ .
$$
Indeed, there is no macroscopic signal from ${\mathcal A}$, only some
microscopic degrees of freedom of ${\mathcal A}$ change due to the interaction
${\mathsf U}_{f1}$. The overwhelming part of the degrees of freedom of
${\mathcal A}$ remain intact, just serve as a backdrop of the process and we
can assume that any dissipation can be neglected. Thus, even if there is a
separation status change, there is no necessity for reduction: one can say
that there is only one channel.

Further evolution is given by ${\mathsf U}_{f2}$ supplemented by
reconstruction of the states in ${\mathbf H}$ and ${\mathbf H}_{\mathcal A}$
as ${\mathcal S}$ is released by ${\mathcal A}$, and we simply obtain: the
formula
\begin{equation}\label{scatt} {\mathsf T}_{\text{end}} = {\mathsf
U}(|\phi\rangle\langle\phi| \otimes {\mathsf T}_0){\mathsf U}^\dagger
\end{equation} of standard quantum mechanics remains valid. Formula
(\ref{scatt}) makes clear that a separation status change alone does not necessarily cause
any reduction.

\subsection{Linear superposition of large quantum systems} In the present
subsection, we shall consider possible linear superpositions of states for
large quantum systems and try to compare the corresponding conclusions of our
theory with those of dynamical state reduction program.

To explain the point, diffraction effects of a beam of molecules $C_{60}$
described in Ref.\ \cite{arndt} can be used. The experiment is conceptually a
simple generalisation of the classic Young's double slit experiment with slits
spaced by 100 nm. Despite the rather large velocity spread in the incoming
beam, the diffracted intensity as a function of angle clearly shows a central
peak flanked by two first-order satellites.

Thus, states of relatively large quantum systems can be superposed and the
superposition can be confirmed by observation. The superposition must not be
destroyed by any state reduction during the whole motion of the molecules from
the source to the detector in order that the diffraction pattern can
appear. Our explanation is that there is no relevant\footnote{There is the
separation status change due to the double slit screen, see Section 2.3.}
separation status change in the studied system state during this motion
independently of the number of particles from which the system is
composed. According to our theory, such a change can occur only at the end of
the motion in a detector, when the system is registered.

The dynamical state reduction program \cite{GRW,pearle} postulates the
existence of a kind of noise. The effect of this noise on all physical objects
is to add to the standard linear time-dependent Schr\"{o}dinger equation
an extra stochastic term that preserves neither linearity nor unitarity. This
term leads to reduction of the linear superposition of different position
eigenstates to particular position eigenstates at random time instants during
any evolution. The term is specified by two adjustable parameters: a length
scale $a$, which determines the minimum difference in positions
necessary to trigger the reduction process and a quantity $\lambda$ that
determines the rate of the reduction process for a single particle. A very
fundamental aspect of the theory is that a superposition state of a complex
body containing $N$ correlated micro-objects is reduced at a rate of
the order of $N\lambda$.

Thus the dynamical reduction scenario leads to reductions occurring during the
whole motion history of the molecule and can explain the absence of any
observational consequences only by a judicious choice of the parameters $a$
and $\lambda$ so that the resulting effect will be very small for sufficiently
small molecules such as $C_{60}$. Of course, the effects must be very strong,
if truly macroscopic systems $N = 10^{23}$ would be used instead of $C_{60}$.

Today, there are many different large (but not yet macroscopic) quantum
systems that can be observed in superposition states, and the analysis above
can be extended to them without much change. Thus, in principle, there is a
difference in observable consequences between our theory and the dynamical
reduction program.

\chapter{Conclusion} Our careful study of difficulties that are met along the
way from quantum mechanics to classical world has lead to a new understanding
of quantum mechanics in a number of aspects. The main points might be
summarised as follows.

A value of an observable of system ${\mathcal S}$ cannot be considered as a
property of ${\mathcal S}$ but only as an indirect piece of information on
such properties. It is well known that each such value is only formed in the
process of interaction between system ${\mathcal S}$ and a suitable
registration apparatus. We have looked, therefore, for another kind of
observable properties that could be ascribed to quantum systems and we have
found them among those that are uniquely determined by preparations. This
simple observation has been developed into a systematic realist
interpretation, the so-called Realist Model Approach to quantum
mechanics. Thus, the myth of quantum mechanics disproving realism has been
shown to be unfounded.

There are two kinds of objective properties: structural, which are common to
the whole class of indistinguishable systems such as mass, charge and spin,
and dynamical, which are different for different dynamical situations such as
states and averages of observables. The space of quantum states is convex and
exhibits a rich face structure. Two kinds of states are distinguished: the
indecomposable ones, which are analogous to points in phase space in Newtonian
mechanics, and the decomposable ones, which are analogous to probability
distributions in Newtonian mechanics. The opposites
"decomposable--indecomposable" is different from
"pure--mixed".

We have also attempted to make our interpretation compatible with a
full-fledged and self-consistent realist philosophy, namely the so-called
Constructive Realism as introduced by Ronald Giere. This is not a naive
realism\footnote{Naive realism is the view that the world is as we perceive
it.} so that it is immune to usual arguments against realism. A practical
aspect of the realist interpretation is not only that the emergence of an
objective classical world from quantum mechanics is not hindered by possible
non-objective character of the latter, but also that it provides explicit
help and guidance in constructing models of classical world.

The aim of our theory of classical properties is a unified approach to all
classical theories, such as Newtonian mechanics, thermodynamics or Maxwellian
electrodynamics. The starting point is that the main assumption of classical
theories, viz. the existence of absolutely sharp trajectories, does not
correspond to reality but is only a practical and productive
idealisation. Indeed, any classical measurement is much fuzzier than the
minimum quantum uncertainty. Thus, we can change the whole aim of
semi-classical approximation: what is to be approximated by quantum models are
reasonably fuzzy classical trajectories, not sharp ones.

This allows to formulate a so-called Modified Correspondence Principle, which
specifies which quantum observables can correspond to important classical ones
for a general classical system as well as what is the form of the
correspondence: the classical property is the average of the corresponding
quantum observable in a particular kind of state, the so-called classicality
one.

The second main hypothesis of our theory is that classicality states are some
quantum states of high entropy. (It follows, in particular, that coherent
states are not classicality states.) This principle is already in use in
thermodynamics. We have shown how it can be applied to classical mechanics by
introducing a new class of states, the so-called ME packets. They maximise von
Neumann entropy for fixed averages and variances of positions and
momenta. Gaussian wave packets are a special case of ME packets for the value
of maximum entropy equal to zero. ME packets approximate classical trajectories
better when  their entropy is higher. As yet, only thermodynamics and
Newtonian mechanics could be unified in this way, but Maxwellian
electrodynamics is hoped to allow an analogous approach. In this way, we
arrive at the natural conclusion that quantum mechanics can be much less fuzzy
than any part of classical reality ever is.

We also stress that macrosystems, as well as generally large composite
systems, have much smaller number of observables than one would expect
according to standard quantum mechanics. There are two reasons why observables
concerning single constituents of such a system can be measured only in
exceptional cases. On the one hand, the constituents may be elements of a
large family of identical systems within the macroscopic system from which
they are not separated by preparation and do not, therefore, possess any
really measurable observables of their own. On the other hand, they are not
individually manipulable by fields and shields and registrable by detectors
and can only be measured by those measurements that use ancillas. The
differences between macroscopic and microscopic systems are thus not due to
inapplicability of quantum mechanics to macroscopic systems. Just the opposite
is true: they result from strict and careful application of standard quantum
mechanics to macroscopic systems.

One of our most important observations is that none of the quantum observables
that are introduced in textbooks is measurable because the form of the
corresponding operators implies that their measurement must be disturbed by
all other systems of the same type in the environment. Only some sufficiently
local kind of observables could could be registered and only if the microsystems to be
measured are sufficiently separated from the set of identical
microsystems. Starting form this point, a new quantum theory of observables
has been constructed and it is rather different from the standard one.

These considerations also lead to an important condition on preparation
procedures: they must give the prepared microsystem a non-trivial separation
status. Only then, it can be viewed as an individual system that can be dealt
with as if its entanglement with other identical particles does not
exist. This explains why quantum mechanics is viable at all. In particular,
the standard rule for composition of identical microsystems must be
weakened. This is justified by the idea of cluster separability. In addition,
a preparation must separate the microsystem from all other microsystems, even
of a different kind, so that it can be individually manipulated by external
fields or matter screens and registered by detectors.

Not only the notion of preparation has been changed, but also registrations
have been given a more specific form than is usually assumed. The necessity to
distinguish systematically between ancillas and detectors has been
justified. The interaction of the registered system with an ancilla is not
considered as the whole registration. The system or the ancilla, or both, has
to be further registered by a detector. Any apparatus that is to register a
microsystem must therefore contain a detector and what is read off the
apparatus is a classical signal from the detector rather than a value of the
observable that is usually called "pointer". Thus, our theory of
classical properties finds an application here. We assume that each detector
contains an active volume of sensitive matter with which the registered
microsystem becomes unified and in which it looses its separation status.

The next important assumption is that standard quantum mechanics does not
provide true information about processes, in which the separation status of
microsystems changes. Preparations and registrations belong to such
processes. Blind application of standard rules to such processes leads to
contradictions with experimental evidence. This justifies adding new rules to
quantum mechanics that govern changes of separation status. We have formulated
such rules and shown that they form a logically coherent whole with the other
rules of quantum mechanics.

A substantial progress has been achieved in the theory of wave function
collapse, more precisely, of the state reduction. This is considered to be a
real physical process. A reason, or justification, for why the state reduction
takes place has been found in the loss of separation status of the registered
quantum microsystem. Our theory replaces the collapse by a more radical
transformation, a change in microsystem kinematic description. The loss of
separation status accompanied by a dissipation can be considered as a kind of
disappearance of a registered object during its registration. The state is
degraded, because the system is more or less lost. This justification comes,
so to speak, from outside of the measurement theory.

Moreover, our theory leads to sharper specification of where and when the
reduction occurs than that given by the von Neumann theory of the
collapse. The place is the sensitive volume of the detector and the time is
that of the detector signal. However, our theory gives only the final change
from the state $|\text{unitary}\rangle$ at some suitable time resulting from
the standard unitary evolution to the reduced state $|\text{reduce}\rangle$
and the question of the detailed time evolution $|\text{unitary}\rangle
\mapsto |\text{reduce}\rangle$, or even the existence of such evolution, is
left open. This can be compared with the quantum decoherence theories or with
the dynamical reduction theories. If the exact state of the environment and
the exact Hamiltonian of its interaction with the registered system plus
apparatus are known, the decoherence theory would give the evolution in all
detail (of course, the desired end of the evolution, a decomposable state,
cannot be obtained by the decoherence theory alone). Similarly, if the two
parameters of the dynamical collapse theory are chosen, again such an
evolution can be calculated. We hope that future improvements in
experimental techniques will allow to address the question of detailed
evolution.

\subsection*{Acknowledgements}The author is indebted to \v{S}tefan
J\'{a}no\v{s} for help with experimental physics and to Heinrich Leutwyler for
useful discussions. Thanks go to Ji\v{r}\'{\i} Tolar for encouragement and
important suggestions.

\addtocontents{toc}{{\bf Bibliography} \dotfill {\bf 168}}


\begin{thebibliography}{999}
\bibitem{AL}Araki, H.; Lieb, E. H., {\em Commun.\ Math.\ Phys.} {\bf 1970},
{\em 18}, 160.
\bibitem{arndt}Arndt, M.; Nairz, O.; Vos-Andreae, J.; Keller, C.; van der
Zouw, G.; Zeilinger, A., {\em Nature} {\bf 1999}, {\em 401}, 680.
\bibitem{AK}Arthurs, E.; Kelly, J. L. Jr., {\em Bell System Tech.\ J.} {\bf
1965}, {\em 44}, 725.
\bibitem{ball}Ballentine, L. E., {\em Rev.\ Mod.\ Phys.} {\bf 1970}, {\em 42},
358.
\bibitem{BR}Barut, A. O.; R\c{a}cka, R., {\em Theory of Group Representations
and Applications}; PWN: Warsaw, 1980.
\bibitem{Ghirardi}Bassi, A.;Ghirardi, G., {\em Phys.\ Letters A} {\bf 2000},
{\em 275}, 373.
\bibitem{bell1}Bell, J. S., {\em Physics} {\bf 1964}, {\em 1}, 195.
\bibitem{Bell3}Bell, J. S., {\em Helv. Phys.~Acta} {\bf 1975}, {\em 48}, 93.
\bibitem{bell4}Bell, J. S., {\em Rev.\ Mod.\ Phys.} {\bf 1966}, {\em 38}, 447.
\bibitem{belt}Beltrametti, E. G.; Cassinelli, G.; Lahti, P. J., {\em J. Math.\
Phys.} {\bf 1990}, {\em 31}, 91.
\bibitem{BvN}Birkhoff, G.; Von Neumann, J. {\em Ann.\ of Math.} {\bf 1936},
{\em 37}, 823.
\bibitem{BEH}Blank, J.; Exner, P.; Havl\'{\i}\v{c}ek, M., {\em Hilbert Space
Operators in Quantum Physics}; Second Edition, Springer: Berlin, 2008.
\bibitem{bohm}Bohm, D., {\em Quantum Theory}; Prentice-Hall: Englewood Cliffs,
1951.
\bibitem{Bona}B\'{o}na, P., {\em Acta Phys.\ Slov.} {\bf 1973}, {\em 23},
149. {\bf 1975}, {\em 25}, 3. {\bf 1977}, {\em 27}, 101.
\bibitem{Born}Born, M., {\em Phys.\ Bl\"{a}tter} {\bf 1955}, {\em 11},
49.
\bibitem{bragin}Braginsky, V. B.; Khalili, F. Ya.\, {\em Rev.\ Mod.\ Phys.}
{\bf 1996} {\em 68}, 1.
\bibitem{brukner}Brukner, \v{C}.; Zukowski, M., Bell's Inequalities:
Foundation and Quantum Communication, Preprint {\bf 2009}, arXiv:0909.2611.
\bibitem{bub}Bub, J., {\em Interpreting the Quantum World}; Cambridge
University Press: Cambridge, UK, 1999.
\bibitem{BLM}Busch, P.; Lahti, P. J.; Mittelstaedt, P., {\em The Quantum
Theory of Measurement}; Springer: Heidelberg, 1996.
\bibitem{cart}Cartwright, N., {\em The Dappled World}; Cambridge University
Press: Cambridge 1999.
\bibitem{coester}Coester, F., {\em Int.\ J. Modern Phys.} {\bf 2003}, {\em
17}, 5328.
\bibitem{EQM}Contributions to {\em J. Phys.: Conference Series} {\bf 2012},
{\em 361}.
\bibitem{DG}Davisson C.; Germer, L., {\em Phys.\ Rev.} {\bf 1927}, {\em 30},
705.
\bibitem{DFT}Dell'Antonio; G.; Figari, R.; Teta, A., {\em J. Math.\ Phys.}
{\bf 2008}, {\em 49}, 042105.
\bibitem{DST}Doebner, H. D.; \v{S}\v{t}ov\'{\i}\v{c}ek, P.; Tolar, J., {\em
Rev.\ Math.\ Phys.} {\bf 2001}, {\em 13}, 799.
\bibitem{EPR}Einstein, A.; Podolsky, B.; Rosen, N., {Phys.\ Rev.} {\bf 1935},
{\em 47}, 777.  705.\bibitem{Exner}Exner, F., {\em Vorlesungen \"{u}ber
die physikalischen Grundlagen der Naturwissenschaften}; Deuticke: Leipzig,
1922.
\bibitem{d'Espagnat}d'Espagnat, B., {\em Veiled Reality}; Addison-Wesley:
Reading, 1995.
\bibitem{fano}Fano, U., {\em Amer.\ J. Phys.} {\bf 1961}, {\em 29}, 536.
\bibitem{fraassen}Van Fraassen, B., {\em Laws and Symmetries}; Clarendon:
Oxford, UK, 1989.
\bibitem{fraass}Van Fraassen, B.,{\em Quantum Mechanics: An Empiricist View};
Clarendon: Oxford, UK, 1991.
\bibitem{gelfand}Gel'fand, J. M.; Vilenkin, N. Ya., {\em Generalized Functions
IV}; Academic Press: New York 1964.
\bibitem{GMM}Gemmer, J.; Michel, M.; Mahler, G., {\em Quantum
Thermodynamics. Emergence of Thermodynamic Behavior Within Composite Quantum
Systems}; LNP 657, Springer: Berlin, 2004.
\bibitem{GRW}Ghirardi, G. C.; Rimini, A.; Weber, T., {\em Phys.\ Rev. D} {\bf
1986}, {\em 34}, 470.
\bibitem{giere}Giere, R. N., {\em Explaining Science: A Cognitive Aproach};
The University of Chicago Press: Chicago, 1988.
\bibitem{Zeh}Giulini, D.; Joos, E.; Kiefer, C.; Kupsch, J.; Stamatescu, I.-O.;
Zeh, H.~D., {\em Decoherence and the Appearance of Classical World in Quantum
Theory}; Springer: Berlin, 1996.
\bibitem{gleason}Gleason, A. M., {\em J. Math.\ Mech.} {\bf 1957}, {\em 6},
885.
\bibitem{goldstein}Goldstein, S.; Lebowitz, J. L.; Tumulka R.; Zanghi, N.,
{\em Phys.\ Rev.\ Letters} {\bf 2006}, {\em 96}, 050403.
\bibitem{GHZ}Greenberger, D. M.; Horne, M. A.; Shimony, A.; Zeilinger, A.,
{Am.\ J. Phys.} {\bf 1990}, {\em 58}, 1131.
\bibitem{haag}Haag, R., {\em Local Quantum Physics. Fields, Particles,
Algebras}; Springer: Berlin 1992.
\bibitem{PHJT}H\'{a}j\'{\i}\v{c}ek, P.; Tolar, J., {\em Found.\ Phys.} {\bf
2009}, {\em 39}, 411.
\bibitem{hajicek}H\'{a}j\'{\i}\v{c}ek, P., {\em Found.\ Phys.} {\bf 2009},
{\em 39}, 1072.
\bibitem{hajicek2}H\'{a}j\'{\i}\v{c}ek, P., {\em Found.\ Phys.} {\bf 2011},
{\em 41}, 640.
\bibitem{survey}H\'{a}j\'{\i}\v{c}ek, P.; Tolar, J., {\em Acta Phys.\ Slovaca}
{\bf 2010}, {\em 60}, 613.
\bibitem{hajicek4}H\'{a}j\'{\i}\v{c}ek, P., {\em Found.\ Phys.} {\bf 2012},
{\em 42}, 555.
\bibitem{HBT}Hanbury Brown, R.; Twiss, R. Q., {\em Nature} {\bf 1956}, {\em
178}, 1046.
\bibitem{hardy}Hardy, L., {\em Phys.\ Rev.\ Letters} {\bf 1992}, {\em 68},
2981.
\bibitem{heisenberg}Heisenberg, W., {\em The physical principles of quantum
mechanics}; University of Chicago Press: Chicago 1930.
\bibitem{Hepp}Hepp, K., {\em Helvetica Phys.\ Acta} {\bf 1972}, {\em 45}, 237.
\bibitem{Jaynes}Jaynes, E.\ T., {\em Probability Theory. The Logic of
Science}; Cambridge University Press: Cambridge, UK, 2003.
\bibitem{KS}Kayser, B.; Stodolsky, L., {\em Phys.\ Lett.} {\bf 1995}, {\em B
395}, 343.
\bibitem{KP}Keister, B. D.; Polyzou, W. N., Relativistic Hamiltonian Dynamics
in Nuclear and Particle Physics. In {\em Advances in Nuclear Physics}, Negele,
J. W., Vogt, E.; Plenum: New York, 2002; Vol 20, p.\ 225--483.
\bibitem{Kittel}Kittel, C., {\em Introduction to Solid State Physics}; Wiley:
New York, 1976.
\bibitem{klyachko}Klyachko, A., Dynamic Symmetry Approach to Entanglement,
Preprint {\bf 2008}, arXiv:0802.4008.
\bibitem{kochen}Kochen, S.; Specker, E. P., {\em J. Math.\ Mech.} {\bf 1967},
{\em 17}, 59.
\bibitem{Kofler}Kofler, J.; Brukner, \v{C}., {Phys.\ Rev.\ Lett.} {\bf 2007},
{\em 99}, 180403.
\bibitem{leggett}Leggett, A. J., {\em J. Phys.: Condens.\ Matter} {\bf 2002},
{\em 14}, R415.
\bibitem{leo}Leo, W. R., {\em Techniques for Nuclear and Particle Physics
Experiments}; Springer: Berlin, 1987.
\bibitem{levy}L\'{e}vy-Leblond, J. M. In {\em Group Theory and Its
Applications}, Vol.\ 2. Loebl, E. M.; Academic Press: New York. 1941.
\bibitem{short2}Linden, N.; Popescu, S.; Short, A. J.; Winter, A., {\em Phys.\
Rev.\ E} {\bf 2009}, {\em 79(6 Pt 1)}:061103.
\bibitem{ludwig1}Ludwig, G., {\em Foundations of Quantum Mechanics I};
Springer: New York, 1983. {\em Foundations of Quantum Mechanics II}; Springer:
New York, 1985.
\bibitem{MW}Mandel, L.; Wolf, E., {\em Optical Coherence and Quantum Optics};
Cambridge University Press: Cambridge, 1995.
\bibitem{merton} {\em Methods of Experimental Physics}, Vol.\ 4, Part A,
Hughes, V. W., Schultz, H. L.; Academic Press: New York, 1967.
\bibitem{MTW}Misner, C. W.; Thorn, K. S.; Wheeler, J. A., {\em Gravitation};
Freeman: San Francisco, 1973.
\bibitem{mott}Mott, N. F., {\em Proc.\ Roy.\ Soc.\ London, Ser.\ A} {\bf
1929}, {\em 126}, 79.
\bibitem{JvN}Von~Neumann, J., {\em Mathematical Foundation of Quantum
Mechanics}; Princeton University Press: Princeton NJ, 1955.
\bibitem{NROBRH}Nogues, G.; Rauschenbeutel, A.; Osnaghi, S.; Brune, M.;
Raimond, J.-M.; Haroche, S., {\em Nature} {\bf 1999}, {\em 400}, 239.
\bibitem{pauli}Pauli, W., {\em Die allgemeinen Prinzipien der Wellenmechanik};
Springer: Berlin, 1990.
\bibitem{pearle}P. Pearle, {\em Phys.\ Rev.\ A} {\bf 1989}, {\em 39}, 2277.
\bibitem{peres}Peres, A., {\em Quantum Theory: Concepts and Methods}; Kluwer:
Dordrecht, 1995.
\bibitem{pierce}Pierce, J. R., {\em An Introduction to Information
Theory. Symbols, Signals and Noise}; Dover: New York, 1980.
\bibitem{Piron}Piron, C., {\em Foundations of Quantum Physics}; Benjamin:
Reading 1976; {\em Found.\ Phys.} {\bf 1972}, {\em 2}, 287.
\bibitem{short1}Popescu, S.; Short, A. J.; Winter, A., The foundations of
statistical mechanics from entanglement: Individual states vs. averages,
Preprint {\bf 2006}, arXiv:quant-ph/0511225.
\bibitem{poulin}Poulin, D., {\em Phys.\ Rev.\ A} {\bf 2005}, {\em 71}, 022102.
\bibitem{Primas}Primas, H., {\em Chemistry, Quantum Mechanics and
Reductionism}; Springer; Berlin, 1983.
\bibitem{PBR}Pursey, M. F.; Barrett, J.; Rudolf, T., {\em Nature Physics},
{\bf 2012}, 2309.
\bibitem{putnam}Putnam, H., {\em Realism and Reason}, Vol.\ 3 of {\em
Philosophical Papers}; Cambridge University Press: Cambridge, USA 1983.
\bibitem{RS}Reed, M.; Simon, B., {\em Methods of Modern Mathematical Physics},
Vol.\ I; Academic Press: New York, 1972.
\bibitem{reichenbach}Reichenbach, H., {\em The Direction of Time}; University
of California Press: Berkeley, 1956.
\bibitem{RSH}Ritchie, N. W. M.; Story, J. G.; Hullet, R. G., {\em Phys.\ Rev.\
Letters} {\bf 1991}, {\em 66}, 1107.
\bibitem{Rutherford}Rutherford, D. E., {\em Proc.\ Roy.\ Soc.\ (Edinburgh),
Ser.\ A} {\bf 1947}, {\em 62}, 229. {\bf 1951}, {\em 63}, 232.
\bibitem{schloss}Schlosshauer, M., {\em Rev.\ Mod.\ Phys.} {\bf 2004}, {\em
76}, 1267.
\bibitem{schroeck}Schroeck, F. E., Jr., {\em Quantum Mechanics on Phase
Space}; Kluwer: Dordrecht, 1996.
\bibitem{Sewell}Sewell, G. L., {\em Quantum Mechanics and its Emergent
Macrophysics}; Princeton University Press; Princeton, USA, 2002.
\bibitem{shannon}Shannon, C. E., {\em Bell Syst.\ Tech.\ J.} {\bf 1948}, {\em
27}, 379, 623.
\bibitem{SAL}Simenel, V.; Avez B.; Lacroix, D., Microscopic approaches for
nuclear Many-Body dynamics: application to nuclear reactions, Preprint {\bf
2008}, arXiv:0806.2714.
\bibitem{sneed}Sneed, J. D., {\em The Logical Structure of Mathematical
Physics}; Reidel: Dordrecht, 1971.
\bibitem{spekkens}Spekkens, R. W., {\em Phys.\ Rev.\ A} {\bf 2007}, {\em
75}(3):032110.
\bibitem{suppes}Suppes, P., {\em Introduction to Logic}; van Nostrand
Princeton, USA 1957.
\bibitem{svensson}Svensson, B. E. Y., New wine in old bottles: Quantum
measurement---direct, indirect, weak---with some applications, Preprint {\bf
2012}, arXiv:1202.5148.
\bibitem{tono}Tonomura, A.; Matsuda,T; Kawasaki, T; Ezawa, H. {\em Am.\ J. Phys.} {\bf 1989}, {\em 57}, 117.
\bibitem{thirring}Thirring, W., {\em Lehrbuch der Mathematischen Physik};
Springer: Berlin, 1980.
\bibitem{stefan}Twerenbold, D., {\em Rep.\ Progr.\ Phys.} {\bf 1996}, {\em
59}, 239.
\bibitem{wan}Wan, K. K.; McLean, R. G. D., {\em J. Phys.\ A: Math.\ Gen.} {\bf
1984}, {\em 17}, 837.
\bibitem{wanb}Wan, K. K., {\em From Micro to Macro Quantum Systems. A Unified
Formalism with Superselection rules and its Applications}; Imperial College
Press: London, 2006.
\bibitem{wehrl}Wehrl, A., {\em Rev.\ Mod.\ Phys.} {\bf 1978}, {\em 50}, 221.
\bibitem{Weinberg}Weinberg, S., {\em The Quantum Theory of Fields} Vol.\ I,
Chap.\ 4; Cambridge University Press: Cambridge, 1995.
\bibitem{wigner}Wigner, E. P., {\em Group Theory}; Academic Press: New York,
1959.
\bibitem{wilczek}Wilczek, F., {\em Phys.\ Rev.\ Letters} {\bf 1982}, {\em 49},
957.
\bibitem{WM}Wiseman H. M.; Milburn, G. J., {\em Quantum Measurement and
Control}; Cambridge University Press: Cambridge, UK, 2010.
\bibitem{Zurek}Zurek, W. H., {\em Rev.\ Mod.\ Phys.} {\bf 2003}, {\em 75},
715.
\end{thebibliography}
\end{document}